\pdfoutput=1

\documentclass[12pt,reqno]{article}
\usepackage{jheppub}
\usepackage{amsmath,amssymb,amsfonts}

\graphicspath{ {./Graphs/} }

\usepackage{float}

\usepackage{mathtools}

\usepackage{blkarray}

\usepackage[usenames,dvipsnames]{xcolor}
\usepackage{epsfig}
\usepackage{epstopdf}
\usepackage{dcolumn}
\usepackage{tikz}
\usetikzlibrary{shapes.geometric, arrows}
\usepackage{upgreek}
\usepackage{setspace}
\usepackage{enumitem}
\usepackage{array,multirow,bigdelim}
\usepackage{cleveref}
\usepackage{bm}

\def\be{\begin{equation}}
\def\ee{\end{equation}}
\def\ba{\begin{eqnarray}}
\def\ea{\end{eqnarray}}

\def\a{\alpha}
\def\b{\beta}

\def\b#1{\overline{#1}}

\def\CP1{\mathbb{CP}^1}
\def\SL2C{\mathrm{SL}(2,\mathbb{C})}

\def\Z2{\mathbb{Z}_2}

\def\su2{{SU(2)}}
\def\eps{{\epsilon}}

\def\a{{\alpha}}

\def\[{\left[}
\def\]{\right]}

\def\L{\Lambda}

\def\s{\sigma}
\def\a{\alpha}
\def\b{\beta}

\def\({\left(}
\def\){\right)}
\def\[{\left[}
\def\]{\right]}

\def\<{\langle}
\def\>{\rangle}

\def\i2{\frac{i}{2}}

\def\2F1{\,_2{\rm F}_1}

\newcommand{\beq}{\begin{equation}}
\newcommand{\eeq}{\end{equation}}
\newcommand{\beqq}{\begin{equation*}}
\newcommand{\eeqq}{\end{equation*}}
\newcommand\beqa{\begin{eqnarray}}
\newcommand\eeqa{\end{eqnarray}}
\newcommand\beqaa{\begin{eqnarray*}}
\newcommand\eeqaa{\end{eqnarray*}}
\newcommand\bea{\begin{array}}
\newcommand\eea{\end{array}}

\newcommand{\ie}{{\it i.e.}}

\title{Scattering Equations and a new Factorization for Amplitudes I: Gauge Theories}

\author{ Humberto Gomez}
\affiliation{Niels Bohr International Academy and Discovery Center, University of Copenhagen\\
Blegdamsvej 17, DK-2100 Copenhagen , Denmark.\\
\\
Facultad de Ciencias Basicas,  Universidad Santiago de Cali,\\
Calle 5 $N^\circ$  62-00 Barrio Pampalinda, Cali, Valle, Colombia.}

\emailAdd{humgomzu@gmail.com}

\abstract{
In this work we show how a double-cover (DC) extension of the  Cachazo, He and Yuan formalism (CHY) can be used to provide a new realization for the factorization of the 
amplitudes involving gluons and scalar fields.
First, we propose a graphic representation for a color-ordered Yang-Mills (YM) and special Yang-Mills-Scalar (YMS) amplitudes   
within the scattering equation formalism.
Using the DC prescription, we are able to obtain an algorithm (integration-rules) which decomposes amplitudes in terms of three-point building-blocks. It is  important to remark that the pole structure of this method is totally different to ordinary factorization (which is a consequence of the scattering equations). Finally, as a byproduct, we show that the soft limit  in the CHY approach, at leading order,  becomes trivial by using the technology described in this paper. 
}

\begin{document}
{\setstretch{1}
\maketitle
}
\onehalfspacing

\section{Introduction}\label{sectionIntro}

The Cachazo, He and Yuan (CHY) formalism, which was motivated by the remarkable work of Witten \cite{Witten:2003nn},  provides an intriguing novel way of computing gauge,
gravity and effective field theories  S-matrix elements~\cite{Cachazo:2013gna,Cachazo:2013hca,Cachazo:2013iea,Mason:2013sva,Berkovits:2013xba}. 

In the usual CHY formalism\footnote{We call this formalism the {\it single-cover} approach.},  amplitudes are contour integrals over the moduli space of $n$-punctured Riemann spheres (${\cal M}_{0,n}$).
This contour integral is localized on the solutions of the so-called scattering equations, $  
S_a \equiv \sum_{b \ne a}\frac{k_a\cdot k_b}{\s_{ab}}=0,  \,  \s_{ab}\equiv \s_a - \s_b$, 
where $(\s_1,\ldots,\s_n)$ denote  local coordinates over ${\cal M}_{0,n}$
and the index ``$a$"  labels the  external particles of 
momentum $k_{a}$ (and polarization vector $\eps_a$) at the puncture ``$\s_{a}$". The prescription for the tree-level S-matrix of any quantum field theory may be given by the expression
\begin{equation}\label{singleC}
A_n = \int_{\Gamma} \,d\mu_n^{\rm CHY}\times  \Delta(pqr)\,\Delta(pqr)  \times {\cal I}_n^{\rm CHY}(\s), 
\end{equation}
where, $\Delta(pqr)\equiv \s_{pq} \s_{qr} \s_{rp}$ (Faddeev-Popov determinant),  $d\mu_n^{\rm CHY}\equiv \prod_{a\neq p,q,r}^{n} \frac{ d \s_a }{   S_a  }$, and the $\Gamma$ contour is defined by the $n-3$ independent equations\footnote{It is simply to verify that, $\sum_{a=1}^n S_a= \sum_{a=1}^n\s_a \,S_a=\sum_{a=1}^n\s_a^2\,S_a=0$, on the support of momentum conservation and on-shell conditions, \ie\, $k_1+\cdots +k_n=0$ and $k_a^2=0$.}, $S_a=0, \, a\neq  p,q,r $.
A different integrand describes a different theory, in particular we focus our attention  in pure Yang-Mills and special Yang-Mills-Scalar theories. For example, for a color-ordered pure Yang-Mills amplitude, $A^{\rm YM}_n(1,2,...,n)$, 
the CHY integrand is given by,
\begin{equation}
{ \cal I}_n^{\rm YM} (1,\ldots, n) = {\rm PT}_{(1,2,...,n)} \times {\rm Pf}^\prime \Psi_n,
\end{equation}
with the Parke-Taylor factor (${\rm PT}$) and the reduced Pfaffian (${\rm Pf}^\prime \Psi_n$)  define as
\begin{equation}\label{PTandPF}
{\rm PT}_{(1,2,...,n)}=
\frac{1}{\s _{12}\, \s_{23}\cdots  \s_{n1}}, \qquad {\rm Pf}^\prime \Psi_n = \frac{ (-1)^{i+j} }{\s_{ij} }\,\,{\rm Pf}[(\Psi_n)^{ij}_{ij}],
\end{equation}
where the  $2n\times 2n$ matrix, $\Psi_n$, is given by the blocks\footnote{Let us recall ourselves that the on-sell polarization vectors,  $\epsilon_a^\mu$, obey the transverse condition, $\eps_a\cdot k_a=0$. The gauge symmetry of the theory is given by the shifting, $\eps_a^\mu \, \rightarrow \,\eps_a^\mu+k_a^\mu$.}
\begin{eqnarray}\label{Pmatrix}
\Psi_n  \equiv  \left( 
\begin{matrix}
 {\cal A} & -{\cal C}^{\rm T} \\
{\cal C}  & {\cal B} 
\end{matrix}
\right)\,\, .
\end{eqnarray}
These blocks are given by the expressions 
\begin{equation}
{\cal A}_{ab} = \frac{k_{a}\cdot k_b}{ \s_{ab}}, \,
 {\cal B}_{ab} = \frac{\eps_{a}\cdot \eps_b}{ \s_{ab}}, \,  {\cal C}_{ab} = \frac{\eps_{a}\cdot k_b}{ \s_{ab}},\,  a\neq b,\, {\cal A}_{aa}={\cal B}_{aa}= 0,\, {\cal C}_{aa}= -\sum_{c=1 \atop c\neq a}^n \frac{\epsilon_a \cdot k_c}{ \s_{ac}}.
\end{equation}
The reduced matrix, $(\Psi_n)^{ij}_{ij}$, is built by removing the rows and columns $(i, j)$ from $\Psi_n$, with $1 \leq i< j \leq n$.

\begin{figure}
\centering
   \includegraphics[scale=0.5]{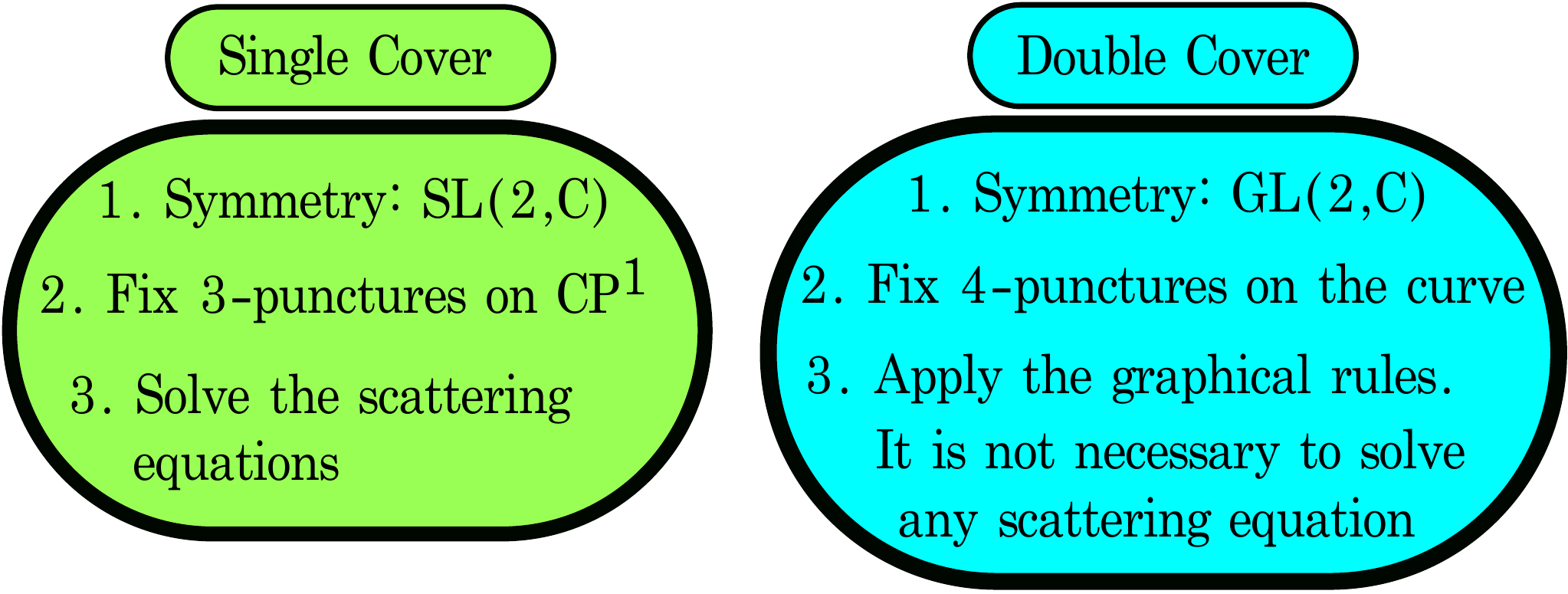}
\caption{General characteristics of the single and double cover approach.} 
 \label{characteristics}
\end{figure}

\begin{figure}
\centering
   \includegraphics[scale=0.4]{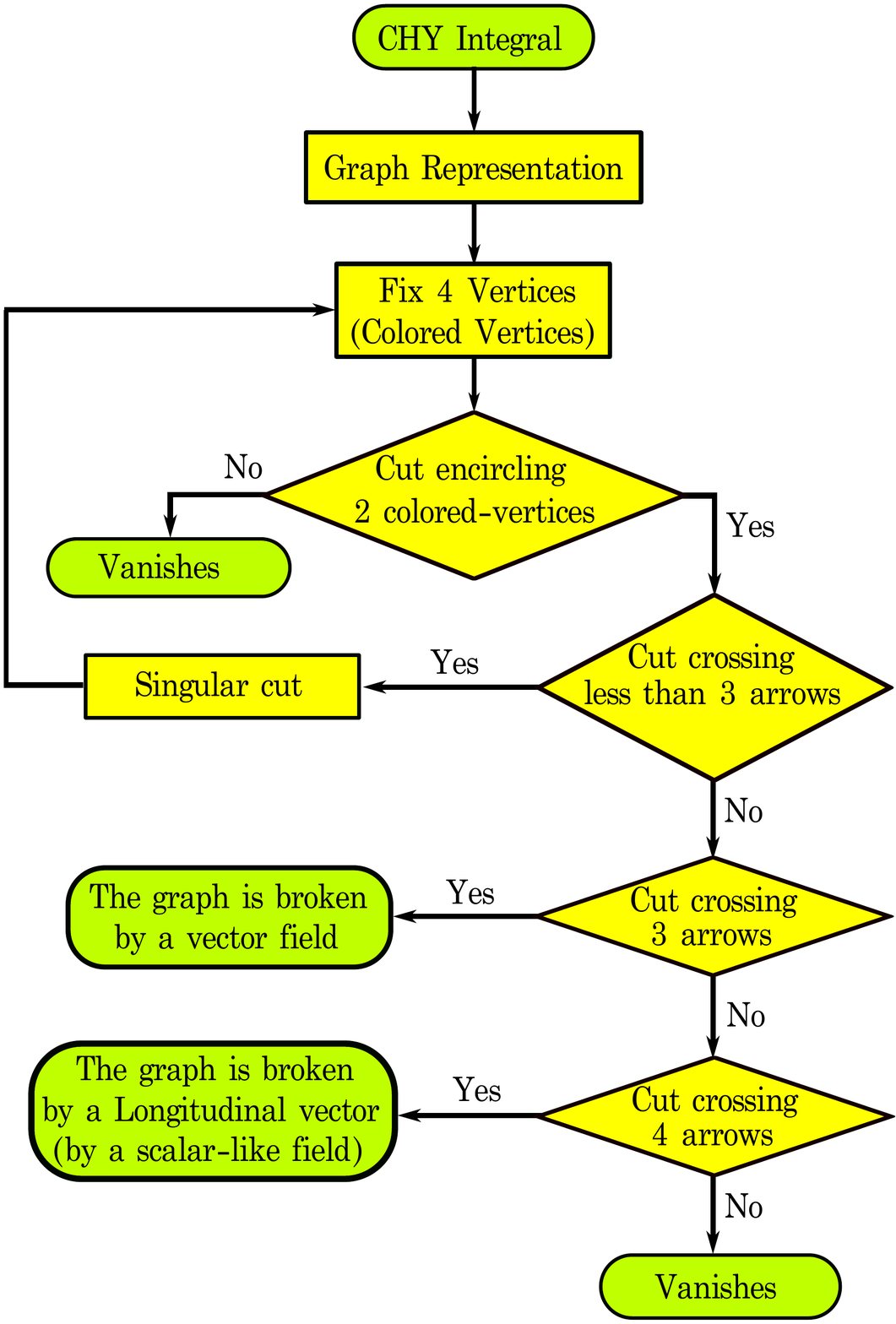}
\caption{\small 
{\bf Integration rules} as a Flowchart.  Given a CHY integral, we build its graph representation,  section \ref{graphR}. Each vertex of the graph represents a puncture on a Riemann sphere, including its physical parameters (momentum and polarization vector). On the other hand, the arrows of the graph represent the factors ``$\frac{1}{\s_{ab}}$" while the reduced Pfaffian is always implicit in it. 
The integration rules give a program to break any graph in terms of two smaller graphs ({\it resulting-graphs}). The first step is to choose four vertices (colored vertices), we call to this process a gauge fixing. The condition over the vertices, to obtain a good splitting of the graph (non-vanishing contribution), is that each split part must contain two colored vertices. After getting a good splitting, we can focus on the arrows. If the graph is split by cutting less than three arrows, then, this is a singular-cut (section  \ref{section-IntR}) and the method can not be applied, therefore, one must come back to choose a new gauge fixing. Solving the singular-cut issue by choosing an appropriate gauge fixing, we have three options: 
{\bf 1}. The  graph is split by cutting three arrows: in this case the two resulting graphs are glued by an off-shell vector field.
{\bf 2}. The graph is split by cutting four arrows: in this case the two resulting graphs are glued by a longitudinal vector field or scalar field. 
{\bf 3}. The graph is split by cutting more than four arrows: then, this contribution vanishes trivially. These rules are going to be clearer after section  \ref{section-IntR}.} 
 \label{flux-diagram}
\end{figure}

Recently, in~\cite{Gomez:2016bmv} (see also ref.~\cite{Cardona:2016bpi}), we showed how the CHY approach can be written in  
a new formulation in which the basic variables 
$\s_a$ live not on $\mathbb{CP}^1$ but on a quadratic algebraic curve embedded in the complex projective plane $\mathbb{CP}^2$. Dubbed the ``$\Lambda$-formalism" in~\cite{Gomez:2016bmv}, 
we shall here refer to it as CHY on a double-cover (DC). At first sight it may seem to be a complication
to extend the CHY formalism in this manner.  For example, the reduced Pfaffian in the double cover representation  is given by  (see sections \ref{reviewDC} and \ref{sectionPfaffian})
\begin{eqnarray}\label{}
{\bf Pf}^\prime\Psi_n^\L= \frac{(-1)^{i+j}}{(y_i+\s_i) - (y_j+\s_j)} \,  \left[ \prod_{a=1}^n   \frac{y_a+\s_a}{y_a}    \right]
 {\rm Pf} \left[(\Psi^\L_n)^{ij}_{ij}\right] , ~ {\rm with}, \, y_a^2=\s_a^2 - \L^2. ~~ ~~
\end{eqnarray}
However, as we shall demonstrate in the present paper the double-cover 
formalism adds a new ingredient to the standard CHY formalism that is much more difficult to extract in the single cover
formulation. Briefly stated, it is this: after integrating the new auxiliary variables, $(y_1,...,y_n,\L)$,
the double-cover formalism naturally expresses the scattering amplitudes so that they appear  factorized into  channels.
Nevertheless, this procedure must be performed with care, 
broadly speaking,
the integral over the auxiliary variables is based on two main points: \\
{\bf 1.} The number of fixed punctures (colored vertices) over each Riemann sheet.\\
{\bf 2.} The number of arrows cut by the branch-cut when it is getting closed.\\
In Fig. \ref{flux-diagram} and section \ref{section-IntR} we will give detail about this subject. 
The propagator that forms the bridge between two factorized pieces arises as the link between two separate $\mathbb{CP}^1$ in the single-cover approach, thus
intuitively explaining why the double-cover naturally expresses amplitudes in a factorized manner. Since in most cases this process can be iterated, then, we will not need to solve any scattering equation, which is one of the virtues of this approach. In Fig. \ref{characteristics} we give a few characteristics of the single and double cover approach, in order to reference some general differences among these two prescriptions.  

It is interesting to remark that, in many cases, the factorizations obtained in this way corresponds directly to the physical channels. Interestingly, there are instances
where, unavoidably, the factorizations proceed in a slightly different manner: some physical channels appear immediately, but others only
resurface after pole-cancelling terms have rearranged the expressions. This seems to lead to an intriguing connection to on-shell
BCFW recursion~\cite{Britto:2005fq}. It turns out that momentum shifts that lead to poles at infinity become evaluated in a quite straightforward
way by means of the double-cover formalism.  

On the other hand, several methods have been developed to compute  the CHY contour integral given  in \eqref{singleC}, most of them are applied to $\phi^3$ or focused on solving the scattering equations \cite{Gomez:2016bmv,Cachazo:2015nwa,Baadsgaard:2015voa,Baadsgaard:2015ifa,Cardona:2016gon,Dolan:2013isa,Farrow:2018cqi,Bjerrum-Bohr:2016juj,Liu:2018brz,Huang:2016zzb,Gomez:2013wza,Kalousios:2013eca,Kalousios:2015fya,Cardona:2015eba,Cardona:2015ouc,Lam:2018tgm}. In this work, from the double-cover representation,  we have been able to achieve a graphic off-shell algorithm to carry out any color-ordered scattering of $n$-gluons and interactions with scalar fields, resulting in an expansion in terms of three-point amplitudes\footnote{Notice that, although the methods presented in \cite{Dolan:2013isa,Roehrig:2017gbt} look somewhat similar to the one developed in this paper, the process and the form of results obtained by us are different.}.  

In view that the algorithm obtained in this work is an off-shell method\footnote{Since all intermediate particles are off-shell, including their polarization vectors, \ie\, $\eps_{[i]}\cdot k_{[i]}\neq 0$, and as it is an iterative program, then, we can consider this algorithm as an off-shell method.}, then a natural question arises, Is there any connection with the method proposed by Berends-Giele in \cite{Berends:1987me}? In order to illustrate the answer,  we focus on the bi-adjoint $\phi^3$ theory (since its integration rules are simpler  \cite{Gomez:2016bmv}). \\
Usually, the double-color partial amplitude for the bi-adjoint $\phi^3$ theory is denoted as $m^{(0)}(\a|\b)$, where $\a$ and $\b$ are two partial orderings \cite{Cachazo:2013iea}.  When $\a=\b$, we denote  $m^{(0)}(\a|\a)$ as $A^{\phi^3}(\a)\equiv m^{(0)}(\a|\a)$. In particular, 
let us analyse the five point example,  $A^{\phi^3}(1,2,3,4,{\color{red} 5})$, where the red label means an off-shell particle, $k_5^2\neq 0$.
Following the recurrence relation obtained by us {\it et al.} from the double cover approach (see equation (C.7) in \cite{Cardona:2016gon}), one has
\begin{eqnarray}\label{Lalgo}
&&
\hspace{-0.6cm}
A^{\phi^3}(1,2,3,4,{\color{red} 5})= \frac{A^{\phi^3} (3,4,[5,1],2) } {\tilde s_{234}}
+ \frac{ A^{\phi^3} ([3,4],5,1,2) } { \tilde s_{34}} 
+ \frac{A^{\phi^3} (3,[4,5],1,2) } { \tilde s_{123}}  \nonumber \\
&& 
\hspace{1.3cm}
= \frac{1}{\tilde s_{234}} \left(    \frac{1}{ \tilde s_{2[5,1]} } + \frac{1}{ \tilde s_{23}  } \right)  +
 \frac{1}{\tilde s_{34} } \left(    \frac{1}{ \tilde s_{12} }  + \frac{1}{  \tilde s_{2[3,4]}  } \right)   
+
 \frac{1}{\tilde s_{123}} \left(    \frac{1}{ \tilde s_{12} } + \frac{1}{  \tilde s_{23 } } \right) \,\, ,
\end{eqnarray}
with,  $ A^{\phi^3} ({\color{red} a},{\color{red} b}, {\color{red} c},d) \equiv \frac{1}{ \tilde s_{dc} } + \frac{1}{ \tilde s_{da}}$, where $\{k_a, k_b, k_c\}$ can be off-shell.  We have also introduced the notation,  $k_{[a_1,..., a_p]} \equiv k_{a_1}+\cdots  + k_{a_p}$,  $\tilde s_{a_1a_2...a_p} \equiv \sum_{i<j} k_{a_i}\cdot k_{a_j} $ and $ s_{a_1a_2...a_p} \equiv {1 \over 2}\, k^2_{[a_1,..., a_p]} $. Pictorially,  \eqref{Lalgo} looks like \cite{Gomez:2016bmv}, 
\begin{eqnarray}\label{phi3-diagrams}
\parbox[c]{20.5em}{\includegraphics[scale=0.35]{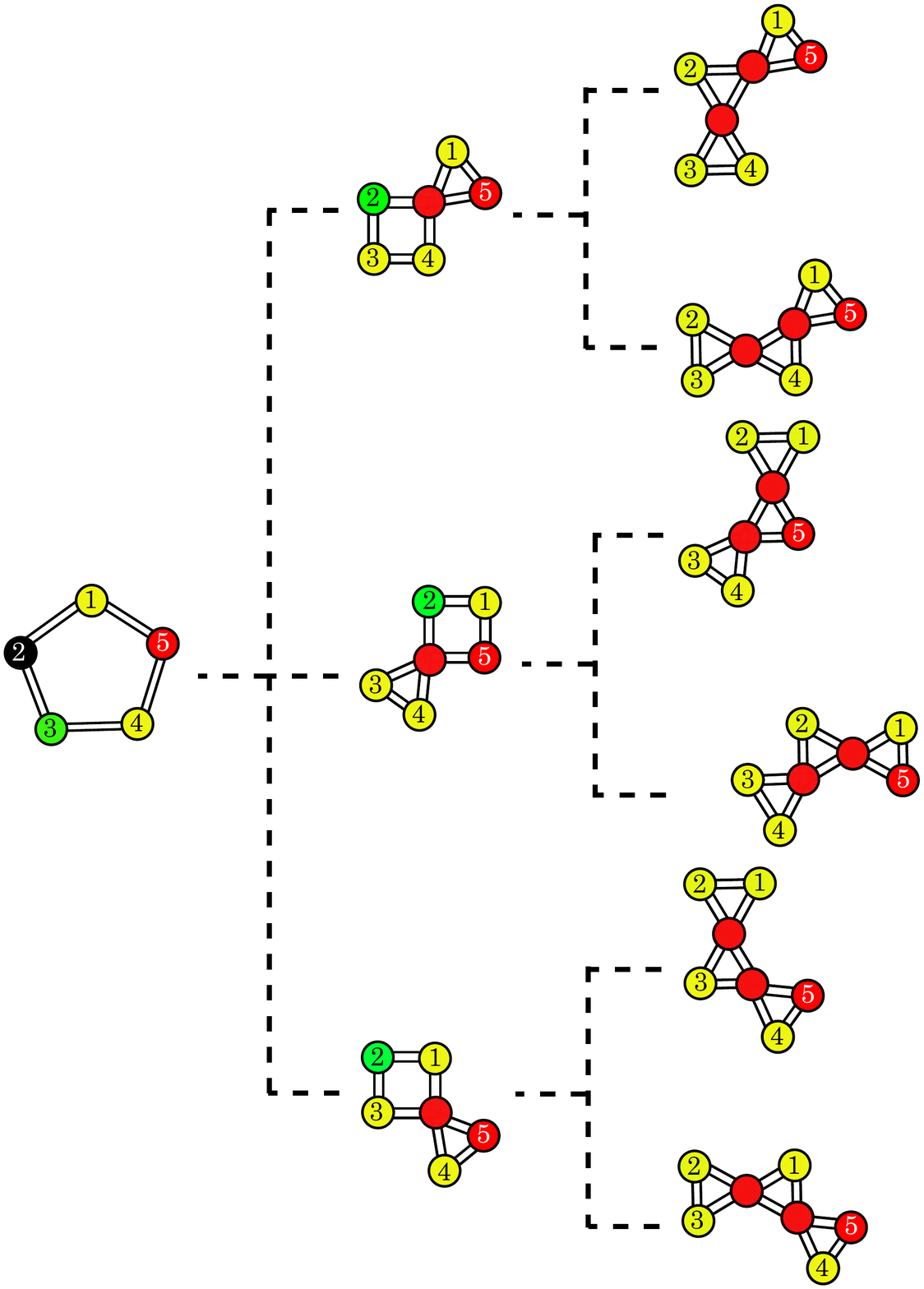}}  .
\end{eqnarray}
Although the color in each vertex has an important meaning, as we will explain later, roughly speaking, the colored vertices symbolize the punctures have been fixed by the global isometry generators (red means off-shell), while the black one means is  unfixed puncture over the double cover sphere\footnote{Notice that the first diagram on the third column in \eqref{phi3-diagrams} looks like the same to the fourth one. However, they are different because during the $\L$-algorithm process appeared spurious poles, as one can see  in \eqref{Lalgo}. For instance, the first diagram is given by the expression, $\frac{1}{\tilde s_{234}\, \tilde s_{2[5,1]}}$, and the fourth one by, $\frac{1}{\tilde s_{34}\, \tilde s_{2[3,4]}}$.}.

Otherwise,  Mafra was able to obtain the Berends-Giele-like currents for the bi-adjoint $\phi^3$ theory  \cite{Mafra:2016ltu}, where he used the Perturbiner method\footnote{Currently, the Perturbiner method has been successfully used for several theories, \cite{Mafra:2016ltu,Mafra:2015vca,Mizera:2018jbh,Garozzo:2018uzj}.}. 
For instance, the $n$-point amplitude, $ A^{\phi^3} ( 1,\ldots, n)$,  is given by the recurrence relation 
\vspace{-0.1cm}
{\small
\begin{eqnarray}\label{maf1}
\hspace{-0.4cm}
A^{\phi^3} ( 1,\ldots, n) = \lim_{k_n^2\rightarrow 0}  s_{12...n-1} \, \phi_{12...n-1} \, \phi_n \, ,      
\end{eqnarray}
}
\vskip-0.5cm\noindent
where, $\phi_{i} = 1$, $\phi_P =\frac{1}{ s_P} \sum_{X Y =P} \phi_X\, \phi_Y$, $X, \, Y \neq \emptyset$. The notation $\sum_{XY =P}$ means a sum over all possible
ways to deconcatenate the word $P$ in two non-empty words $X$ and $Y$. For example, $\phi_{1234}=\frac{1}{ s_{1234}}\sum_{X Y =1234} \phi_X\, \phi_Y = \frac{1}{ s_{1234}} \left( \phi_{1}\,\phi_{234}+\phi_{12}\,\phi_{34}+\phi_{123}\,\phi_{4} \right)$. Therefore, from  \eqref{maf1}, it is straightforward to see
 \vspace{-0.1cm}
 {\small
\begin{eqnarray}\label{mafraF}
A^{\phi^3} ( 1,2,3,4,5) &=& \lim_{k_5^2\rightarrow 0}  s_{1234} \, \phi_{1234} \, \phi_5 = \phi_{1}\,\phi_{234}+\phi_{12}\,\phi_{34}+\phi_{123}\,\phi_{4}   \nonumber\\  
&=&
\frac{1}{ s_{234}}\left(  \frac{1}{ s_{34}} + \frac{1}{ s_{23}} \right)  
+\frac{1}{ s_{12} \,  s_{34}} 
+ \frac{1}{ s_{123}} \left(  \frac{1}{ s_{12}} + \frac{1}{ s_{23}} \right) \,\,.
\end{eqnarray}
}
\vskip-0.3cm\noindent
Graphically,  \eqref{mafraF} is represented by the diagram, 
\vspace{-0.3cm}
{\small
\begin{eqnarray}\label{phi3-BG}
\hspace{-1.5cm}
\parbox[c]{22.6em}{\includegraphics[scale=0.19]{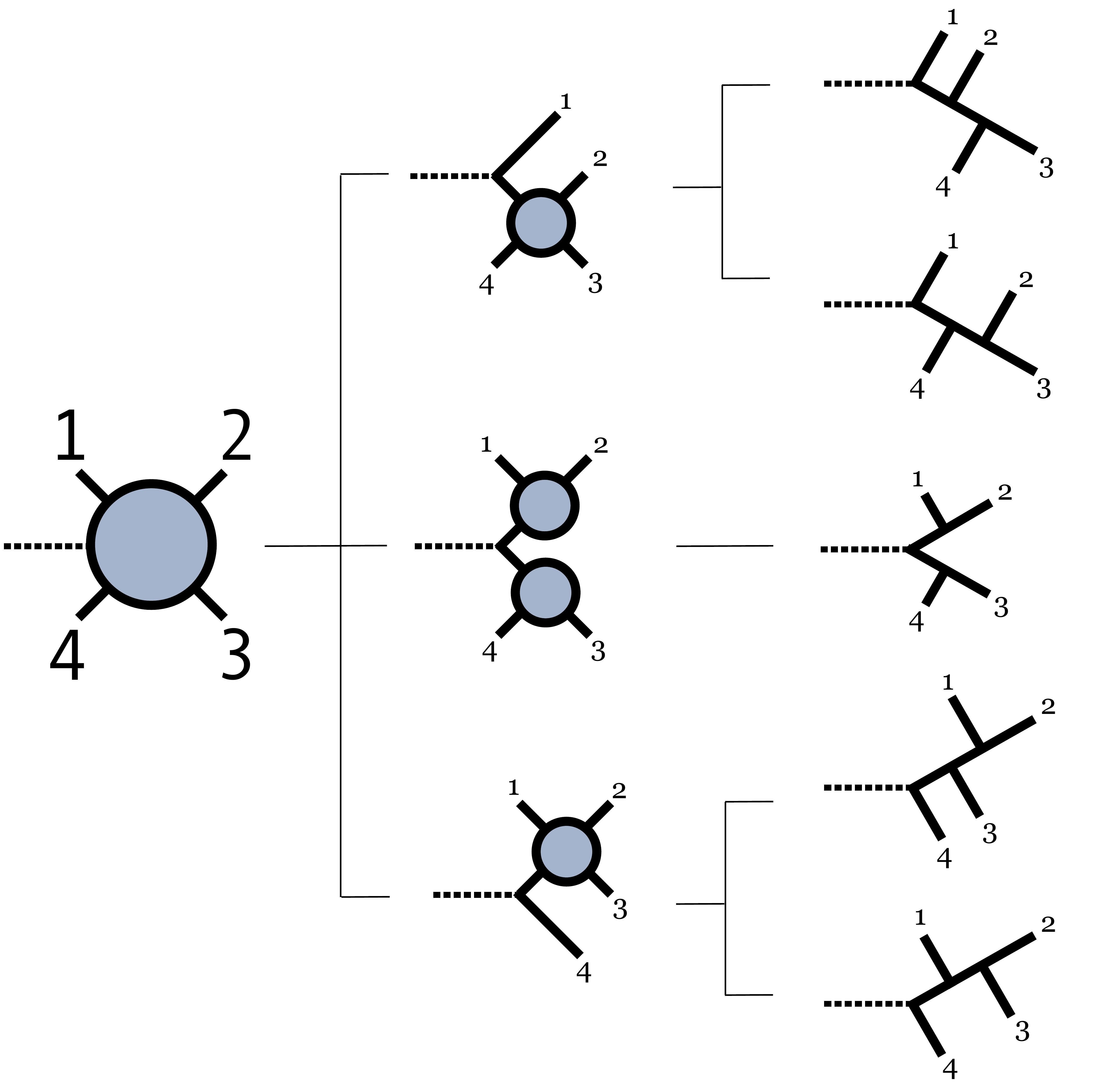}}  ,
\end{eqnarray}
}
\vskip-0.6cm\noindent
where the dashed line is an off-shell particle ($k_5^2\neq 0$). 

Clearly,  the results obtained in \eqref{Lalgo}  and \eqref{mafraF}  are related by the partial fraction decomposition
\begin{eqnarray}\label{partialFrac}
\frac{1}{ s_{234}\,  s_{34}} = \frac{1}{ s_{234} ( s_{34} -  s_{234})} + \frac{1}{ s_{34} ( s_{234}-  s_{34}) }  = \frac{1}{\tilde s_{234}\, \tilde s_{2[5,1]}} + \frac{1}{\tilde s_{34}\, \tilde s_{2[3,4]}},
\end{eqnarray}
which is the same phenomenon found at loop-level \cite{Casali:2014hfa,Geyer:2016wjx,Geyer:2015jch,Geyer:2015bja,Cardona:2016wcr,Gomez:2016cqb,Cardona:2016bpi,Gomez:2017cpe,Ahmadiniaz:2018nvr}.

Nevertheless, although in Yang-Mills theory the double-cover approach also produces spurious poles, such as those on the right-hand side in \eqref{partialFrac} (it can be checked in section \ref{sectionEX5P} in the five-point example), we have no idea how to relate our method with the one developed by  Berends-Giele in \cite{Berends:1987me}. At four-point, the relation among these two methods is straightforward, but at five-point, we do not know how to get it. 
This could be a possible future direction in order to understand better the double-cover representation. 

It is important to mention that in most of the case we will consider on-shell external particles. Nevertheless, our method is able to support up to two off-shell gluons, such as  the example at  four-point shows in section \ref{sectionEX5P}. 

On the other hand, on the support of the scattering equations, the reduced Pfaffian can write as a linear combination of Parke-Taylor factors  \cite{Cachazo:2013iea,Mafra:2011kj,Teng:2017tbo,Fu:2017uzt}
\begin{eqnarray}\label{Psi-PT}
{\rm Pf}^\prime \Psi_n= \sum_{\rho \in S_{n-2}} {\rm PT}_{(1,\rho(2,...,n-1),n)} \, N_{1|\rho(2,...,n)|n}^{\rm tree} ,
\end{eqnarray}
where the $N_{1|\rho(2,...,n)|n}^{\rm tree}$ terms are the tree-level Bern-Carrasco-Johansson (BCJ) numerators \cite{Bern:2008qj} and $S_{n-2}$ is the group of all possible permutations of the labels  $(2,...,n-1)$. Clearly, from the expansion in  \eqref{Psi-PT}, the $n$-point ordered YM amplitude, $A_n^{\rm YM} (1,...,n)$, is written in terms of the $(n-2)!$ $\phi^3$-amplitudes, $m^{(0)}(1,...,n|1,\rho(2,...,n-1),n)$,  times its corresponding BCJ factor,   $N_{1|\rho(2,...,n)|n}^{\rm tree} $. Although this formula looks simple, notice that the expansion grows quickly with the number of points, additionally, when the number of points is large, the BCJ numerators are not straightforward to carry out. Conversely,  in the proposal given by us here, the $n$-point color-ordered YM amplitude is represented just by one graph, and, after using iteratively the integration rules from Fig. \ref{flux-diagram}, the final answer is expressed as a product of the three-point amplitude, $A_3^{\rm YM}$.

To end, it is interesting to remember that  the usual CHY approach of the $n$-point color-ordered YM amplitude can be seen as the $d$-dimensional version of the Roiban, Spradlin and Volovich  formula (RSV)  \cite{Roiban:2004yf} (which is called the connected prescription). Otherwise, Cachazo, Svrcek and Witten (CSW) proposed an alternative\footnote{Historically speaking, the CSW approach was formulated before than the RSV.} formulation that describes the same physical object  given by the RSV formula (disconnected prescription) \cite{Cachazo:2004kj}. The relation between the RSV and CSW was shown in Refs. \cite{Gukov:2004ei,ArkaniHamed:2009sx,ArkaniHamed:2009dg}.  Since the double-cover representation is able to express the $n$-point amplitude as a product of three-point building-blocks, then, we think this is the disconnected version of the usual CHY approach. It would be very interesting to obtain a relation between the CSW prescription and the CHY double-cover representation.

{\bf Outline:}

The present work is organized as follows: 

In section \ref{reviewDC}, we give a simple review of the double-cover prescription for the double color-ordered $\phi^3$ theory, for detail see \cite{Gomez:2016bmv}. 

In section \ref{sectionPfaffian}, we introduce a new deferential form given by, $T_{ab}\, d\s_a\equiv \frac{d\s_a}{(y_a+\s_a) - (y_b+\s_b)}$, which is the key to define the CHY matrices in the double-cover approach.  In the original paper where the double-cover prescription was introduced \cite{Gomez:2016bmv}, the fundamental object is given by the expression, $\tau_{(a,b)} d\s_a \equiv \frac{1}{2y_a}\left( \frac{y_a+y_b}{\s_{ab}}+1  \right) d\s_a$, and  
although the relationship between $T_{ab}$ and $\tau_{(a,b)}$ looks pretty simple,
{\small
\begin{eqnarray}
\tau_{(a,b)}= \left(\frac{y_a+\s_a}{y_a} \right) \times T_{ab}\,\, ,
\end{eqnarray}
}
\hspace{-0.15cm}
\noindent
on the support of  $y_a^2=\s_a^2-\L^2$ and $y_b^2=\s_b^2-\L^2$, this is non-trivial.

Now, since $T_{ab}$ is antisymmetric, we define the mapping, $\frac{1}{\s_{ab}} \rightarrow T_{ab}$,  to translate the most familiar CHY matrices in the double-cover language. However, the $\Pi_{\a_1,...,\a_p}$ matrix \cite{Cachazo:2014xea} must be treated with care \cite{inpreparation}.

To end the section,  we propose the DC integrands to compute the scattering amplitudes of $n$-gluons and interactions with scalar fields.

Section \ref{graphR} is conceptually important, since here we introduce a graphical representation for  CHY integrals, as much as in the double-as in the single-cover. These graphics are essentials in this work, because all our computations are performed over them. Additionally, we clarify the color notation of the vertices. 

In  section \ref{section-IntR}, we formulate the {\bf integration rules}, perhaps the central part of the paper.
We perform an extensive and careful analysis of the double cover representation and the integration over the auxiliary variables, $(y_1,...,y_n,\L)$. 

First, we study a general situation, i.e. for any CHY integrand. In that case, we obtain the {\bf rule-I}, which basically claims that if any given cut does not encircle two colored-vertices, then, that cut vanishes trivially. Next, we go to a particular case, the Yang-Mills integrand. Thus, studying  the Pfaffian, we obtain three more {\bf integration rules}. Finally, in Fig. \ref{flux-diagram}, we have summarized  all rules in a flowchart. 

In sections \ref{sectionBB} and \ref{sectionEXS}, we compute the fundamental three-point building blocks. This method is able to break a big graph as a product of three-point amplitudes,  similar as it was done in Fig. \eqref{phi3-diagrams}. Subsequent, we give the simplest example, the four-point  Yang-Mills amplitude. Applying the algorithm schematized in Fig. \ref{flux-diagram} over this four-point graph,  we obtain two standard factorization cuts and an additional (apparently) non-physical contribution (which we call  a {\it strange-cut}).

Next, in section \ref{LongContributions}, following the Pfaffian identities  given  in appendix \ref{appendix}, we interpret  the {\it strange-cuts} as longitudinal contributions from some off-shell Yang-Mills amplitude. To be more precise, we obtain the equality (up to overall sign),
{\small
\begin{eqnarray}
&& \int d\mu_{(n-1)}^{\rm CHY} \times \Delta([1,2]\, 3\, 4)^2\times {\rm PT}_{([1,2],3,\ldots, n)}\times \frac{\s_{34}}{\s_{[1,2]3} \, \s_{4[1,2]}}\times {\rm Pf}\left[ \left(  \Psi_{(n-1)} \right)^{[1,2]}_{[1,2]}  
\right] \nonumber \\
=&& 
 \int d\mu_{(n-1)}^{\rm CHY} \times \Delta([1,2]\, 3\, 4)^2\times {\rm PT}_{([1,2],3,\ldots, n)}\times \frac{1}{ \s_{[1,2]4}}\times {\rm Pf}\left[ \left(  \Psi_{(n-1)} \right)^{[1,2]\,4}_{[1,2]\,4}  
\right] \nonumber \\
=&&  \left. A_{(n-1)}^{\rm YM} ( [1,2] ,3, 4,\ldots, n)\right|_{\eps^{\mu}_{[1,2]} \rightarrow k^\mu_{[1,2]}} \, ,
\end{eqnarray}
}
\hspace{-0.15cm}\noindent
where the integrand on the first line is  a generic {\it strange-cut}, and
the  fixed puncture, $\s_{[1,2]}$, is an off-shell  vertex, (i.e. $k^\mu_{[1,2]}= k^\mu_1+k^\mu_2,\,  \eps^\mu_{[1,2]}\cdot k^\mu_{[1,2]}\neq 0 $). 

We also analyze the standard factorization cuts ({\it standard-cuts}), and the different ways to glue their resulting-graphs. It is crucial in order to obtain a recursive method. 

In section \ref{sectionMOFF}, we basically generalize the results found previously in section \ref{LongContributions} to more than one off-shell particle.  Additionally, we use reverse engineering to compute (applying the algorithm in Fig. \ref{flux-diagram}) some non-trivial standard factorization contributions. 

In section \ref{sectionEX5P} we apply all technology developed up to this point. Here, we carry out analytically (and in a simple way)  the five-point amplitude, $A^{\rm YM}_5(1,2,3,4,5)$.

In sections \ref{sectionYMS} and \ref{examplesYMS}, we see that the same ideas developed for Yang-Mills theory are naturally extended to the special Yang-Mills-Scalar theory. We give some simple examples,  for instance, two gluons interacting with two scalars, and four and six scalars. Additionally, in this section we examine as the ${\cal A}$ matrix (the kinematic matrix) is factored into a product of two matrices that involve scalars and gluons (usually denoted by $\Psi_{\rm g,s:g}$).

In section  \ref{strange-YMS}, we obtain a connection between the Yang-Mills-Scalar amplitudes and the {\it strange-cuts}. In other words, a {\it strange-cut} factorizes the reduced Pfaffian, ${\rm Pf}^\prime \Psi_n$, into a product of two Pfaffians that involve the matrix $\Psi_{\rm g,s:g}$, which contains both gluons and scalars.

Finally, in section \ref{sectionSoft}, we see as the soft limit, at leading order,  becomes simple to carry out after using the ideas developed in this paper.

Some conclusions are presented in section \ref{sectionConc}, and in appendix \ref{notation} and \ref{appendix}, we give a small glossary and  some  properties of the off-shell Pfaffians, ${\rm Pf}\left[ \left(\Psi_p \right)^{i}_{i} \right]$, ${\rm Pf}\left[ \left(\Psi_p \right)^{ij}_{ij} \right]$, ${\rm Pf}\left[ \left(\Psi_p \right)^{ijk}_{ijk} \right]$.

{\it In this work we describe carefully the results found by the author  et al. in the recent paper \cite{Bjerrum-Bohr:2018lpz}}.

\section{A Brief Review of the Double Cover Representation }\label{reviewDC}

The double cover representation of the CHY construction is given as a contour integral on $n$-punctured double-covered Riemann sphere. Restricted to the curves,
$ {\rm C}_a \equiv y_a^2 - \sigma_a^2 + \Lambda^2=0$ for $a = 1,\ldots,n$, the pairs, $(\sigma_1,y_1), (\sigma_2,y_2),\ldots, (\sigma_n,y_n)$,
provide a set of doubled variables  (a translation table has been worked out in detail in ref. \cite{Gomez:2016bmv}).

As a fast overview, in the DC approach, a CHY-like integrand is built using the third kind form,
$\tau_{(a,b)}\,d\s_a\equiv \left[ \frac{1}{2}\left(  \frac{ y_b}{y_a}+1\right) \frac{1}{\s_{ab}} + \frac{1  }{2\,y_a} \right] d \s_a  $, on the support ${\rm C}_a={\rm C}_b=0$, and the integration measure is given by
\vspace{-0.1cm}
{\small
\begin{eqnarray}\label{measureGF}
&& 
\hspace{-0.8cm}
d\mu_n^{\L}\times \Delta_{(pqr)}\,\Delta_{(pqr|m)}= 
\left[\frac{1}{2^2}\times \frac{d\L}{\L}  \times  \prod_{a=1}^n \frac{y_a\, dy_a}{{\rm C}_a}  
\times\hspace{-0.2cm} \prod_{d=1 \atop  d\neq p,q,r,m}^n
\hspace{-0.2cm}
\frac{d\s_d}{ S^{\tau}_d}\right] 
\times\Delta_{(pqr)}\, \Delta_{(pqr|m)}\,,
\end{eqnarray}
}
\hspace{-0.15cm}\noindent
where the Faddeev-Popov determinants, $\Delta_{(pqr)}\,\Delta_{(pqr|m)} $, 
and the scattering equations, $S^\tau_a$'s, are defined as (we use the notation $(y\s)_a\equiv y_a+\s_a$)
\vspace{-0.1cm}
{\small
\begin{eqnarray}
\hspace{-0.4cm}
\Delta_{(pqr)} =  \frac{y_p\, y_q\, y_r}{(y\s)_p\,(y\s)_q\,(y\s)_r}\times  \left|
                            \begin{array}{ccc}
                              1 & (y\s)_p\, &  \left[ (y\s)_p \right]^{2}  \\
                              1 &  (y\s)_q \,\, &   \left[ (y\s)_q \right]^{2}\\
                              1  &  (y\s)_r\,\,&  \left[ (y\s)_r \right]^{2} \\
                            \end{array}
                          \right| =  \left(\tau_{(p,q)} \, \tau_{(q,r)} \, \tau_{(r,p)} \right)^{-1}
                       \label{Delta-3}
\end{eqnarray} 
}
\vskip-0.4cm\noindent
\vspace{-0.8cm}
{\small
\begin{eqnarray}
\hspace{-0.4cm}
\Delta_{(pqr|m)} =
\Delta_{(pqr)} \, \s_m -  \Delta_{(mpq)} \, \s_r +  \Delta_{(rmp)} \, \s_q -\Delta_{(qrm)} \, \s_p,  \label{Delta-4}
\end{eqnarray} 
}
\vskip-0.5cm\noindent
\vspace{-0.8cm}
{\small
\begin{eqnarray}
\hspace{-0.4cm}
S_a^{\tau} =
\sum_{b\neq a}^n \,k_a\cdot k_b \,\tau_{(a,b)} \, \, . \label{Stau}
\end{eqnarray}
}
\vskip-0.2cm\noindent

Here is  important to remind ourselves where the factors, $\Delta_{(pqr)}$ and $\Delta_{(pqr|m)}$, come from.  Such as in the single cover approach, the number of independent scattering equations, $S_a^{\tau}=0$, is ``$n-3$", which is straightforward to see of the identities (on the support $ {\rm C}_a=0$), 
\begin{eqnarray}\label{se-symm}
\sum_{a=1}^n y_a\, S_a^\tau = \sum_{a=1}^n y_a\,(y_a+\s_a) S_a^\tau  = \sum_{a=1}^n \frac{y_a}{y_a+\s_a}\, S_a^\tau  =0.
\end{eqnarray}
From these identities, we can read  the ${\rm SL}(2,\mathbb{C})$ generators given by global vectors
\begin{equation}
L_{\pm 1} = \L^{\pm 1} \sum_{a=1}^n y_a \, (y_a+\s_a)^{\mp 1}\, \partial_{\s_a}, ~ L_{0} =  \sum_{a=1}^n y_a \,  \partial_{\s_a},~[L_{\pm 1},L_0]=\pm L_{\pm1}, ~ [L_{1},L_{-1}]=2 L_{0}
\end{equation}
The Faddeev-Popov determinant to fix the \eqref{se-symm} redundancy is the factor defined in \eqref{Delta-3} that we denoted by $\Delta_{(pqr)}$. 

When $\Lambda $ is promoted as a variable, the double cover formulation is invariant under  the scale transformation, $(\s_1,...,\s_n,y_1,...,y_n,\L) \rightarrow  (\rho\, \s_1,...,\rho\, \s_n,y_1,...,y_n,\rho\,\L), \, \rho\in \mathbb{C}^\ast$, which is generated by the vector field  (on the support $ {\rm C}_a=0$)
\begin{equation}
D= \L \,\partial_{\L} + \sum_{a=1}^n \s_a \, \partial_{\s_a}.
\end{equation}
The global vectors, $\{L_{\pm1},\, L_0,\, D  \}$, satisfy a $gl(2,\mathbb{C})$ algebra, therefore, these generators can be used to gauge four punctures, for example $(\s_p,\s_q,\s_r,\s_m)$,  and the integrand must be multiplied by  the Faddeev-Popov determinant  defined in \eqref{Delta-4} and denoted by $\Delta_{(pqr|m)}$ (in the $\Delta_{(pqr|m)}$ factor, the label ``$m$" is referred to the scale generator). We call this process a {\it gauge fixing} (or {\it initial setup}). 

The amplitudes are derived from the integral
\vspace{-0.1cm}
\begin{equation}\label{LprescriptionGFF}
\hspace{-0.1cm}
A_n=   \int_{\Gamma}  d\mu^{\L}_n    \times  \frac{(-1) \,\Delta_{(pqr)}\times \Delta_{ (pqr|m)}} {S^\tau_m } \times {\cal I}^\L_n(\sigma,y)  \, ,
\end{equation}
where the  $\Gamma$  contour is defined by the equations\footnote{The rewriting of the amplitude in terms of this contour ``$\Gamma$", which does
not encircle the scattering equation $S^{\tau}_m$, follows from the global residue theorem.},
$\L=0,~ S^{\tau}_d=0,~{\rm for}~ d\neq \{p,q,r,m\},~\, {\rm C}_a =0,~ \forall\, a$
.

Like in the single-cover approach, the precise form of the integrand ${\cal I}^\L_n(\s,y)$ defines the theory. 
For example, the double color-ordered partial amplitude $\phi^3$-theory, usually denoted by $m^{(0)}=(\a|\b)$, corresponds to the integrand\footnote{At loop level see \cite{Gomez:2017lhy,Gomez:2017cpe,Ahmadiniaz:2018nvr}.}
{\small 
\begin{eqnarray}\label{integrandP}
{\cal I}^{\phi^3}_n(\a|\b) ~=~  {\rm PT^\tau}_{(\a_1,\a_2,\ldots \a_n)} \times {\rm PT^\tau}_{(\b_1,\b_2,\ldots \b_n)} \,,
\end{eqnarray}}
\hspace{-0.15cm}\noindent
with the Parke-Taylor factors,
$
{\rm  PT^\tau}_{(\a_1,\a_2,\ldots \a_n)} \equiv \tau_{(\a_1,\a_2)}\,\tau_{(\a_2,\a_3)}\cdots \tau_{(\a_n,\a_1)} \,,
$
where \\
$\a=(\a_1,\ldots , \a_n)$ and 
$\b=(\b_1,\ldots , \b_n)$ are two partial orderings
(note that although $\tau_{(a,b)}$ is the equivalent to $\frac{1}{\s_{ab}}$ in the single-cover approach, this is neither antisymmetric nor symmetric).

Similarly, other theories correspond to products of such modified Parke-Taylor factors with additional expressions, much like in the original
CHY formalism. Again, the integrands for these other theories can be broken down to products of shuffled Parke-Taylor expressions.%

\section{Matrices in The DC prescription}\label{sectionPfaffian}

Since $\tau_{(a,b)}\neq - \tau_{(b,a)}$, it is not immediately obvious how to define the CHY anti-symmetric matrices to describe
the double-cover analog of the pure Yang-Mills, Gravity, NLSM theory, among others.
In order to obtain  a double-cover version for the CHY matrices, we rewrite $\tau_{(a,b)}$ as
{\small
\begin{eqnarray}
\hspace{-0.1cm}
\tau_{(a,b)} =\frac{(y\s)_a}{y_a} \times \, T_{ab} \,, \quad {\rm with \,\,} ~~  T_{ab}  = 
\frac{1}{(y\s)_a-(y\s)_b} ~~~ {\rm on} ~~ {\rm C}_a={\rm C}_b=0\, .
\end{eqnarray}
}
\hspace{-0.15cm}\noindent
Clearly, $T_{ab}$ is an anti-symmetric form, $T_{ab}= -T_{ba}$, thus, we can establish the simple map, $\displaystyle\frac{1}{\s_{ab}} \, \rightarrow \, T_{ab}$, in order to define the matrices in the double-cover representation.  For example, the ${\cal A}^\L$ matrix is obtained from the ${\cal A}$ matrix by the replacement, ${\cal A}^\L\equiv {\cal A}\Big|_{\frac{1}{\s_{ab}}\, \rightarrow \, T_{ab}}$, i.e.  ${\cal A}_{ab}= k_a\cdot k_b\, T_{ab}$, $a\neq b$ and  ${\cal A}_{aa}=0$. The same replacement is made for the other matrices, ${\cal B}^\L\equiv {\cal B}\Big|_{\frac{1}{\s_{ab}}\, \rightarrow \, T_{ab}}$, ${\cal C}^\L\equiv {\cal C}\Big|_{\frac{1}{\s_{ab}}\, \rightarrow \, T_{ab}}$ and $\Psi^{\L}_n\equiv \Psi_n\Big|_{\frac{1}{\s_{ab}}\, \rightarrow \, T_{ab}}$. An identical correspondence can be done for more matrices, for instance, ${\cal X}_{\rm s}^\L$ and\footnote{We will come back to the $\Psi^{\L}_{\rm g,s:g}$ matrix later.}  $\Psi^{\L}_{\rm g,s:g}$ (see \cite{Cachazo:2014xea}), but,  there is one matrix that must be handled with care, the $\Pi_{\a_1,...,\a_m}$ matrix. 
This matrix has elements such as, $\sum_{i^\prime\in\a_{p^\prime}, j\in\a_q} \frac{\s_{i^\prime} k_{i^\prime} \cdot k_j }{\s_{i^\prime j}}$, and we  in \cite{inpreparation} will explain how to deal with that type of terms.

Before analyzing the pure and scalar Yang-Mills theories in the double-cover formalism, it is useful to see how $\phi^3$-integrands may be rewritten in terms of $T_{ab}$,
\small{
\begin{eqnarray}\label{phi3T}
{\cal I}^{\phi^3}_n(\a|\b) &=&  {\rm PT^\tau}_{(\a_1,\ldots \a_n)} \times 
\left[ \prod_{a=1}^n    \frac{(y\s)_a}{y_a}  \times
{\rm PT}^T_{(\b_1,\ldots \b_n)} \right] ,  \nonumber \\
&=& \left( \prod_{a=1}^n    \frac{(y\s)_a}{y_a} \right)^2 \times
{\rm PT^T}_{(\a_1,\ldots \a_n)} \times 
{\rm PT}^T_{(\b_1,\ldots \b_n)} ,
\end{eqnarray}
}
\hspace{-0.15cm}\noindent
where
$
{\rm  PT}^T_{(a_1,a_2,\ldots a_n)} \equiv T_{a_1 a_2}\, T_{a_2 a_3 }\cdots T_{a_n a_1} \,.
$

Following the CHY program developed in \cite{Cachazo:2013iea,Cachazo:2014xea}, the DC prescription for the color-ordered scattering amplitudes of the pure Yang-Mills theory can be obtained from \eqref{phi3T} by replacing, ${\rm  PT}^T_{(\b_1,\ldots , \b_n)} \rightarrow (-1)^{i+j}\,T_{ij}\,{\rm Pf}[(\Psi^\L_n)^{ij}_{ij}]$,  i.e.
{\small
\begin{eqnarray}\label{}
{\cal I}^{\rm YM}_n(\a) =  {\rm PT}^\tau_{(\a)} \times 
{\bf Pf}^{\prime} \Psi^\L_n , 
\quad
{\bf Pf}^{\prime} \Psi_{n}^\L \equiv   
(-1)^{i+j}\, T_{ij}\left(  \left[ \prod_{a=1}^n   \frac{(y\s)_a}{y_a}    \right]
 {\rm Pf} \left[(\Psi^\L_n)^{ij}_{ij}\right] \right) ,   ~~~\label{Pfdef}
\end{eqnarray}
}
\hspace{-0.15cm}\noindent
where $(\Psi^\L_n)^{ij}_{ij}$ is  built by removing the rows/columns $i, j$  from $\Psi^\L_n$, with $1 \leq i< j \leq n$. In a similar way, the DC integrand for the special Yang-Mills-Scalar theory is given by
{\small
\begin{eqnarray}\label{}
{\cal I}^{\rm YMS}_n (\a)=  {\rm PT^\tau}_{(\a)} \times 
\hspace{-0.7cm}
\sum_{~~~\{a,b\}\in {\rm p.m. (s)}}
\hspace{-0.6cm}
{\rm sgn}_{(\{a,b\})}\, \delta^{I_{a_1},I_{b_1}} \cdots \delta^{I_{a_m},I_{b_m}}
\, T_{a_1b_1}\cdots T_{a_m b_m} \,
{\bf Pf}^{\prime} \Psi^\L_{\rm g,s:g} \, ,  ~~ ~~~\label{PfYMS}
\end{eqnarray}
}
\hspace{-0.15cm}\noindent
where, ${\bf Pf}^{\prime} \Psi_{\rm g,s:g}^\L \equiv   
(-1)^{i+j}\, T_{ij}\left(  \left[ \prod_{a=1}^n   \frac{(y\s)_a}{y_a}    \right]
 {\rm Pf} \left[(\Psi^\L_{\rm g,s:g})^{ij}_{ij}\right] \right)$ and $\Psi^\L_{\rm g,s:g} \equiv \Psi_{\rm g,s:g}\Big|_{\frac{1}{\s_{ab}} \rightarrow T_{ab}}$. Here, the set of gluons is denoted as ``${\rm g}$" (${\rm g}=\{  g_1,...,g_p\}$) while that the set of scalars\footnote{We apologise for the abuse of notation between Maldenstand variables and the scalar particles. 
Additionally, notice that, $p+2m=n$, the total number of particles.} as ``${\rm s}$" (${\rm s}=\{s_1,,..., s_{2m}\}$). The symbol ``${\rm p.m.}$" means perfect matching and, $\{a_1, b_1, . . . , a_m, b_m\} = {\rm s}$. 
Since the \eqref{PfYMS} expansion comes from the pfaffian, ${\rm Pf}({\cal X}_{\rm s}^\L)$ \cite{Cachazo:2014xea}, then 
 ${\rm sgn}_{(\{a,b\})}$ means the corresponding signature (more details about this theory in section \ref{sectionYMS}).

Finally, the pure Yang-Mills and the special Yang-Mills-Scalar amplitudes at tree-level in the DC language are given by the integrals
{\small
\begin{eqnarray}
&& 
A_n^{\rm YM}(\a)  = \int_\Gamma d\mu^\L_n \, 
 \frac{(-1)\,\Delta_{(pqr)} \, \Delta_{(pqr|m)}} {S^\tau_m } \times
 {\cal I}^{\rm YM}_n (\a)\, 
 , 
 \label{YMgeneric} \\
&&
A_n^{\rm YMS}(\a) =  
\int_\Gamma d\mu^\L_n \, 
 \frac{(-1)\,\Delta_{(pqr)} \, \Delta_{(pqr|m)}} {S^\tau_m } \times
{\cal I}^{\rm YMS}_n (\a)
 \, ,  \label{YMSgeneric}
\end{eqnarray}
}
\hspace{-0.15cm}\noindent
where $(\a)\equiv (\a_1,\a_2,\ldots ,\a_n)$ is a partial (generic) ordering.

The following sections will be dedicated to the pure Yang-Mills theory. The generalization to the special Yang-Mills-Scalar theory is straightforward, and we will come back to this model in section \ref{sectionYMS}.

{\bf It is important to remark that due to the normalization of the kinematic parameters and the polarization vectors  chose in these notes, we are going to obtain an extra overall factor, $(2)^{\frac{(4-n)}{2}}$, compared
with color-ordered Feynman rules given in\footnote{In order to obtain the normalization given by Dixon in \cite{Dixon:1996wi} for color ordered amplitudes,  we just need to perform the replacement, $k^\mu_a\rightarrow \sqrt{2}\, k^\mu_a$.} \cite{Dixon:1996wi}.}

\subsection{Graphical Representations of DC Integrands}\label{graphR}

Since the method that we are going to describe in this work is based on so called the $\L-$algorithm, which is given by graphic rules  \cite{Gomez:2016bmv}, we  introduce a simple graph representation for the amplitude $A_n^{\rm YM}(\a) $  in \eqref{YMgeneric}.

Like in $\phi^3$, a Parke-Taylor factor is drawn by lines join the vertices in a sequence way. In order to specify the ordering we replace the lines by arrows, for example 
\vspace{-0.45cm}
\begin{eqnarray}\label{PTgraph}
\hspace{0.1cm}
{\rm  PT}^\tau_{(1,\ldots, n)} 
=
\hspace{-0.32cm}
\parbox[c]{4.9em}{\includegraphics[scale=0.14]{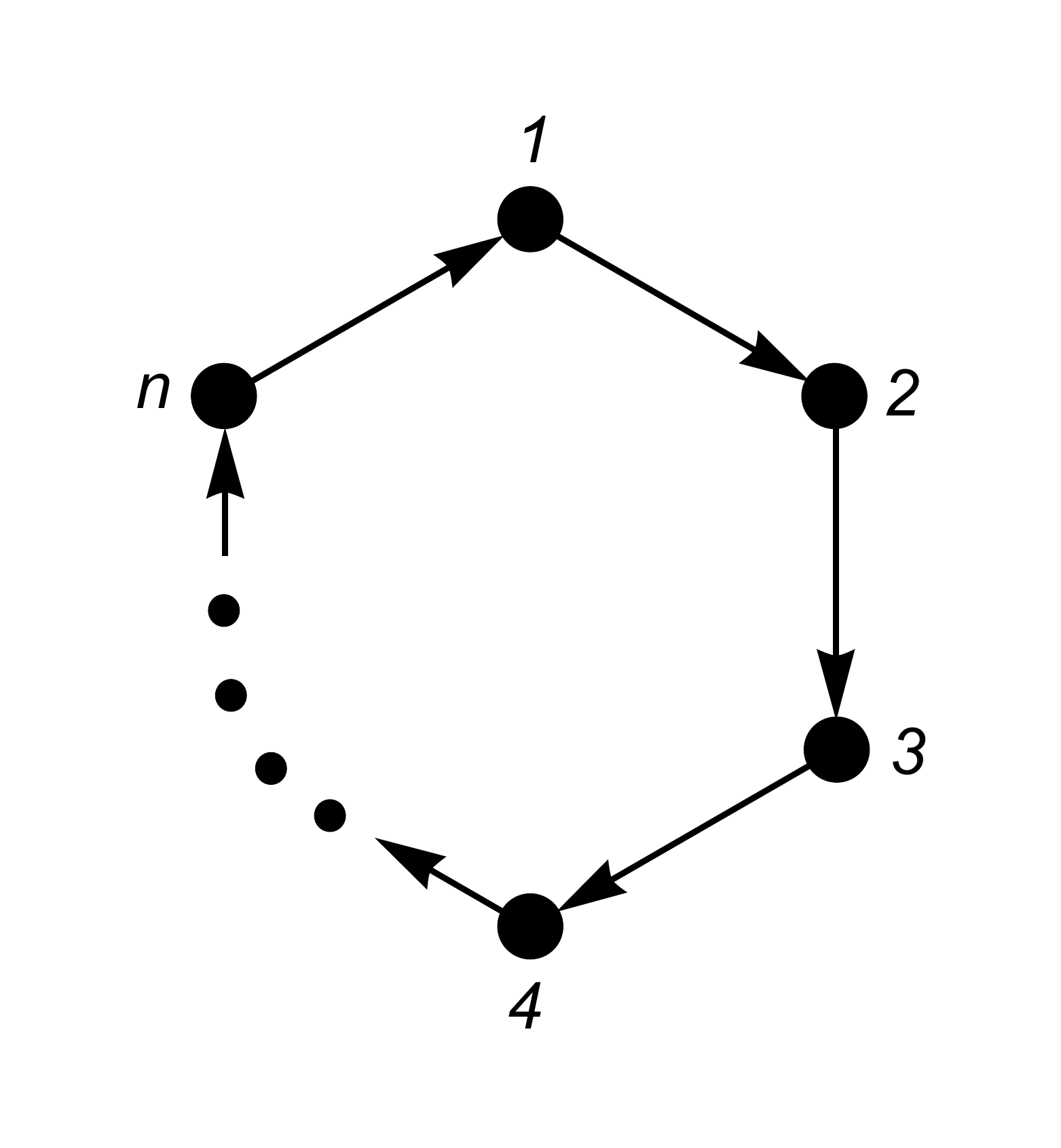}}
= (-1)^n \times
\hspace{-0.25cm}
\parbox[c]{4.9em}{\includegraphics[scale=0.14]{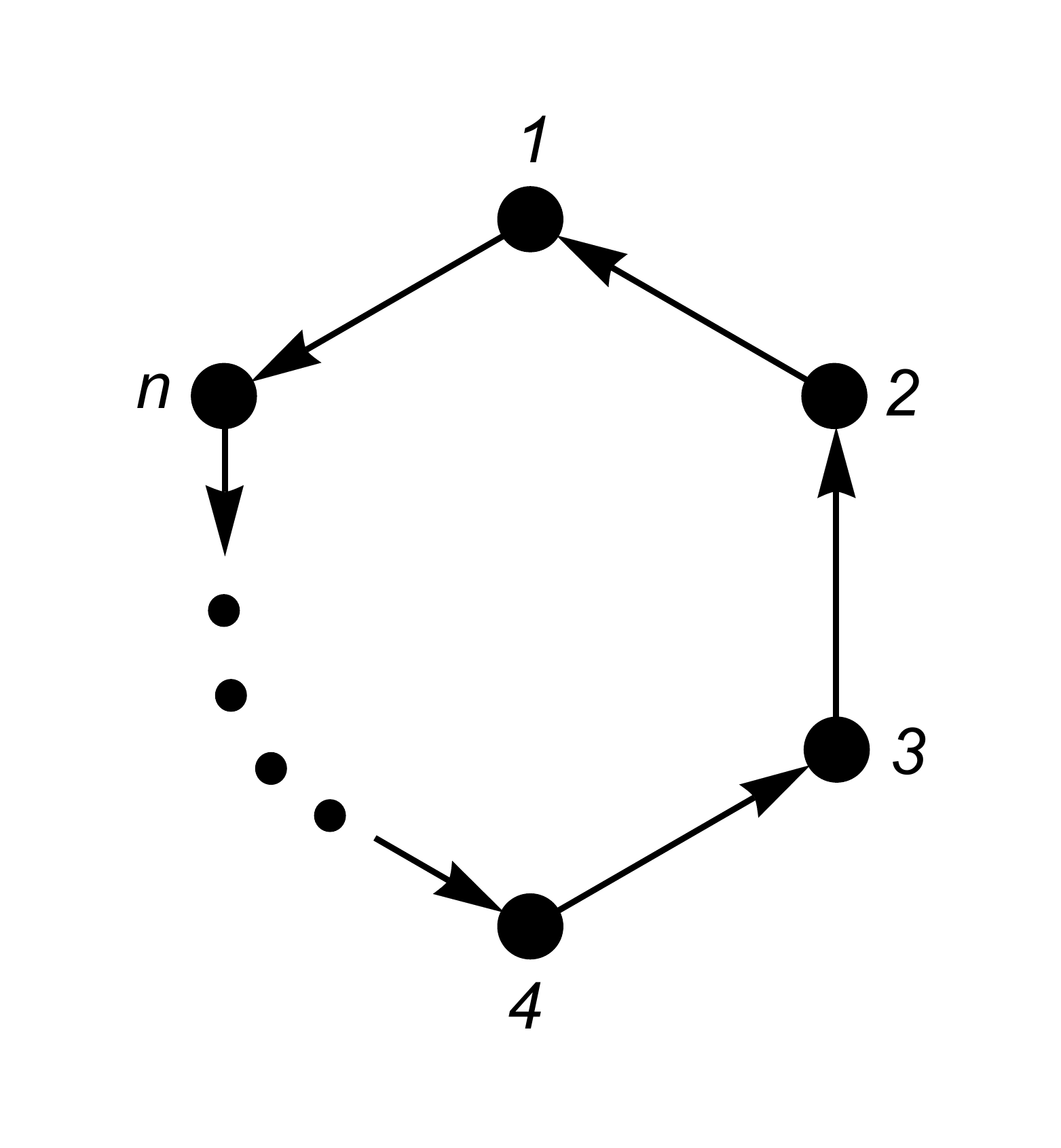}}=  (-1)^n \times {\rm  PT}^\tau_{(n,\ldots, 1)} \, ,\,\,
\end{eqnarray}
\vskip-0.45cm\noindent

On the other hand, due to the factor, $ {\rm Pf} \left[(\Psi^\L_n)^{ij}_{ij}\right]$, is a non-ordered object, we do not know how to build a graph representation for it. Nevertheless, there is a term in ${\bf Pf}^{\prime} \Psi^\L_n $ that one can draw, specifically $T_{ij}$, which we  sketch by red arrow, namely,
$T_{ij} \equiv i \, {\color{red} \rightarrow} \, j$.  Finally, so as in \cite{Gomez:2016bmv}, the factor, $ \frac{(-1)\,\Delta_{(pqr)}\, \Delta_{ (pqr|m)}} {S^\tau_m }$, is symbolized by yellow vertices for $(\s_p,\s_q,\s_r)$ and a green one for the $\s_m$-puncture. Therefore, the complete graphs for the YM integrand in the DC representation is\footnote{From this graph representation we can conclude that, a vertex with two black arrow is a gluon and, a vertex with two black arrows and a red one represents a gluon such that its row/column (among 1 and n) must be removed from the matrix.}
\vspace{-0.4cm}
{\small
\begin{eqnarray}
 \frac{(-1)\,\Delta_{(123)}\, \Delta_{ (123|4)}} {S^\tau_4 } \times  {\rm  PT}^\tau_{(1,\ldots, n)} \times
 (-1)^{2+p}T_{2p} 
 \left( \left[ \prod_{a=1}^n  \frac{(y\s)_a}{y_a}\right]  
\,
 {\rm Pf} \left[(\Psi^\L_n)^{2p}_{2p}\right] \right)
\equiv
\hspace{-0.3cm}
\parbox[c]{5.3em}{\includegraphics[scale=0.14]{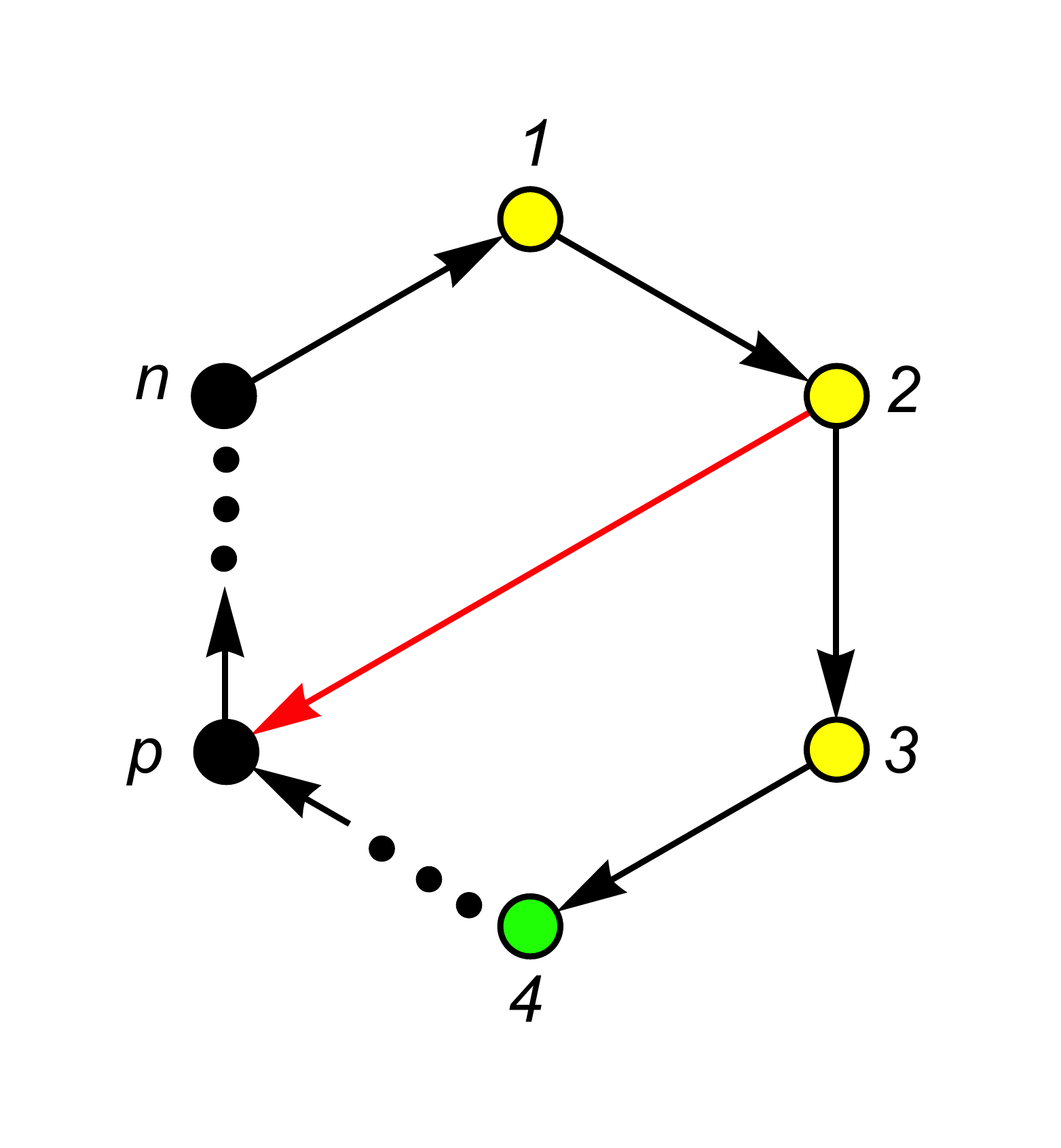}} ,\qquad
\end{eqnarray}
}
\vskip-0.5cm\noindent
where, without loss of generality, we have chosen $(pqr|m)=(123|4)$ and $(i,j) = (2,p)$. The amplitude, $ A_n^{\rm YM}(1,2,...,n)$, is given by 
\vspace{-0.3cm}
{\small
\begin{eqnarray}\label{YMgraph}
\hspace{-0.2cm}
A_n^{\rm YM}(1, ..., n)=
A_n^{(2,p)}(1, ..., n)
= \int d\mu_n^\L
\hspace{-0.1cm}
\parbox[c]{5.8em}{\includegraphics[scale=0.14]{YM1-n.pdf}} ,
\end{eqnarray}
}
\vskip-0.7cm\noindent
where the superscript denotes the $(i,j)$ choosing, always with $i<j$ in order to give the direction to the red arrow, this is import in order to obtain the cyclic property.
We call to \eqref{YMgraph} a {\it YM-graph}.

Obviously,  when all polarization vectors are transverse and all particles are on-shell  (i.e.  $k_a\cdot \eps_a=k_a^2=0$), the above expression is independent of the red arrows and colored vertices \cite{Cachazo:2013hca,Cachazo:2014xea} (physical amplitudes). However, when there is one off-shell particle,  this notation becomes important and that expression depends on the colored vertices and red arrows, such as it will be shown later. 

The generalization of the graph representation from the DC prescription to the single-cover approach is  simple, it is just to replace the green vertex  by a black one, since the single-cover is not scale invariant, namely 
\vspace{-0.5cm}
{\small
\begin{eqnarray}\label{}
\int d\mu_n^\L
\hspace{-0.1cm}
\parbox[c]{5.3em}{\includegraphics[scale=0.14]{YM1-n.pdf}} 
=
\hspace{-0.01cm}
\int d\mu_n^{\rm CHY}
\hspace{-0.1cm}
\parbox[c]{5.0em}{\includegraphics[scale=0.14]{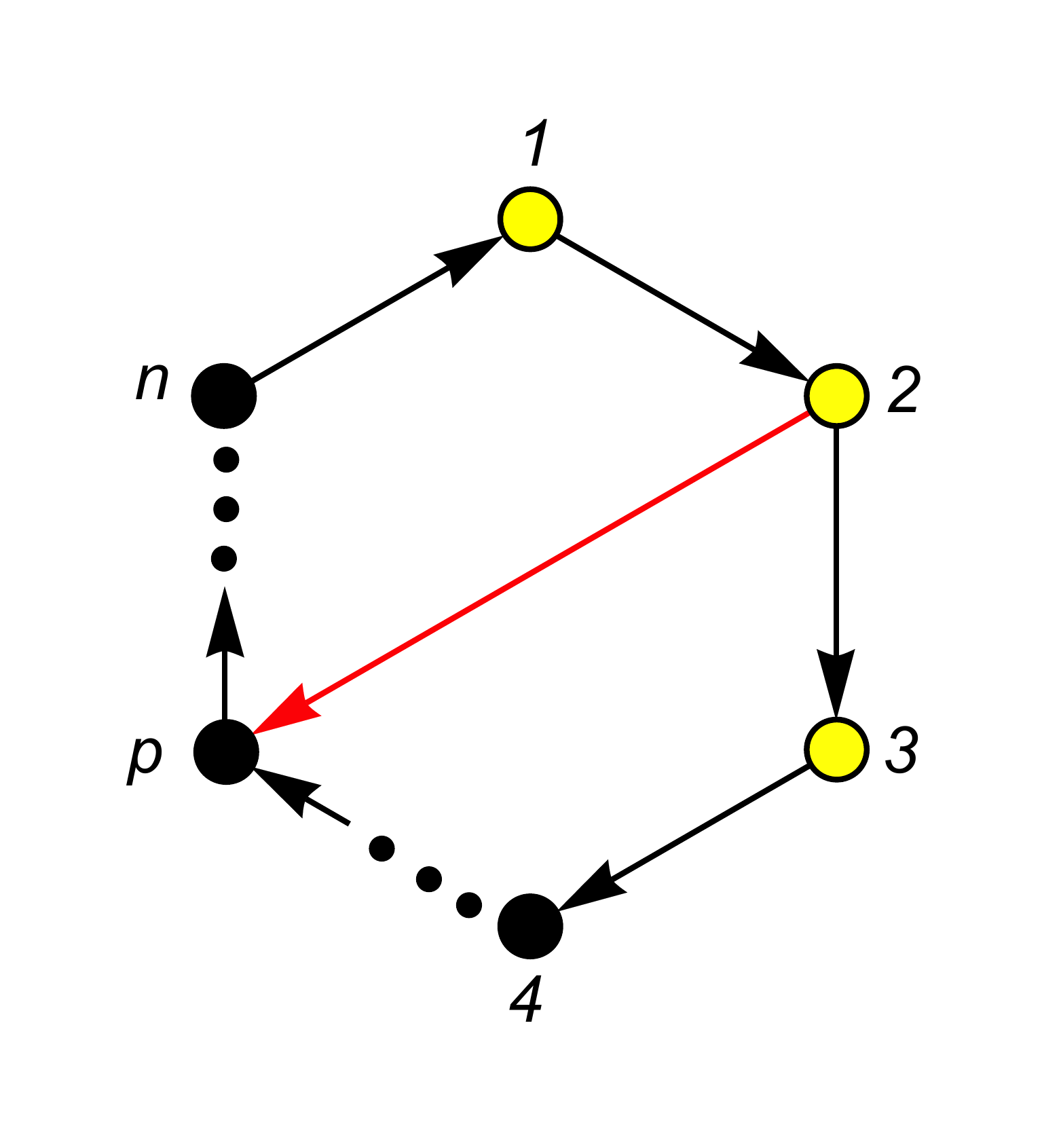}} .
\nonumber
\end{eqnarray}
}
\vskip-0.4cm\noindent
On the right hand side, the colored vertices mean the punctures $(\s_1,\s_2,\s_3)$ are fixed, i.e. the Faddeev-Popov determinant, $\Delta(123)^2=(\s_{pq}\s_{qr}\s_{rp})^2 $.  The map from the double-cover to the single one can be seen in \cite{Gomez:2016bmv}

At last,  it is useful to observe the following two properties 
\vspace{-0.1cm}
\begin{eqnarray}\label{properties}
&& 
\hspace{-0.2cm}
 A_n^{(2,p)} (1,2, \ldots, n)=
A_n^{(2,p)}(2,\ldots, n,1) = A_n^{(p,2)}(3,\ldots,n,1,2) = \cdots =
A_n^{(2,p)}(n,1\ldots,n-1)
, 
\quad
\nonumber \\
&&
\hspace{3.0cm}
A_n^{(2,p)}(1,2, \ldots, p, \dots , n) = 
(-1)^n\,
 A_n^{(p,2)}(n,\ldots , p, \ldots ,2,1)\,,~~~~~
\end{eqnarray}
Notice the flipping in the superscript, $(2,p) \rightarrow (p,2)$,
which is because the position of the label "$p$" is first that the label "$2$" in the partial ordering.  
These identities are satisfied even if the particles are off-shell.

We would like to clarify that, although right now the notation, $ A_n^{(i,j)} (1,2, \ldots, n)$, looks useless, it will become important since the method developed here depends of the gauge fixing, i.e. of the choosing of $(pqr|m)$ and $(i,j)$.

\section{The double-cover integration rules}\label{section-IntR}

In this section, we  schematize how the double cover formalism works. Following, we obtain some integration rules, which are applied  to the {\it YM-graphs}, in order to solve the CHY integral. 

Since the amplitudes are computed in a manner that differs in detail substantially
from the original CHY prescription, we will provide a few explicit examples.
Let us start the discussion with the integration measure of the double cover prescription, namely
{\small
\begin{eqnarray}\label{}
d\mu_n^{\L}~=\underbrace{ \left[ \frac{1}{2^2} \right]}_{\rm symmetry \atop factor}
\times
~
\underbrace{\left[\prod_{a=1}^n \frac{y_a\, dy_a}{{\rm C}_a}\right]}_{\rm sum \,\,over \atop all\,\,configurations  } 
~~
\times 
\hspace{-0.1cm}
\underbrace{\left[\frac{d\L}{\L} \right]}_{\rm split \,\,the \,\,sphere \atop in\,\, two \,\, pieces  }
\hspace{-0.3cm}
\times
~
\left[\prod_{d\neq 1,2,3,4}^n\frac{d\s_d}{ S^{\tau}_d}\right]\, ,
\end{eqnarray}
}
\hspace{-0.15cm}\noindent
where we have fixed $(pqr|m)=(123|4)$.
Like it was shown in \cite{Gomez:2016bmv},  after integrating the $y_a$ coordinates around the solutions, ${\rm C}_a=0 \Rightarrow \, y_a= \pm \sqrt{\s_a^2-\L^2}, \, \forall\,a$, we obtain a sum over all possible configurations (cuts), i.e. $2^n$ possibilities,  where we call the sign, $+/-$, the {\it upper$/$lower sheet} \footnote{Since the $\mathbb{Z}_2$ symmetry, $y_a \,\rightarrow\, -y_a,\, a =1,2,\ldots, n$, there are $2^{n-1}$ nonequivalents configurations. In other words, the upper and lower spheres are indistinguishable.}. 
For example, at six-point one has $2^6=2\times 32$ possibilities  given, schematically, by\footnote{  
The factor $(1/2^2)$ fixes the discrete symmetry, see~\cite{Gomez:2016bmv}.}
\vspace{-0.3cm}
{\small
\begin{eqnarray}
\hspace{-0.2cm}
A_6^{\rm YM } (1,2,3,4, 5,6)
=
\hspace{-1.1cm}
\parbox[c]{6.0em}{\includegraphics[scale=0.21]{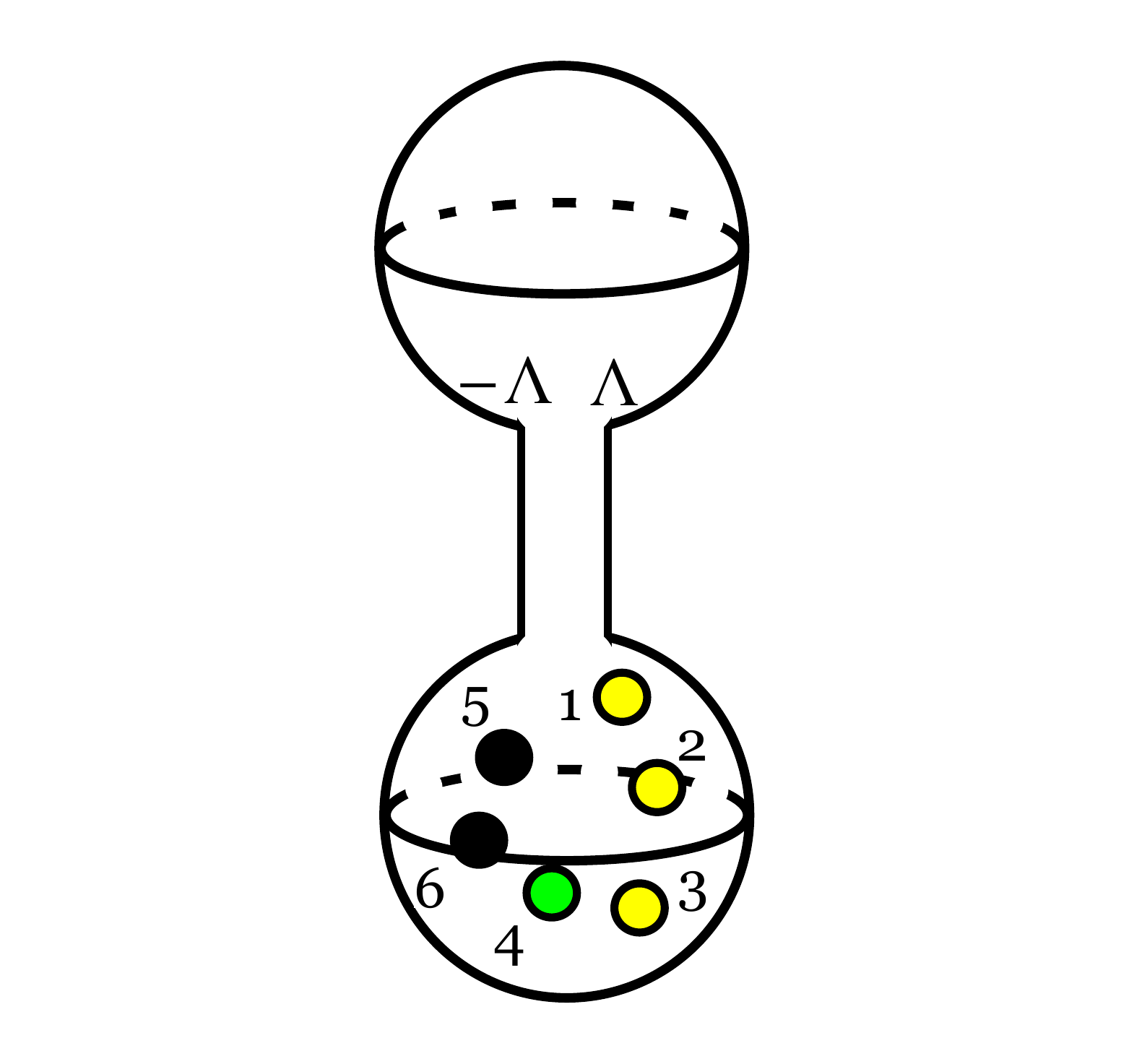}}
+ 
\hspace{-1.1cm}
\parbox[c]{6.0em}{\includegraphics[scale=0.21]{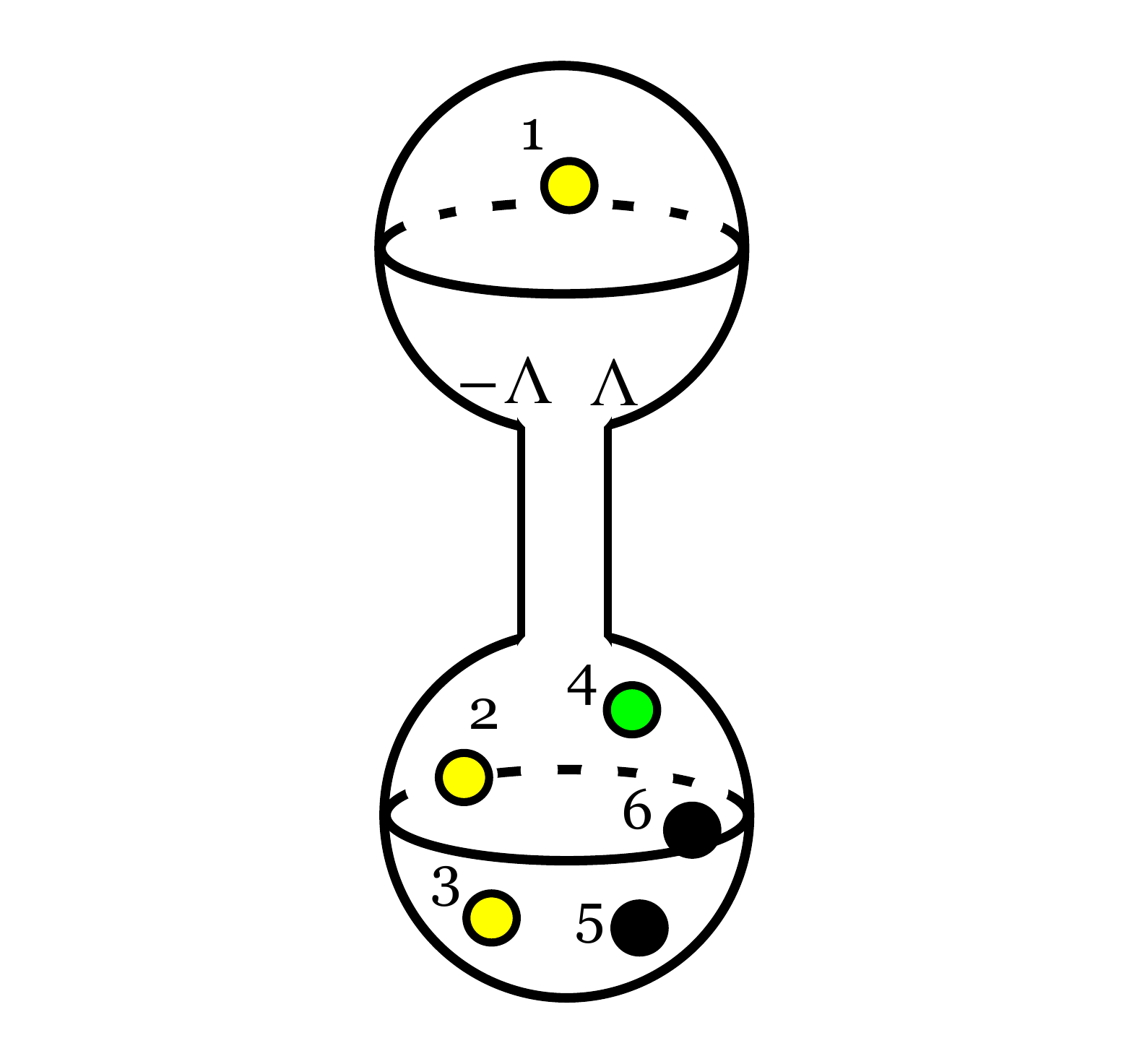}}
+ 
\hspace{-1.1cm}
\parbox[c]{6.0em}{\includegraphics[scale=0.21]{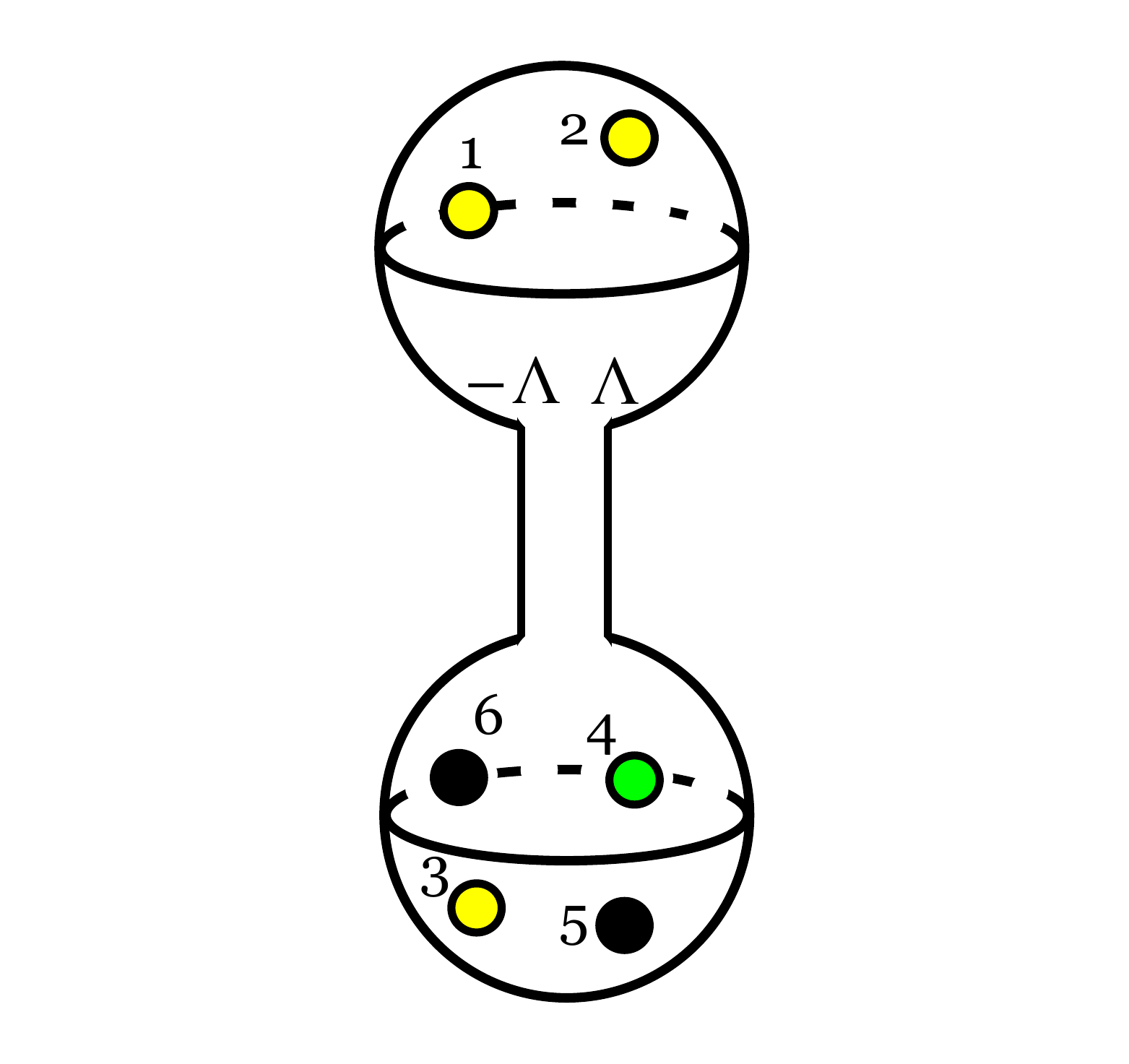}}
+ \cdots +
\hspace{-1.1cm}
\parbox[c]{6.0em}{\includegraphics[scale=0.21]{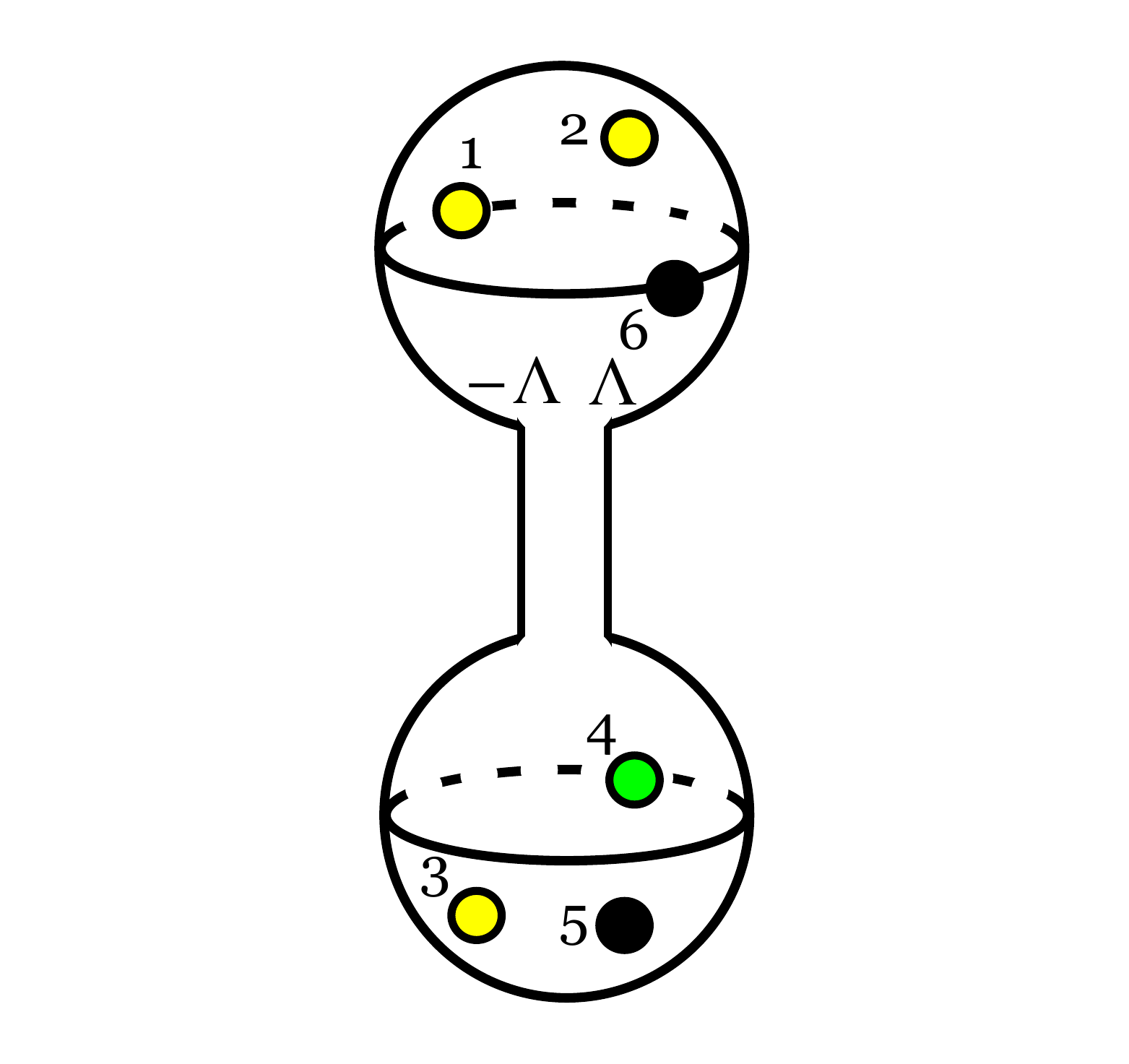}} +\cdots
+
\hspace{-1.1cm}
\parbox[c]{6.0em}{\includegraphics[scale=0.21]{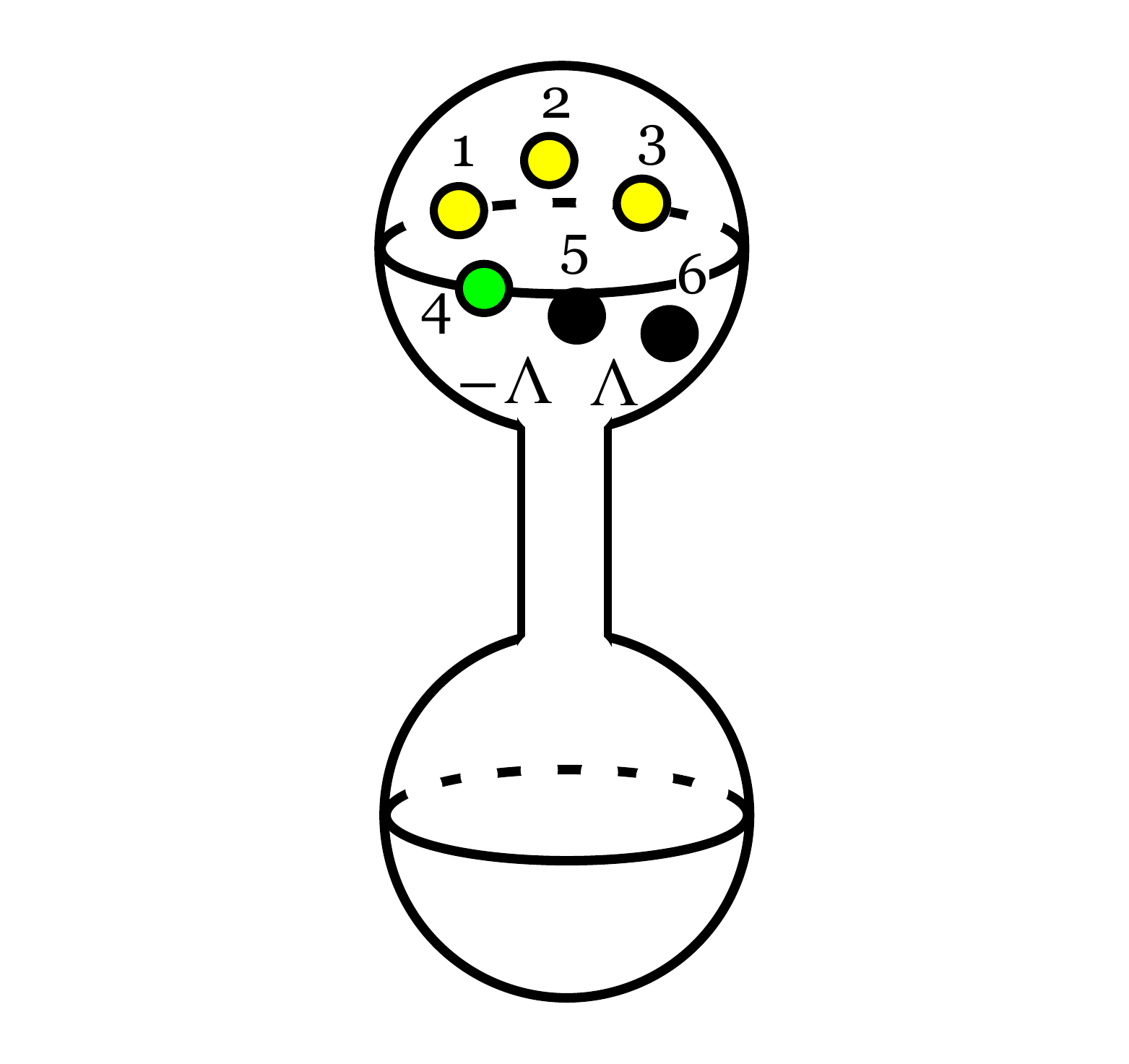}} .
\nonumber
\end{eqnarray}
}
\vskip-0.1cm\noindent
where  upper/lower spheres are represent by the local coordinates, $(y_a=\sqrt{\s_a^2-\L^2},\s_a)$ and 
$(y_a=-\sqrt{\s_a^2-\L^2},\s_a)$, respectively. The neck that joins the spheres represents the branch-cut and the branch-points, $-\L$ and $\L$,  give the width of the neck. 

Integrating $\L$ (keeping $\L\sim \mathcal{O}(0)$),  the double cover formulation factorized into two single covers attached by an off-shell propagator (the free scattering equation in the DC formalism reduce to the propagator when one performs the integration over $\L$). 
For instance, let us consider the punctures $\{\s_p+1, \ldots, \s_n,\s_1,\s_2\}$ on the upper sheet and $\{\s_3,\s_4, \ldots, \s_p\}$, on the lower one. The measure and the Faddeev-Popov determinant turn into\footnote{Let us remember that in this work we are using the momentum notation, $k_{[a_1,a_2,\dots, a_p]}\equiv k_{a_1}+\cdots + k_{a_p}$, $\tilde s_{a_1a_2\ldots a_p}\equiv \sum _{i<j}^p k_{a_i}\cdot k_{a_j}$ and $ s_{a_1a_2\ldots a_p}\equiv k_{[a_1,a_2,\dots, a_p]}^2$.} (expanding around $\L=0$)
\vskip-0.5cm
\begin{eqnarray}\label{fpL0}
&&\hspace{-0.4cm}
d\mu_n^\L \Big|^{p+1,\ldots,1,2}_{3,\, 4,\ldots ,p}  
= \frac{d\L}{\L} \times  \left[ \frac{d\s_{p+1}}{S_{p+1}}\cdots  \frac{d\s_{n}}{S_n} \right] \times  \left[\frac{d\s_{5}}{S_5}\cdots \frac{d\s_{p}}{S_p} \right]
+ {\cal O}(\L) \nonumber\\
&&
\hspace{1.8cm}
=
\frac{d\L}{\L} \times
d\mu^{\rm CHY}_{n-(p-2)+1} \times d\mu^{\rm CHY}_{(p-2)+1} + {\cal O}(\L) , \nonumber \\
&&
\hspace{-0.4cm}
\frac{(-1)\Delta_{(123)}  \Delta_{(123|4)}}{ S^{\tau}_4}
\Big|^{p+1,\ldots,1,2}_{3,\, 4,\ldots ,p}  
\hspace{-0.2cm}
 =  \frac{2^5}{\L^4} (\s_{12}\, \s_{2[{\rm up}]} \,  \s_{[{\rm up}]1}  )^2 
\left[ \frac{ 1 }{\tilde s_{34 \ldots p}} 
\right] 
(\s_{[{\rm down}]3} \, \s_{34} \, \s_{4[{\rm down}]})^2
 +  {\cal O}\left(\L^{-2} \right). \nonumber\\
\end{eqnarray}
\vskip-0.2cm\noindent
The two new punctures,  $\s_{[{\rm up}]}\equiv\s_{[p+1,\ldots,n,1,2]}$ and  $\s_{[{\rm down}]}\equiv\s_{[3,4,\ldots,p]}$,  are fixed at the point, $``\s_{[{\rm up}]}=\s_{[{\rm down}]}=0"$, on  the upper and lower sphere, respectively. It is import to remind  that the sub-index at the puncture is related with the momentum of the particle, {\it e.g.} the punctures $\s_{[p+1,\ldots,n,1,2]}$ and  $\s_{[3,4,\ldots,p]}$  are particles with momenta, $k_{[p+1,\ldots,n,1,2]}$ and  $k_{[3,4,\ldots,p]}$, respectively (off-shell particles). This process is exemplified in the following figure, 
\vspace{-0.3cm}
{\small
\begin{eqnarray}\label{cuts-app}
\hspace{-1.0cm}
\parbox[c]{6.1em}{\includegraphics[scale=0.21]{cut-1-apdx.pdf}}
{\underrightarrow{~~ \L\sim 0  ~~}}
\hspace{-1.2cm}
\parbox[c]{5.8em}{\includegraphics[scale=0.21]{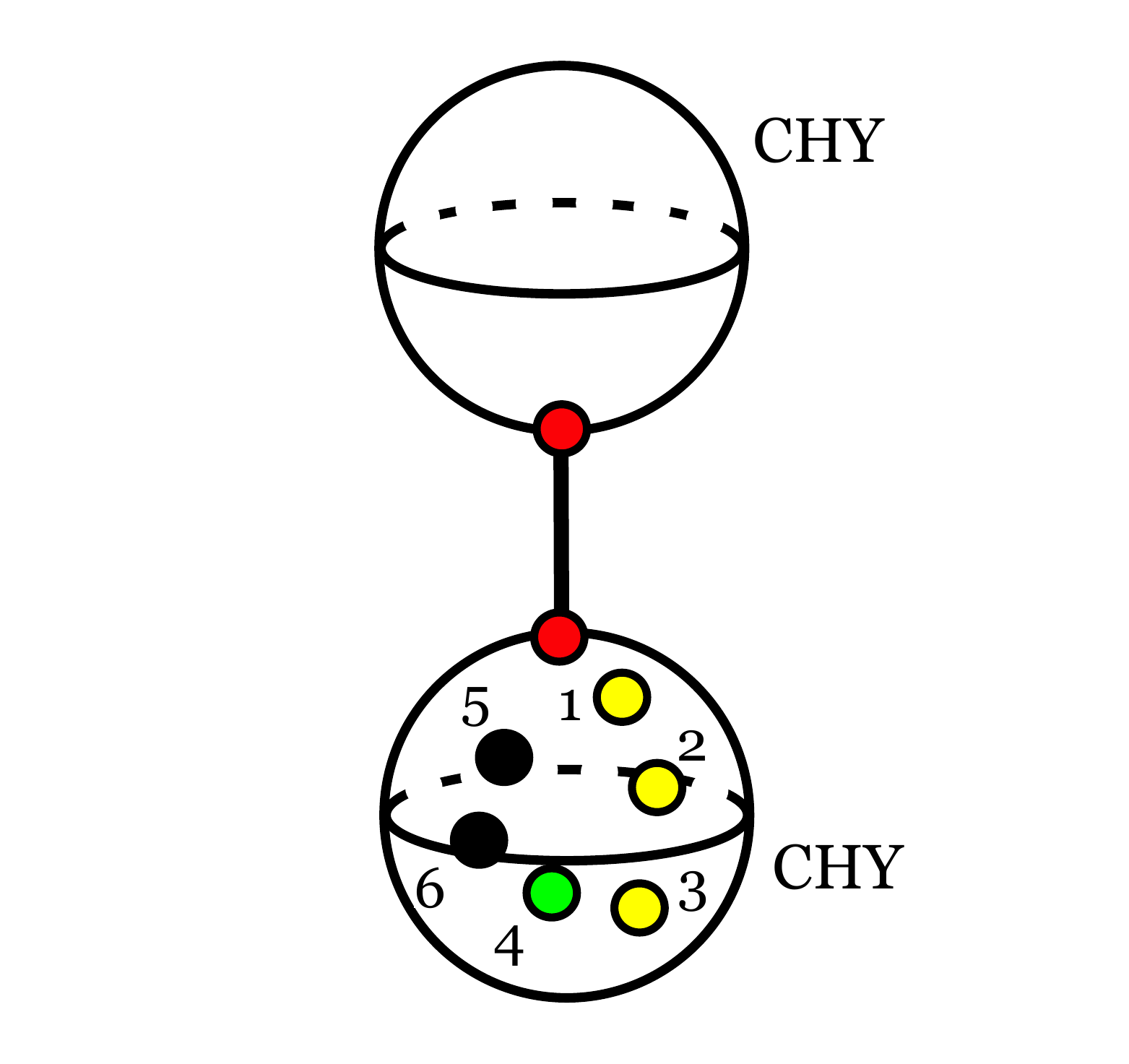}}
=0, 
\parbox[c]{6.1em}{\includegraphics[scale=0.21]{cut-4-apdx.pdf}}
{\underrightarrow{~~ \L\sim 0  ~~}}
\hspace{-0.8cm}
\parbox[c]{5.5em}{\includegraphics[scale=0.21]{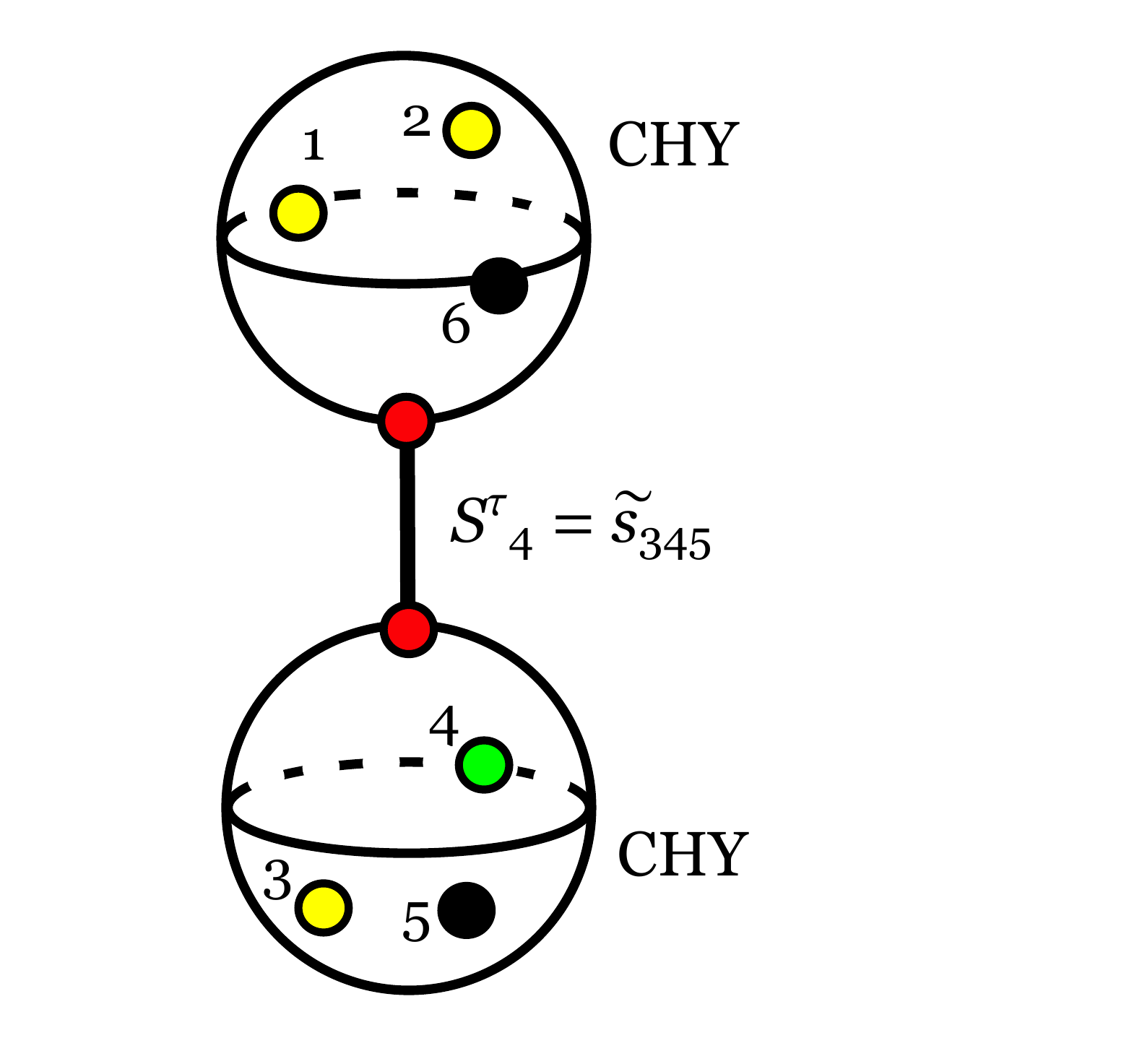}}
=
\begin{matrix}
 \text{Could be} \\ 
 \text{non-zero},
\end{matrix} 
~~~
\end{eqnarray}
}
\vskip-0.4cm\noindent
where we have introduced the red vertices to point out they are off-shell  particles ($\s_{[{\rm up}]}$ and $\s_{[{\rm down}]}$). Since over each sphere there is a single-cover prescription (see \eqref{fpL0}), then, three punctures must be fixed by
the ${\rm PSL}(2,\mathbb{C})$ redundancy. Thus, this is the reason why the first graphic in \eqref{cuts-app} vanish trivially (the ${\rm PSL}(2,\mathbb{C})$ symmetry has not been completely fixed on the upper sphere).
 
This analysis gives us the first integration rule \cite{Gomez:2016bmv}
\begin{itemize}
\item {\bf Rule-I}. {\it All configurations (or cuts) where there are less (or more) than two colored vertices (yellow or green) on each branch, vanish trivially} .
\end{itemize}
For simplicity, we represent a cut (or configuration) over a {\it YM-graph} by a dashed red line, which separates (encircles) the punctures localized on the upper (or lower) sheet.  For example,
\vspace{-0.8cm}
\begin{eqnarray}\label{Gen-examples}
\hspace{-0.4cm}
\parbox[c]{5.2em}{\includegraphics[scale=0.14]{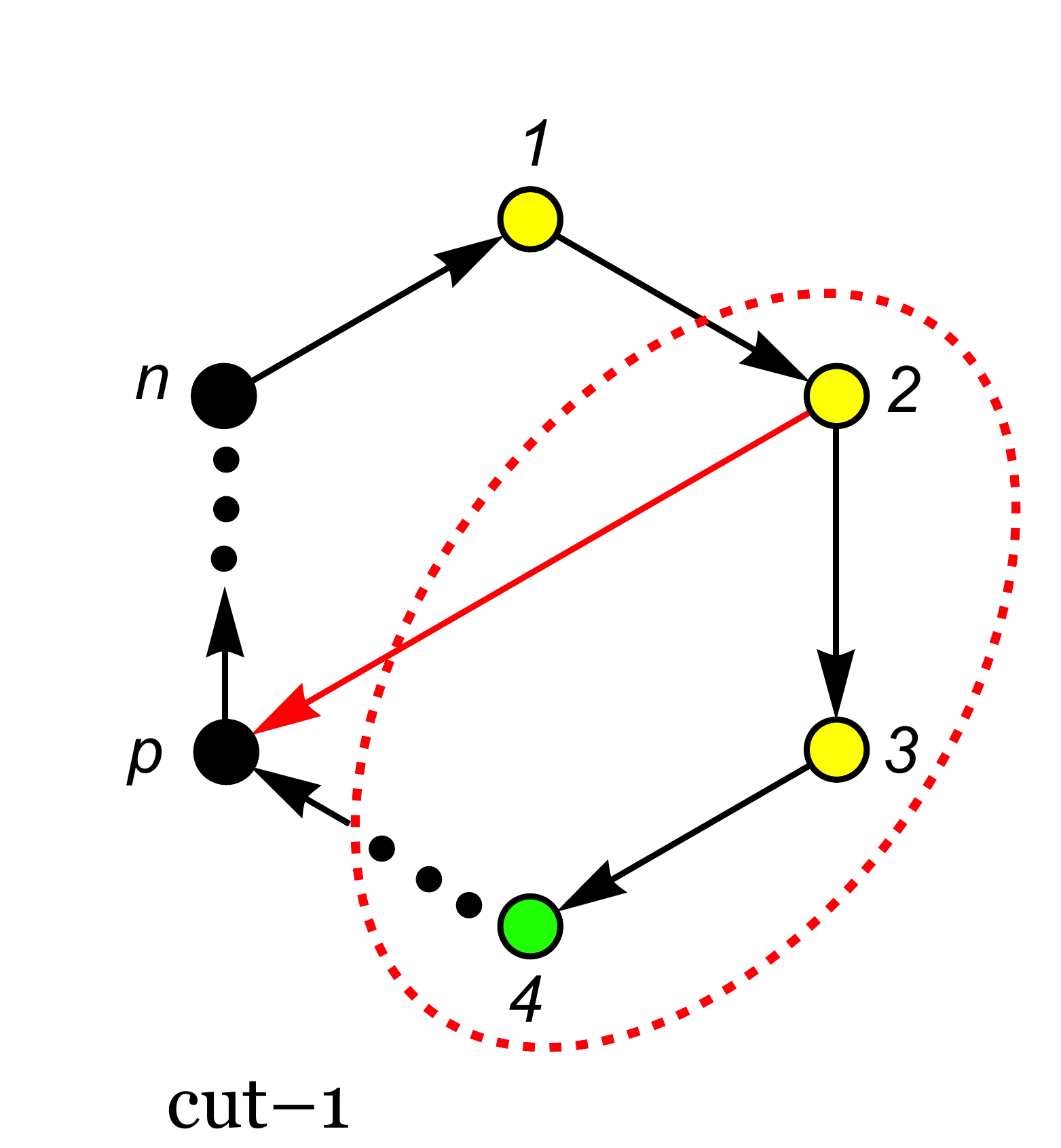}}
= \,0\,, \qquad 
\parbox[c]{6.0em}{\includegraphics[scale=0.14]{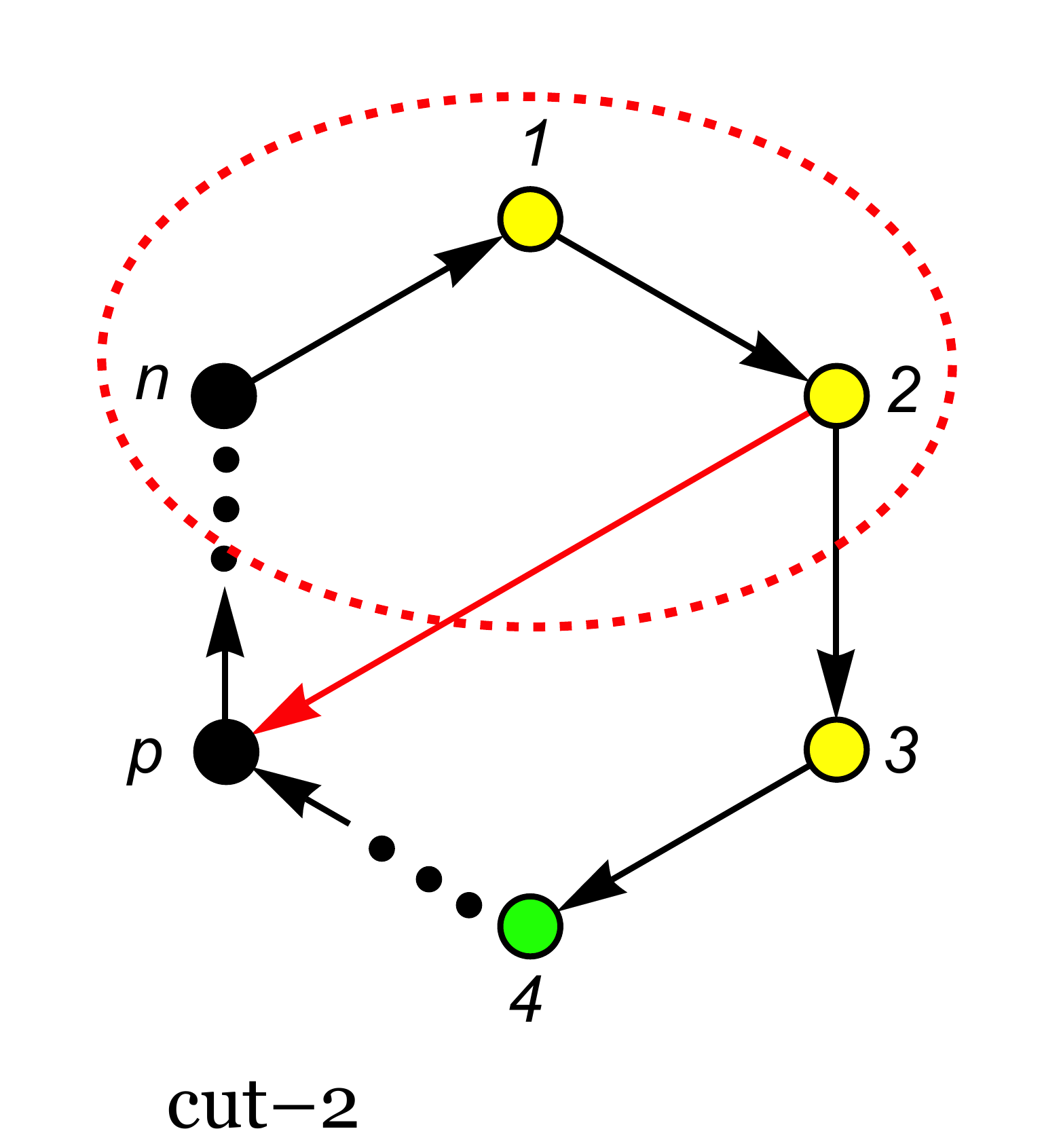}}
,
\end{eqnarray}
\vskip-0.2cm\noindent
where {\it cut-1} vanished trivially by {\bf rule-I}, but {\it cut-2} could give a non-zero contribution.

\subsection{Specific rules}\label{specificrules}

The {\bf rule-I} is a general rule which works for any integrand ${\cal I}^\L(\s,y)$. In this section, we are going to formulate some integration rules which depend on the integrand similar to those discussed in \cite{Gomez:2016bmv} for  $\phi^3$.

Following the {\bf rule-I}, it is enough to focus on all possible non-vanishing configurations. Expanding the YM integrand of each possible non-zero cut around $\L=0$, we notice that
{\small
\begin{eqnarray}\label{Lbehavior}
&&{\rm PT}^\tau_{(1,\ldots,n)} \times  {\bf Pf}^\prime \Psi^{\L}_n \sim  {\cal O}(\L^6),\,\quad\qquad\, \textit{The dashed red line cuts more than four arrows.}\, \nonumber \\
&&{\rm PT}^\tau_{(1,\ldots,n)} \times  {\bf Pf}^\prime \Psi^{\L}_n \sim  \L^4  + {\cal O}(\L^2),\quad\, \textit{The dashed red line cuts three or four arrows.}\,\,~~~ \nonumber \\
&&{\rm PT}^\tau_{(1,\ldots,n)} \times  {\bf Pf}^\prime \Psi^{\L}_n \sim  \L^2  + {\cal O}(\L^0),\quad\, \textit{The dashed red line cuts two arrows (singular cut).}\,\,~~~ \nonumber  \\
\end{eqnarray}
}
\vskip-0.1cm\noindent
Therefore, considering the expansion obtained in \eqref{fpL0}, the next rule is obvious
\begin{itemize}
\item {\bf Rule-II.} {\it If the dashed red line cuts less than three arrows over the YM-graph, the integrand must be expanded next to leading order. If the dashed red line cuts three or four arrows, the leading order expansion  is sufficient. Otherwise, the cut is zero.} 
\end{itemize}
This rule is equivalent to the $\L-$theorem given in \cite{Gomez:2016bmv}. 

Finally, the last rule we are going to formulate is only applied when the leading order expansion is sufficient, i.e. when the dashed red line cuts three or four arrows. 

First, let us consider the case when the dashed red line cuts three arrows
over a {\it YM-graph} (see {\it cut-2} in \eqref{Gen-examples}).  The only way to break the Pfaffian into two pieces is by introducing new polarization vectors, one for each new puncture  (one for $\s_{[{\rm up}]}$ and one for $\s_{[{\rm down}]}$) \cite{Cachazo:2013hca,Dolan:2013isa}.  For example\footnote{Here, we are assuming the dashed red line is cutting two arrows of the Parke-Taylor factor, i.e. ${\rm PT}^\tau_{(1,2,3,4)}\Big|^{1,2}_{3,4}   \sim  \L^{2}$. }, let us consider  the four point matrix, $(\Psi_4^\L)^{13}_{13}$, and let us expand its Pfaffian around $\L=0$ when the punctures, $(\s_1,\s_2)$, are on the upper sphere (while,  $(\s_3,\s_4)$, are on the lower one).  It is straightforward to check the leading order expansion
\vspace{-0.1cm}
{\small
\begin{eqnarray}
&&
{\rm Pf}\left[
(\Psi_4^\L)^{13}_{13}
\right]\Big|^{1,2}_{3,4} = \frac{\s_{4[1,2]}^2 \, \s_{3[1,2]}  }{2\, \L^2} \times \sum_r  \,\,
{\rm Pf}\left[
{\small
\begin{matrix}
0 & - \frac{\eps^r_{[3,4]}\cdot k_{2}}{\s_{[3,4]2}} & - \frac{\eps_{1}\cdot k_{2}}{\s_{12}} & - {\cal C}_{22}\\
 \frac{\eps^r_{[3,4]}\cdot k_{2}}{\s_{[3,4]2}} & 0 & \frac{\eps^r_{[3,4]}\cdot \eps_{1}}{\s_{[3,4]1}} & \frac{\eps^r_{[3,4]}\cdot \eps_{2}}{\s_{[3,4]2}}\\
  \frac{\eps_{1}\cdot k_{2}}{\s_{12}} &  \frac{\eps_{1}\cdot \eps^r_{[3,4]}}{\s_{1[3,4]}} &  0 & \frac{\eps_{1}\cdot \eps_{2}}{\s_{12}} \\
{\cal C}_{22} &   \frac{\eps_{2}\cdot \eps^r_{[3,4]}}{\s_{2[3,4]}} &  \frac{\eps_{2}\cdot \eps_{1}}{\s_{21}} &  0  \\
\end{matrix}}
\right] \times
 \nonumber \\
&&
{\rm Pf}\left[
{\small
\begin{matrix}
0 & - \frac{\eps^r_{[1,2]}\cdot k_{4}}{\s_{[1,2]4}} & - \frac{\eps_{3}\cdot k_{4}}{\s_{34}} & - {\cal C}_{44}\\
 \frac{\eps^r_{[1,2]}\cdot k_{4}}{\s_{[1,2]4}} & 0 & \frac{\eps^r_{[1,2]}\cdot \eps_{3}}{\s_{[1,2]3}} & \frac{\eps^r_{[1,2]}\cdot \eps_{4}}{\s_{[1,2]4}}\\
  \frac{\eps_{3}\cdot k_{4}}{\s_{34}} &  \frac{\eps_{3}\cdot \eps^r_{[1,2]}}{\s_{3[1,2]}} &  0 & \frac{\eps_{3}\cdot \eps_{4}}{\s_{34}} \\
{\cal C}_{44} &   \frac{\eps_{4}\cdot \eps^r_{[1,2]}}{\s_{4[1,2]}} &  \frac{\eps_{4}\cdot \eps_{3}}{\s_{43}} &  0  \\
\end{matrix}}
\right]  + {\cal O}(\L^0)
= \frac{\s_{4[1,2]}^2 \, \s_{3[1,2]}  }{2\, \L^2} \, \sum_r
{\rm Pf}\left[
\left(\Psi_3 \right)^{[3,4]1}_{[3,4]1}
\right] \, 
{\rm Pf}\left[
\left( \Psi_3 \right)^{[1,2]3}_{[1,2]3}
\right]  
 \nonumber\\
 \label{pfexpansion}
\end{eqnarray}
}
\vskip-0.7cm\noindent
where\footnote{It is useful to recall that, $\eps_{2}\cdot k_{[3,4]} = -\eps_{2}\cdot k_{1}$ and $\eps_{4}\cdot k_{[1,2]} = -\eps_{4}\cdot k_{3}$.}, ${\cal C}_{22} =-\left( \frac{\eps_2\cdot k_1}{\s_{21}} + \frac{\eps_2\cdot k_{[3,4]}}{\s_{2[3,4]}}   \right) $, ${\cal C}_{44} =-\left( \frac{\eps_4\cdot k_3}{\s_{43}} + \frac{\eps_4\cdot k_{[1,2]}}{\s_{4[1,2]}}   \right) $, $\s_{[1,2]}= \s_{[3,4]}=0 $, and 
 the new polarization vectors, $(\eps_{[3,4]}^{r,\mu}$,  $\eps_{[1,2]}^{r,\mu})$, must satisfy the identity, $\sum_r\eps_{[3,4]}^{r,\mu} \, \eps_{[1,2]}^{r,\nu}=\eta^{\mu \nu}$. The same phenomenon is observed to higher number of points. Thus, we have one more rule, 
\begin{itemize}
\item {\bf Rule-IIIa.} {\it If the dashed red line cut three arrows over a YM-graph, there is an off-shell vector field (gluon) propagating among the two resulting graphs ({\bf standard-cut}).  These two resulting graphs must be glued by the identity, $\sum_{r}\epsilon^{r, \mu}_{[{\rm up}]} \, \epsilon_{[{\rm down}]}^{r,\nu}= \eta^{\mu \nu}$.} 
\end{itemize}

On the other hand, when the dashed red line cuts four arrows, the Pfaffian (${\rm Pf} \left[(\Psi^\L_n)^{ij}_{ij}\right]$)  breaks spontaneously into two pieces. For instance\footnote{Here, we are assuming the dashed red line is cutting four arrows of the Parke-Taylor factor, i.e. ${\rm PT}^\tau_{(1,2,3,4)}\Big|^{1,2}_{3,4}   \sim  \L^{4}$.}, 
let us consider again the matrix, $(\Psi_4^\L)^{13}_{13}$, and let us expand its Pfaffian around $\L=0$, but now, when the punctures, $(\s_1,\s_3)$, are on the upper sphere (and  $(\s_2,\s_4)$, are on the lower one). It is simple to show that the leading order contribution is given by
\vspace{-0.2cm}
{\small
\begin{eqnarray}
&&{\rm Pf}\left[
(\Psi_4^\L)^{13}_{13}
\right]\Big|^{1,3}_{2,4} = 
- \frac{ 2\, \s_{4[1,3]}^2 \, \s_{2[1,3]}^2  }{\L^4} \times
\frac{(\eps_1 \cdot \eps_3)}{\s_{13}}
 \times 
\frac{(\eps_2 \cdot \eps_4) \, (k_2\cdot  k_4)}{\s_{24}^2} + {\cal O}(\L^{-2}) 
\nonumber\\
 &&= 
- \frac{ 2\, \s_{4[1,3]}^2 \, \s_{2[1,3]}^2  }{\L^4} \times
{\rm Pf}\left[
{\small
\begin{matrix}
0 &  \frac{\eps_{1}\cdot \eps_{3}}{\s_{13}}  \\
\frac{\eps_{3}\cdot \eps_{1}}{\s_{31}} &  0  \\
\end{matrix}}
\right] 
\times {\rm Pf}\left[
{\small
\begin{matrix}
 0 &    \frac{ k_2\cdot k_4  }{\s_{24}}  & - {\cal C}_{22} & -{\eps_4\cdot k_2 \over \s_{42} }  \\
 \frac{ k_4\cdot k_2 }{\s_{42}} & 0 &- {\eps_2\cdot k_4 \over \s_{24} } & - {\cal C}_{44}  \\
{\cal C}_{22} & {\eps_2\cdot k_4 \over \s_{24} } & 0 & \frac{\eps_2\cdot \eps_4}{\s_{24}}  \\
{\eps_4\cdot k_2 \over \s_{42} } & {\cal C}_{44} &  \frac{\eps_4\cdot \eps_2}{\s_{42}} &0  \\
\end{matrix}}
\right] 
+ {\cal O}(\L^{-2})\, ,   \quad
 \label{pfexpansion2}
\end{eqnarray}
}
\vskip-0.1cm\noindent
where, ${\cal C}_{22}=-\left( \frac{\eps_2\cdot k_4}{\s_{24}} +\frac{\eps_2\cdot k_{[1,3]}}{\s_{2[1,3]}} \right)$, ${\cal C}_{44}=-\left( \frac{\eps_4\cdot k_2}{\s_{42}} +\frac{\eps_4\cdot k_{[1,3]}}{\s_{4[1,3]}} \right)$ and $\s_{[2,4]}=\s_{[1,3]}=0$. The same behavior can be checked at higher number of points. Finally, notice that the matrices in  \eqref{pfexpansion2} do not have any rows/columns associated with the new punctures, $\s_{[2,4]}$ and $\s_{[1,3]}$. So, we have the last integration rule
\begin{itemize}
\item {\bf Rule-IIIb.} {\it When the dashed red line cut four arrows, the YM-graph breaks spontaneously into two resulting graphs, which are written in the single-cover language (times a propagator given by \eqref{fpL0}). 
All rows/columns related to the new resulting vertices  (punctures with four arrows) must be removed from the resulting  matrices.}
\end{itemize}
We call to this type of cut a {\bf strange-cut}, since it produces spurious poles, such as we will show in the next section.

In general, a puncture with four arrows represents a scalar particle  \cite{Gomez:2016bmv} (we will come back to this point later). Nevertheless, in pure Yang-Mills a puncture with four arrows can be interpreted as a longitudinal gluon, it will be explained in detail later. Therefore,  this means that two new punctures, $\s_{[{\rm up}]}$ and $\s_{[{\rm down}]}$,  are longitudinal off-shell gluons when the dashed red line cuts four arrows.

We would like to draw attention to the importance of the green vertex. This is differentiated since its scattering equation is responsible for generating the propagator of a given cut, via equation \eqref{fpL0}.

As a final observation, in order to formulate a well-defined method, we remark that the {\bf integration rules} obtained in this section are independent of the embedding. The only thing that one must keep in mind is the following additional rule
\begin{itemize}
\item {\bf Rule-IV.} {\it The number of intersection points among the dashed red-line and the arrows is given mod 2.}
\end{itemize}
This means that  when the dashed red-line cuts an even number of times an arrow, it is always possible to find an embedding such that the dashed red-line does not cut with that arrow. In a similar way, when the dashed red-line cuts an odd number of times an arrow, it is always possible to find an embedding such that the dashed red-line cuts just one time with that arrow.

\section{Three-point building-block}\label{sectionBB}

Before giving simple examples, it is going to be useful to introduce the three-point functions that will work as building blocks., Additionally, in appendix \ref{notation} we give a small glossary in order to remember the notation.

The first three-point function, which is important to remark its normalization, is the  biadjoint $\phi^3$ 
computation
\vspace{-0.7cm}
\begin{eqnarray}\label{3pt-Nor}
\int d\mu_3^{\rm CHY} \times (\s_{[a][b]}\, \s_{[b][c]}\, \s_{[c][a]})^2 \times {\rm PT}_{([a],[b],[c])}^2
=
\int d \mu_3^{\rm CHY}
\hspace{-0.7cm}
\parbox[c]{5.1em}{\includegraphics[scale=0.17]{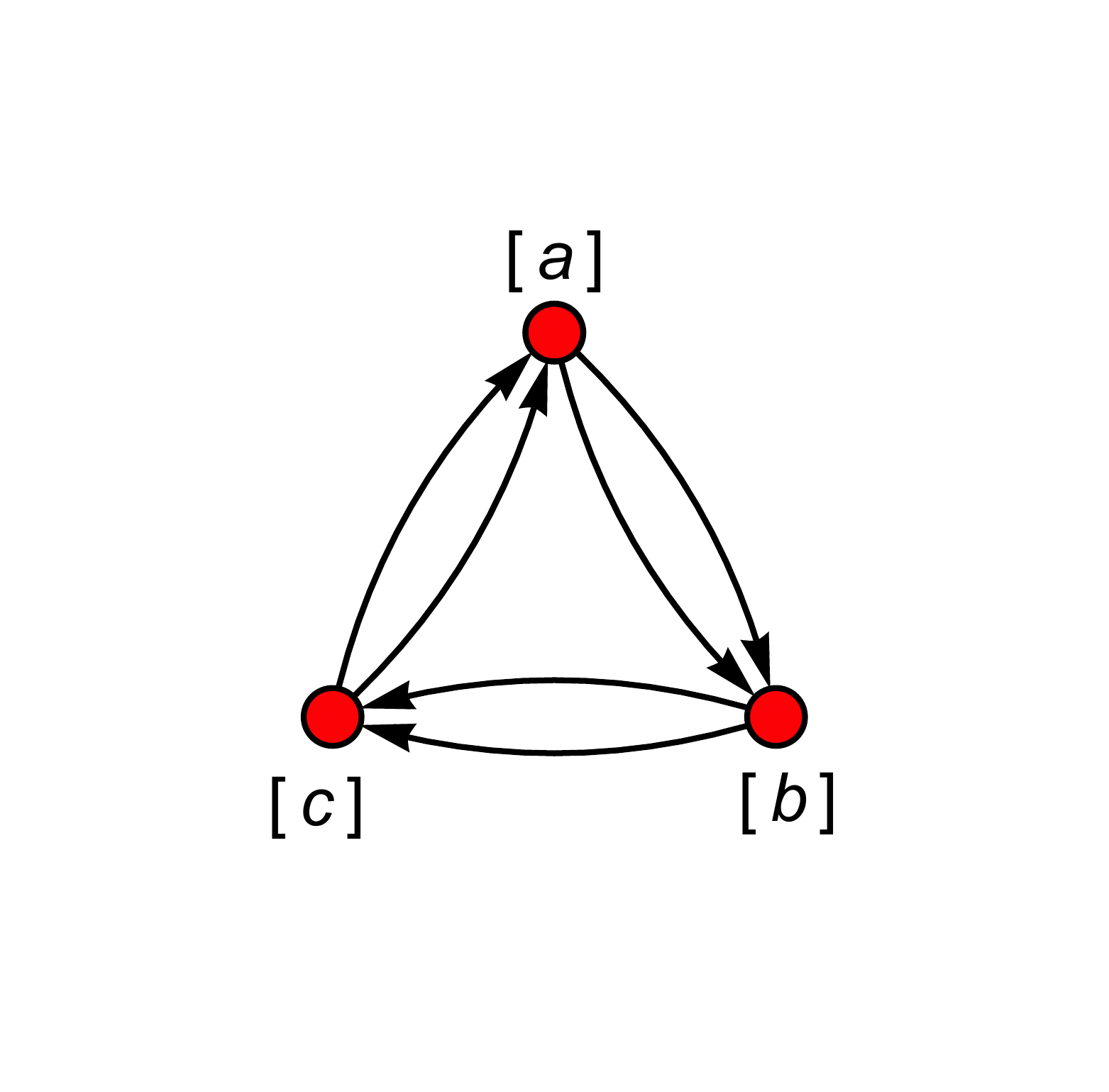}}
= \,1\, ,
\end{eqnarray}
\vskip-0.5cm\noindent
where the punctures are off-shell, $k^2_{[i]}\neq 0$ and $k_{[a]}+k_{[b]}+k_{[c]}=0$. 

It is obvious to note that any three-point computation is just algebraic, i.e. there is no an integral  (its integration  measure is trivial,  $d\mu_3^{\rm CHY}=1$). Therefore, for the rest of the paper, we will always omit the symbol $\int d\mu_3^{\rm CHY}$.

Additionally to the $\phi^3$ normalization, the off-shell ($k_{[i]}^2\neq 0$ and $k_{[a]}+k_{[b]}+k_{[c]}=0$) three-point building-block for the Yang-Mills amplitudes is given by the expression  
\vspace{-0.3cm}
\begin{eqnarray}\label{YMabc}
{ A}_3^{([a],[b])}([a],[b],[c])  = 
\hspace{-0.4cm}
\parbox[c]{5.3em}{\includegraphics[scale=0.19]{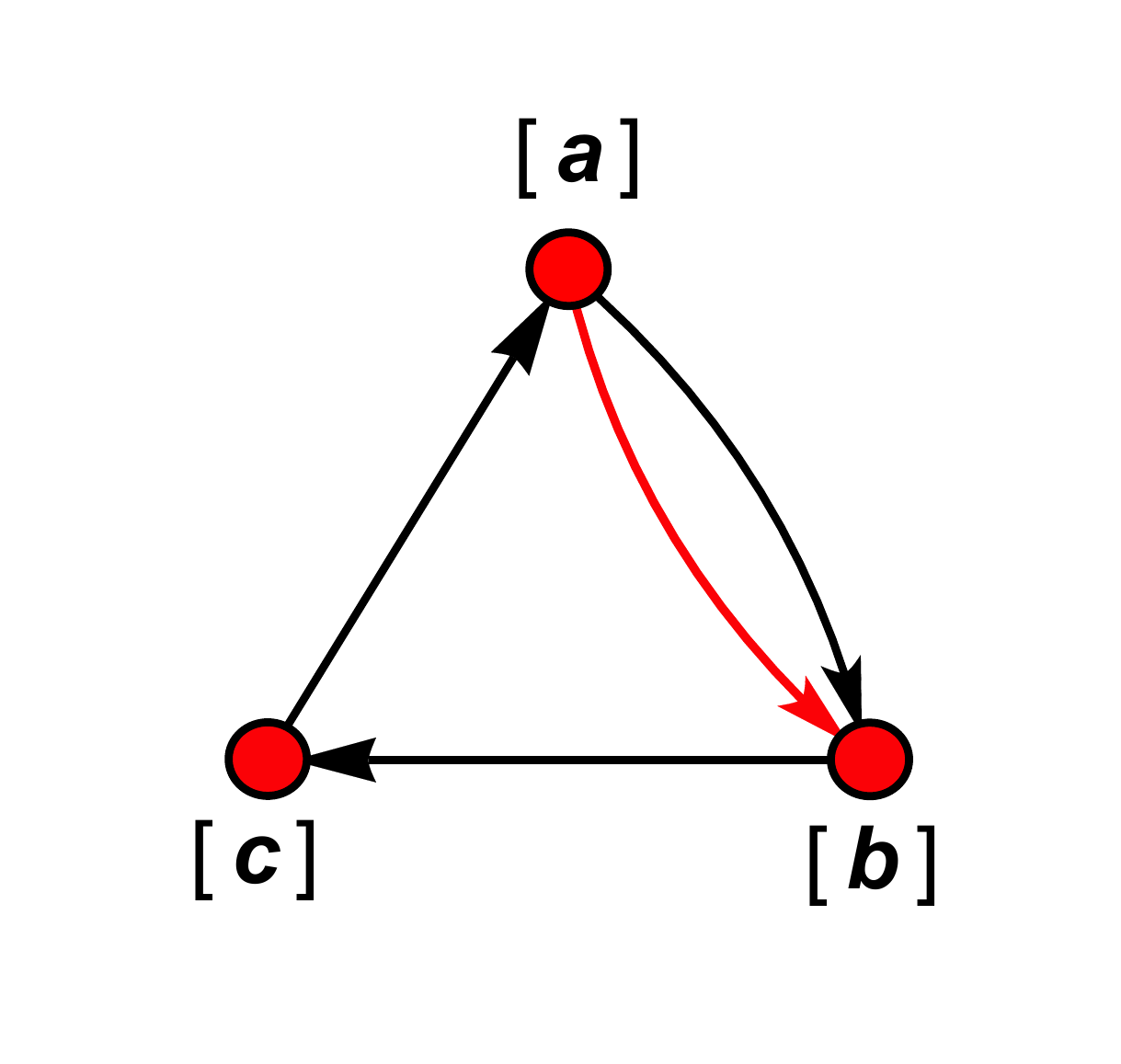}}
=
(\s_{[a][b]}\,\s_{[b][c]}\,\s_{[c][a]})^2\times {\rm PT}_{([a],[b],[c])}
\nonumber
\end{eqnarray}
\vskip-0.9cm\noindent
\begin{eqnarray}\label{YM-BB}
&&
\hspace{0.9cm}
\times
\frac{(-1)}{\s_{[a][b]}}
\times
{\rm Pf}\left[
{\small
\begin{matrix}
0 & - \frac{\eps_{[a]}\cdot k_{[c]}}{\s_{[a][c]}} & - \frac{\eps_{[b]}\cdot k_{[c]}}{\s_{[b][c]}} & - {\cal C}_{[c][c]}\\
 \frac{\eps_{[a]}\cdot k_{[c]}}{\s_{[a][c]}} & 0 & \frac{\eps_{[a]}\cdot \eps_{[b]}}{\s_{[a][b]}} & \frac{\eps_{[a]}\cdot \eps_{[c]}}{\s_{[a][c]}}\\
  \frac{\eps_{[b]}\cdot k_{[c]}}{\s_{[b][c]}} &  \frac{\eps_{[b]}\cdot \eps_{[a]}}{\s_{[b][a]}} &  0 & \frac{\eps_{[b]}\cdot \eps_{[c]}}{\s_{[b][c]}} \\
{\cal C}_{[c][c]} &   \frac{\eps_{[c]}\cdot \eps_{[a]}}{\s_{[c][a]}} &  \frac{\eps_{[c]}\cdot \eps_{[b]}}{\s_{[c][b]}} &  0  \\
\end{matrix}}
\right] 
\nonumber 
\end{eqnarray}
\vskip-0.3cm\noindent
\begin{eqnarray}\label{YM-BB}
&&
\hspace{-0.1cm}
=  (\eps_{[a]}\cdot \eps_{[b]}) ( \eps_{[c]} \cdot k_{[a]} ) - (\eps_{[b]}\cdot \eps_{[c]}) ( \eps_{[a]} \cdot k_{[c]} )
+  (\eps_{[c]}\cdot \eps_{[a]}) ( \eps_{[b]} \cdot k_{[c]} )  \,\,
\nonumber\\
&&
\hspace{-0.1cm}
=  \eps_{[a]}^\mu \eps_{[b]}^\nu \eps_{[c]}^\rho \left\{ \frac{1}{2} \left[
\eta_{\mu\nu} ( k_{[a]} -  k_{[b]} )_{\rho} +\eta_{\nu\rho} ( k_{[b]} - k_{[c]} )_{\mu}
+  \eta_{\rho\mu} ( k_{[c]} - k_{[a]} )_{\nu}  
\right]
\right\}
\,\,
\nonumber\\
&&
+ \,   \frac{1}{2} \left[\,
(  \eps_{[a]} \cdot k_{[a]}  )\, (\eps_{[b]} \cdot \eps_{[c]}) - ( \eps_{[b]} \cdot k_{[b]})  \, (\eps_{[a]} \cdot \eps_{[c]}) \,
\right] 
\end{eqnarray}
where ${\small {\cal C}_{[c][c]}=-\left(  \frac{\eps_{[c]}\cdot k_{[a]}}{\s_{[c][a]}} +\frac{\eps_{[c]}\cdot k_{[b]}}{\s_{[c][b]}}  \right)}$ and $\eps_{[c]}\cdot k_{[c]}=0$. Notice that the transverse constraint, $\eps_{[c]}\cdot k_{[c]}=0$, is a necessary  and sufficient condition to obtain an expression independent of $\s_{[i]}$'s (the ${\rm PSL}(2,\mathbb{C})$ symmetry). In addition, the polarization vectors, $\eps_{[a]}$ and $\eps_{[b]}$,  are not necessarily transverse, i.e. $\eps_{[a]}\cdot k_{[a]}\neq 0$, $\eps_{[b]}\cdot k_{[b]}\neq 0$. This is an important fact since we will need to apply the {\bf rule-IIIa}  to glue {\it YM-graphs} ($\sum_r \eps^{r,\mu}_{\rm [up]}\, \eps^{r,\nu}_{\rm [up]} = \eta^{\mu\nu}$).

Clearly, the three-point building block obtained in \eqref{YM-BB} is not the three-point Feynman vertex (it has a correction which depends on the transversality of $\epsilon_{[a]}$ and  $\epsilon_{[b]}$). Therefore, this means that the integration rules proposed in this paper are neither the Berends-Giele method nor
the usual Feynman rules.

It is trivial to see that under the transversality conditions, $\epsilon_{[a]}\cdot k_{[a]}=0$ and $\epsilon_{[b]}\cdot k_{[b]}=0$,  \eqref{YM-BB} turns into the very well known three-point amplitude (the three-point Feynman vertex), $A_3^{([a],[b])}([a],[b],[c])= (\eps_{[a]}\cdot \eps_{[b]}) ( \eps_{[c]} \cdot k_{[a]} ) + (\eps_{[b]}\cdot \eps_{[c]}) ( \eps_{[a]} \cdot k_{[b]} )+(\eps_{[c]}\cdot \eps_{[a]}) ( \eps_{[b]} \cdot k_{[c]} )$. 

Finally, notice that although we have chosen a particular gauge in \eqref{YM-BB}, i.e. $(i,j) = ([a],[b])$, by the properties in \eqref{properties} one can always carry any off-shell three-point amplitude to the form, ${ A}_3^{([a],[b])}([a],[b],[c])$.

\section{Simple examples}\label{sectionEXS}

In this section, we show simple examples to understand the DC integration rules. 
First, we start with the simple amplitude, $A^{\rm YM}(1,2,3,4)$. Next,  we schematize the five-point computation in order to introduce new concepts. The Yang-Mills amplitude at five-point will be computed explicitly in section \ref{sectionEX5P}. 

Before going explicitly to the computations, it is useful to understand which vertices are fixed after using the 
{\bf integration rules}, i.e. over the resulting graphs. By the {\bf rule-I}, a resulting graph inherits two fixed punctures from the gauge-fixing set, $\{ p,q,r,m \}$.  Additionally, as it was explained in section \ref{section-IntR},  the two new emerging punctures, $(\s_{[\rm up]}, \s_{[\rm down]})$, are also fixed, therefore, we can conclude that over a resulting the three fixed-vertices (since those graphs are in the single-cover representation) are given by the set
{\small
\begin{eqnarray}\label{fixed-punctures}
\hspace{-0.3cm}
\{\textbf{Fixed  punctures} \} =\left( \{\textbf{All  punctures  in the graph}\} \cap \{p,q,r,m \} \right) \cup\{\textbf{off-shell punctures}\}. \nonumber\\
\end{eqnarray}
}
\vskip-0.7cm\noindent
It is important to always keep this expression in mind, because our algorithm depends of the gauge fixing. 

\subsection{Four-point}\label{four-point}

First, we set the gauge fixing, $(pqr | m)=(123 | 4)$. So,
in order to avoid singular configurations (see \eqref{Lbehavior}),  
we choose the red arrow to join the vertices, $(i,j)=(1,3)$. 
Applying the {\bf rule-I} one has
\vspace{-0.45cm}
\begin{eqnarray}\label{example-4p}
A_4^{(1,3)}(1,2,3,4) =
\int d\mu_4^{\L}
\hspace{-0.55cm}
\parbox[c]{5.4em}{\includegraphics[scale=0.18]{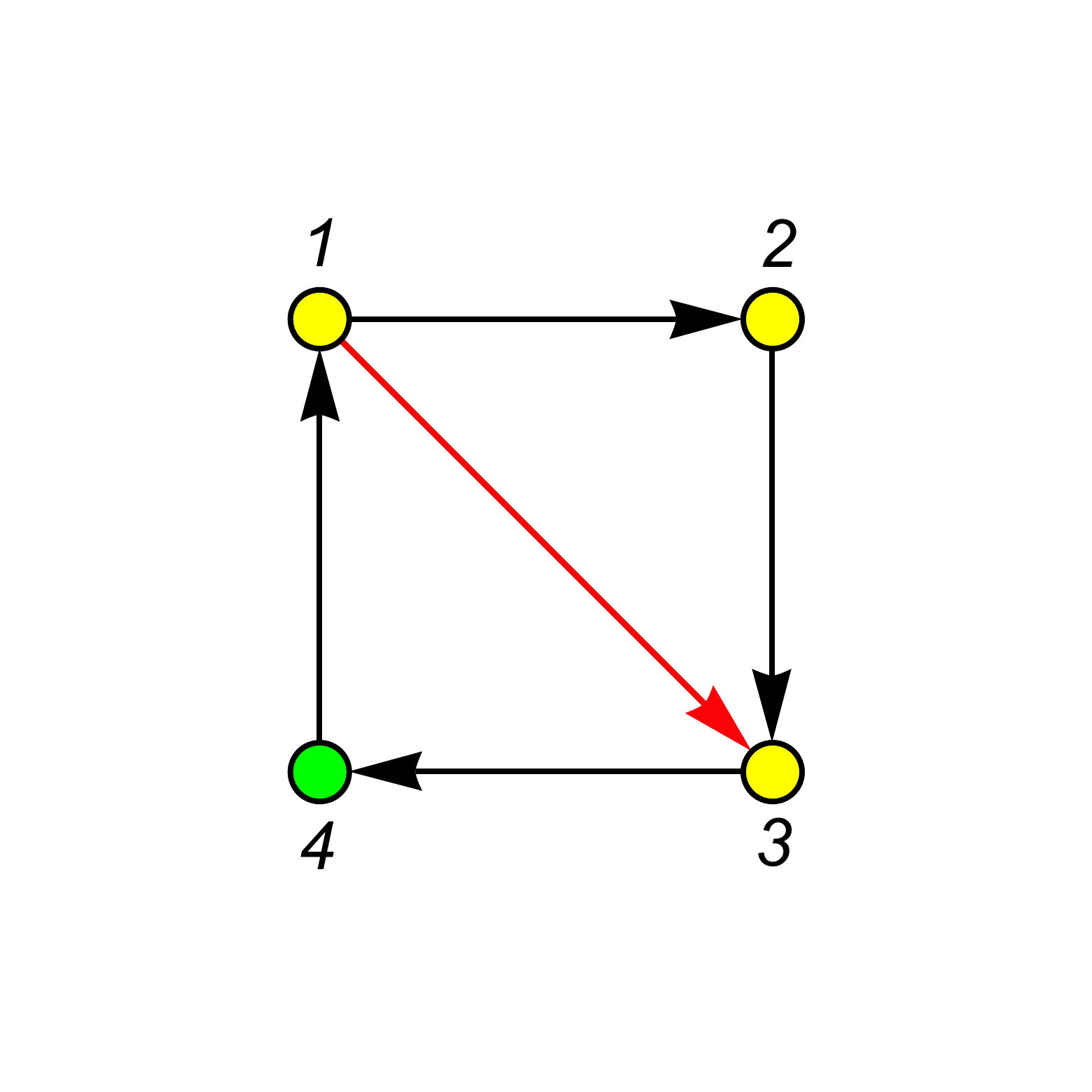}}
= 
\hspace{-0.6cm}
\parbox[c]{5.4em}{\includegraphics[scale=0.18]{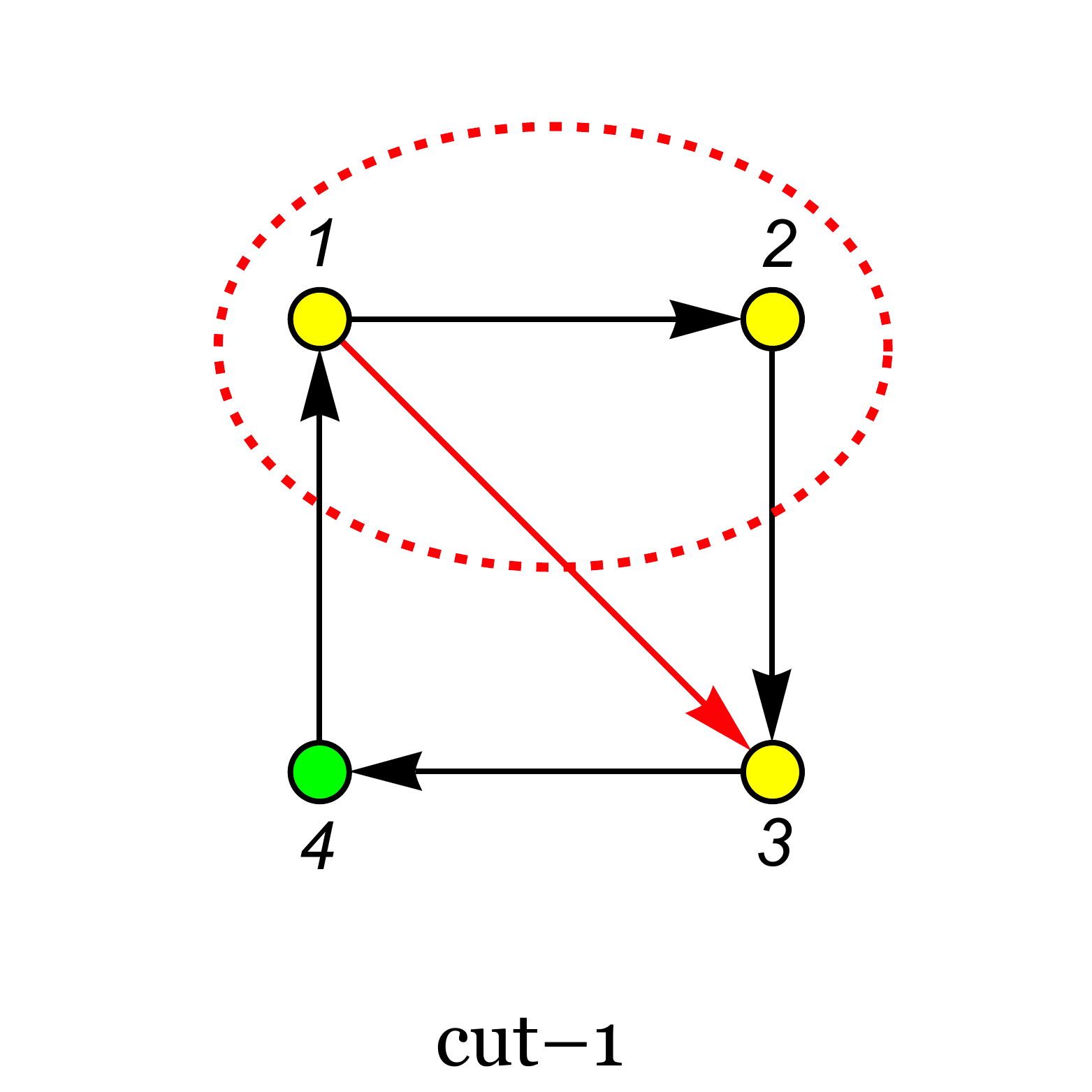}} +
\hspace{-0.6cm}
\parbox[c]{6.2em}{\includegraphics[scale=0.18]{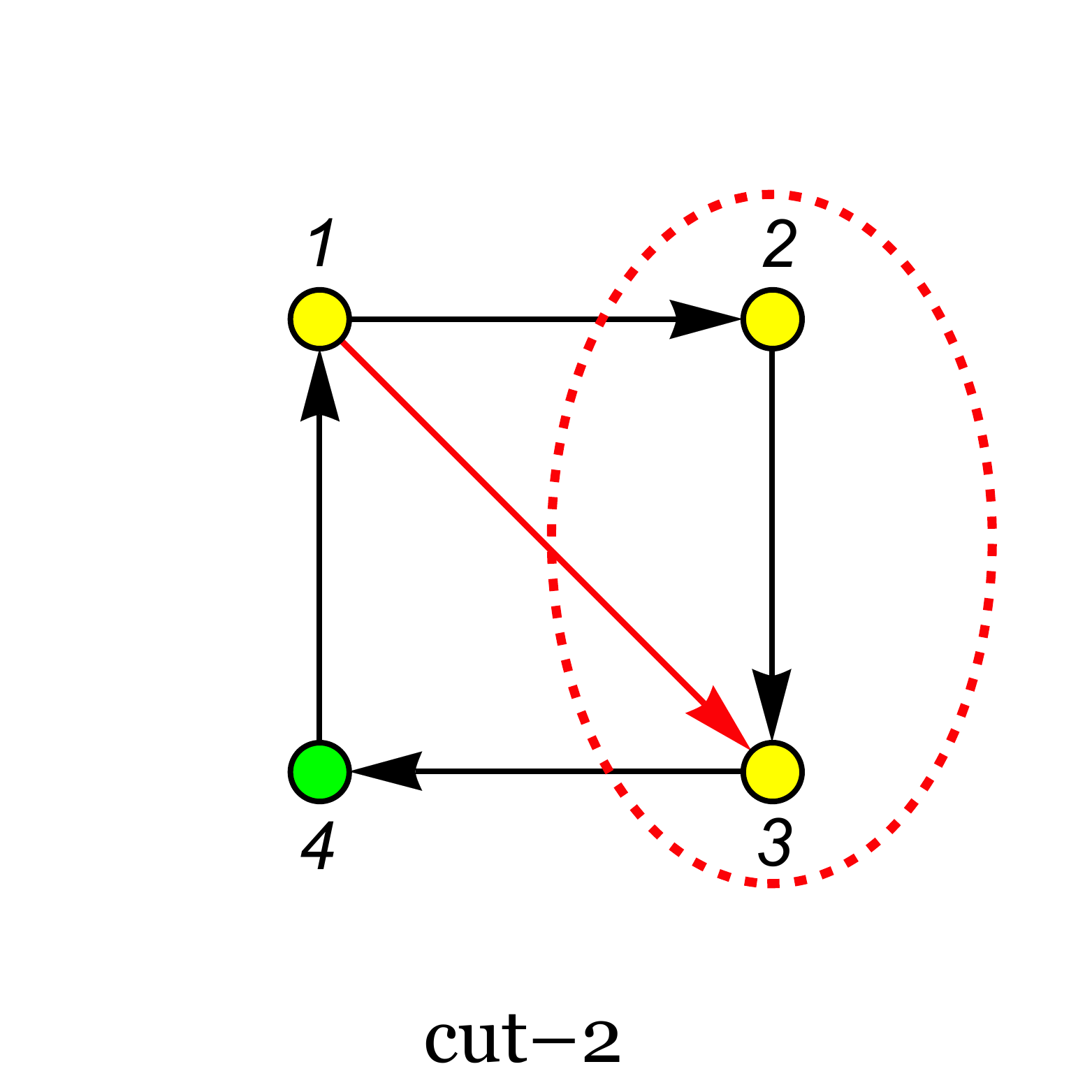}} +
\hspace{-0.6cm}
\parbox[c]{5.8em}{\includegraphics[scale=0.18]{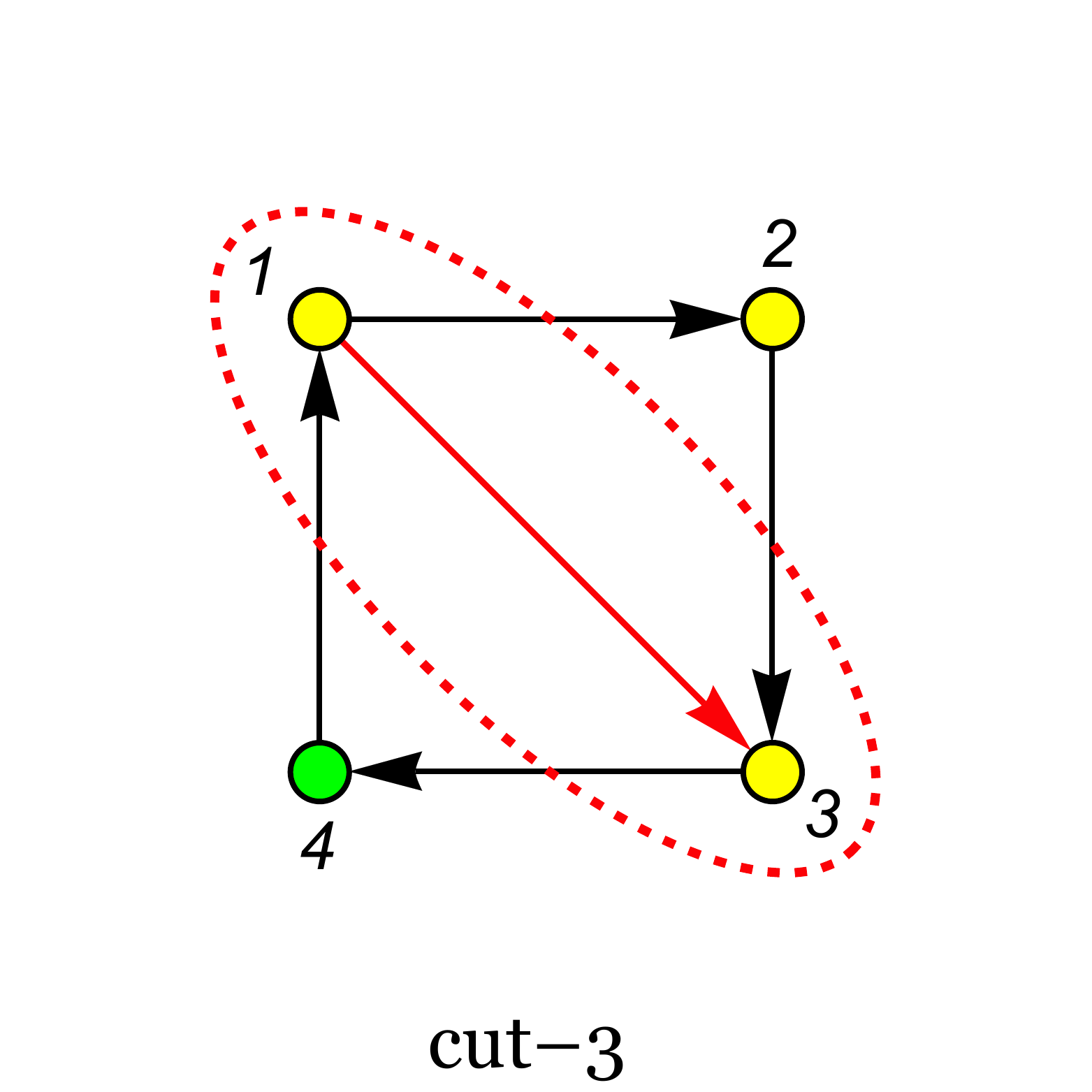}}  \,\, .
\end{eqnarray}
\vskip-0.3cm\noindent
Using the {\bf rules-II, III} and the expansion in \eqref{fpL0}, the {\it cuts} become
\vspace{-0.4cm}
{\small
\begin{eqnarray}\label{Rcut1}
\hspace{-0.4cm}
\parbox[c]{5.0em}{\includegraphics[scale=0.15]{4pt-cut1.pdf}}
&= &
\sum_r
\hspace{-0.5cm}
\parbox[c]{4.8em}{\includegraphics[scale=0.15]{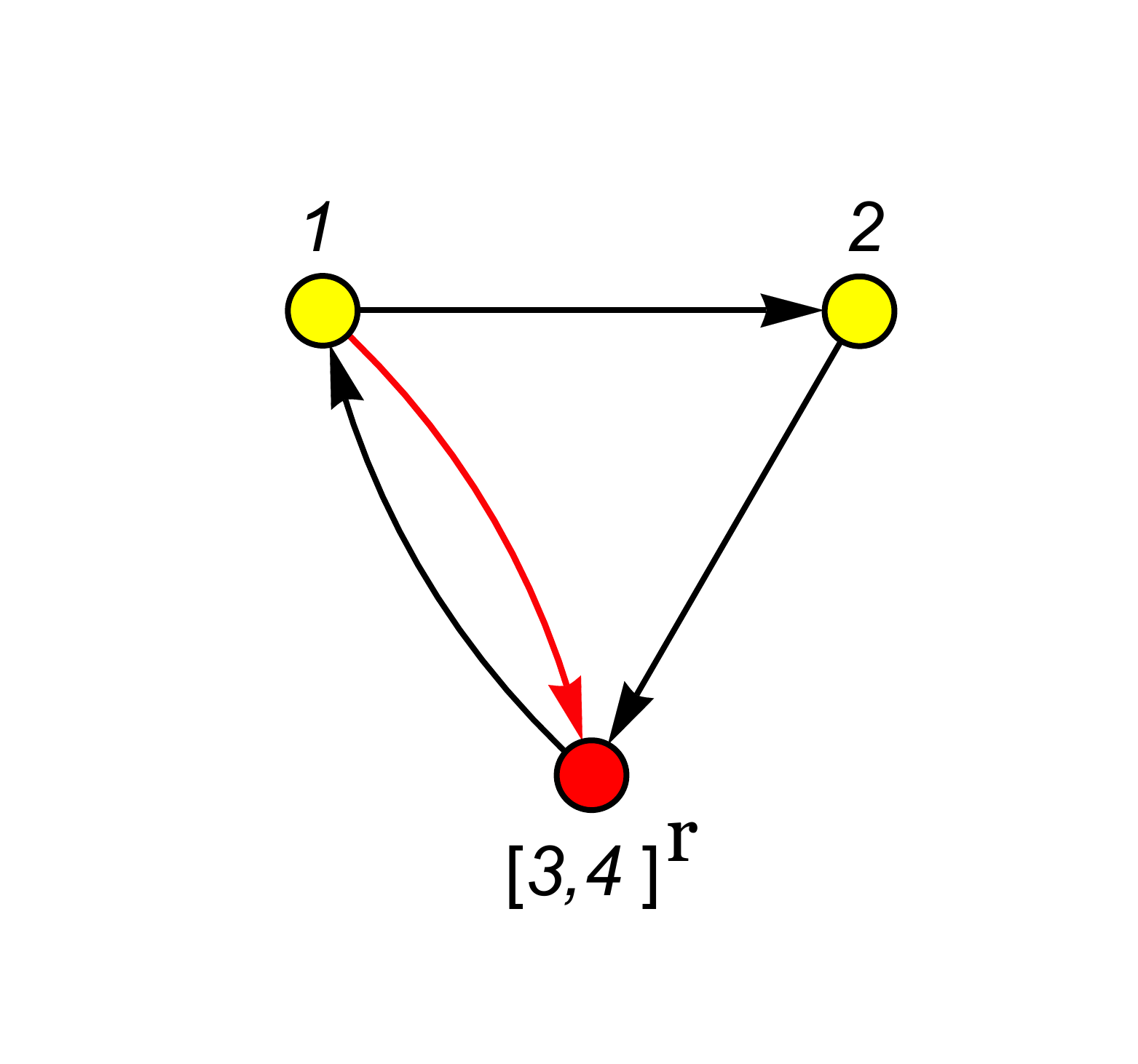}} 
\times  \left(
\frac{1}{\tilde s_{34}} \right)
\times
\hspace{-0.7cm}
\parbox[c]{5.1em}{\includegraphics[scale=0.15]{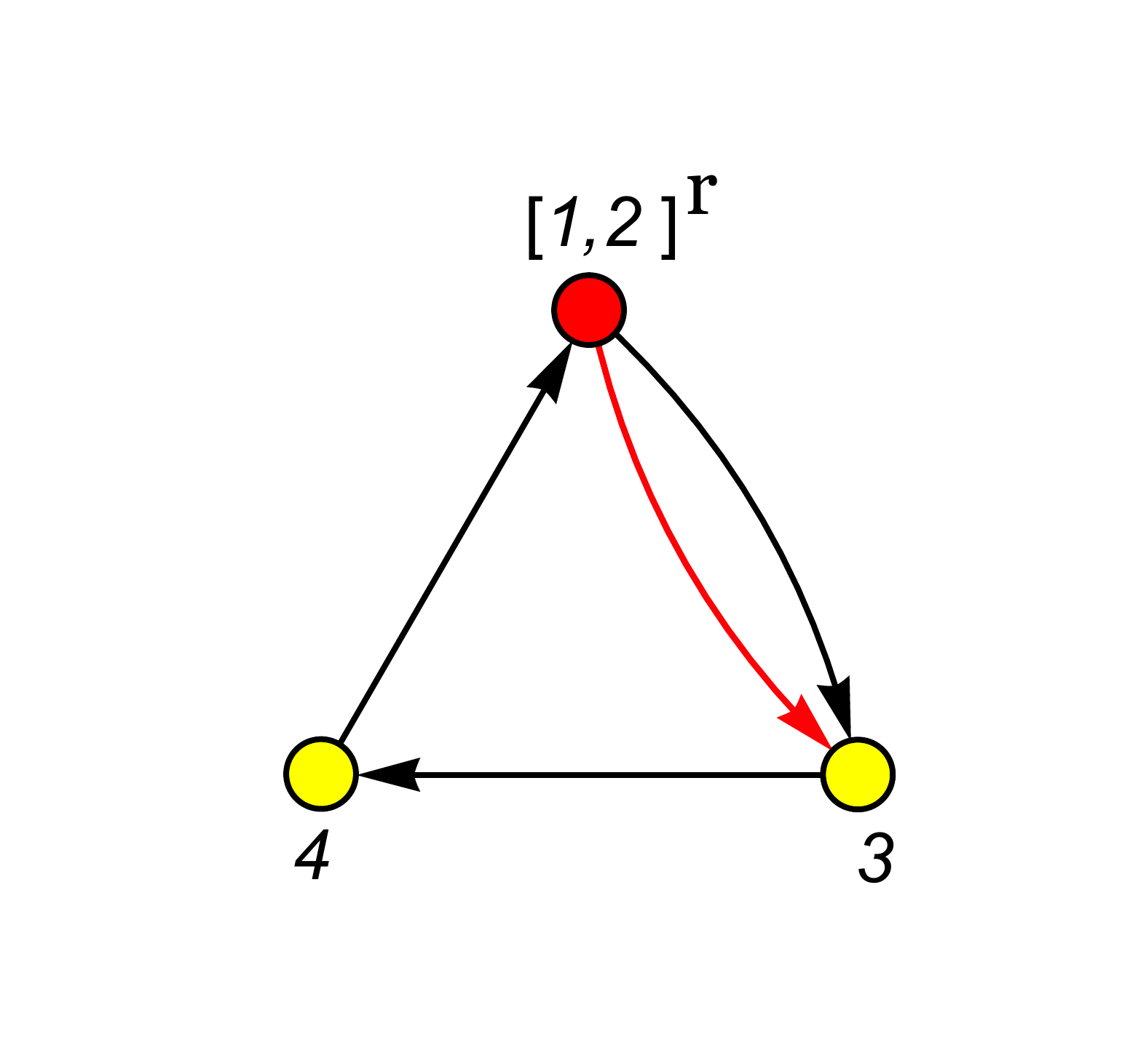}} =
 \frac{ \sum_r \, 
A_3^{([3,4],1)}([3,4]^r,1, 2 ) \times A_3^{([1,2],3)}([1,2]^r,3,4)    }{\tilde s_{34}}   \, , \nonumber
\end{eqnarray}
}
\vskip-0.7cm\noindent
\vspace{-0.9cm}
{\small
\begin{eqnarray}\label{Rcut2}
\hspace{-0.35cm}
\parbox[c]{5.5em}{\includegraphics[scale=0.15]{4pt-cut2.pdf}}
&=& 
\sum_r
\hspace{-0.6cm}
\parbox[c]{5.6em}{\includegraphics[scale=0.15]{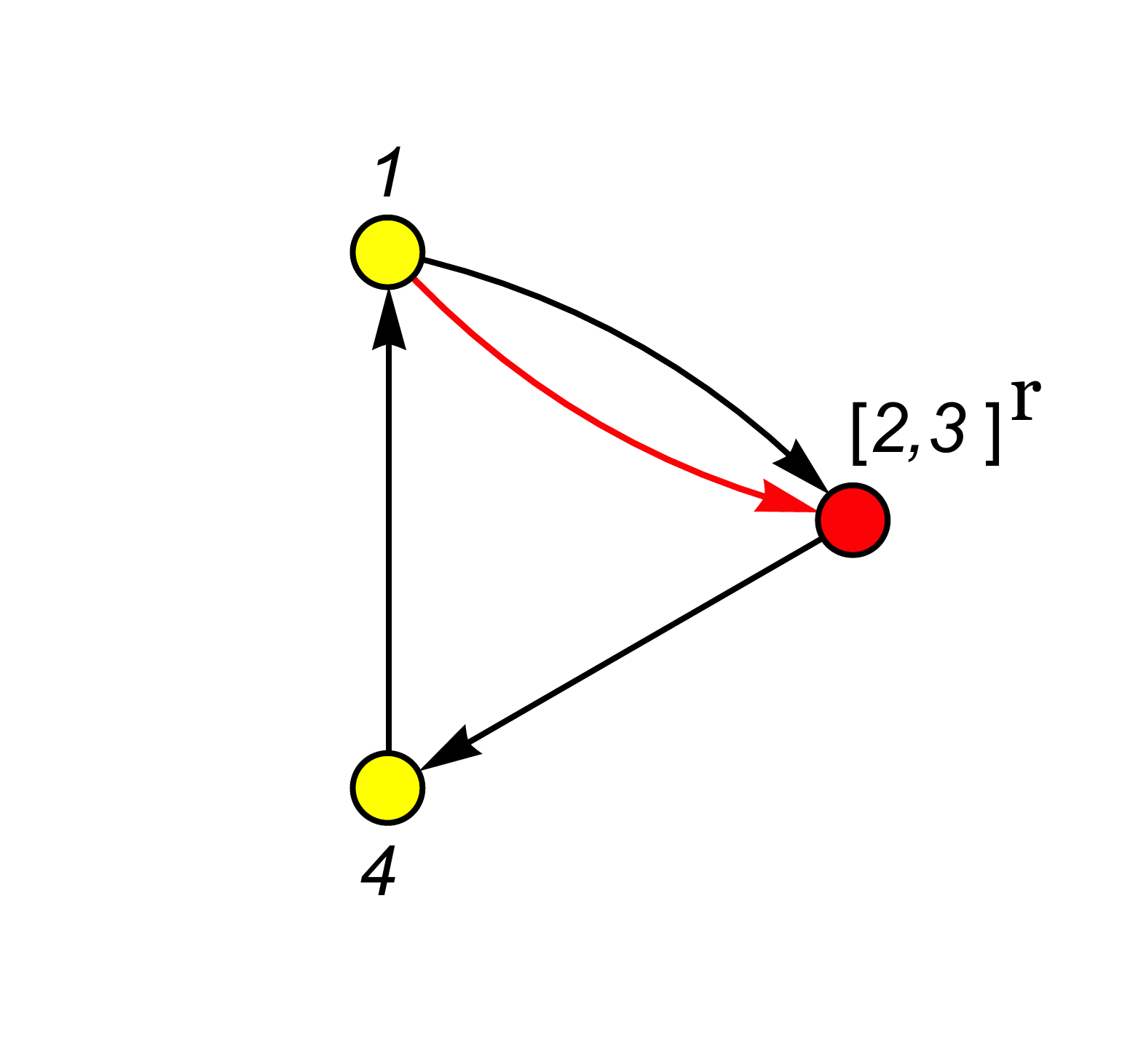}} 
\times
\left(
\frac{1}{\tilde s_{14}} \right)
\times
\hspace{-0.4cm}
\parbox[c]{5.0em}{\includegraphics[scale=0.15]{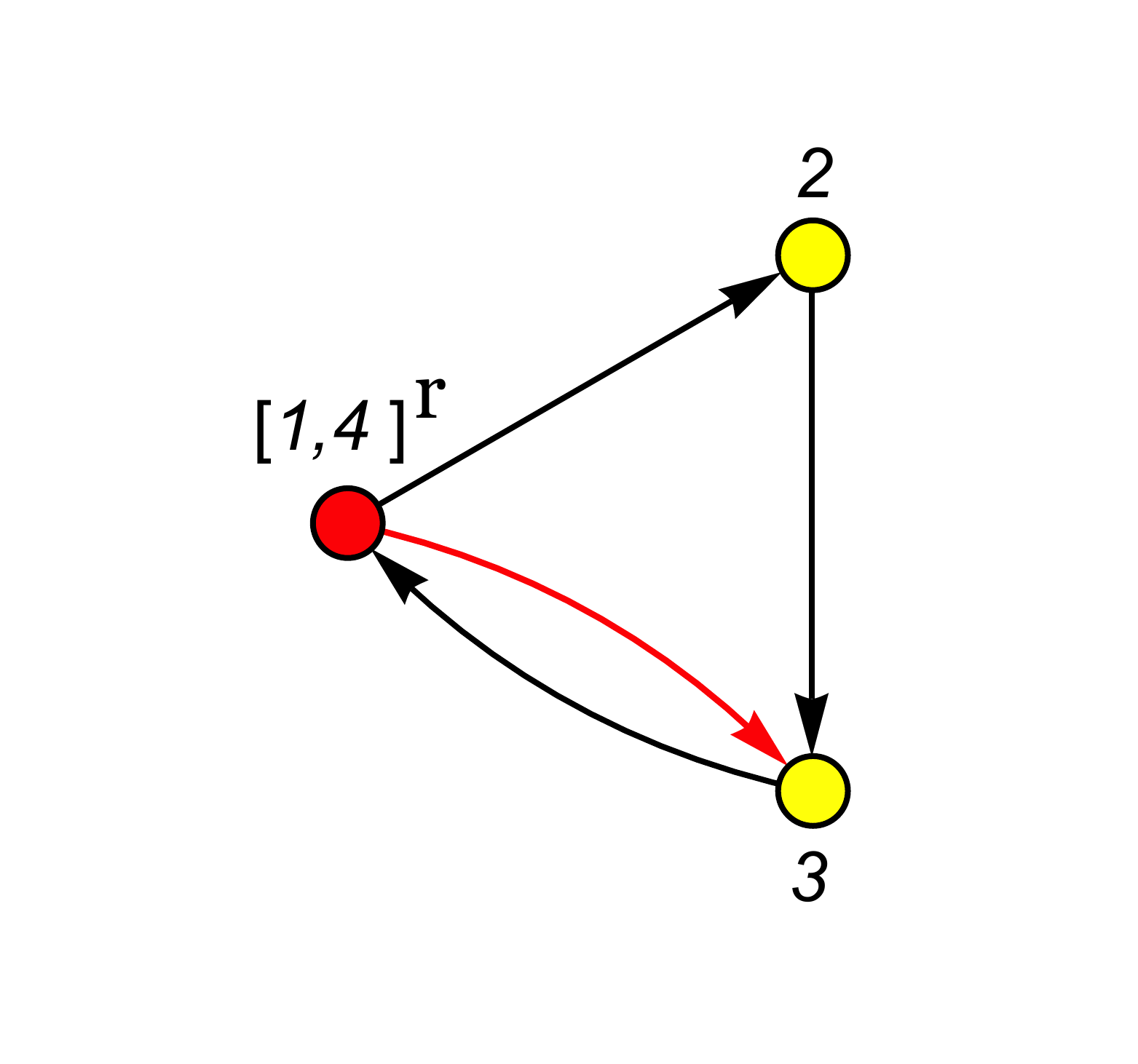}}  =
 \frac{ \sum_r \, 
A_3^{(1,[2,3])}(1,[2,3]^r,4) \times  A_3^{(3,[4,1])}(3,[4,1]^r,2) }{\tilde s_{14}}   \, , \nonumber
\end{eqnarray}
}
\vskip-0.7cm\noindent
\vspace{-0.9cm}
{\small
\begin{eqnarray}\label{Rcut3}
\hspace{-8.0cm}
\parbox[c]{5.2em}{\includegraphics[scale=0.15]{4pt-cut3.pdf}}
= 
\hspace{-0.4cm}
\parbox[c]{5.1em}{\includegraphics[scale=0.15]{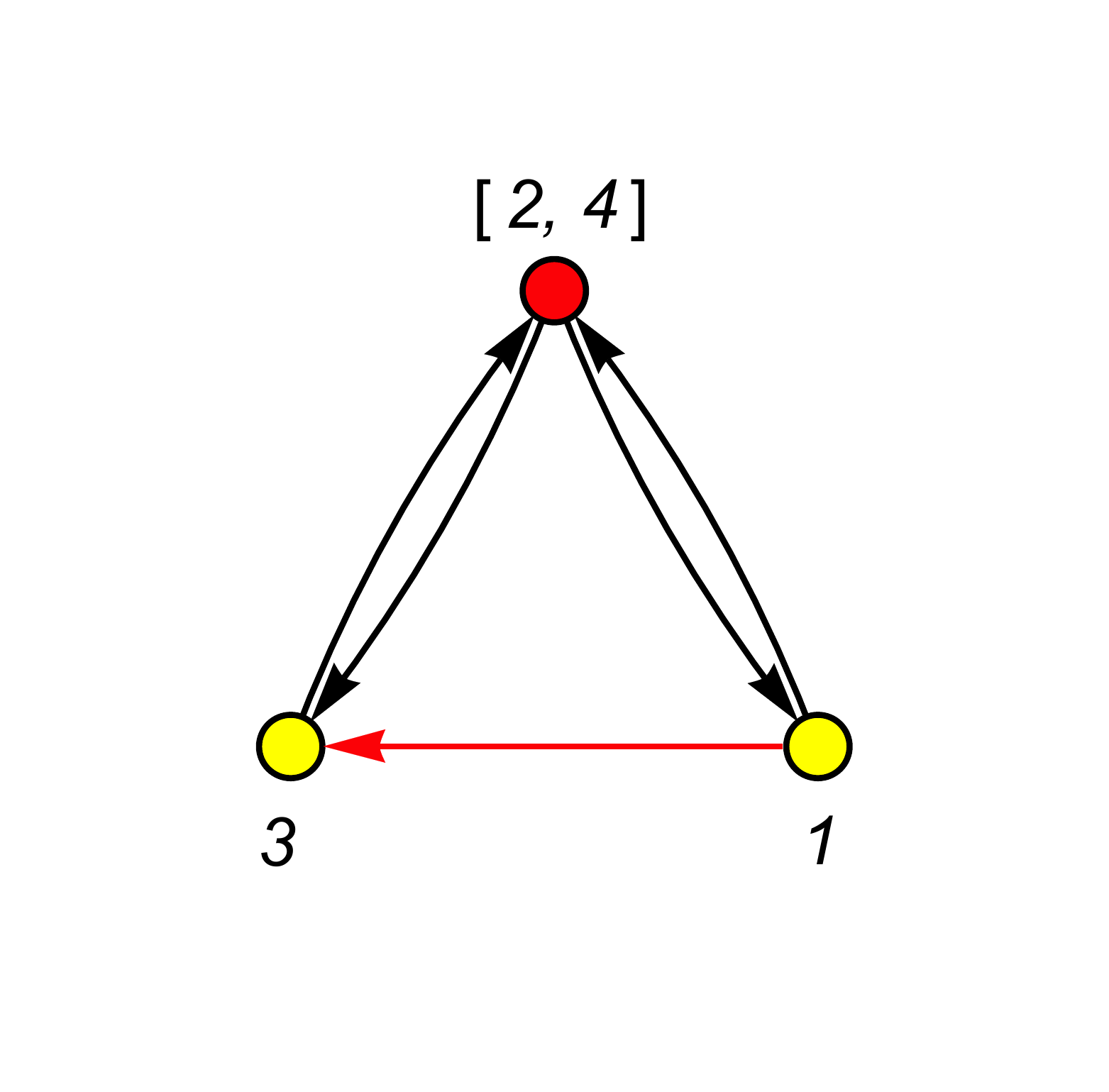}} 
\times  \left(
\frac{1}{\tilde s_{24}} \right)
\times
\hspace{-0.5cm}
\parbox[c]{5.7em}{\includegraphics[scale=0.15]{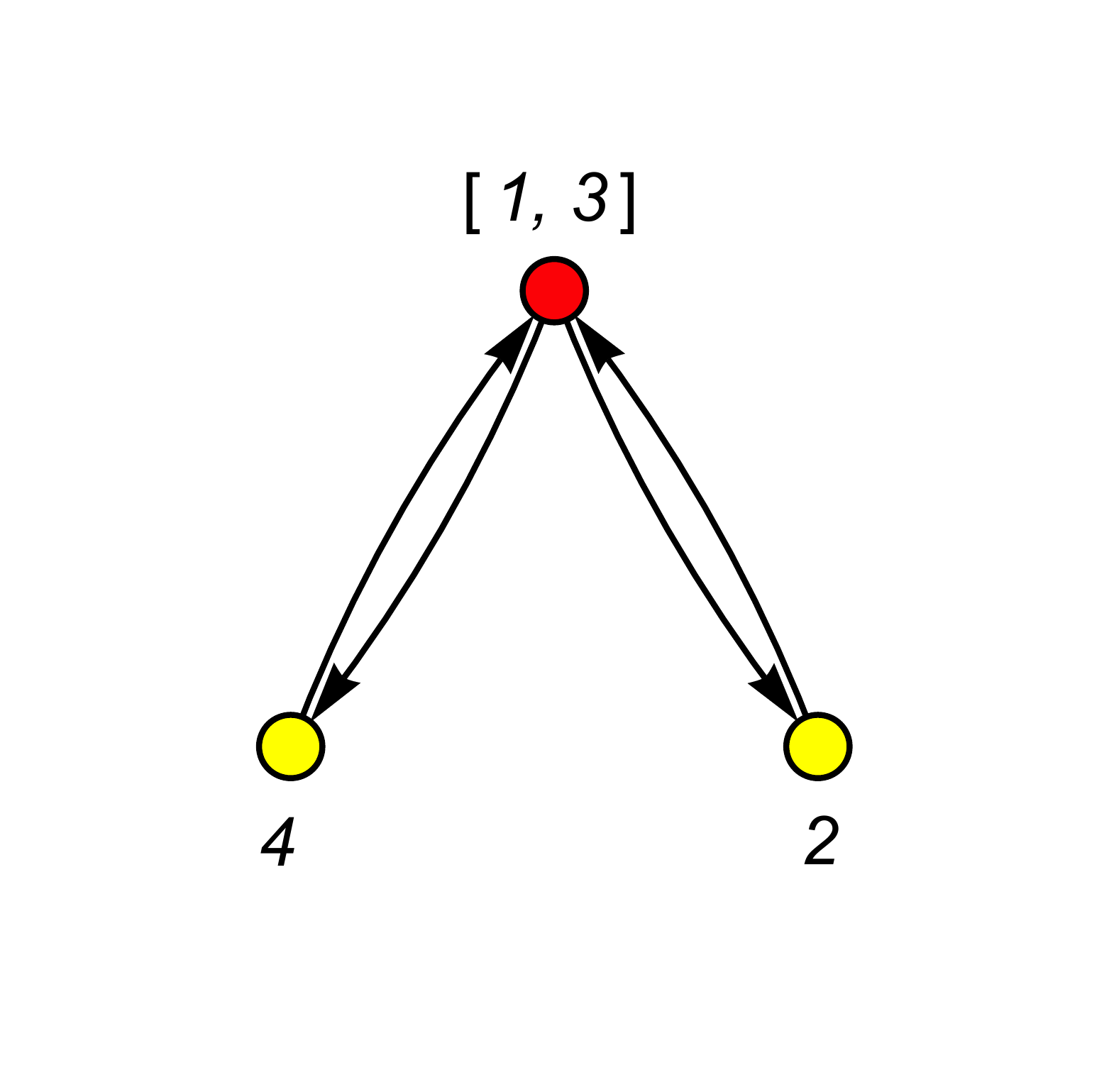}} , 
\end{eqnarray}
}
\vskip-0.5cm\noindent
where let us remind the {\bf red vertices mean they are fixed and off-shell punctures}. Notice the upper index ``$r$" over  the red vertices, for instance $[a_1,\ldots, a_i]^r$, means the off-shell punctures,  $\s_{[a_1,\ldots, a_i]}$, have as associated  polarization vector, $\epsilon^{r,\mu}_{[a_1,\ldots, a_i]}$. 

As it was said above, the four black-arrows on the off-shell punctures over the resulting graphs in {\it cut-3}  
mean all rows/columns related with them must be removed from $\Psi$ matrix ({\bf rule-IIIb}).
In other words,  these off-shell vertices have an associated  polarization vector proportional to their momentum, i.e.  longitudinal gluons.
The  explicit computation will be performed in \eqref{strangeC1}, and in the next section we will give more details about this issue.

Using the three-point off-shell building-block, $A_3^{([a],[b])}([a],[b],[c])$  given in \eqref{YM-BB}, and  the gluing  identity, $\sum_{r} \eps^{r,\mu}_{[3,4]} \, \eps^{r,\nu}_{[1,2]} = \eta^{\mu\nu}$, it is simple to compute  {\it cut-1}
{\small
\begin{eqnarray}\label{4p-cut1}
\hspace{-0.1cm}
 {\it cut\text{-}1} &=& \frac{ \sum_r \, A_3^{([3,4],1)}( [3,4]^r,1,2) \times A_3^{([1,2],3)}([1,2]^r,3,4)  }{\tilde s_{34}}  \nonumber\\ 
\hspace{-0.5cm}
&= &  \left(\frac{ 2}{ s_{12}}\right) \times   \left\{ 
-(\eps_1 \cdot k_2) (\eps_2 \cdot \eps_3)(k_3 \cdot \eps_4) -{\rm cyc}_{(1,2,3,4)} \, + 
(\eps_1 \cdot k_2) (\eps_2 \cdot \eps_4)(k_4 \cdot \eps_3) + {\rm cyc}_{(1,2,4,3)}
\right. \nonumber\\
&&
\hspace{1.7cm}
\left.
+\frac{s_{13}}{2} (\eps_1 \cdot \eps_2)(\eps_3\cdot \eps_4) \right\}.
\nonumber
\end{eqnarray}
}
\vskip-0.3cm\noindent
Analogously,
\vspace{-0.5cm}
{\small
\begin{eqnarray}\label{4p-cut2}
\hspace{-0.7cm}
{\it cut\text{-}2}  &= &\frac{   \sum_r \, A_3^{(1,[2,3])}(1, [2,3]^r ,4) \times A_3^{(3,[4,1])}(3 ,[4,1]^r,2)  }{\tilde s_{14}}  \nonumber \\
&=&
{cut\text{-}1}\Big|_{(1,2,3,4)\, \rightarrow \, (4,1,2,3)} ~~ , \nonumber
\end{eqnarray}
}
\vskip-0.5cm\noindent
with, $\sum_{r} \eps^{r,\mu}_{[2,3]} \, \eps^{r,\nu}_{[4,1]} = \eta^{\mu\nu}$. 

Finally, to compute {\it cut-3}, we just read the resulting graphs, namely
\vspace{-0.3cm}
{\small
\begin{eqnarray}\label{}
\hspace{-1.7cm}
\hspace{-0.5cm}
\parbox[c]{5.3em}{\includegraphics[scale=0.17]{R1-cut3.pdf}} 
= \frac{(\s_{1[2,4]} \s_{[2,4]3} \s_{31})^2}{(\s_{1[2,4]}\s_{[2,4]1})\, (\s_{[2,4]3}\s_{3[2,4]})} 
\times \frac{1}{\s_{13}}
\,  {\rm Pf}\left[
{\small
\begin{matrix}
 0 & {\eps_1\cdot \eps_3 \over \s_{13} } \\
 {\eps_3\cdot \eps_1 \over \s_{31} } & 0
\end{matrix}}
\right] =  (\eps_1 \cdot \eps_3)  ,
\nonumber
\end{eqnarray}
}
\vskip-0.6cm\noindent
\vspace{-0.8cm}
{\small
\begin{eqnarray}\label{strangeC1}
\parbox[c]{5.2em}{\includegraphics[scale=0.17]{R2-cut3.pdf}} 
= \frac{(\s_{2[1,3]} \s_{[1,3]4} \s_{42})^2}{(\s_{[1,3]2}\s_{2[1,3]})\, (\s_{[1,3]4}\s_{4[1,3]})}
\times {\rm Pf}\left[
{\small
\begin{matrix}
 0 &    \frac{\tilde s_{24}}{\s_{24}}  & - {\cal C}_{22} & -{\eps_4\cdot k_2 \over \s_{42} }  \\
 \frac{\tilde s_{42}}{\s_{42}} & 0 &- {\eps_2\cdot k_4 \over \s_{24} } & - {\cal C}_{44}  \\
{\cal C}_{22} & {\eps_2\cdot k_4 \over \s_{24} } & 0 & \frac{\eps_2\cdot \eps_4}{\s_{24}}  \\
{\eps_4\cdot k_2 \over \s_{42} } & {\cal C}_{44} &  \frac{\eps_4\cdot \eps_2}{\s_{42}} &0  \\
\end{matrix}}
\right] = \tilde s_{24}\,  (\eps_2 \cdot \eps_4), 
\nonumber\\ 
\end{eqnarray}
}
\vskip-0.5cm\noindent
where, ${\cal C}_{22}=-\left( \frac{\eps_2\cdot k_4}{\s_{24}} +\frac{\eps_2\cdot k_{[1,3]}}{\s_{2[1,3]}} \right)$ and ${\cal C}_{44}=-\left( \frac{\eps_4\cdot k_2}{\s_{42}} +\frac{\eps_4\cdot k_{[1,3]}}{\s_{4[1,3]}} \right)$. Therefore
\vspace{-0.1cm}
\begin{eqnarray}\label{4p-cut3}
{\it cut\text{-}3} = (\eps_1 \cdot \eps_3)(\eps_2\cdot \eps_4)\, . 
\end{eqnarray}
\vskip-0.2cm\noindent
It is straightforward to check that, in fact, $A_4^{\rm YM}(1,2,3,4) =$ {\it cut-1}$+${\it cut-2}$+${\it cut-3}.

As a final remark, it is interesting to see that the {\bf strange-cut 3} is related with the quartic vertex, ${\rm Tr} \left(  \left[A_\mu,A_\nu \right]^2   \right)$.  First, notice that the {\it cut-3} can be rewritten as
\vspace{-0.2cm}
\begin{eqnarray}\label{}
{\it cut\text{-}3} = \eps_1^\mu \, \eps_2^\nu \, \eps_3^\rho \, \eps_4^\delta \,   \left[ \eta_{\mu \rho} \eta_{\nu \delta} \right] \, . 
\end{eqnarray}
\vskip-0.1cm\noindent
On other hand, the color-ordered contact vertex is given by\footnote{See Dixon normalization in \cite{Dixon:1996wi}.} 
\vspace{-0.1cm}
{\small
\begin{eqnarray}\label{}
\parbox[c]{3.2em}{\includegraphics[scale=0.1]{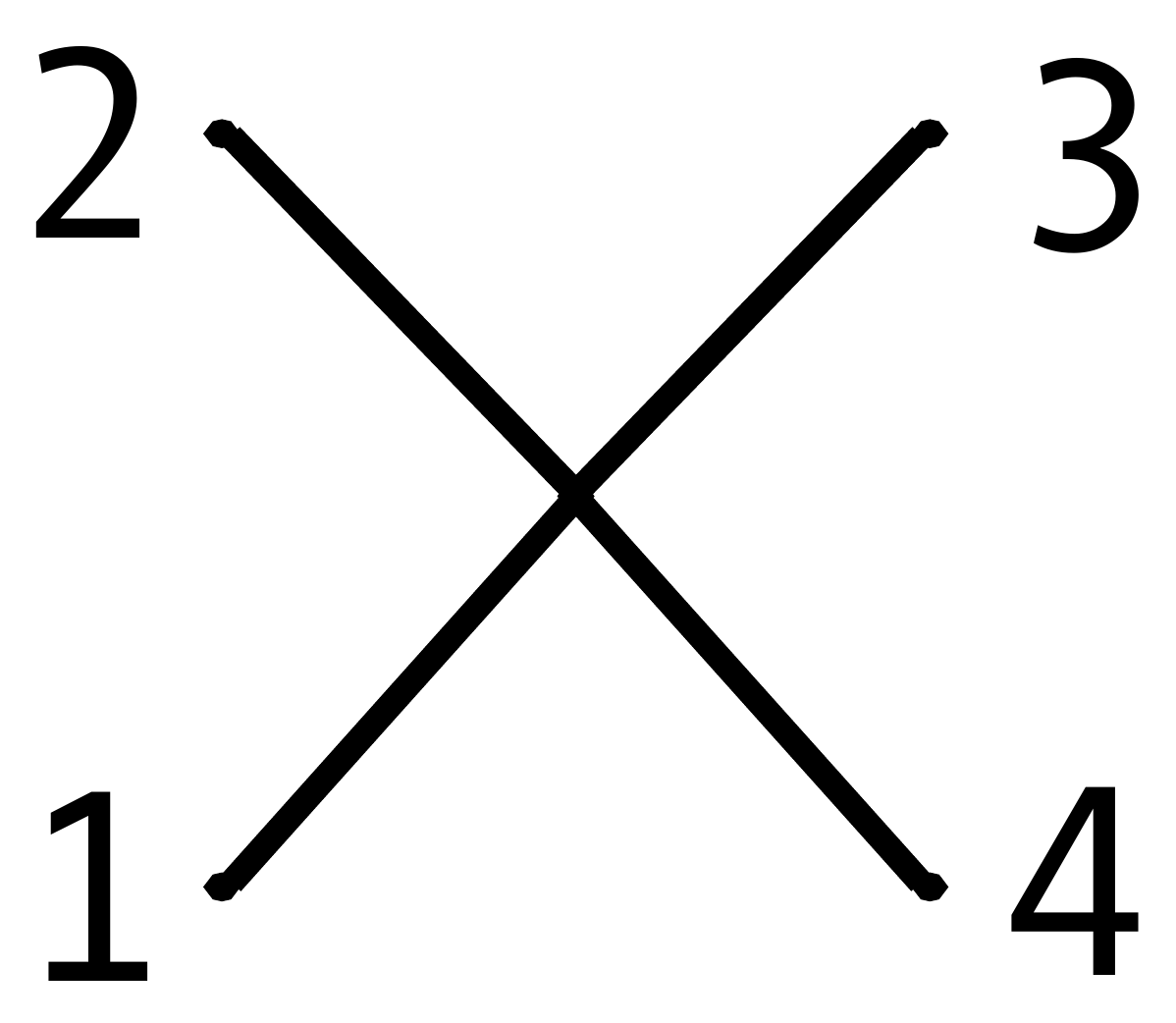}} 
=\eps_1^\mu \, \eps_2^\nu \, \eps_3^\rho \, \eps_4^\delta \,   \left[ \eta_{\mu \rho} \eta_{\nu \delta}
-\frac{1}{2}\left( \eta_{\mu\nu} \eta_{\rho\delta} + \eta_{\mu\delta} \eta_{\nu\rho}
\right)
 \right]
\end{eqnarray}
}
\vskip-0.3cm\noindent
Clearly, the first term matched perfectly, but, the others two are not present in {\it cut-3}. This fact confirms that the {\bf integration rules} proposed in this paper are not the Feynman rules, as a consequence, we obtain spurious poles.

\subsection{Five-Point}

Like in the previous example, we choose the gauge fixing $(pqr | m)=(123 | 4)$. Additionally, 
 to avoid singular cuts (see \eqref{Lbehavior}), we pick out the red arrow among the vertices,   
 $(i,j)=(1,3)$. Thus, applying the {\bf integration  rules}, one has the cutting expansion 
\vspace{-0.6cm}
\begin{eqnarray}
\hspace{0.2cm}
A_5^{(1,3)}(1,2,3,4,5) =
\int  d\mu_5^{\L}
\hspace{-0.2cm}
\parbox[c]{5.9em}{\includegraphics[scale=0.17]{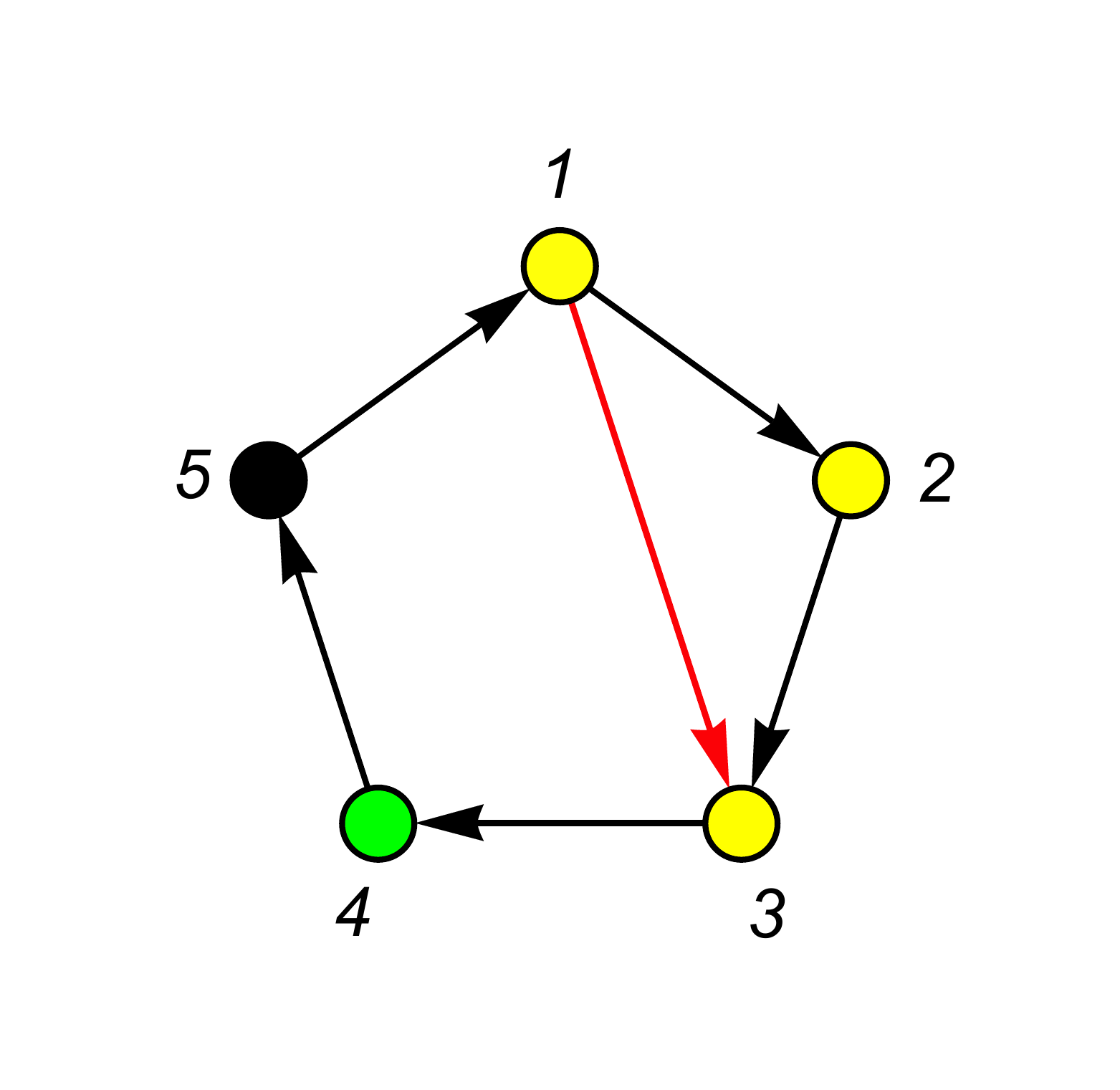}}
=
\hspace{-0.27cm}
\parbox[c]{6.1em}{\includegraphics[scale=0.17]{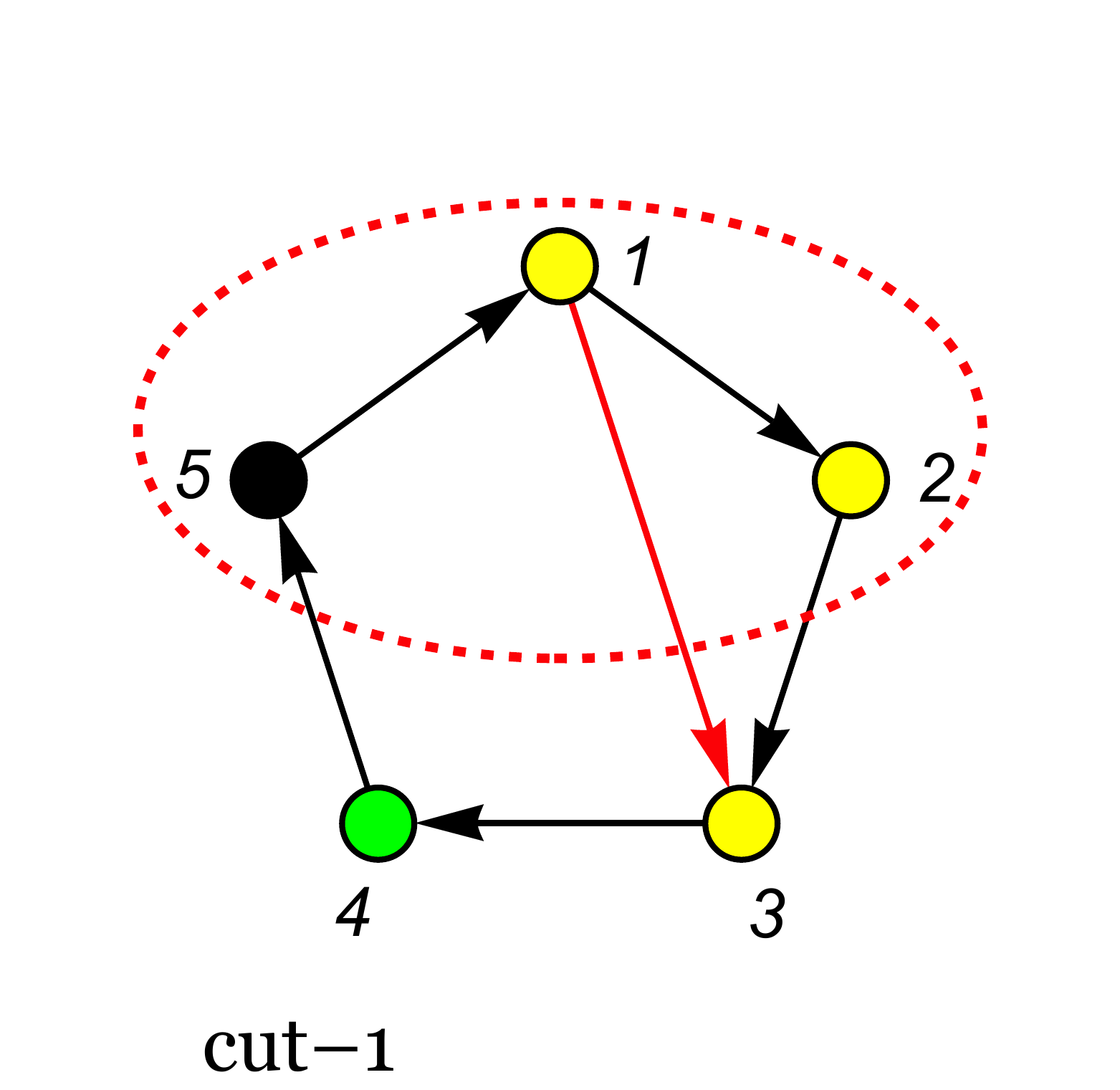}} +
\hspace{-0.3cm}
\parbox[c]{8.0em}{\includegraphics[scale=0.17]{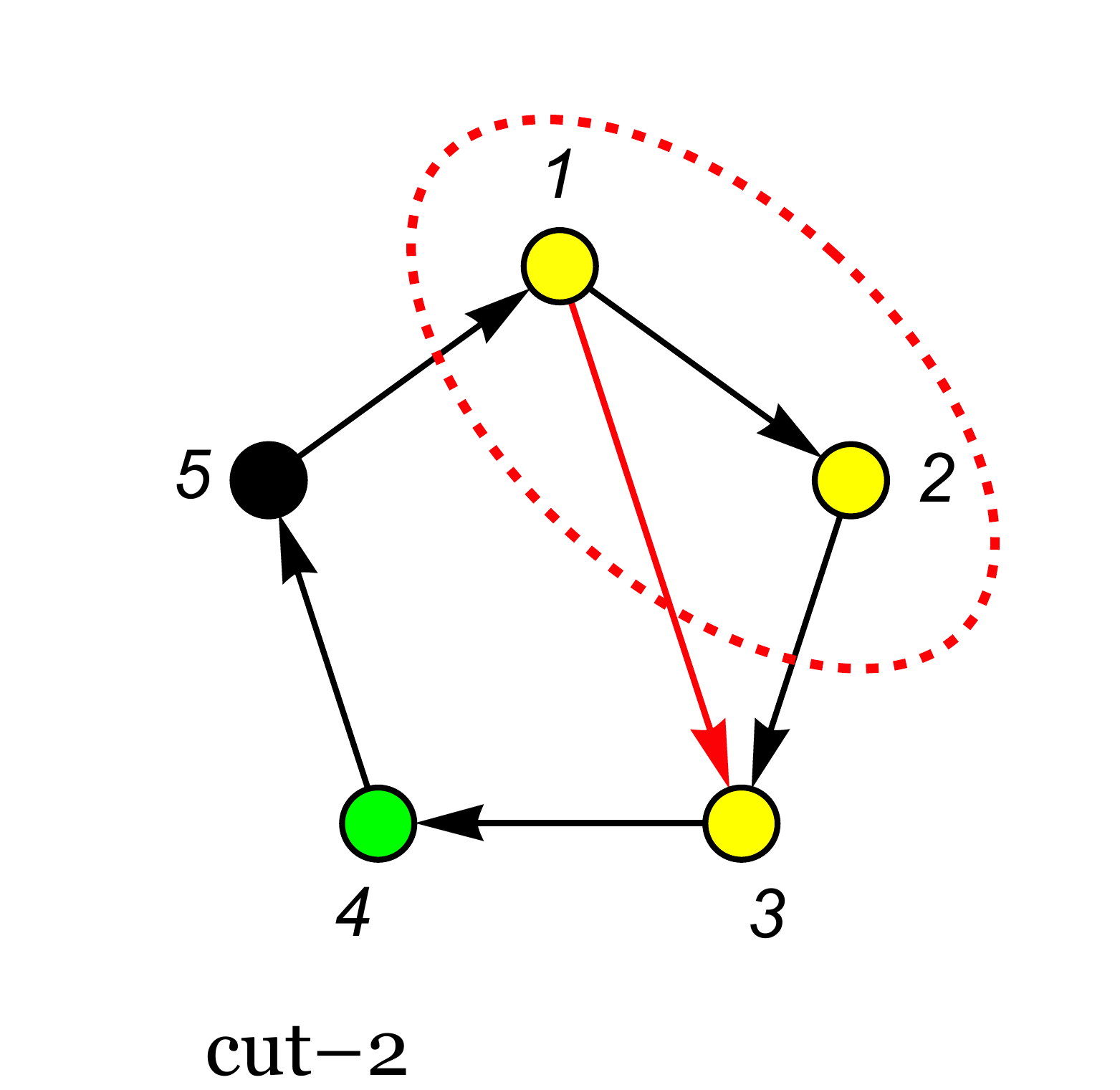}} 
\hspace{-0.7cm}
+
\hspace{-0.3cm}
\parbox[c]{6.1em}{\includegraphics[scale=0.17]{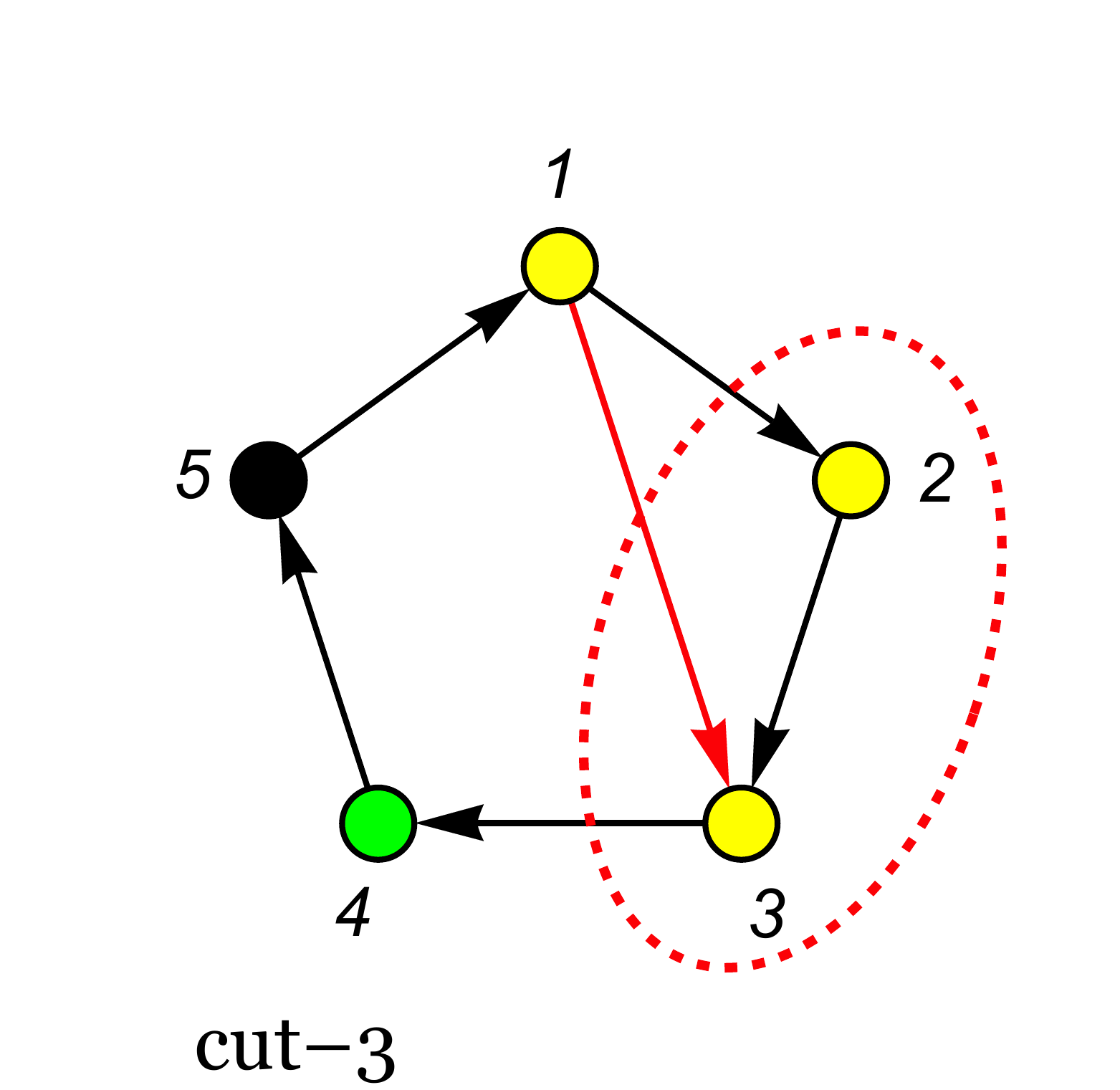}} 
\nonumber
\end{eqnarray}
\vskip-0.2cm\noindent
\vspace{-0.9cm}
\begin{eqnarray}\label{fivePcuts}
\hspace{5.6cm}
+\hspace{-0.1cm}
\parbox[c]{6.1em}{\includegraphics[scale=0.17]{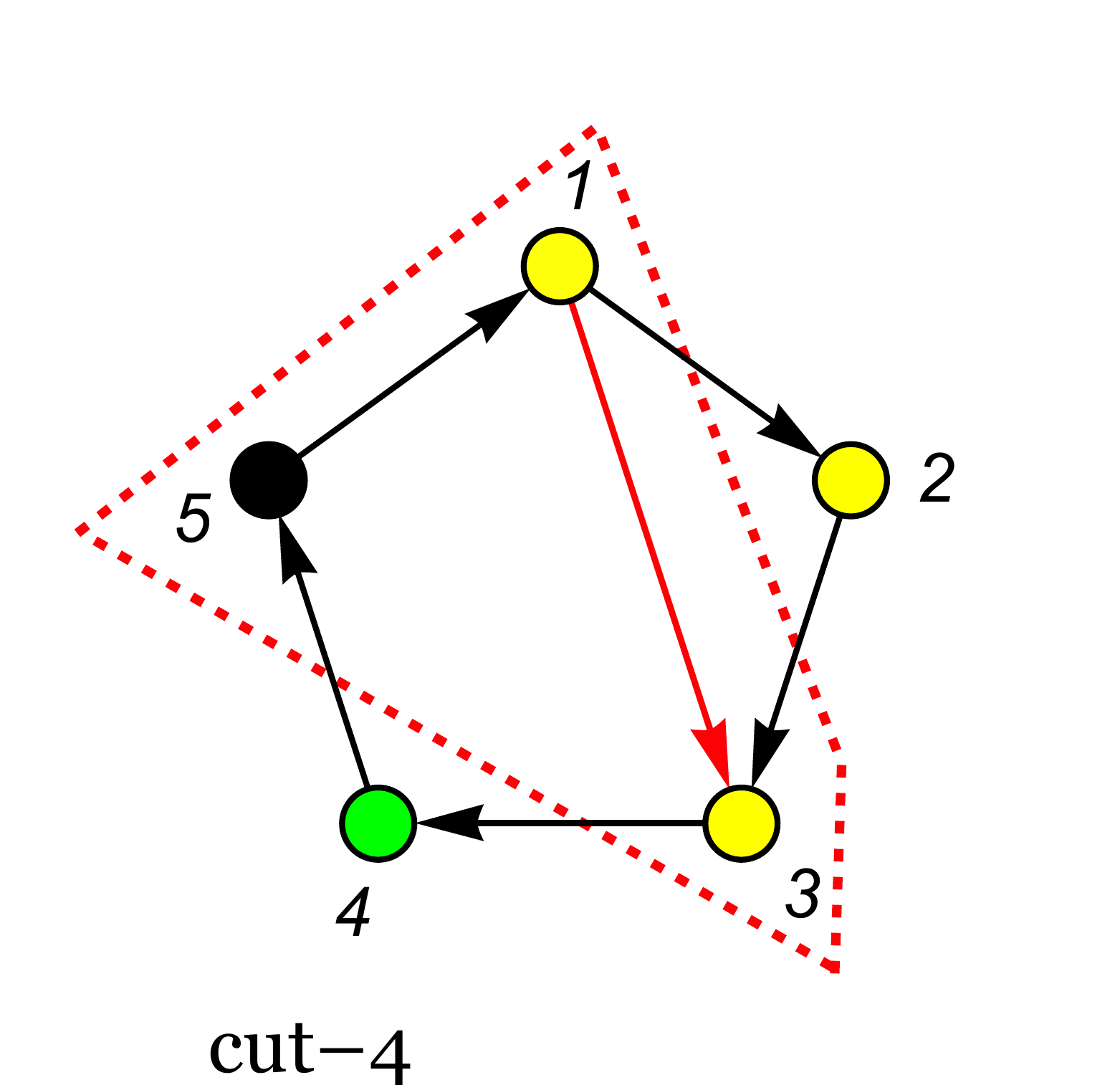}} 
+\hspace{-0.3cm}
\parbox[c]{6.3em}{\includegraphics[scale=0.17]{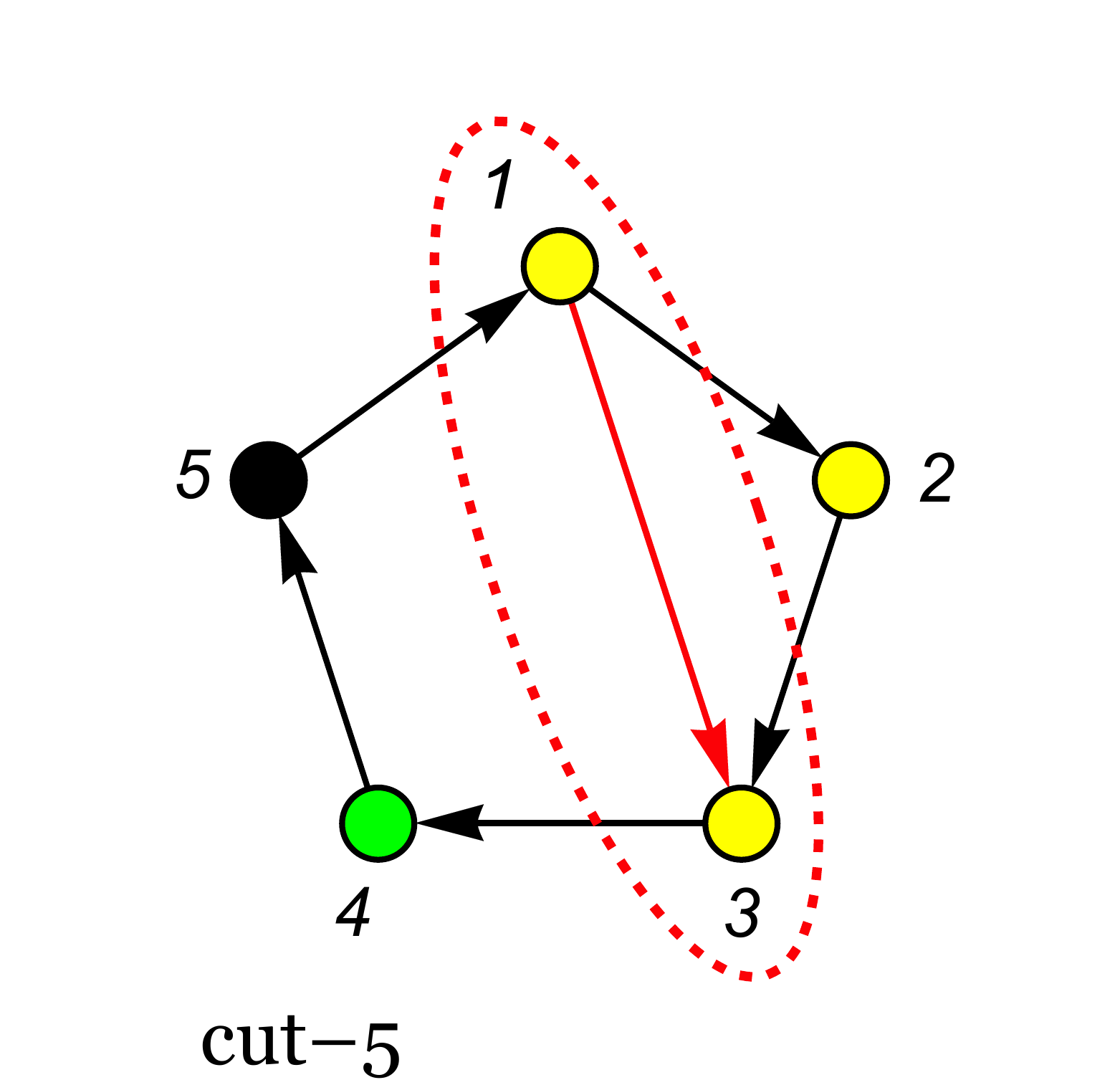}} .\,~~~
\end{eqnarray}
\vskip-0.1cm\noindent
This expansion has been verified numerically, up to an overall sign, and we will compute it explicitly in the last section.

Since one of our objectives is to describe a Yang-Mills CHY-algorithm, we need to understand how to apply the {\bf integration rules} over the resulting graphs. For example, in the $\phi^3$  algorithm obtained in \cite{Gomez:2016bmv}, {\bf the resulting gauge fixing can not be modified on the go to compute the resulting graphs}. As an illustration, let us consider {\it cut-2} in fig. \eqref{fivePcuts}, 
\vspace{-0.5cm}
{\small
\begin{eqnarray}\label{4ptsA}
&&
\hspace{-1.2cm}
\parbox[c]{6.3em}{\includegraphics[scale=0.17]{5pt-cut2.pdf}}
= 
\sum_r
\hspace{-0.5cm}
\parbox[c]{5.3em}{\includegraphics[scale=0.17]{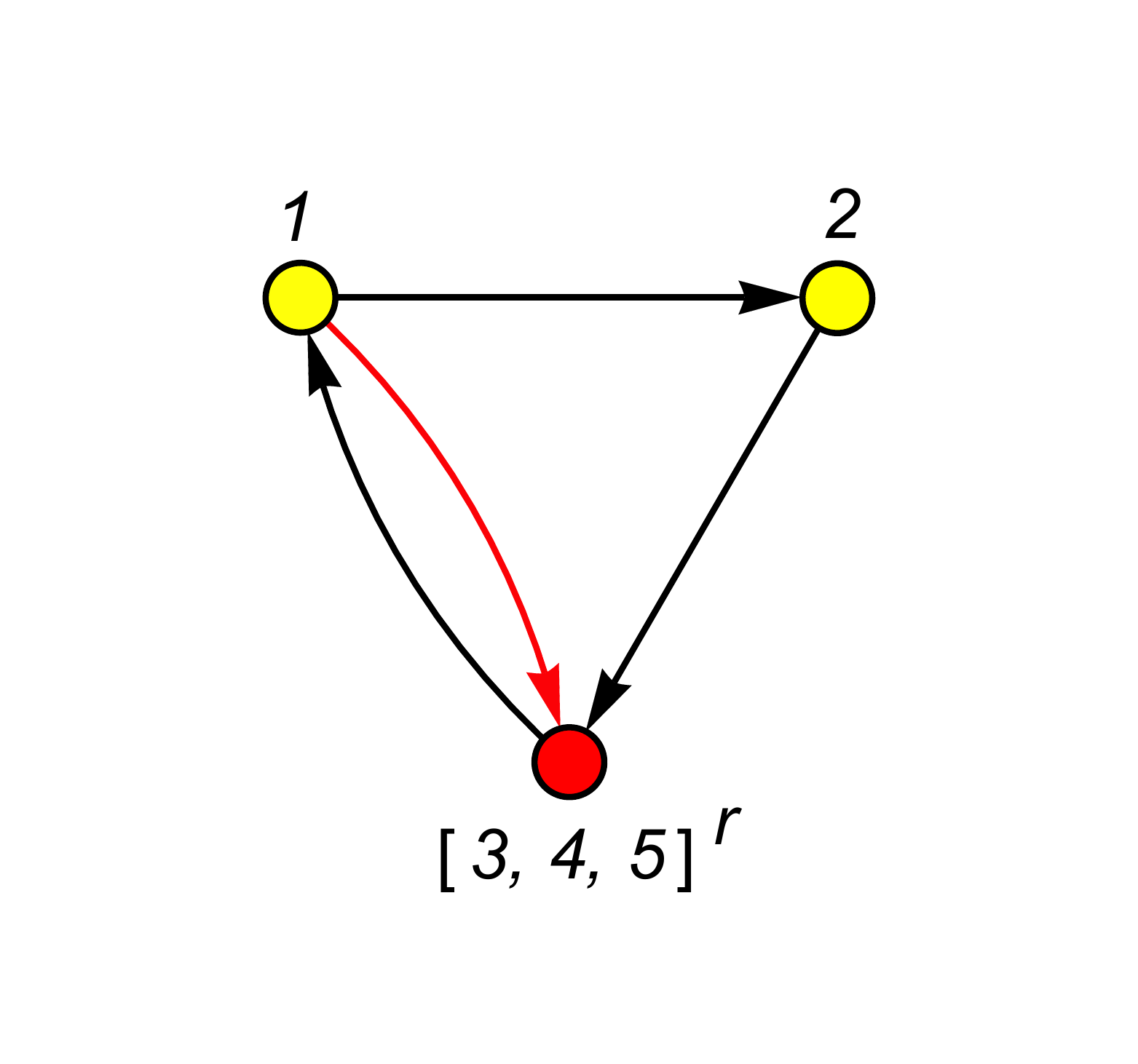}} 
\times
\left(
\frac{1}{\tilde s_{345}} \right)
\times
\int d\mu_4^{\rm CHY}
\hspace{-0.5cm}
\parbox[c]{5.8em}{\includegraphics[scale=0.17]{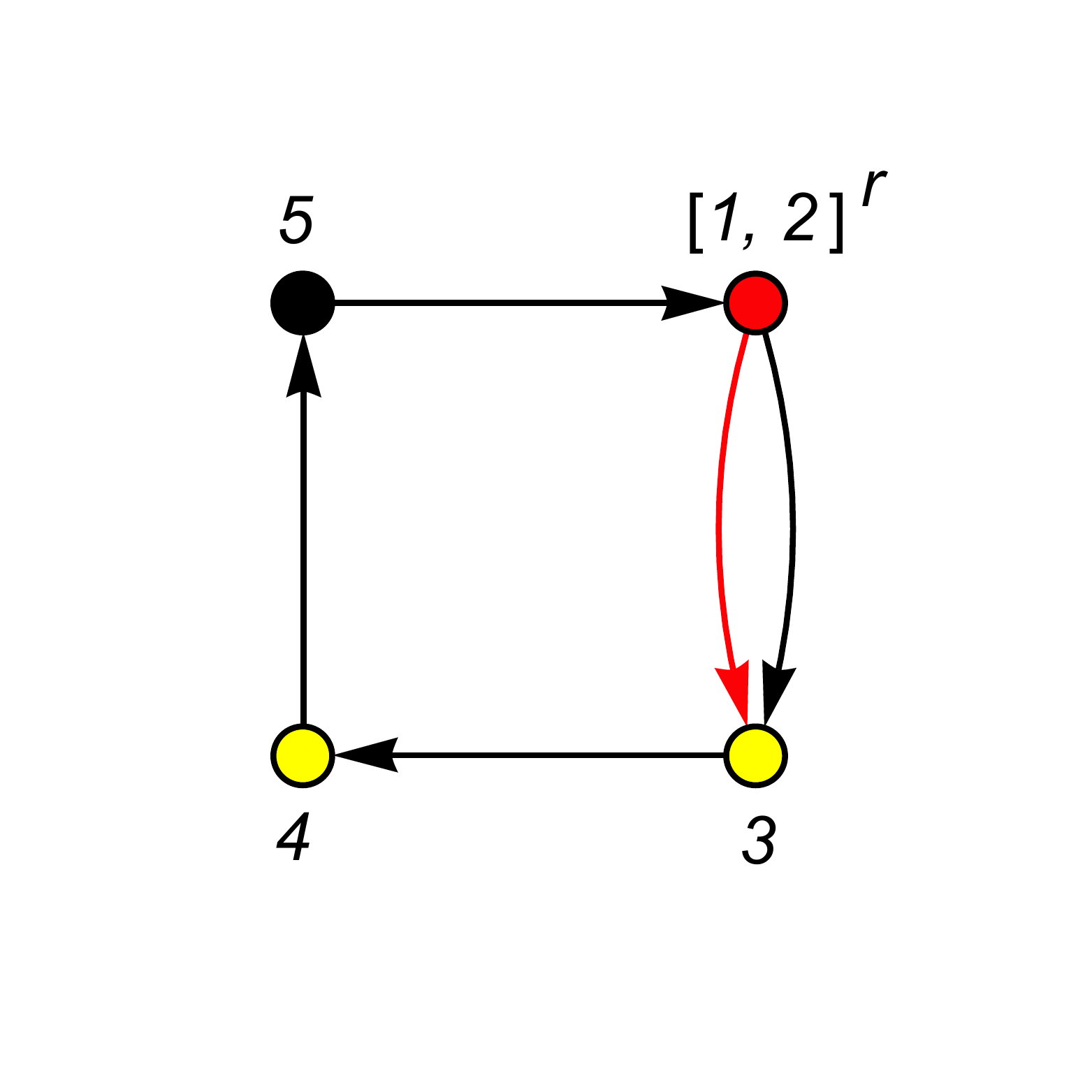}} \quad
\end{eqnarray}
}
\vskip-0.7cm\noindent
{\small
\begin{eqnarray}
\hspace{1.2cm}
=
\frac{\sum_r  A_3^{([3,4,5],1)}([3,4,5]^r,1,2) \times  
A_4^{([1,2],3)}([1,2]^r,3,4,5)
}{\tilde s_{345} }
 \, , \nonumber
\end{eqnarray}
}
\vskip-0.5cm\noindent
where the graphs are glued by the identity,  $\sum_r \eps^{r,\mu}_{[3,4,5]}\, \eps^{r,\nu}_{[1,2]}=\eta^{\mu\nu}$.
Clearly, the {\bf integration rules} can not be used over the resulting four-point graph because there is a singular cut (a configuration that cuts two arrows). A naive solution would be just to change its resulting setup, e.g. by moving the red arrow,  $(i,j)=([1,2],3)$, to the one that joins the vertices, $(i,j)=([1,2],4)$. However, a simple numerical computation shows a mismatch,
\vspace{-0.5cm}
{\small
\begin{eqnarray}\label{nonequality}
\hspace{-0.5cm}
\sum_r
\hspace{-0.5cm}
\parbox[c]{5.3em}{\includegraphics[scale=0.17]{cut2-R1.pdf}} 
\,
\int d\mu_4^{\rm CHY}
\hspace{-0.5cm}
\parbox[c]{5.5em}{\includegraphics[scale=0.17]{cut2-R2.pdf}} 
\neq
\sum_r
\hspace{-0.5cm}
\parbox[c]{5.3em}{\includegraphics[scale=0.17]{cut2-R1.pdf}} 
\,
\int d\mu_4^{\rm CHY}
\hspace{-0.5cm}
\parbox[c]{5.8em}{\includegraphics[scale=0.17]{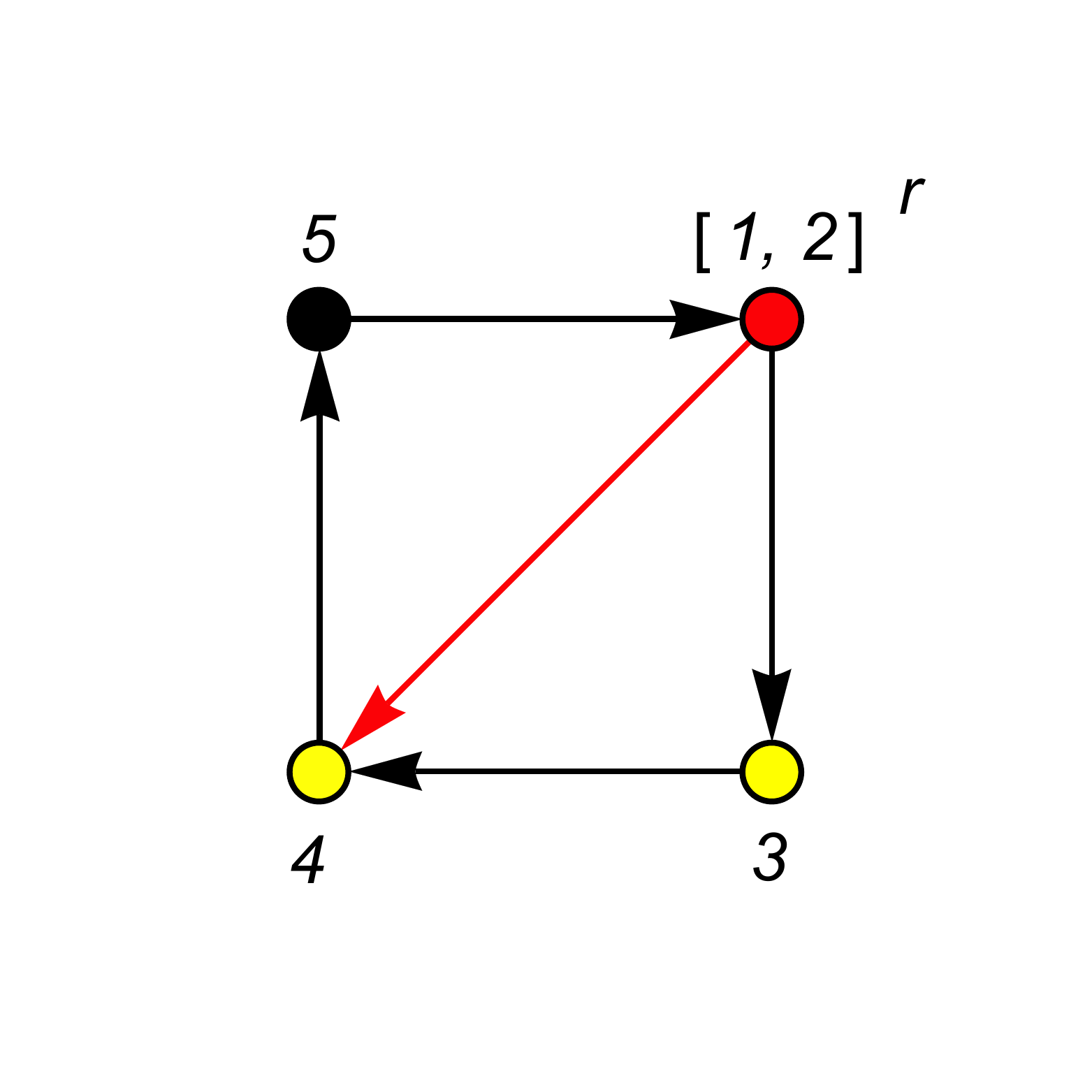}} 
,
\end{eqnarray}
}
\vskip-0.7cm\noindent
with $\sum_r \eps^{r,\mu}_{[3,4,5]}\, \eps^{r,\nu}_{[1,2]}=\eta^{\mu\nu}$.
This fact is a consequence that the polarization vectors, $\eps^{r,\mu}_{[3,4,5]}$ and $\eps^{r,\nu}_{[1,2]}$, are not transverse ($\eps_{[3,4,5]}^{r} \cdot k_{[3,4,5]}\neq 0$, $\eps_{[1,2]}^{r} \cdot k_{[1,2]}\neq 0$). In the next section, we are going to solve this drawback.

\section{Longitudinal and Transverse  Gluons}\label{LongContributions}

Additionally to the standard factorization cuts, where a YM-graph is splitting in two smaller ones with an off-shell gluon propagating among them ({\it standard-cuts}), 
we have also obtained some strange contributions ({\it strange-cut}), with no obvious physical interpretation. In this section, we study these {\it strange-cuts}.

Under the gauge fixing, $(pqr|m)=(123|4)$ and with the red arrow over, $(i,j)=(1,3)$, a generic {\it strange-cut} 
encircles the vertices, $(1,3,p+1,...,n)$,
produces the following two types of resulting graphs
\vspace{-0.6cm}
{\small
\begin{eqnarray}\label{strangec-YMS1}
\int
d\mu_{(n-p+3)}^{\rm CHY}
\hspace{-0.6cm}
\parbox[c]{6.0em}{\includegraphics[scale=0.17]{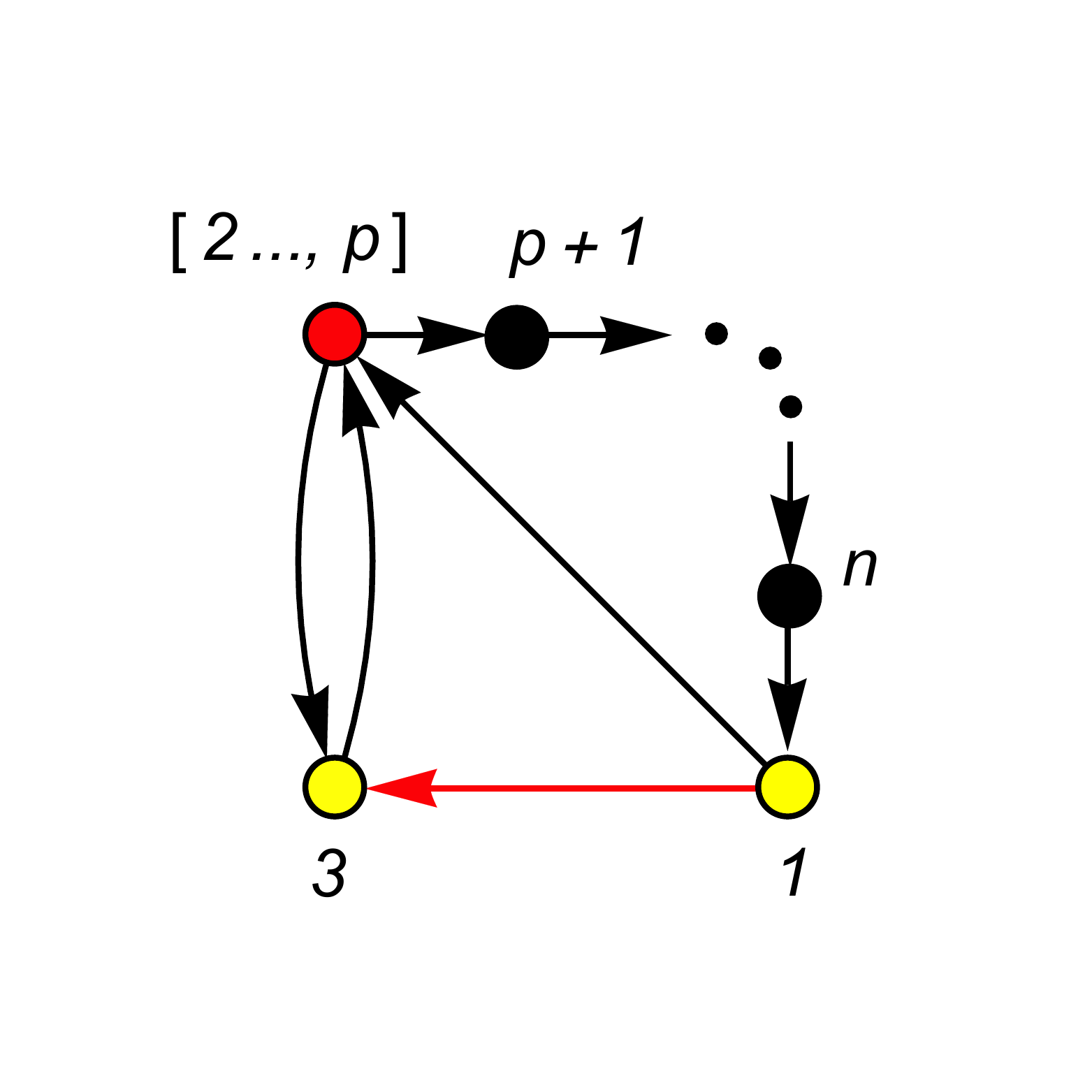}} ,~
~~~
\hspace{-0.1cm}
\int
d\mu_{(p-1)}^{\rm CHY}
\hspace{-0.5cm}
\parbox[c]{5.5em}{\includegraphics[scale=0.17]{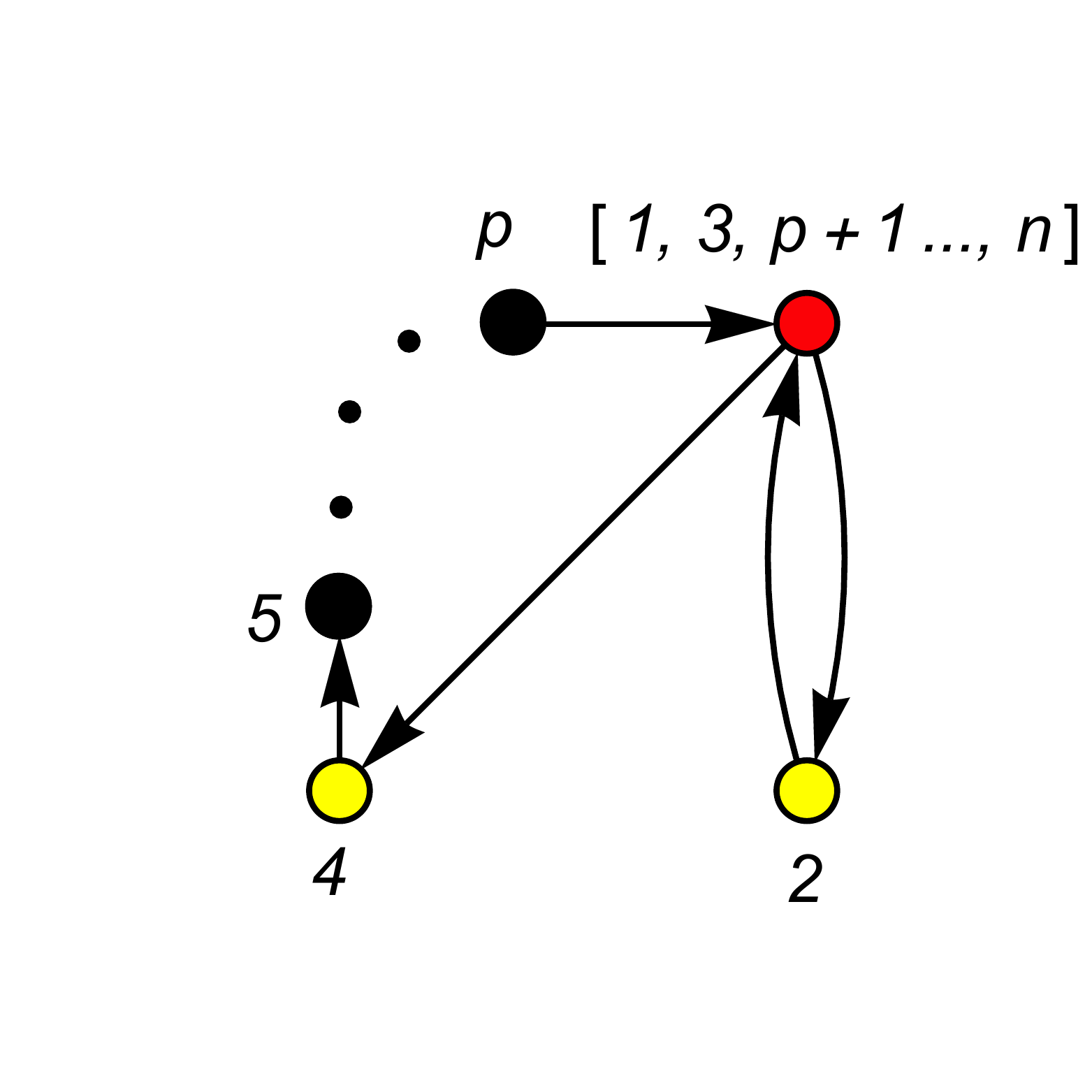}} ~~=~~
\hspace{-0.1cm}
\frac{k^2_{[1,3,p+1,...,n]}}{2}\,\,
\int
d\mu_{(p-1)}^{\rm CHY}
\hspace{-0.5cm}
\parbox[c]{6.0em}{\includegraphics[scale=0.17]{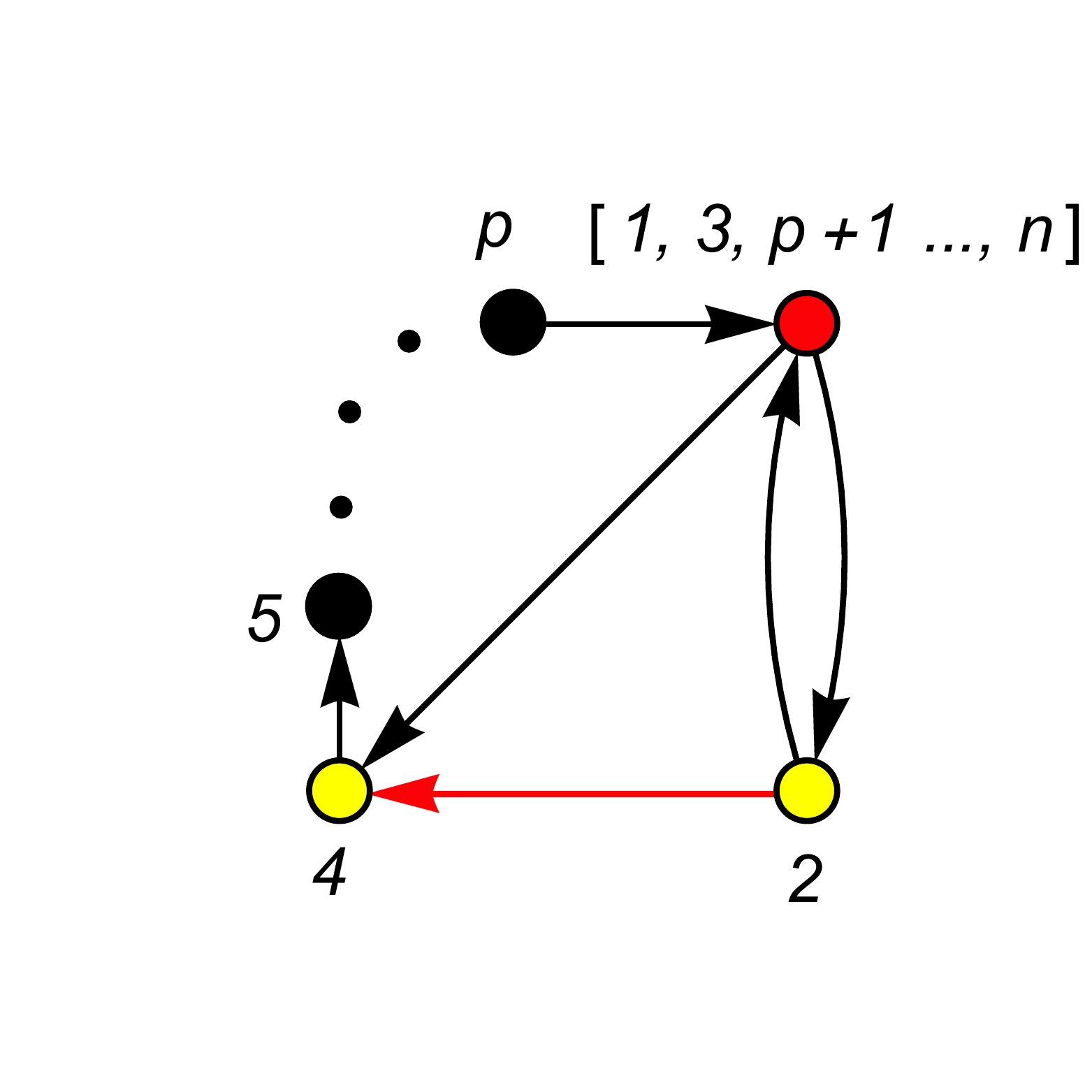}},
\qquad~
\end{eqnarray}
}
\vskip-0.8cm\noindent
where we  used the {\bf property-I} (appendix \ref{appendix}) to obtain the equality. Notice that from the {\bf rule-IIIa}, the  associated matrices of these two resulting graphs are given by, $\left( \Psi_{\rm g,s:g}\right)^{1\,3\,[2,4,...,p]}_{1\,3\,[2,4,...,p]}$ and $\left(\Psi_{\rm g,s:g}\right)^{[1,3,p+1,...,n]}_{[1,3,p+1...,n]}$,  where the gluon and scalar sets are given by the particles, ${\rm g}=\{ 1,3,p+1,...,n  \}$, ${\rm g}=\{ 2,4,5,...,p  \}$, ${\rm s}=\{ [2,4,...,p]  \}$ and ${\rm s}=\{ [1,3,p+1,...,n]  \}$, respectively. In section \ref{strange-YMS}, we will come back to this point.

Using \eqref{strangec-YMS1} and the {\bf property-II} of the appendix \ref{appendix}, we obtain the identity
\vspace{-0.2cm}
{\small
\begin{eqnarray}\label{generalone}
\int
d\mu_{(n-p+3)}^{\rm CHY}
\hspace{-0.6cm}
\parbox[c]{5.5em}{\includegraphics[scale=0.17]{R1-cutp.pdf}} 
~~
\hspace{-0.1cm}
\int
d\mu_{(p-1)}^{\rm CHY}
\hspace{-0.5cm}
\parbox[c]{5.5em}{\includegraphics[scale=0.17]{R2-cutp.pdf}} 
= 2\,
\sum_L
\int
d\mu_{(n-p+3)}^{\rm CHY}
\hspace{-0.6cm}
\parbox[c]{5.4em}{\includegraphics[scale=0.17]{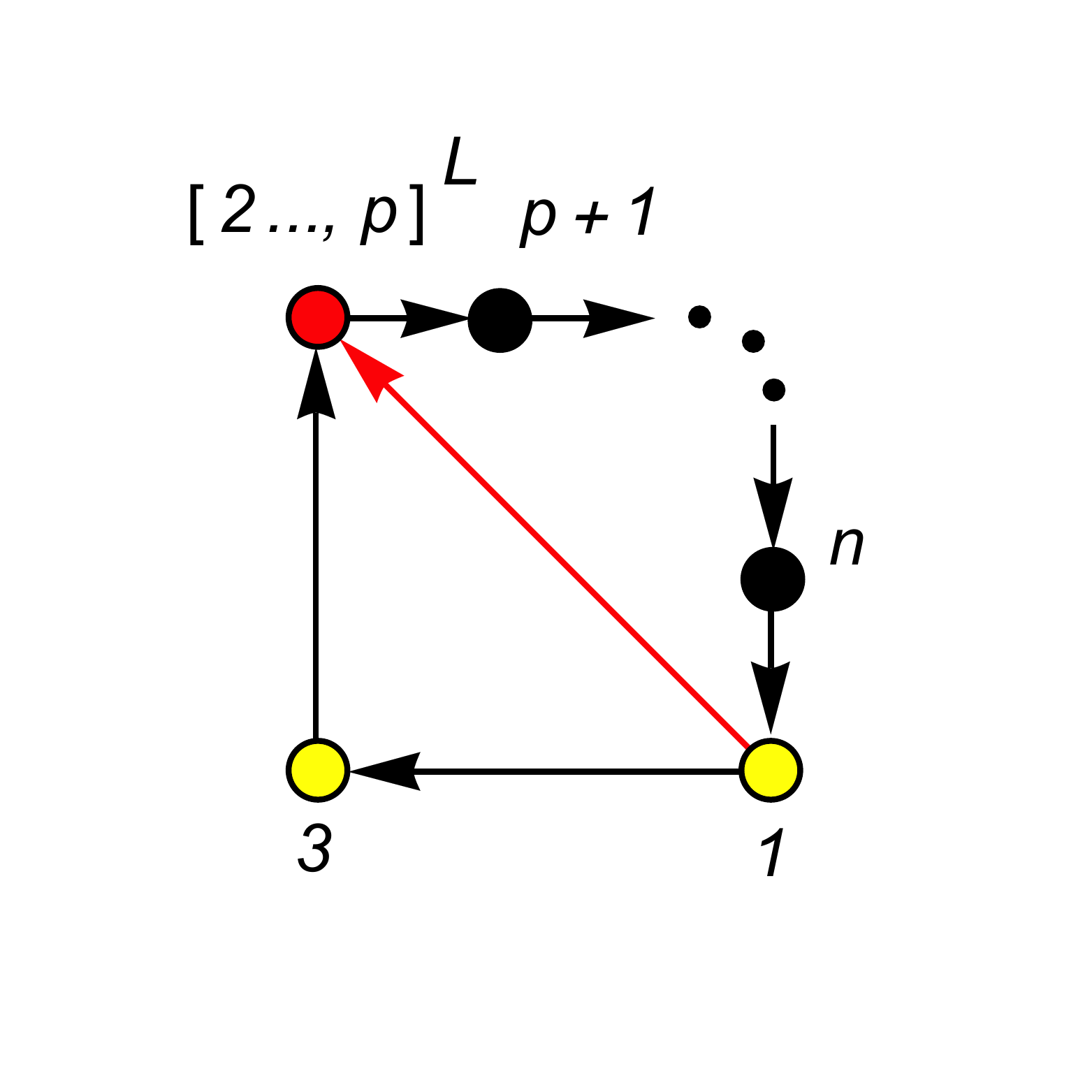}} 
~~
\hspace{-0.1cm}
\int
d\mu_{(p-1)}^{\rm CHY}
\hspace{-0.4cm}
\parbox[c]{5.3em}{\includegraphics[scale=0.17]{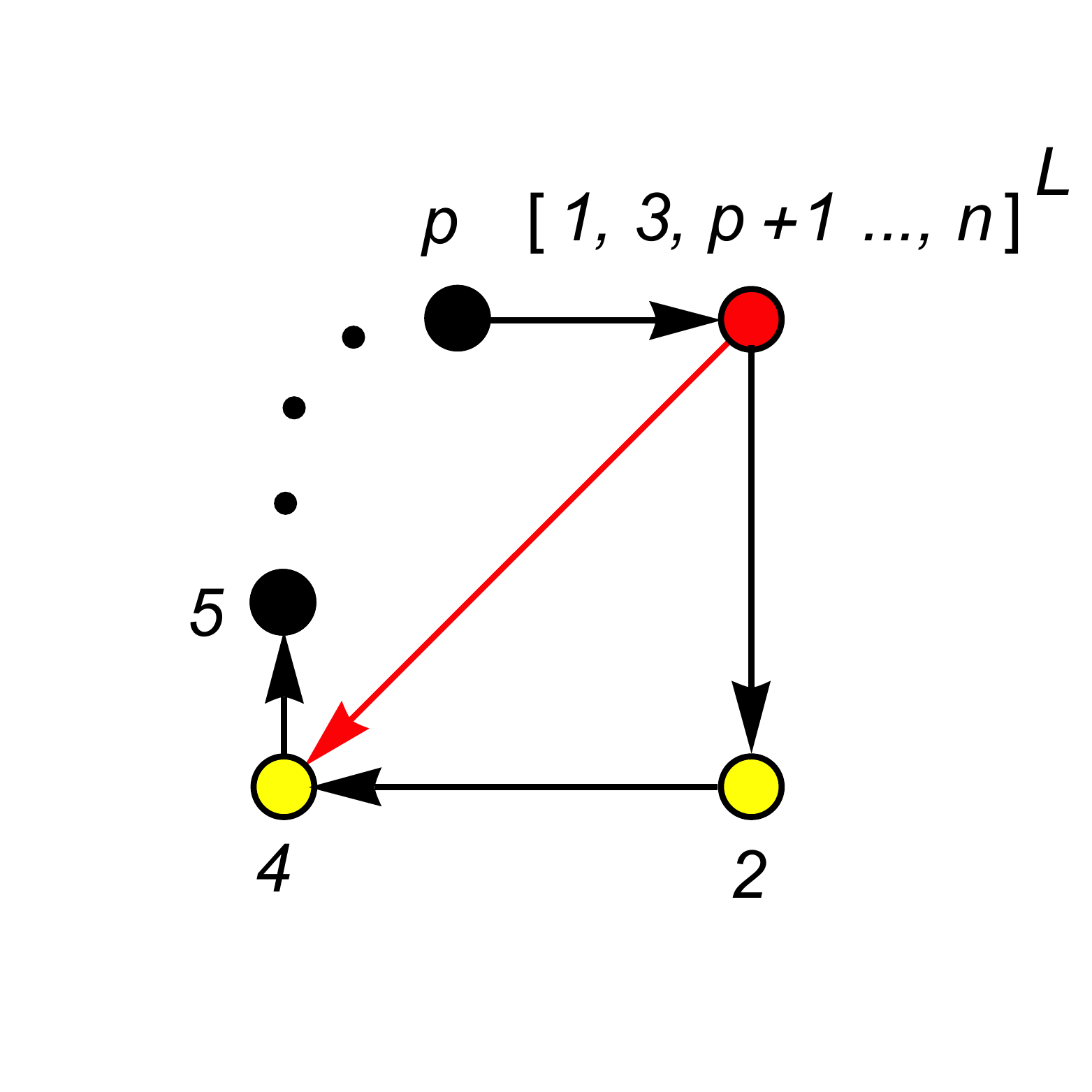}} ,\,\,~~~ \nonumber\\
\end{eqnarray}
}
\vskip-1.3cm\noindent
with\footnote{This identity can be easily extended to another setup.}, 
$\sum_{L}  \eps_{ [2,4,5,...,p]}^{L,\mu} \, \eps^{L,\nu}_{ [1,3,p+1,...,n]} = \frac{k^{\mu}_{[2,4,5,...,p]}\, k^{\nu}_{[1,3,p+1,...,n]}}{ k_{[2,4,5,...,p]}\cdot k_{[1,3,p+1,...,n]}}$.

It is clear that this equality give us the following physical interpretation for the {\it strange cuts}
\vspace{-0.1cm}
\begin{itemize}
\item {\it All strange-cuts can be rewritten as a product of two YM-graphs, which must be glued by a longitudinal gluon.
}
\end{itemize}
\vspace{-0.0cm}\noindent
For example,  it is trivial to verify that the result found in \eqref{4p-cut3} for  {\it cut-3} can be rewritten as,
{\it cut-3} $=\left(\frac{2}{\tilde s_{24}} \right) \sum_{L} A_3^{([2,4],1)}([2,4]^L,1,3)\times  A_3^{(4,[1,3])}(4,[1,3]^L,2) =  (\eps_1\cdot \eps_3)  (\eps_2\cdot \eps_4)$, where the three-point building blocks are gluing by the identity, $\sum_{L}  \eps_{\rm [2,4]}^{L,\mu} \, \eps^{L,\nu}_{\rm [1,3]} = \frac{k^{\mu}_{[2,4]}\, k^{\nu}_{[1,3]}}{k_{[2,4]}\cdot k_{[1,3]}}$ ({\bf longitudinal gluons}).

\subsection{Transverse Gluons}

On the other hand, despite to the result obtained in \eqref{generalone}, we still do not know how to deal with some resulting graphs, for example, the four-point in \eqref{4ptsA}.

Nevertheless, from the {\bf properties-III, IV} in appendix \ref{appendix}, it is straightforward to obtain the following identities for a general {\it standard cut}
\vspace{-0.1cm}
{\small
\begin{eqnarray}\label{}
\hspace{-2.3cm}
\left.
\int
d\mu^{\rm CHY}_{(p-1)}
\hspace{-0.5cm}
\parbox[c]{6.8em}{\includegraphics[scale=0.17]{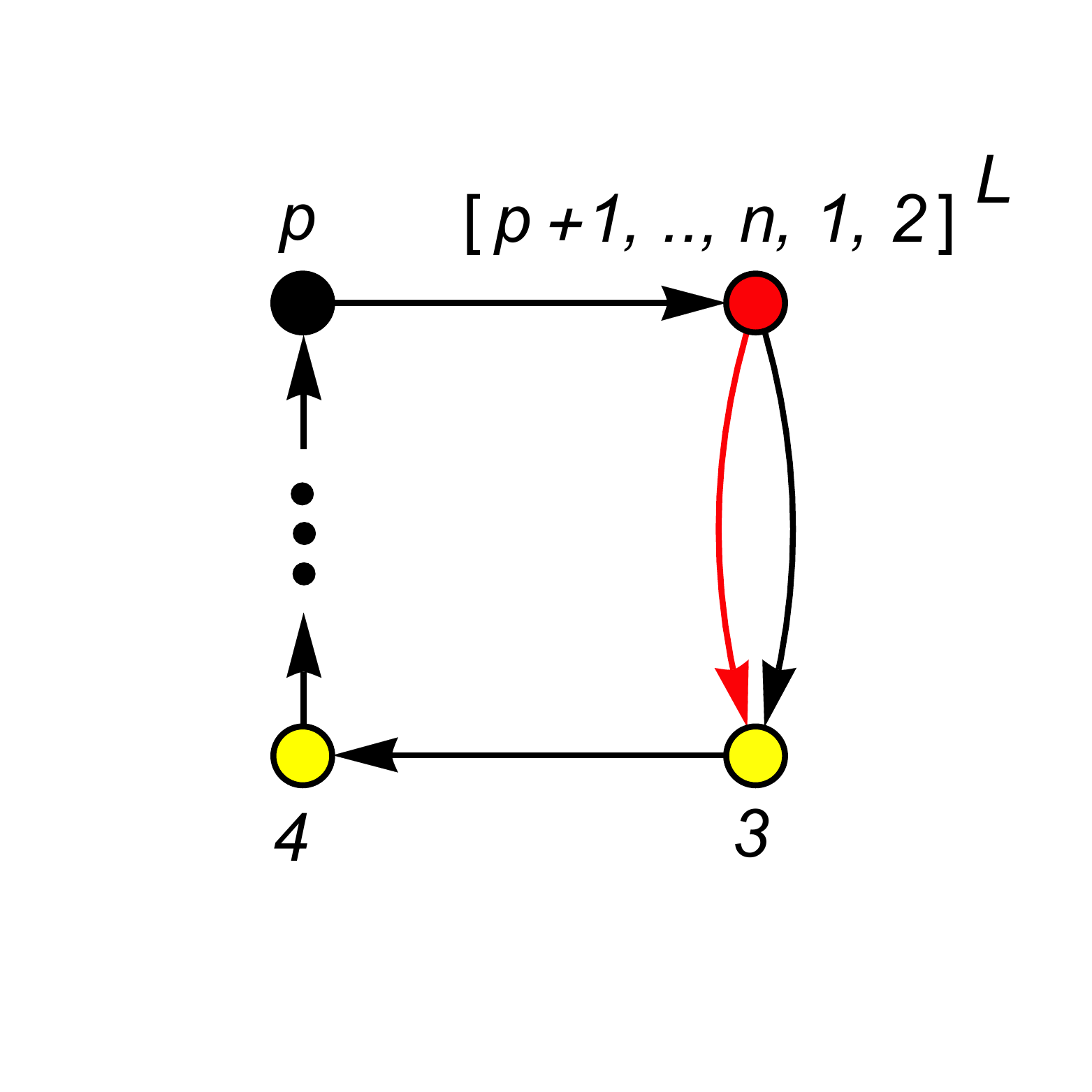}} 
\right|_{\eps^{L,\mu}_{[p+1...n,1,2]} \rightarrow k_{[p+1...n,1,2]}^\mu }
\hspace{-2.9cm}
=\,
(-)
\left.
\int
d\mu^{\rm CHY}_{(p-1)}
\hspace{-0.5cm}
\parbox[c]{6.9em}{\includegraphics[scale=0.17]{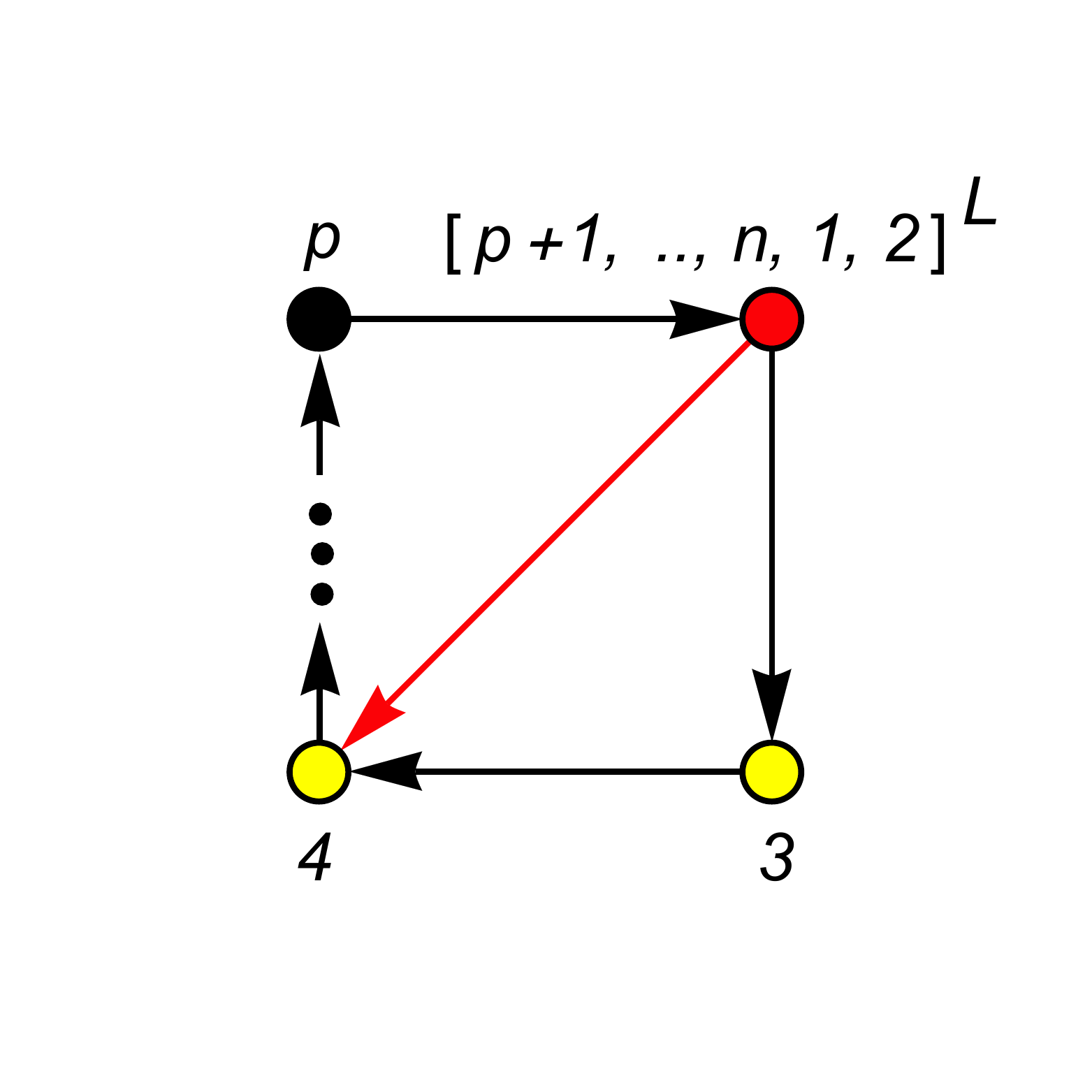}} 
\right|_{\eps^{L,\mu}_{[p+1...n,1,2]} \rightarrow k_{[p+1...n,1,2]}^\mu }
\hspace{-2.8cm} ,
\,\,\quad
\nonumber
\end{eqnarray}
}
\vskip-0.4cm\noindent
\vspace{-0.8cm}
{\small
\begin{eqnarray}\label{}
\hspace{-0.1cm}
\sum_T
\int d\mu^{\rm CHY}_{(n-p+3)}
\hspace{-0.5cm}
\parbox[c]{4.9em}{\includegraphics[scale=0.17]{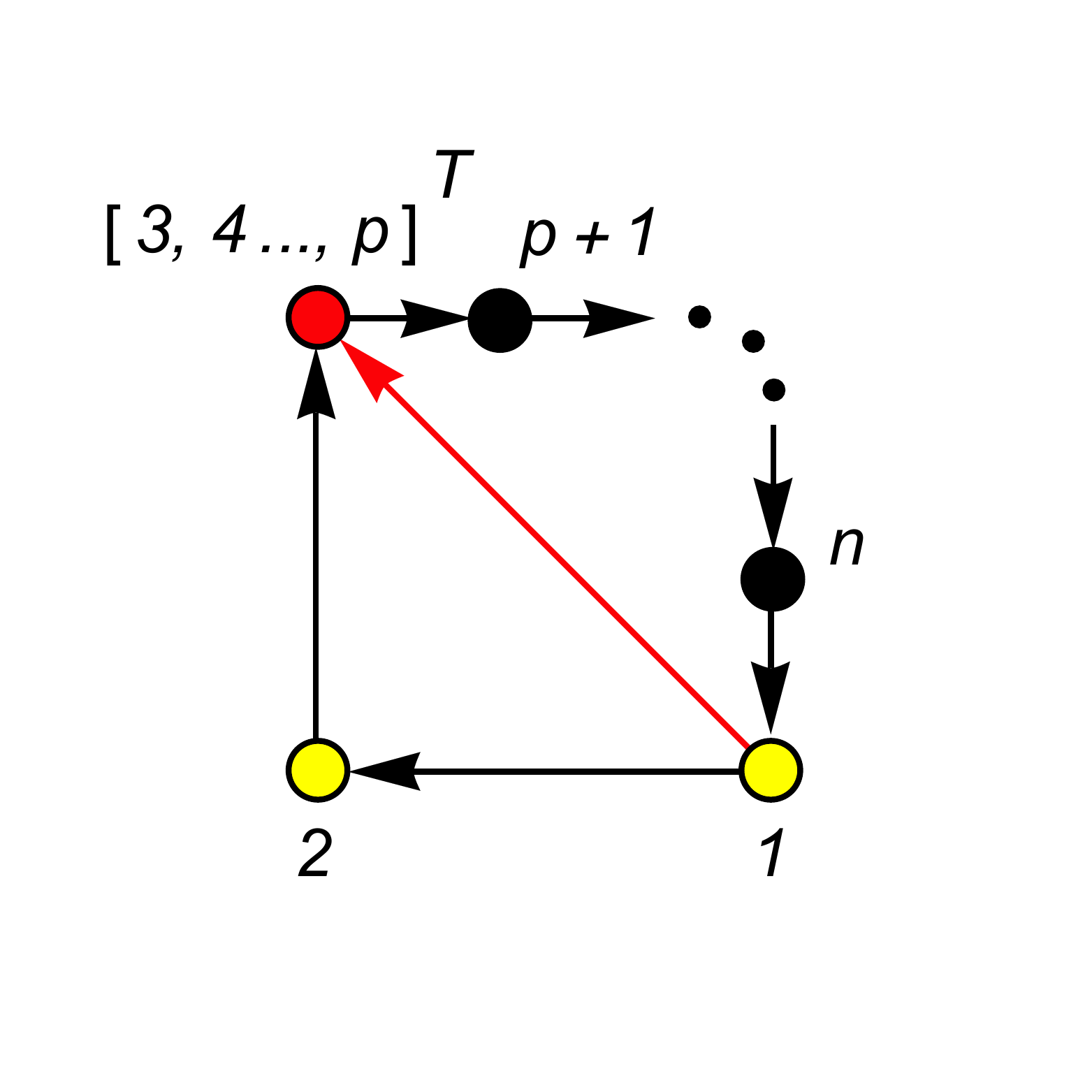}} 
~~~
\hspace{-0.1cm}
\int d\mu^{\rm CHY}_{(p-1)}
\hspace{-0.5cm}
\parbox[c]{5.0em}{\includegraphics[scale=0.17]{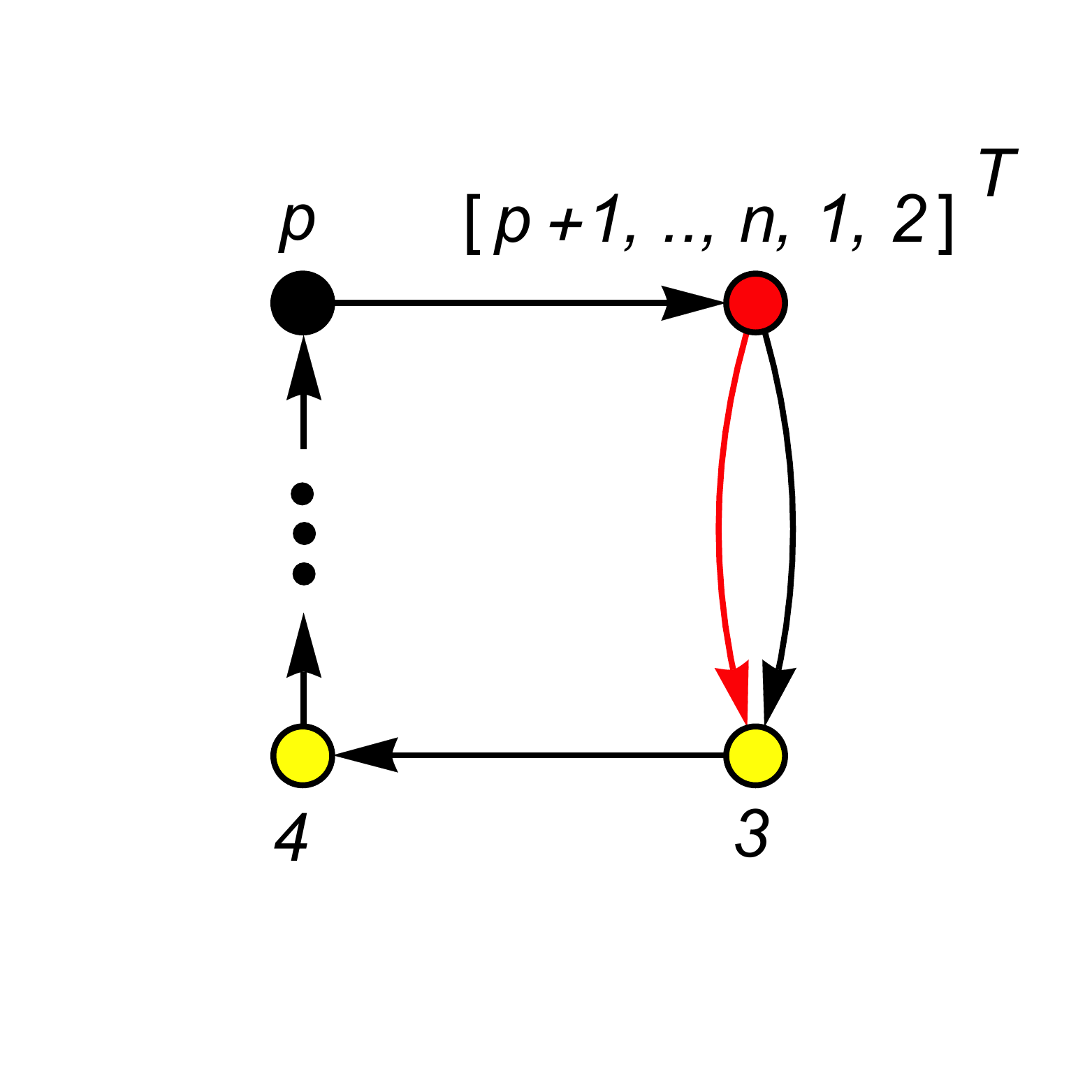}} 
=\,
\sum_T
\int d\mu^{\rm CHY}_{(n-p+3)}
\hspace{-0.5cm}
\parbox[c]{4.8em}{\includegraphics[scale=0.17]{R1A-cutnT.pdf}} 
~~~
\int d\mu^{\rm CHY}_{(p-1)}
\hspace{-0.5cm}
\parbox[c]{5.6em}{\includegraphics[scale=0.17]{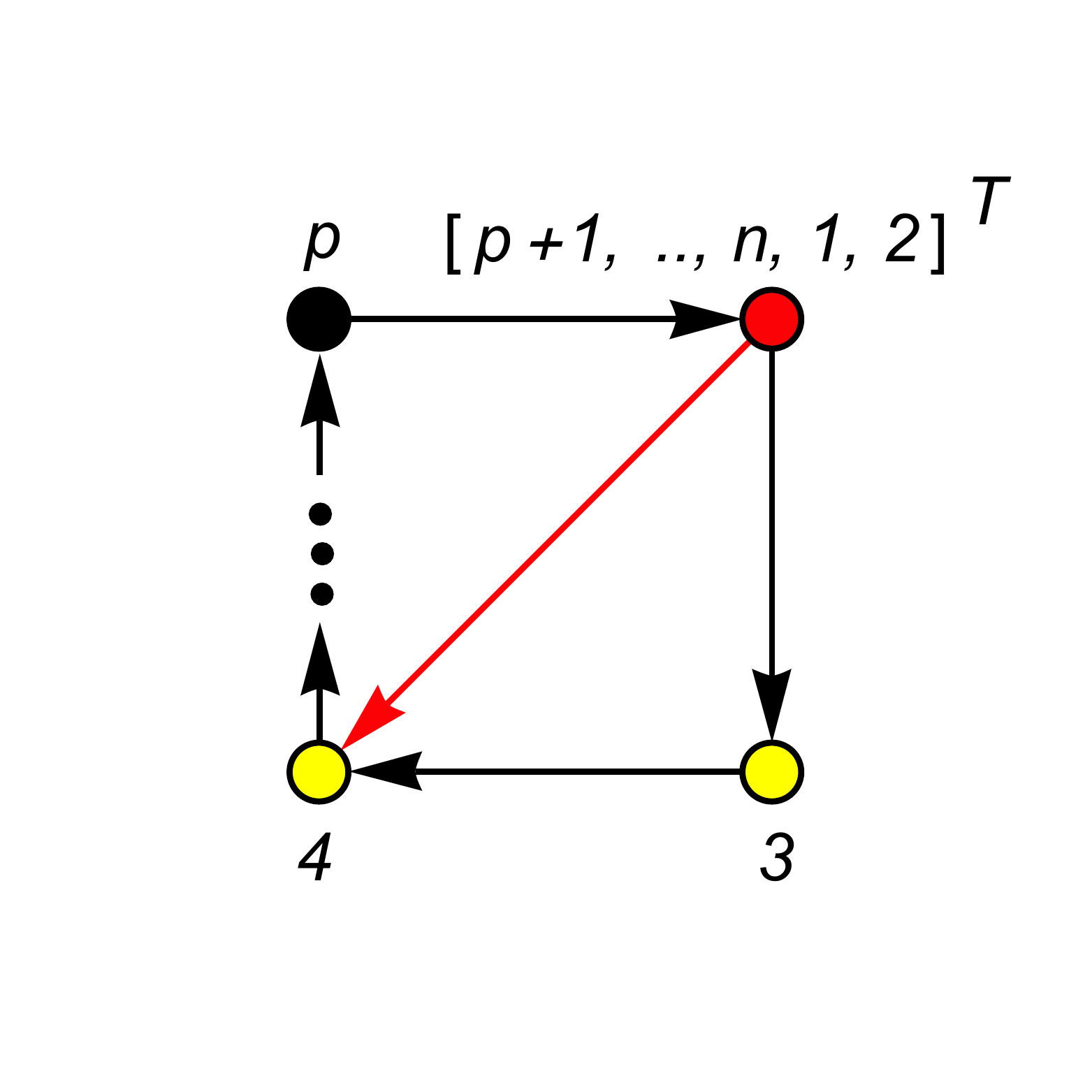}} ,\,\,\quad
\nonumber
\end{eqnarray}
}
\vskip-0.9cm\noindent
\vspace{-0.9cm}
{\small
\begin{eqnarray}\label{generaltwo}
\hspace{-0.1cm}
\sum_r
\int d\mu^{\rm CHY}_{(n-p+3)}
\hspace{-0.5cm}
\parbox[c]{4.9em}{\includegraphics[scale=0.17]{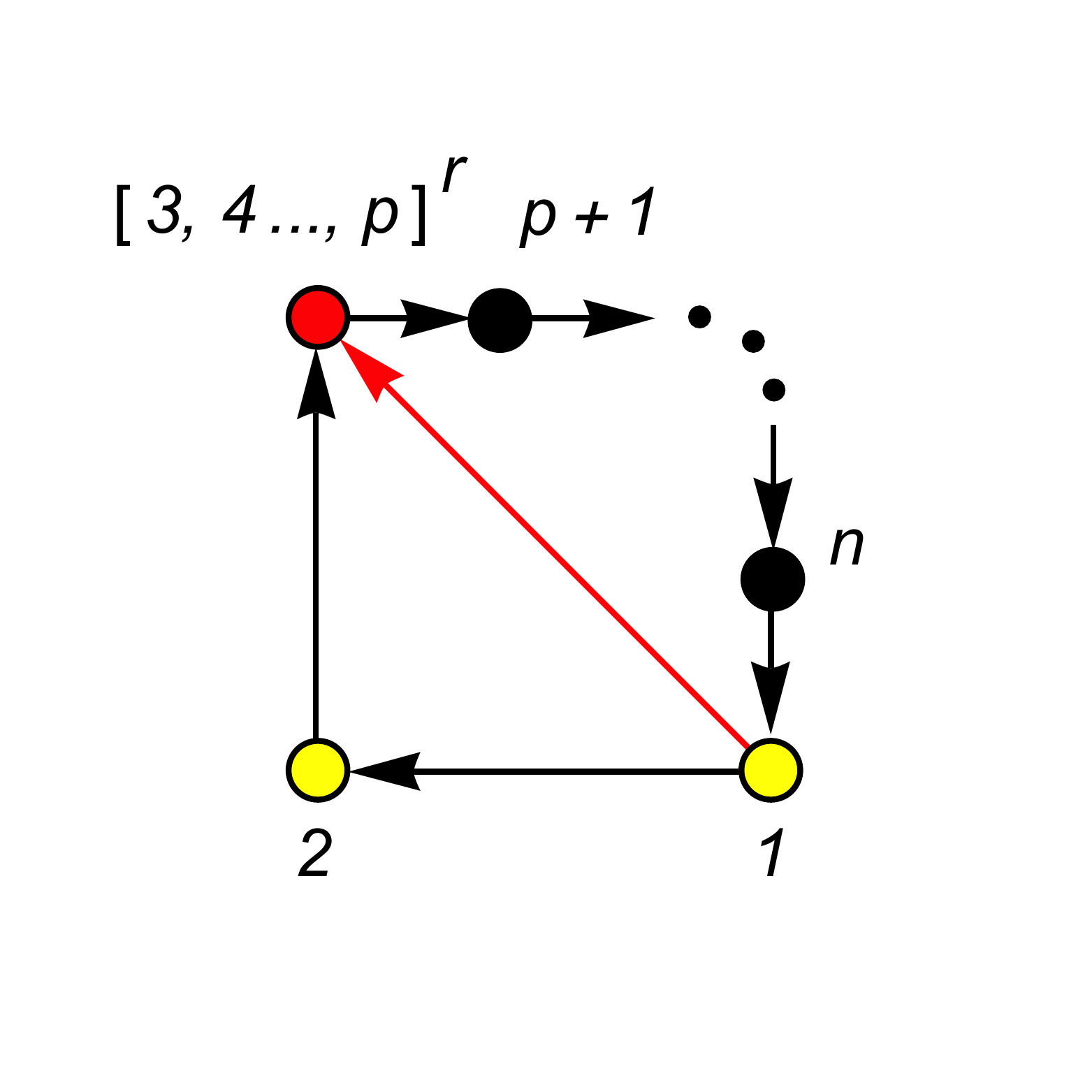}} 
~~~
\hspace{-0.1cm}
\int d\mu^{\rm CHY}_{(p-1)}
\hspace{-0.5cm}
\parbox[c]{5.0em}{\includegraphics[scale=0.17]{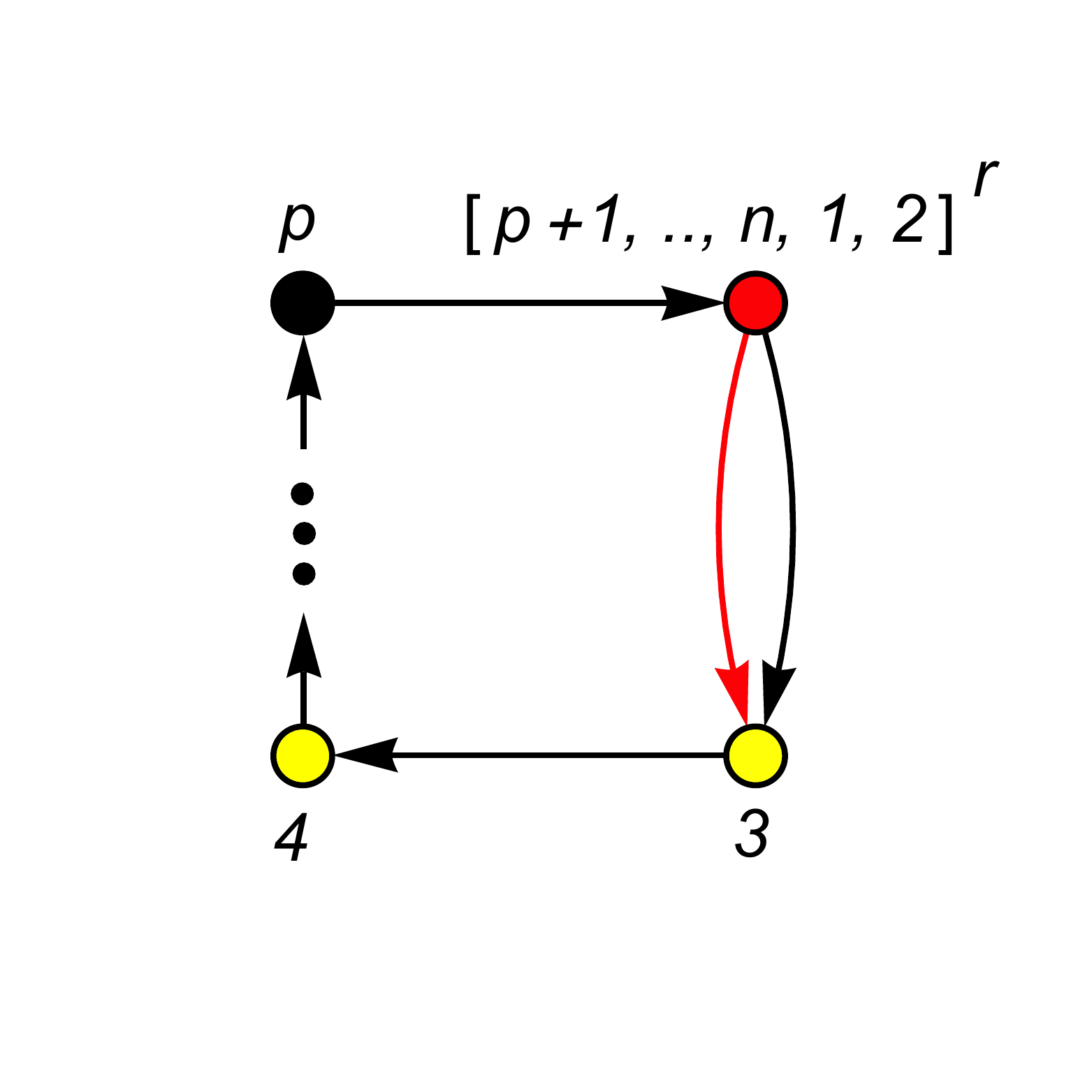}} 
=\,
\sum_A
\int d\mu^{\rm CHY}_{(n-p+3)}
\hspace{-0.5cm}
\parbox[c]{4.8em}{\includegraphics[scale=0.17]{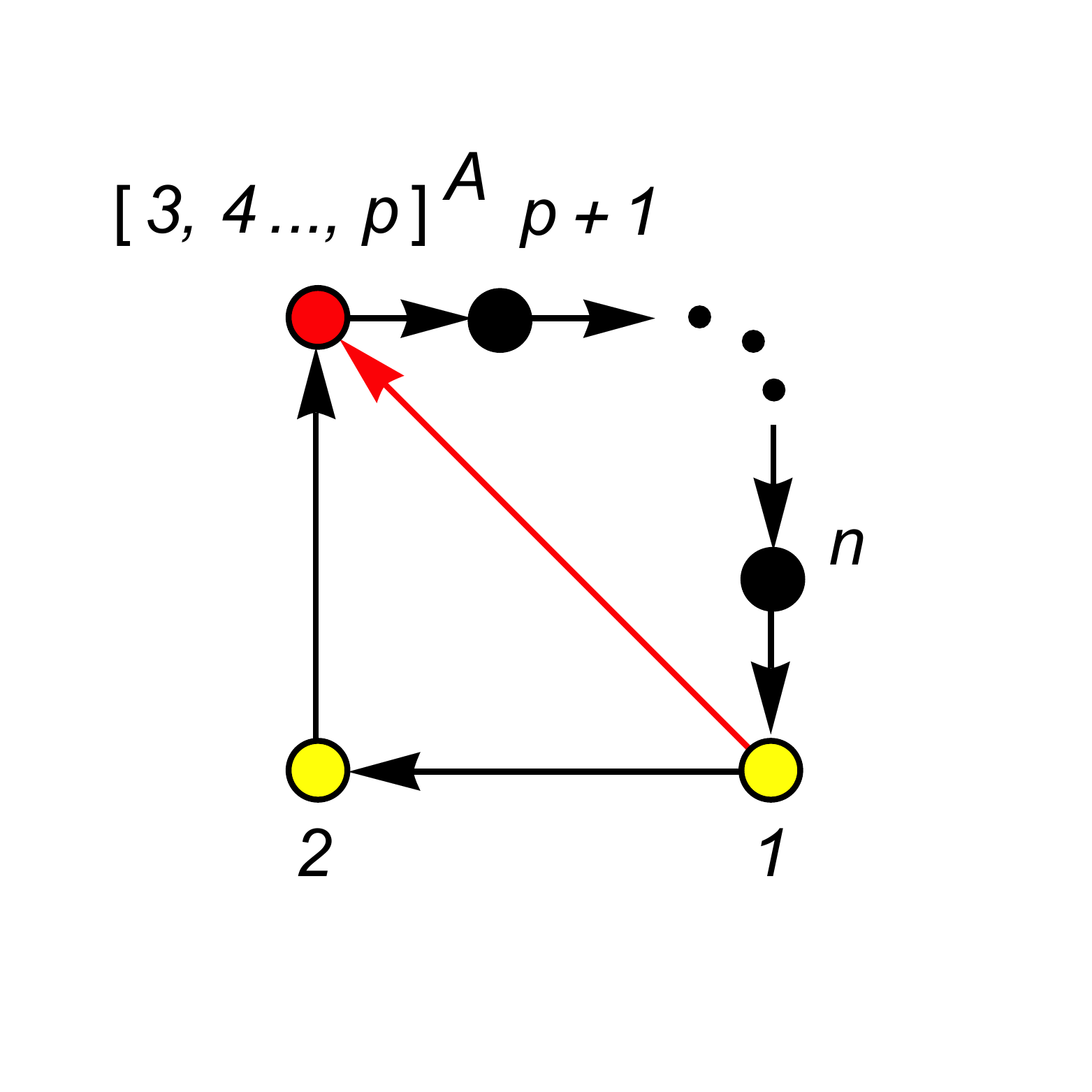}} 
~~~
\int d\mu^{\rm CHY}_{(p-1)}
\hspace{-0.5cm}
\parbox[c]{5.6em}{\includegraphics[scale=0.17]{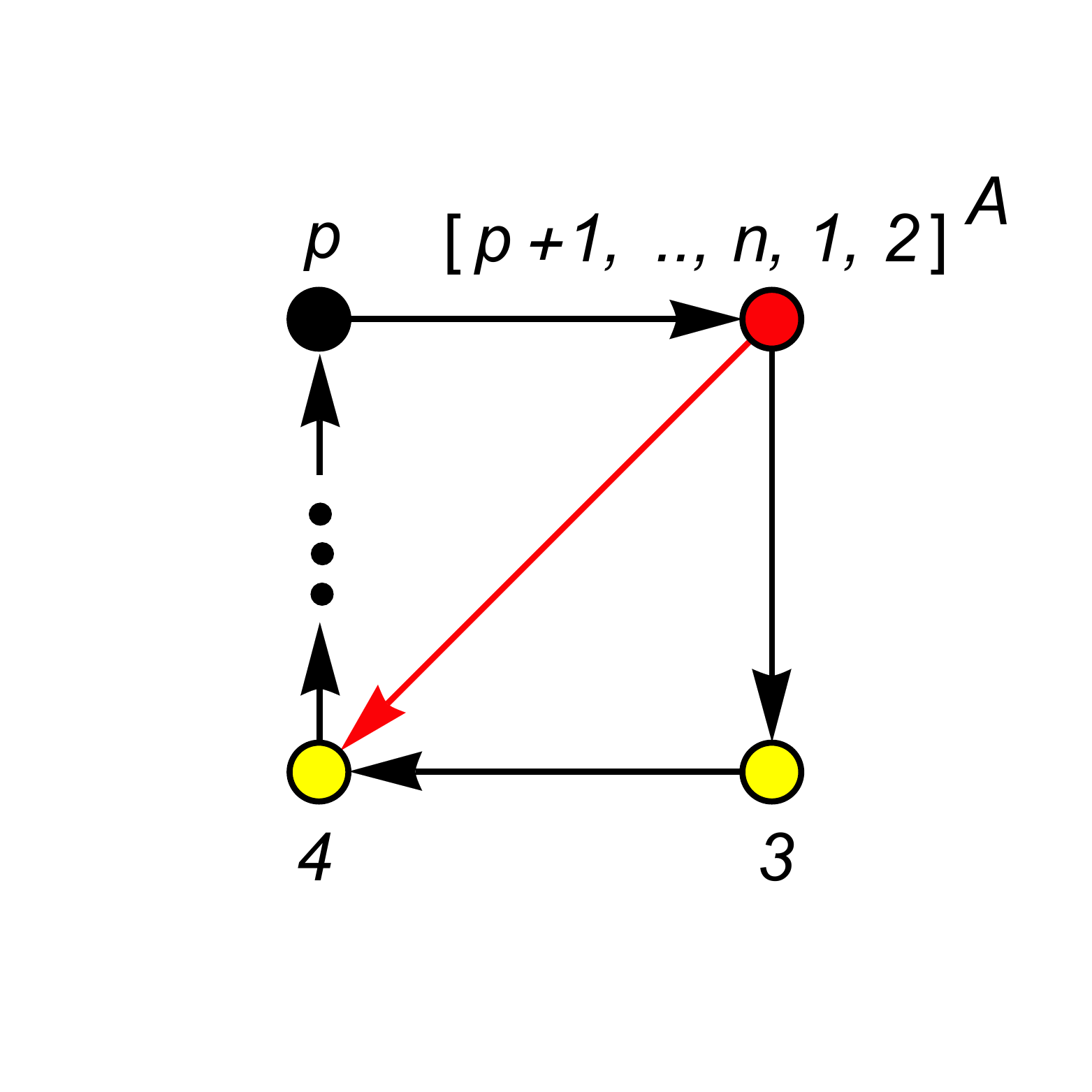}} ,\,\,\quad
\nonumber\\
\end{eqnarray}
}
\vskip-1.2cm\noindent
where we have picked up as the initial setup, $(pqr|m)=(123|4)$, the red arrow from, $(i,j)=(1,3)$, and
the gluing identities are given by, 
$\sum_{r}  \eps_{\rm [3,4,...,p]}^{r,\mu} \, \eps^{r,\nu}_{\rm [p+1,...,n,1,2]} = \eta^{\mu\nu} $, 
$\sum_{T}  \eps_{\rm [3,4,...,p]}^{T,\mu} \, \eps^{T,\nu}_{\rm [p+1,...,n,1,2]} = \eta^{\mu\nu} -\frac{k^{\mu}_{[3,4,...,p]}\, k^{\nu}_{[p+1,...,n,1,2]}}{k_{[3,4,...,p]}\cdot k_{[p+1,...,n,1,2]}}$ 
and
$\sum_{A}  \eps_{\rm [3,4,...,p]}^{A,\mu} \, \eps^{A,\nu}_{\rm [p+1,...,n,1,2]} = \eta^{\mu\nu} -\frac{2\,k^{\mu}_{[3,4,...,p]}\, k^{\nu}_{[p+1,...,n,1,2]}}{k_{[3,4,...,p]}\cdot k_{[p+1,...,n,1,2]}}$. Notice that
the last equality is a consequence of the first two ones.

For example, it is not hard to check (numerically) the equality
\vspace{-0.4cm}
{\small
\begin{eqnarray}\label{conjec2}
\hspace{-0.1cm}
\sum_{r}
\hspace{-0.5cm}
\parbox[c]{4.8em}{\includegraphics[scale=0.17]{cut2-R1.pdf}} 
~~
\int
d\mu_4^{\rm CHY}
\hspace{-0.5cm}
\parbox[c]{5.5em}{\includegraphics[scale=0.17]{cut2-R2.pdf}} 
=
\sum_{A}
\hspace{-0.5cm}
\parbox[c]{4.8em}{\includegraphics[scale=0.17]{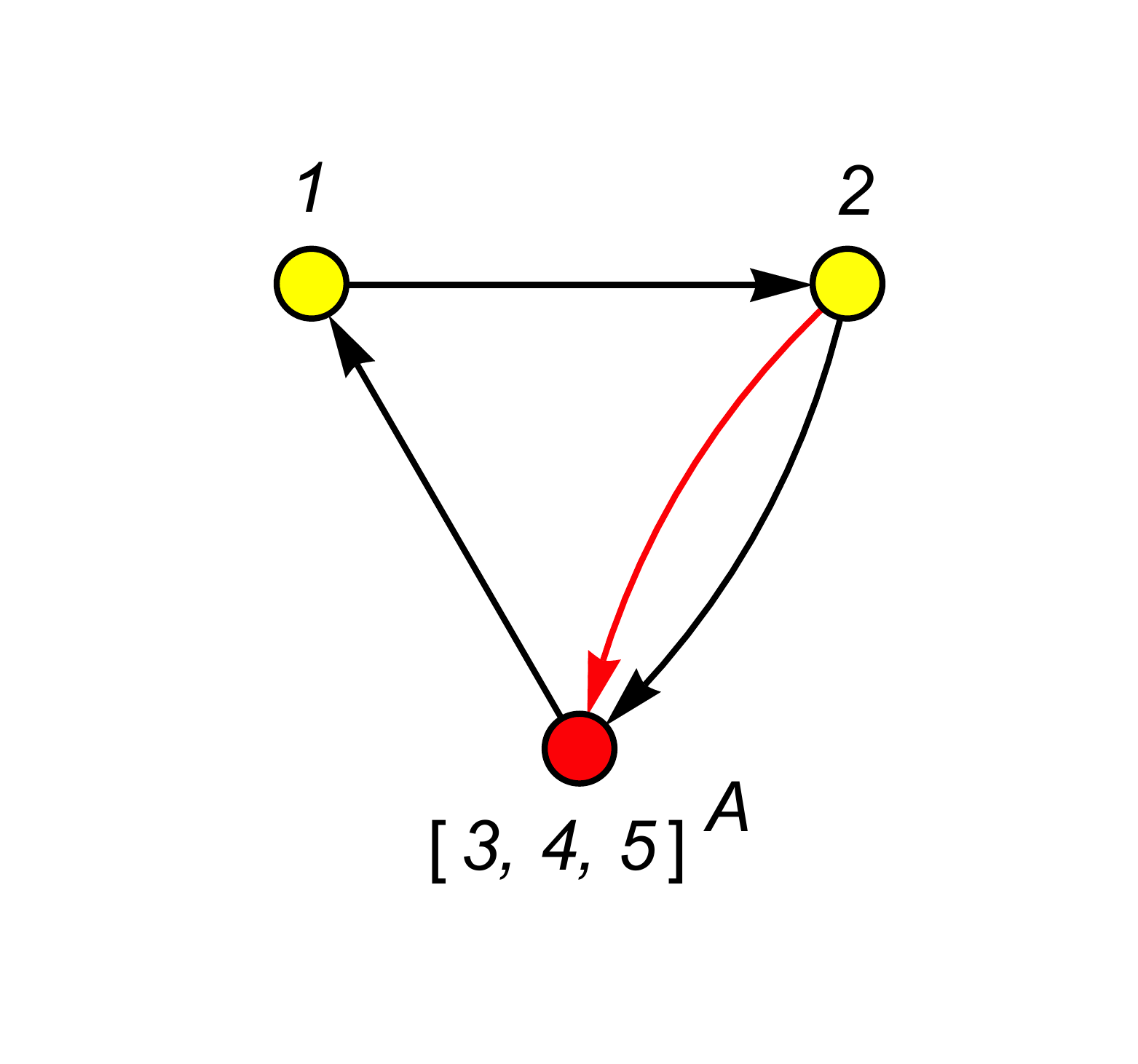}} 
~~
\int
d\mu_4^{\rm CHY}
\hspace{-0.5cm}
\parbox[c]{5.9em}{\includegraphics[scale=0.17]{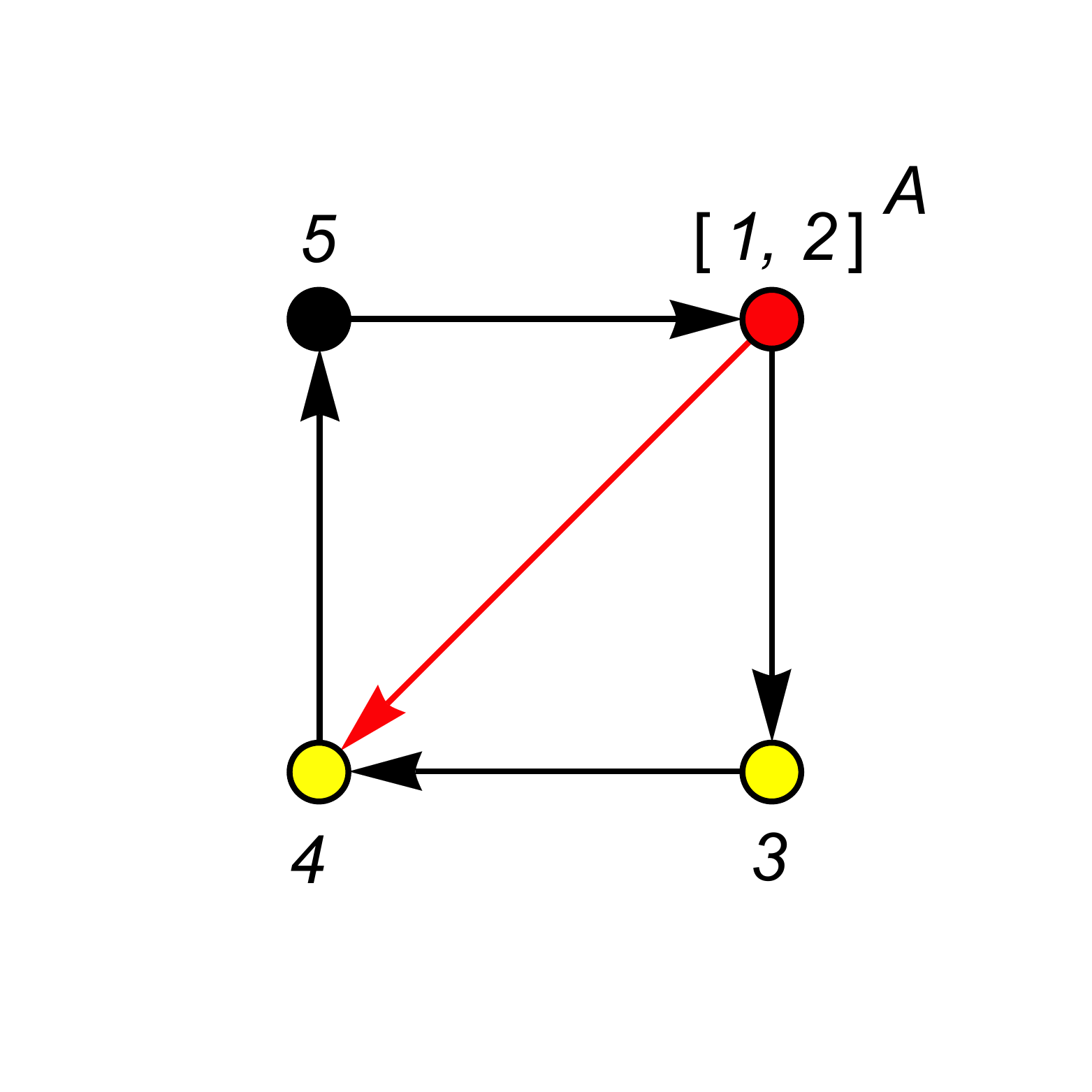}} 
 ,\qquad~
\end{eqnarray}
}
\vskip-0.5cm\noindent
where, $\sum_{r}  \eps_{\rm [3,4,5]}^{r,\mu} \, \eps^{r,\nu}_{\rm [1,2]} = \eta^{\mu\nu} $ and 
$\sum_{A}  \eps_{\rm [3,4,5]}^{A,\mu} \, \eps^{A,\nu}_{\rm [1,2]} = \eta^{\mu\nu} -\frac{2\, k^{\mu}_{[3,4,5]}\, k^{\nu}_{[1,2]}}{k_{[3,4,5]}\cdot k_{[1,2]}}$. This is very important to observe that the {\bf integration rules} can not be applied on the left-hand side, while that on the right-hand side they work perfectly.

Up to this point, under the identities set down in this section, we are able  to write an ordered on-shell YM-amplitude as a sum of the product of two smaller partial off-shell YM-amplitudes, which must be glued by off-shell gluons. However,  it is important to remark that those graphical identities proposed here involve only one  red-vertex over each resulting graph (off-shell gluon).  In the next section, we will refine the gluing process to go beyond more than one off-shell puncture.

\subsection{More off-shell gluons}\label{sectionMOFF}

Since the above  identities can only be applied to graphs with one off-shell particle, we need to discuss what happens when there is more than one red puncture; this will help us to  develop a  graph-algorithm for more general cases. In other words, we would like to generalize the properties given in appendix \ref{appendix}.

Let us start by considering the following simple example, the {\it cut-2} given in \eqref{4ptsA}. Using  \eqref{generaltwo}, it is enough just to focus on the graph
\vspace{-0.6cm}
\begin{eqnarray}\label{off-ex1}
\hspace{-0.1cm}
\int d\mu_4^{\rm CHY}
\hspace{-0.5cm}
\parbox[c]{5.3em}{\includegraphics[scale=0.17]{cut2-R2A.pdf}} 
\rightarrow
\int d\mu_4^{\L}
\hspace{-0.5cm}
\parbox[c]{6.3em}{\includegraphics[scale=0.17]{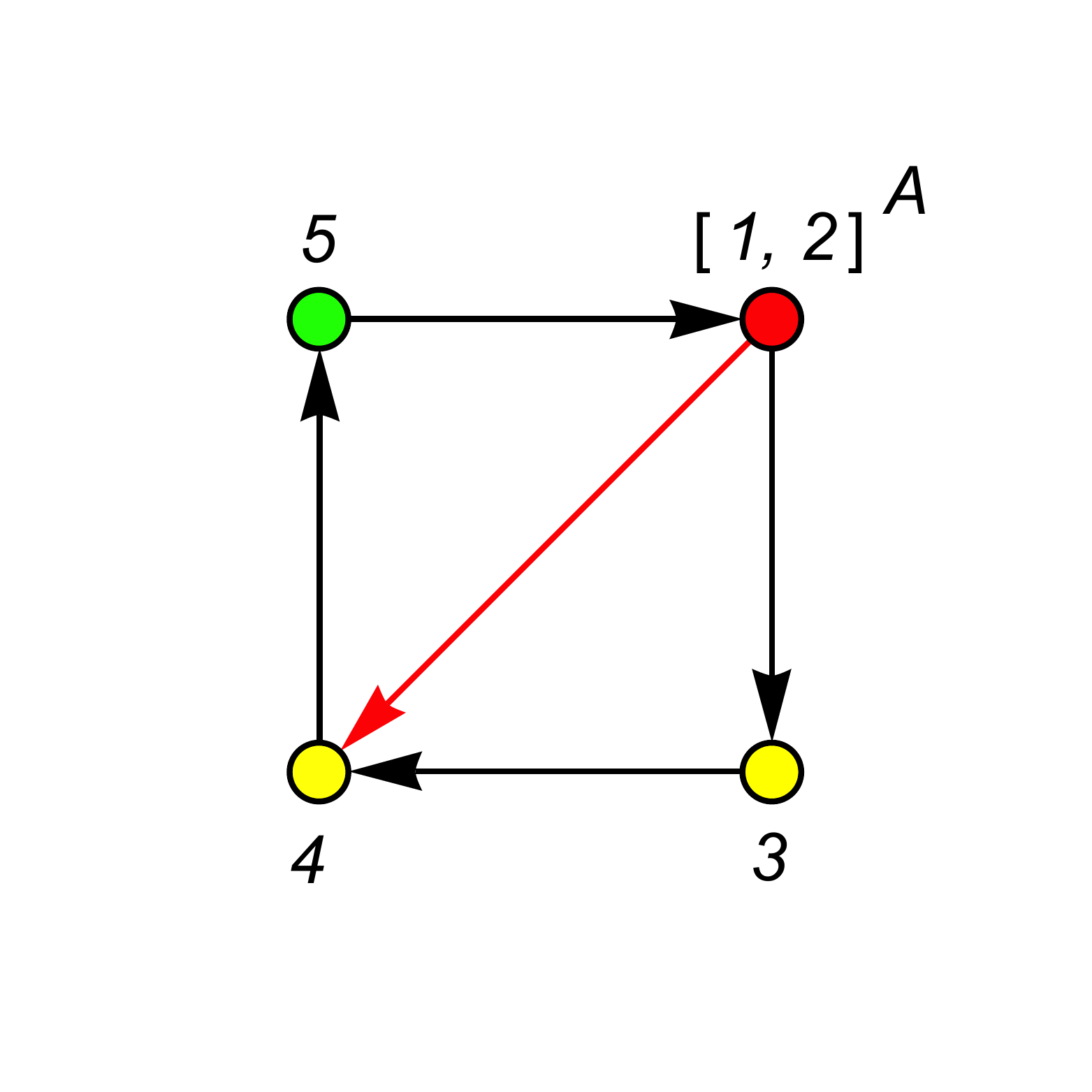}} .
\end{eqnarray}
\vskip-0.5cm\noindent
Such as in \eqref{Rcut1}, applying the {\bf integration  rules} and from the expansion in \eqref{fpL0}, this graph turns into 
\vspace{-0.4cm}
{\small
\begin{eqnarray}\label{}
\int d\mu_4^{\L}
\hspace{-0.4cm}
\parbox[c]{5.5em}{\includegraphics[scale=0.17]{cut2-R2AG.pdf}} 
=\sum_r\left[\frac{A_{3}^{([1,2],[3,4])}([1,2]^A, [3,4]^r,5)\, A_{3}^{(4,[5,1,2])}( 4, [5,1,2]^r,3) }{\tilde s_{5[1,2]}} 
\right.
 \nonumber 
\end{eqnarray}
}
\vskip-1.6cm\noindent
{\small
 \begin{eqnarray}\label{strangec-2off}
\left.
+ \frac{A_{3}^{([1,2,3],4)}([1,2,3]^r,4,5) \times  A_{3}^{([4,5],[1,2])}([4,5]^r,[1,2]^A,3)  }{\tilde s_{54}} 
 \right]
 +
\hspace{-0.5cm}
\parbox[c]{5.8em}{\includegraphics[scale=0.17]{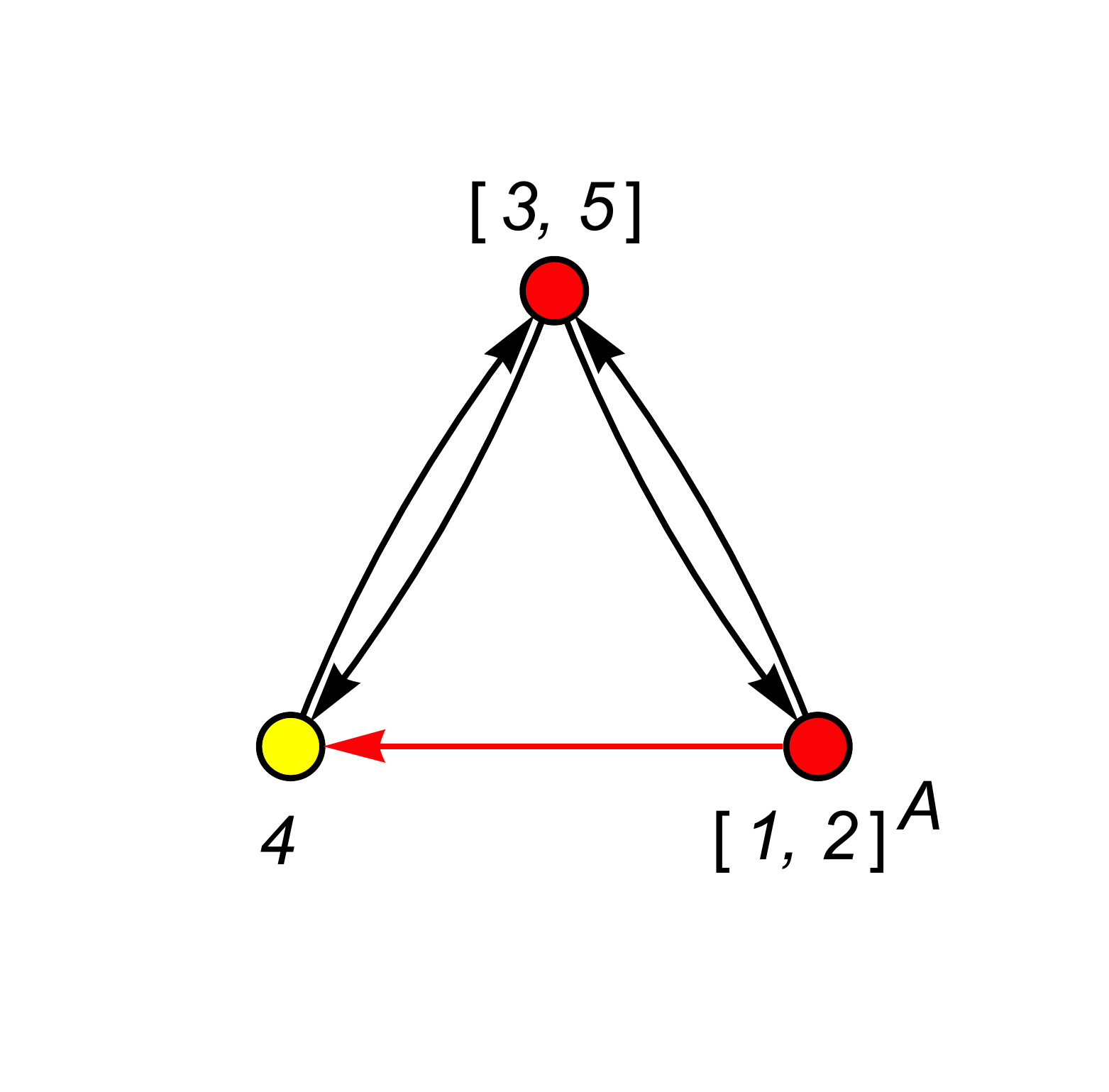}} 
\times
\left(
 \frac{1}{\tilde s_{53}} 
 \right)
 \times
\hspace{-0.55cm}
\parbox[c]{5.6em}{\includegraphics[scale=0.17]{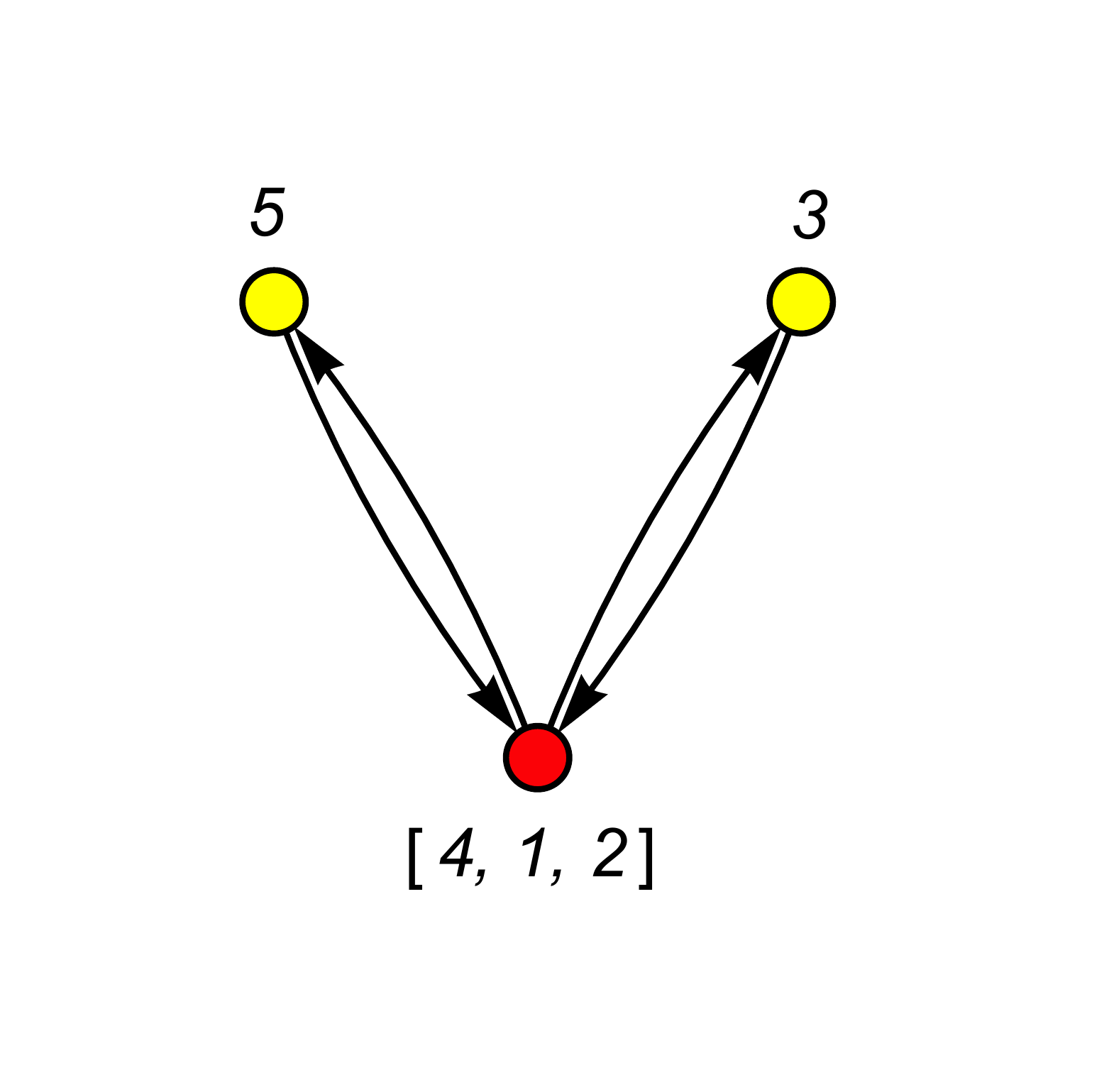}} ,
~~~~~
\end{eqnarray}
}
\vskip-0.55cm\noindent
with, $\sum_r \eps_{[3,4]}^{r,\mu} \eps_{[5,1,2]}^{r,\nu} = \eta^{\mu\nu}$ and  $\sum_r \eps_{[1,2,3]}^{r,\mu} \eps_{[4,5]}^{r,\nu} = \eta^{\mu\nu}$.  The computation of these terms is straightforward and the final result has been checked numerically.
Note the emergence of the spurious pole, $\tilde s_{5[1,2]}=k_5\cdot (k_1+k_2)$, which does not appear in any known method before\footnote{These kind of poles are a direct consequence of the scattering equations, similar to the linear propagators that appear at loop level \cite{Casali:2014hfa,Geyer:2015bja,Cardona:2016bpi,Cachazo:2015aol,Baadsgaard:2015hia,Cardona:2016wcr,Gomez:2016cqb,Feng:2016nrf,Geyer:2015jch,Geyer:2016wjx}.}. 

This example showed us that the {\bf integration rules} work perfectly over a YM-graph with one off-shell particle ($k_{[1,2]}^2\neq 0$ and $\eps_{[1,2]}^A\cdot k_{[1,2]}\neq 0$). In fact, we have successfully tested them over bigger graphs and with more than one off-shell particles\footnote{In this  approach  is enough  to consider up three off-shell particles (the ${\rm PSL}(2,\mathbb{C})$ symmetry). In order to extend these ideas to more off-shell particles, we must introduce the off-shell scattering equations.}. Therefore, we claim that our graph-method is recursive over YM-graphs. Now, we want to know what happens with the {\it strange-cuts}, i.e. Can the conjecture in \eqref{generalone} be generalized to more than one off-shell puncture?  

In order to answer this question, let us consider the first {\it strange-graph} in \eqref{strangec-2off},
\vspace{-0.5cm}
\begin{eqnarray}\label{offex1}
\hspace{-0.6cm}
\parbox[c]{5.7em}{\includegraphics[scale=0.17]{R1-cut35.pdf}} 
=\,
(\eps_{[1,2]}^A\cdot \eps_4)\,\, .
\end{eqnarray}
\vskip-0.5cm\noindent
Following the {\bf properties-I,II} (appendix \ref{appendix}), we are interested to compute the graph
\vspace{-0.2cm}
{\small
\begin{eqnarray}\label{epsAtrans}
\left.
\hspace{-0.6cm}
\parbox[c]{6.5em}{\includegraphics[scale=0.17]{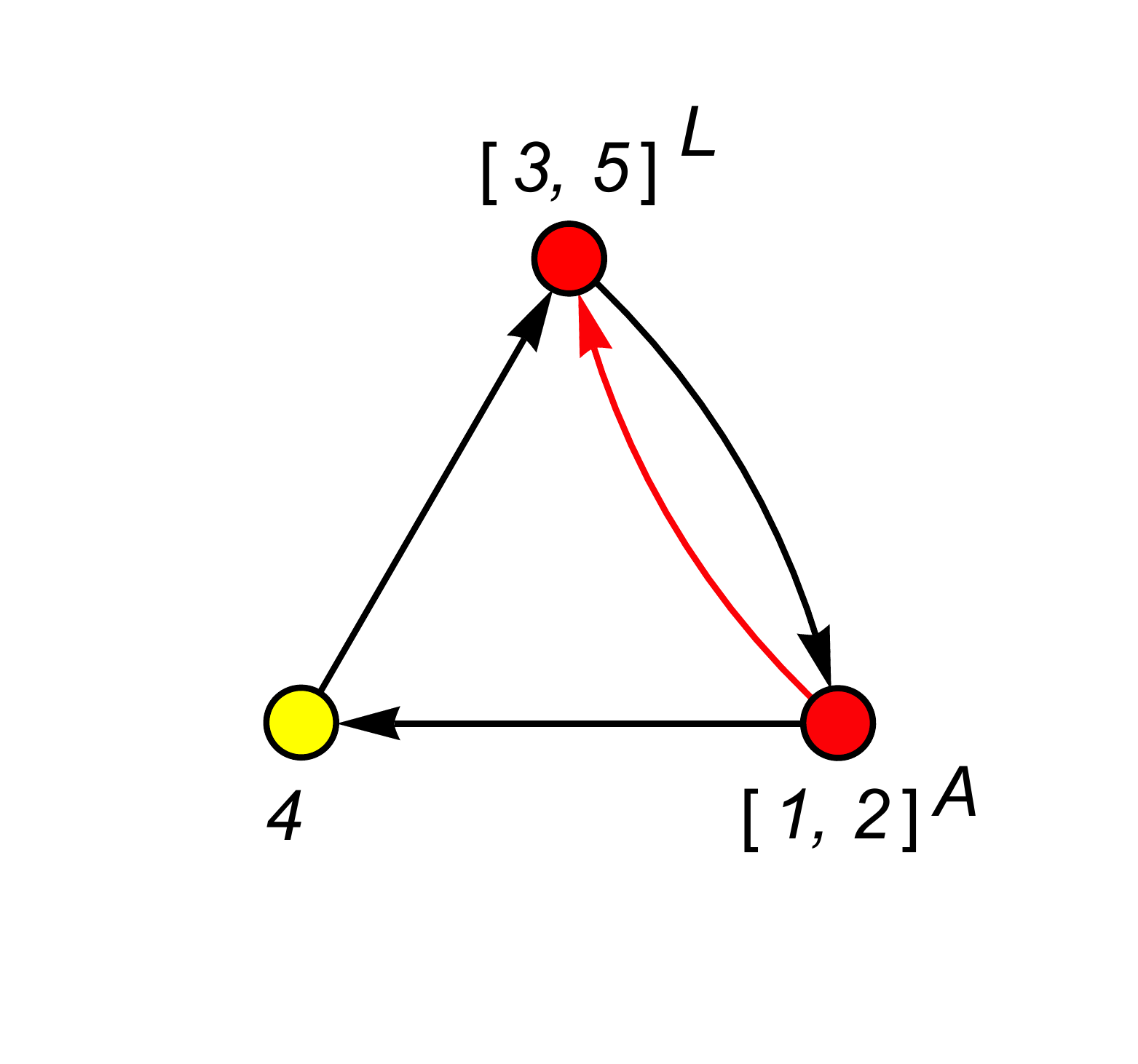}} \right|_{\eps_{[3,5]}^{L,\mu}\rightarrow k^\mu_{[3,5]}}
\hspace{-1.4cm}
=\,
(\eps_{4}\cdot k_{[3,5]})\,(\eps^A_{[1,2]}\cdot k_{[3,4,5]})
+
\frac{k_{[3,5]}^2 - k_{[1,2]}^2}{2}\,(\eps_{[1,2]}^A\cdot \eps_4) .
\end{eqnarray}
}
\vskip-0.25cm\noindent
Notice that, although $\epsilon_{[1,2]}^{A,\mu}$ is not necessarily a transverse polarization vector, 
we can impose the condition, $\epsilon_{[1,2]}^{A}\cdot k_{[1,2]}=0$, 
since the desired result is independent of terms with the form, $\epsilon_{[1,2]}^{A}\cdot k_i$. So, we now are able to reproduce \eqref{offex1} from \eqref{epsAtrans}.  This result is easily generalized to three off-shell punctures
\vspace{-0.3cm}
{\small
\begin{eqnarray}\label{off-str-3}
\hspace{-0.6cm}
\parbox[c]{5.5em}{\includegraphics[scale=0.17]{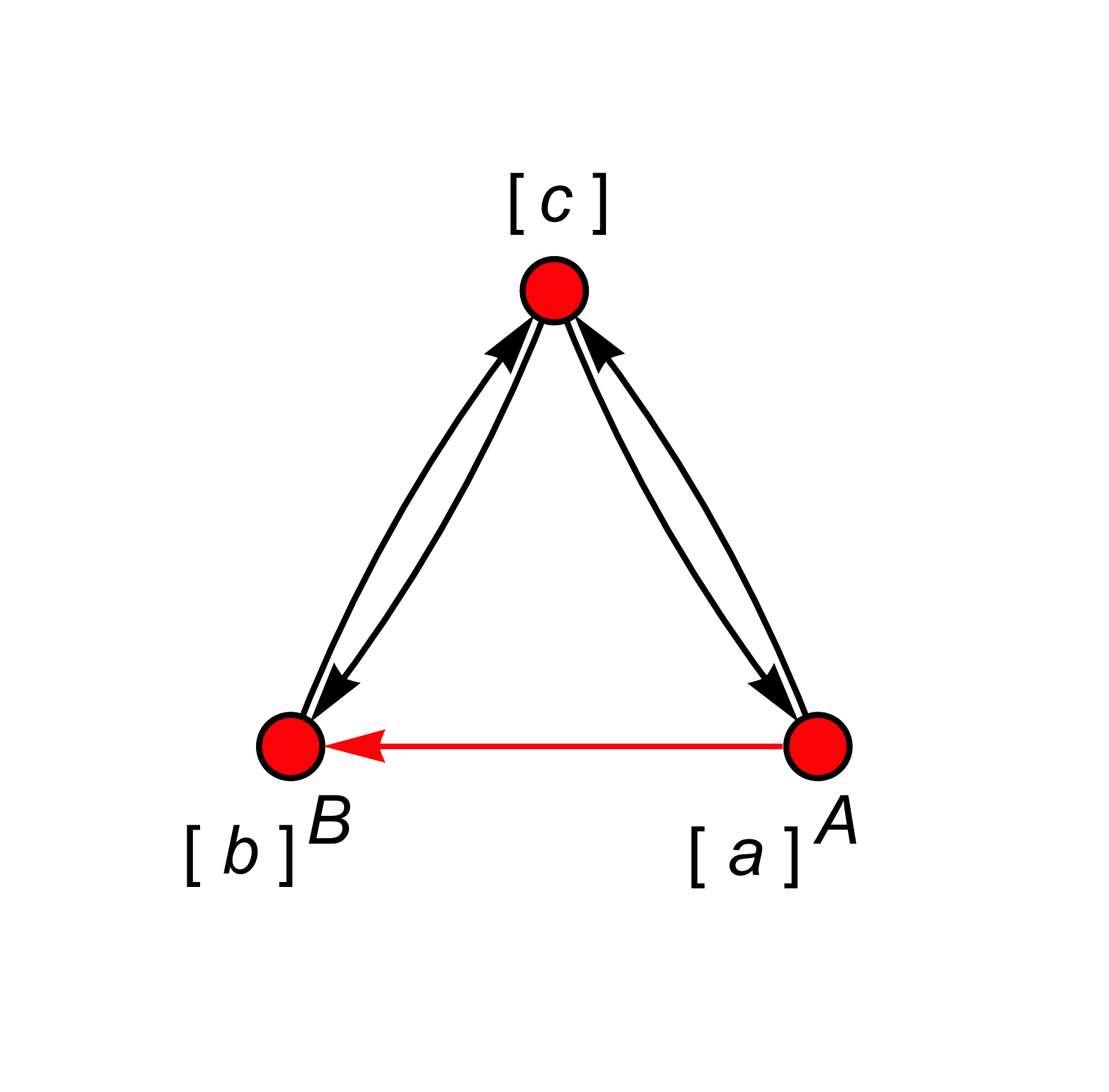}} 
=
\frac{2}{k_{[c]}^2 + k_{[b]}^2 - k_{[a]}^2}
\times
\hspace{-0.6cm}
\left.
\parbox[c]{6.5em}{\includegraphics[scale=0.17]{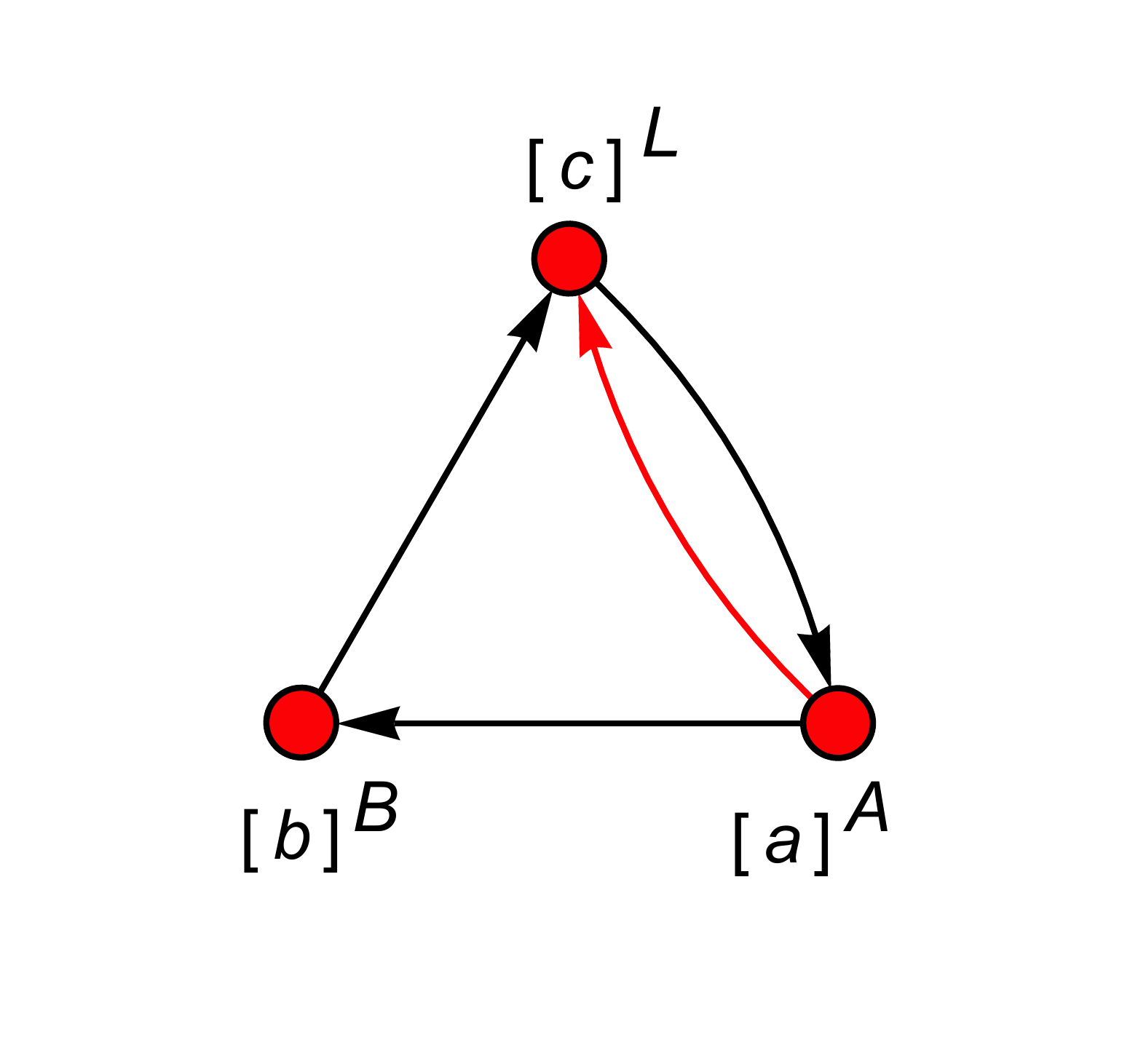}} \right|_{\hspace{-1.0cm} \eps_{[c]}^{L,\mu}\rightarrow k^\mu_{[c]} \atop \,\, \eps_{[a]}^A\cdot k_{[a]}=\eps_{[b]}^B\cdot k_{[b]}=0 }
\hspace{-2.0cm}
= \,\, (\eps_{[a]}^A\cdot \eps_{[b]}^B) \,\, , \, \qquad\,\,
\end{eqnarray}
}
\vskip-0.25cm\noindent
with $k_{[a]}+k_{[b]}+k_{[c]}=0$ and $k_{[i]}^2\neq 0$. Obviously, when the punctures ``$[a]$" and ``$[b]$" are on-shell ($k_{[a]}^2=k_{[b]}^2=\eps_{[a]}\cdot k_{[a]}=\eps_{[b]}\cdot k_{[b]}=0$), we obtain identity found in the previous section. The same behavior has been seen over bigger graphs, so, we propose the generalization 
\vspace{-0.3cm}
{\small
\begin{eqnarray}\label{conj1-mG}
\hspace{-0.55cm}
\int d\mu_{n}^{\rm CHY}
\hspace{-0.55cm}
\parbox[c]{6.5em}{\includegraphics[scale=0.17]{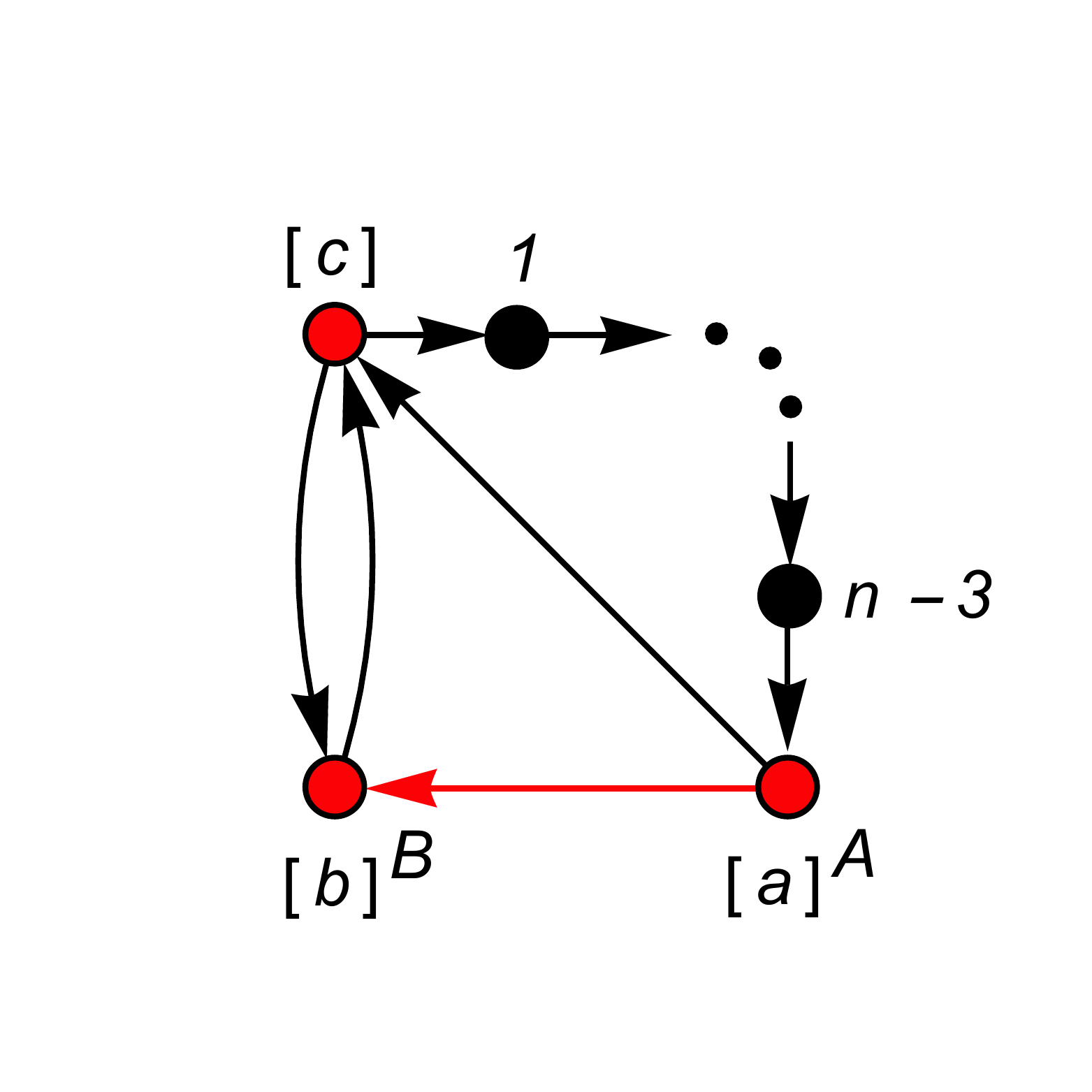}} 
=\, 
\frac{2}{k_{[c]}^2 + k_{[b]}^2 - k_{[a]}^2}
\times
\int d\mu_{n}^{\rm CHY}
\hspace{-0.53cm}
\left.
\parbox[c]{6.6em}{\includegraphics[scale=0.17]{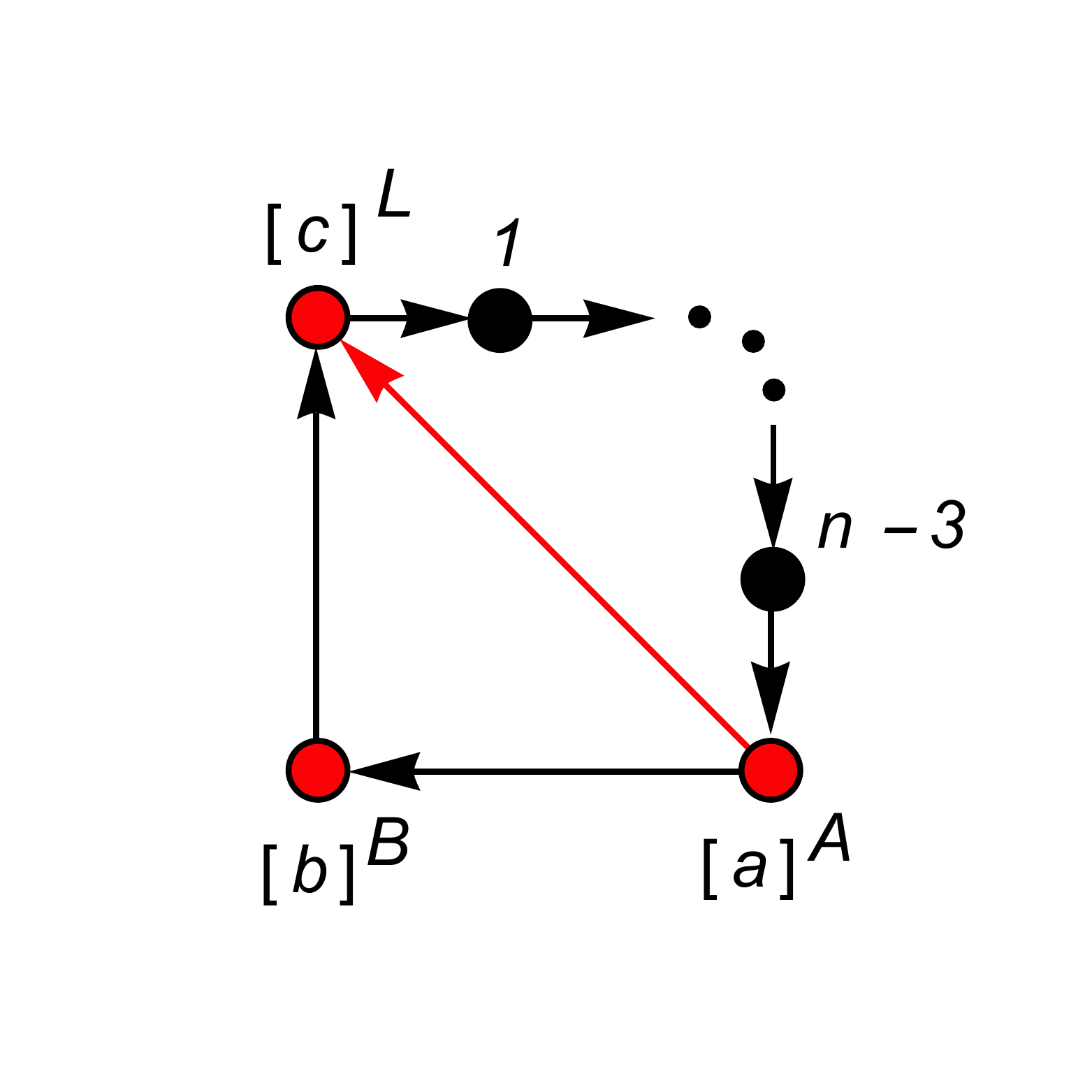}} \right|_{\hspace{-1.0cm} \eps_{[c]}^{L,\mu}\rightarrow k^\mu_{[c]} \atop 
\eps_{[a]}^A\cdot k_{[a]}=\eps_{[b]}^B\cdot k_{[b]}=0 }  
\hspace{-1.6cm} ,\,\,\,\,\,
\end{eqnarray}
}
\vskip-0.3cm\noindent
with, $k_{[a]}+k_{[b]}+k_{[c]}+k_1+\cdots +k_{(n-3)}=0$ and $k_{[i]}^2\neq 0$, $[i]\in\{[a], [b], [c] \}$.
Let us remember ourselves that on the left-hand side  the polarization vectors, $\eps_{[a]}^{A,\mu}$ and $\eps_{[b]}^{B,\mu}$, are not necessarily transverse, however, on the right-hand side, we impose the transversality condition to carry out the computation. 

On the other hand, the generalization of the second {\it strange-graph} in \eqref{strangec-2off} is given by 
\vspace{-0.35cm}
\begin{eqnarray}\label{s-str-graph}
\hspace{-0.6cm}
\parbox[c]{5.7em}{\includegraphics[scale=0.17]{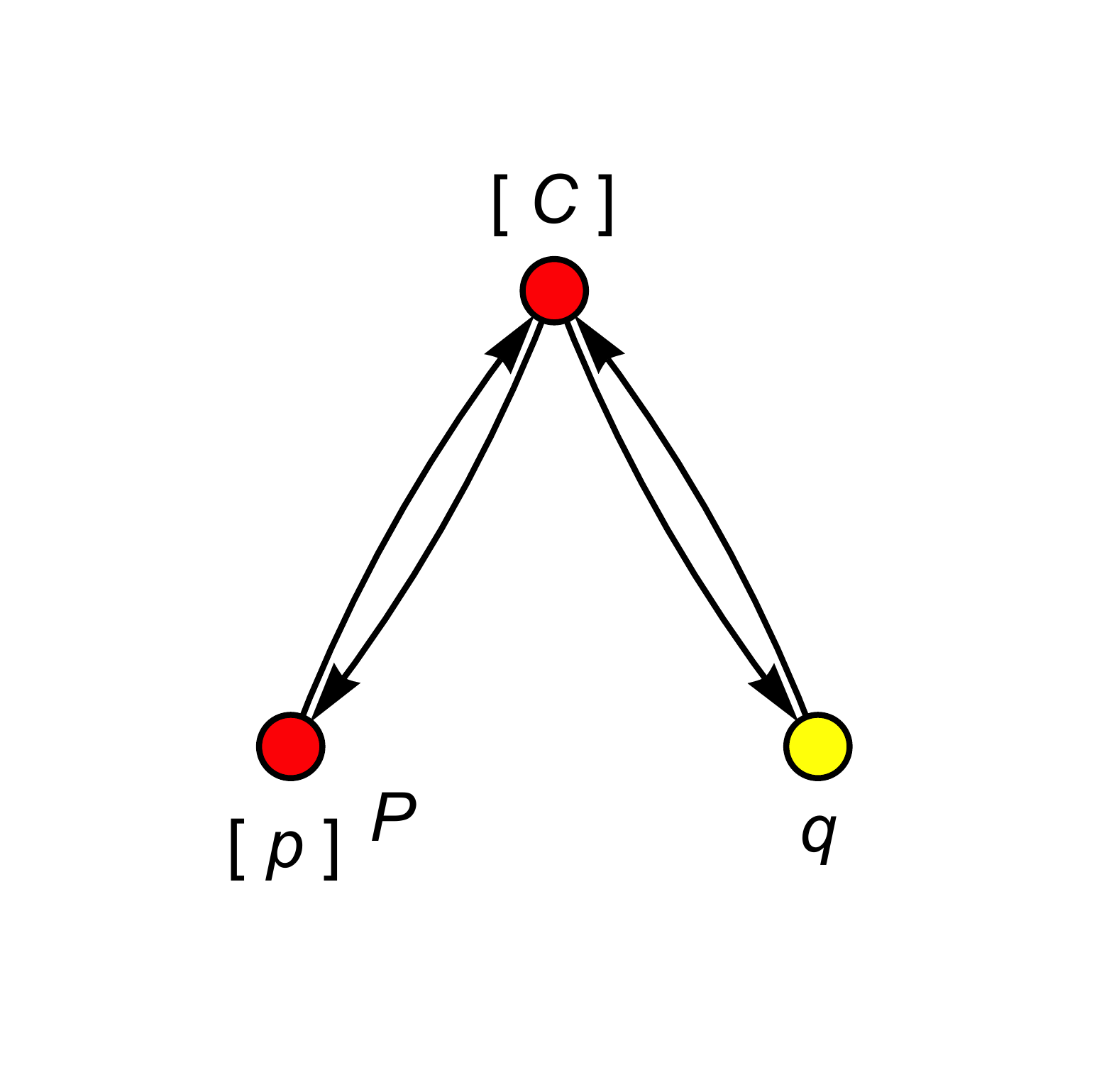}} 
=
\frac{k^2_{[C]} - k^2_{[p]}}{2} \times (\eps_{[p]}^P\cdot \eps_q)\,\, ,
\end{eqnarray}
\vskip-0.5cm\noindent
where, unlike to the graph in \eqref{off-str-3}, the polarization vector $\eps_{[p]}^{P,\mu}$ must be transverse ($\eps_{[p]}^{P}\cdot k_{[p]}=0$), this in order for the computation to not depend on $\s'$s (${\rm PSL}(2,\mathbb{C})$ symmetry). Note that  we have only considered two off-shell punctures, $k_{[C]}^2\neq 0$ and $k_{[p]}^2\neq 0$, which is  enough since after gluing two strange graphs, such as those given in \eqref{conj1-mG} and  \eqref{s-str-graph}, 
one must obtain a cut from a YM-graph, who can just have up to three off-shell particles.  

Following the {\bf properties-I,II} in appendix, we should focus on the graph (let us remind that $\eps_{[p]}^{P}\cdot k_{[p]}=0$ )
\vspace{-0.4cm}
{\small
\begin{eqnarray}\label{}
\hspace{-0.6cm}
(-)
\hspace{-0.5cm}
\left.
\parbox[c]{5.8em}{\includegraphics[scale=0.17]{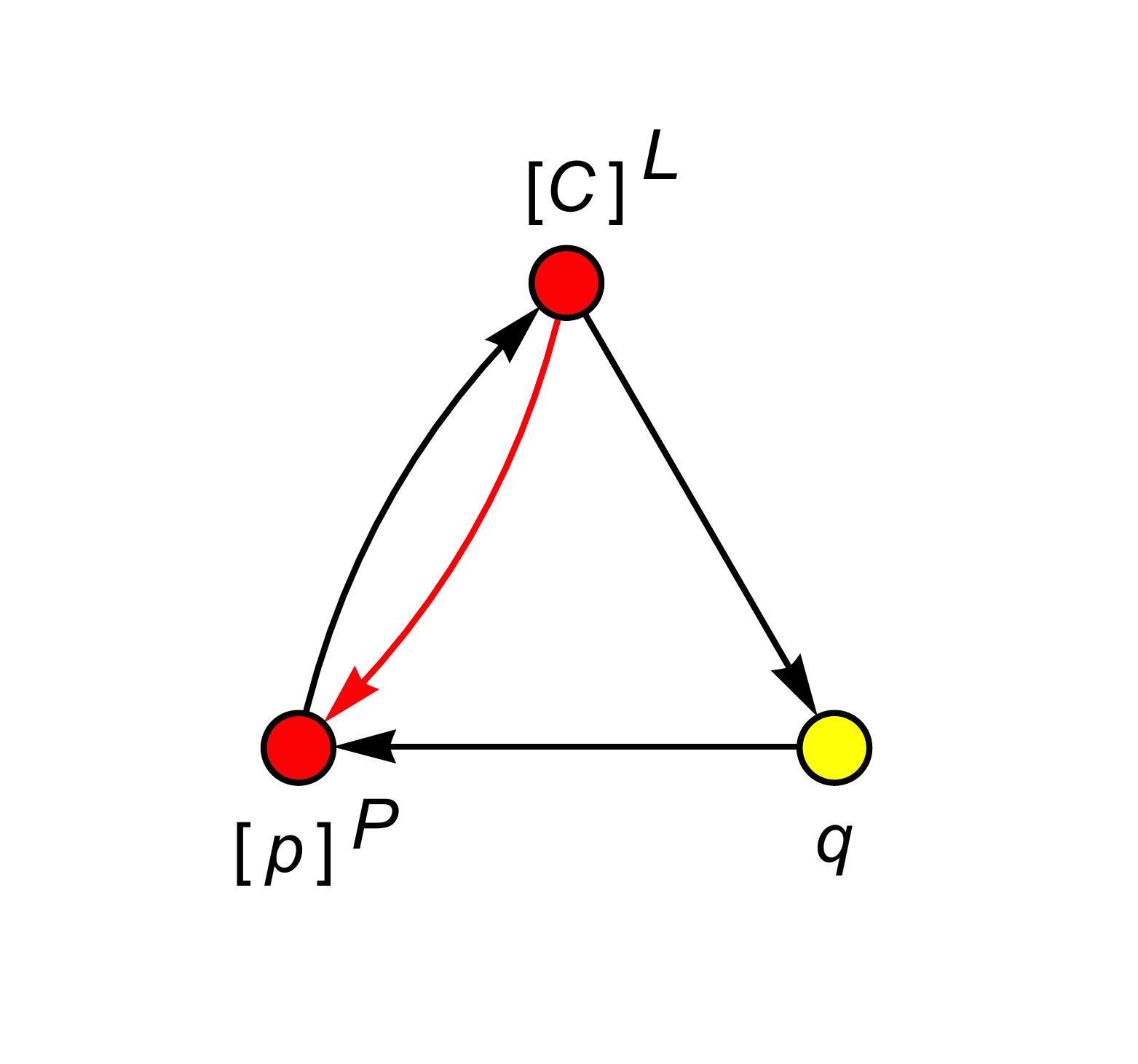}} 
\right|_{\eps_{[C]}^{L,\mu}\rightarrow k^\mu_{[C]}}
\hspace{-1.0cm}
=
\frac{k^2_{[C]} - k^2_{[p]}}{2} \times (\eps_{[p]}^P\cdot \eps_q) \,
=
\hspace{-0.45cm}
\parbox[c]{5.8em}{\includegraphics[scale=0.17]{str2-offG.pdf}} .
 \,\,\,
\end{eqnarray}
}
\vskip-0.25cm\noindent
Visibly, we obtained a matching with \eqref{s-str-graph}. 

The same behavior is observed at four and five points, therefore, the 
generalization of the identity obtained in \eqref{generalone} is direct
\vspace{-0.4cm}
{\small
\begin{eqnarray}\label{conj1-general}
\int d\mu_n^{\rm CHY} 
\hspace{-0.57cm}
\parbox[c]{6.1em}{\includegraphics[scale=0.17]{c1-offG.pdf}} 
\,\,
\int d\mu_m^{\rm CHY} 
\hspace{-0.5cm}
\parbox[c]{5.3em}{\includegraphics[scale=0.17]{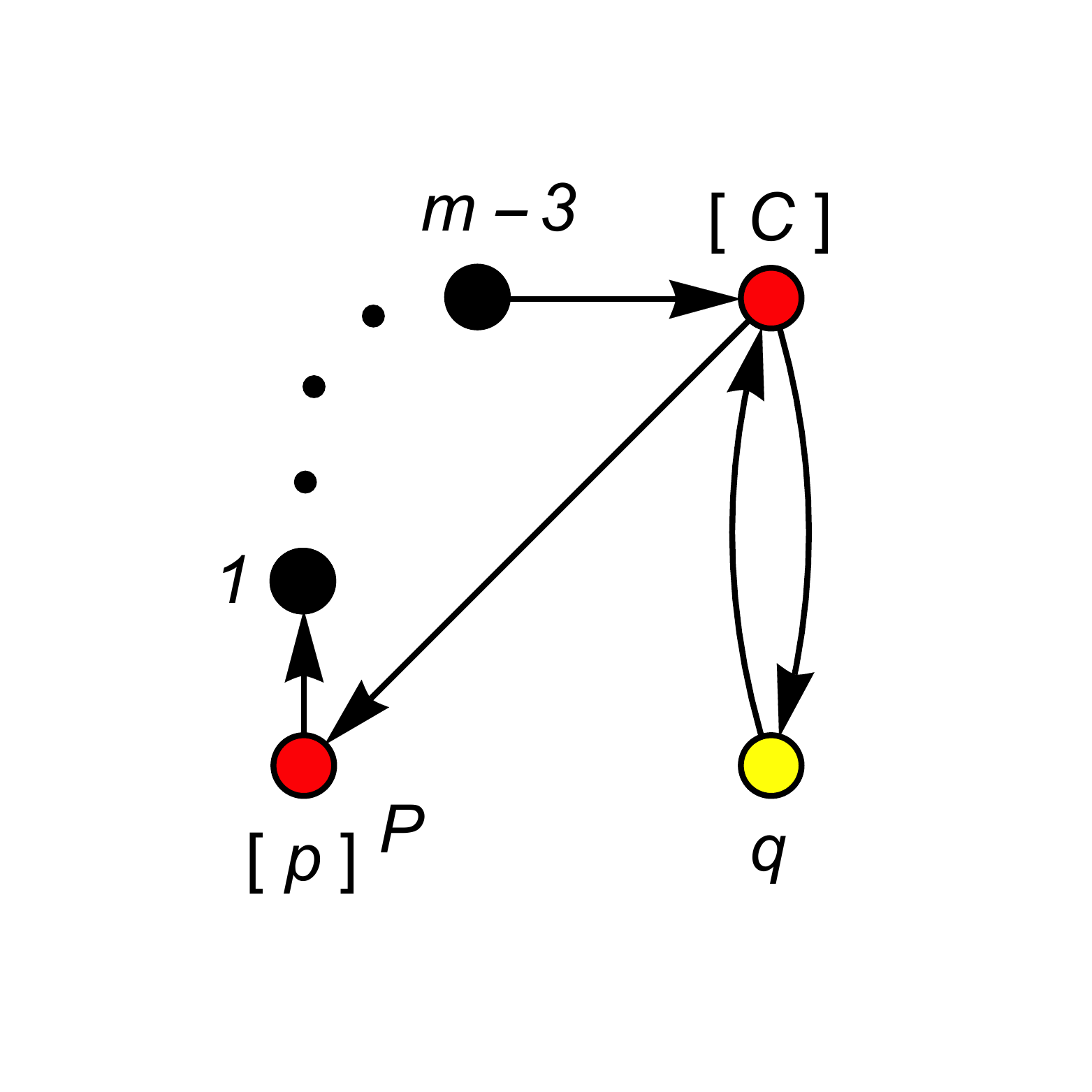}} 
=
2\,
\sum_L
\left.
\int d\mu_n^{\rm CHY} 
\hspace{-0.57cm}
\parbox[c]{6.5em}{\includegraphics[scale=0.17]{YMc1-offG.pdf}} 
\right|_{
\hspace{-0.02cm}
\eps_{[a]}^A\cdot k_{[a]}= 
\eps_{[b]}^B\cdot k_{[b]}=0 } 
\hspace{-2.3cm} 
\hspace{-0.15cm}
\int d\mu_m^{\rm CHY} 
\hspace{-0.5cm}
\parbox[c]{5.7em}{\includegraphics[scale=0.17]{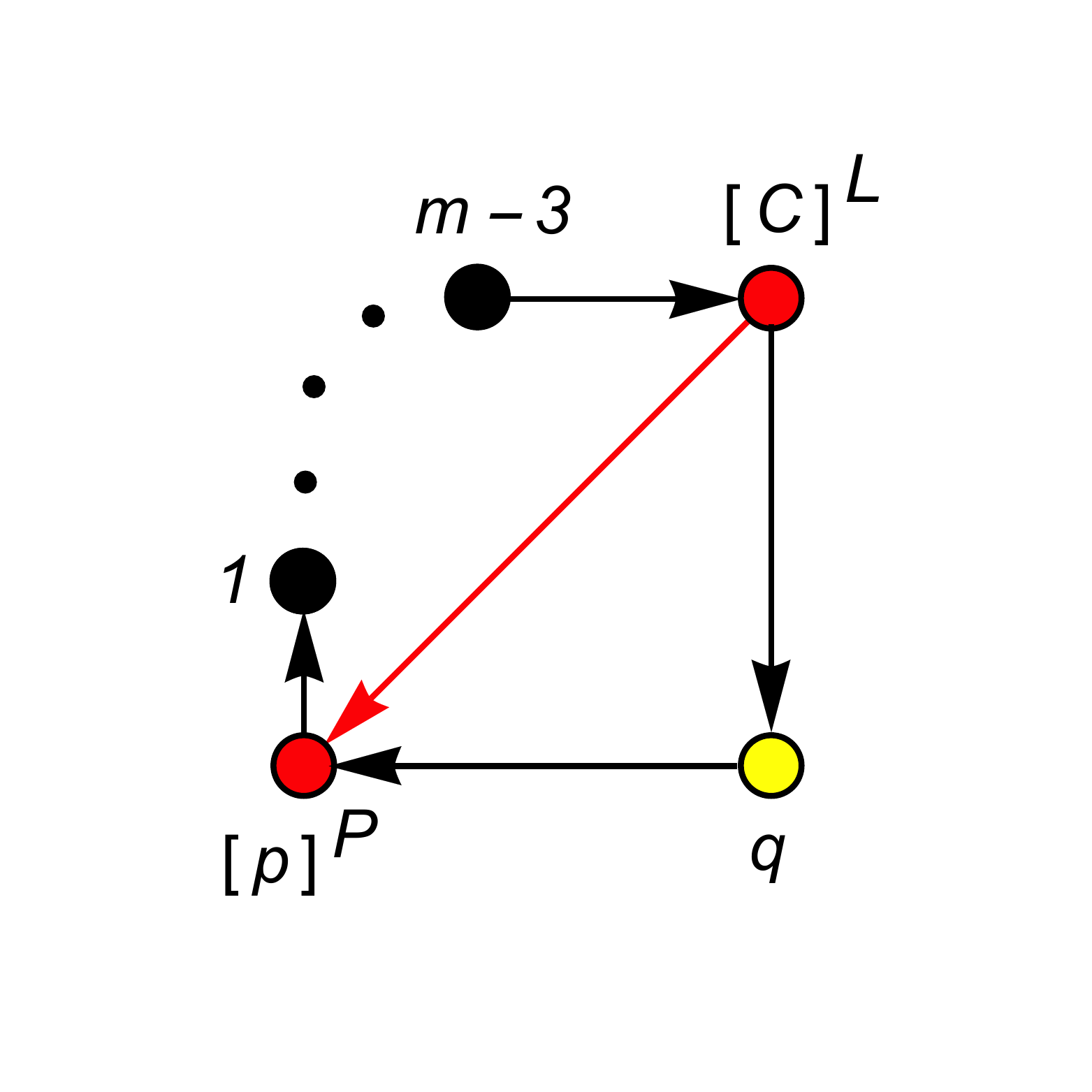}} ,
\quad \nonumber \\
\end{eqnarray}
}
\vskip-0.9cm\noindent
where, $\sum_{L}\eps_{[c]}^{L,\mu}\, \eps_{[C]}^{L,\nu}=-\frac{k_{[c]}^\mu k_{[C]}^\nu  }{k_{[c]}^2 + k_{[b]}^2 - k_{[a]}^2 }$, and with $\eps_{[p]}^P\cdot k_{[p]}=0$. It is useful to remember that 
each graph satisfies the momentum conservation condition, $k_{[a]}+k_{[b]}+k_{[c]}+k_1+\cdots + k_{n-3}=k_{[p]}+k_{[q]}+k_{[C]}+k_1+\cdots + k_{m-3}=0$, additionally, the forward limit, $k_{[c]}=-k_{[C]}$, 
must be imposed in order to glue the graphs.

\subsubsection{Standard-cuts}\label{sectionScuts}

Naively, one can think to achieve a recursive method the  identities in \eqref{generaltwo} should be generalized. However, when there is more than one off-shell particles with non-transverse polarization vectors in a YM-graph, this generalization is not possible.  Furthermore, notice that the \eqref{generaltwo} relationships are not enough to carry out the four-point resulting graph
obtained from the {\it cut-3} in the five-point amplitude. For instance, applying the third identity given in \eqref{generaltwo}  over the  {\it cut-3} in \eqref{fivePcuts}, one arrives
\vspace{-0.5cm}
\begin{eqnarray}\label{example-CR}
\hspace{-0.1cm}
 \text{{\it cut-3}} =
 \frac{1}{\tilde s_{23}}\times
\sum_{r}
\hspace{-0.6cm}
\parbox[c]{5.5em}{\includegraphics[scale=0.17]{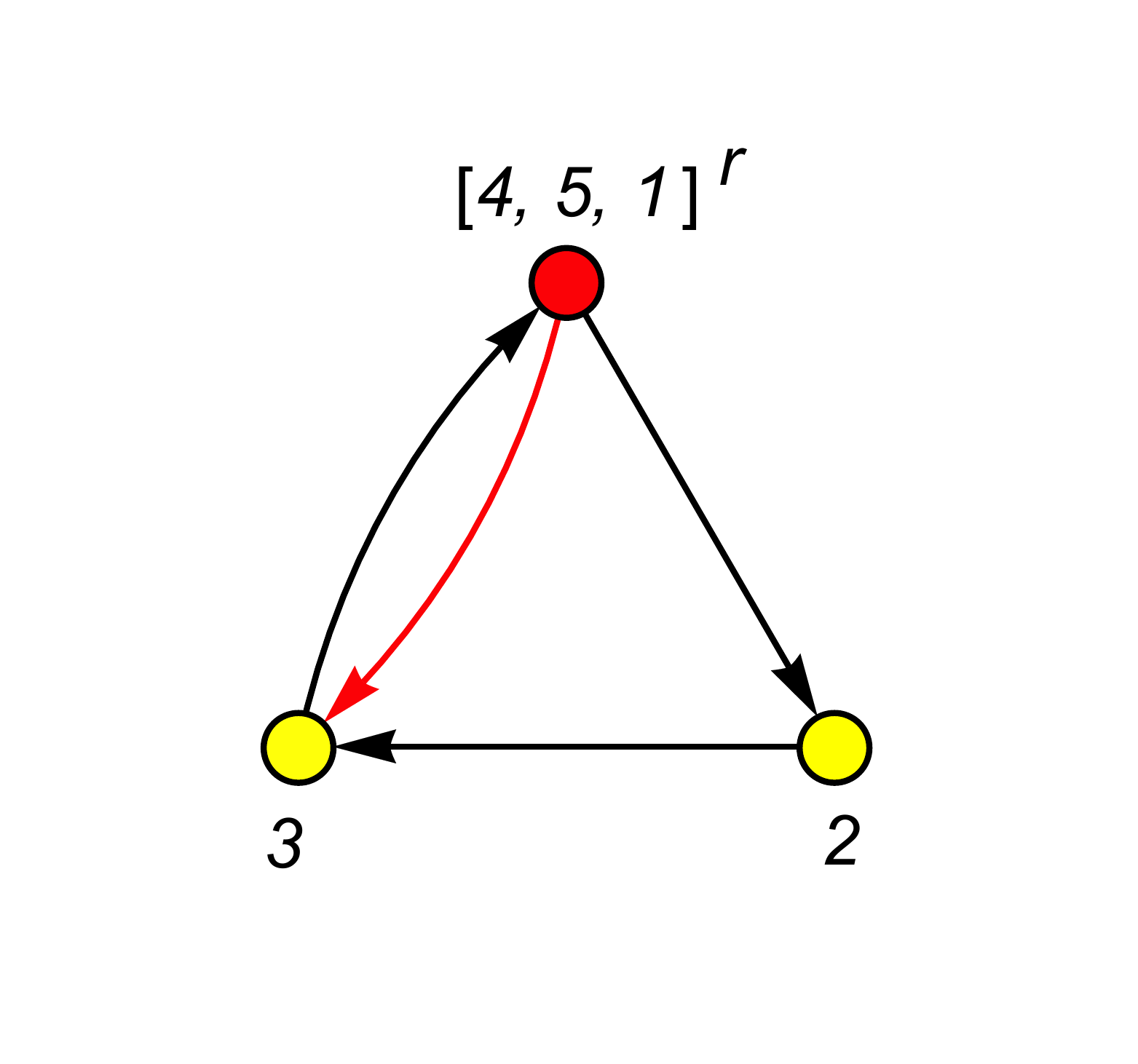}} 
\int d\mu_4^{\rm CHY} 
\hspace{-0.6cm}
\parbox[c]{5.1em}{\includegraphics[scale=0.17]{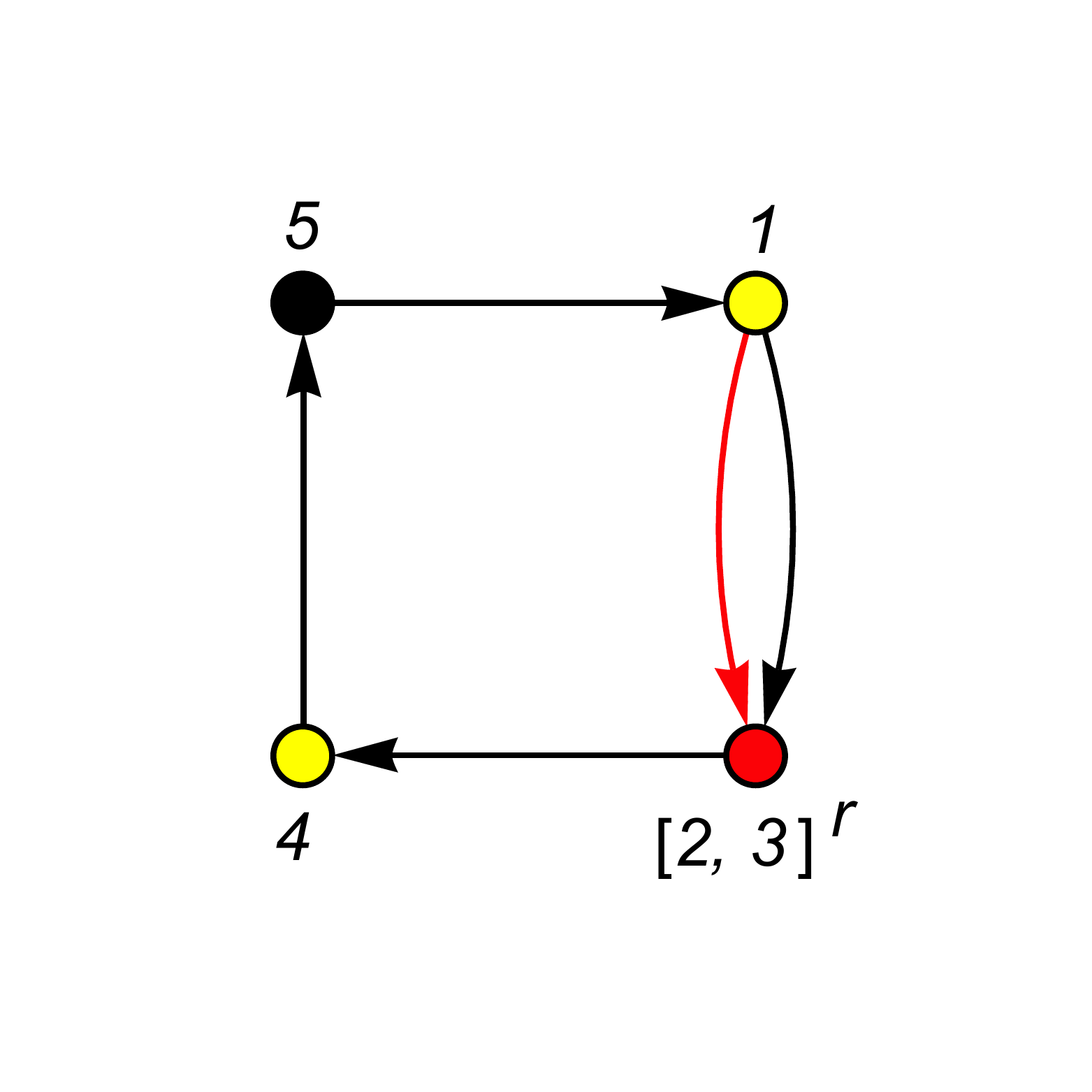}} 
=
\frac{1}{\tilde s_{23}}
\times
\sum_{A}
\hspace{-0.6cm}
\parbox[c]{5.5em}{\includegraphics[scale=0.17]{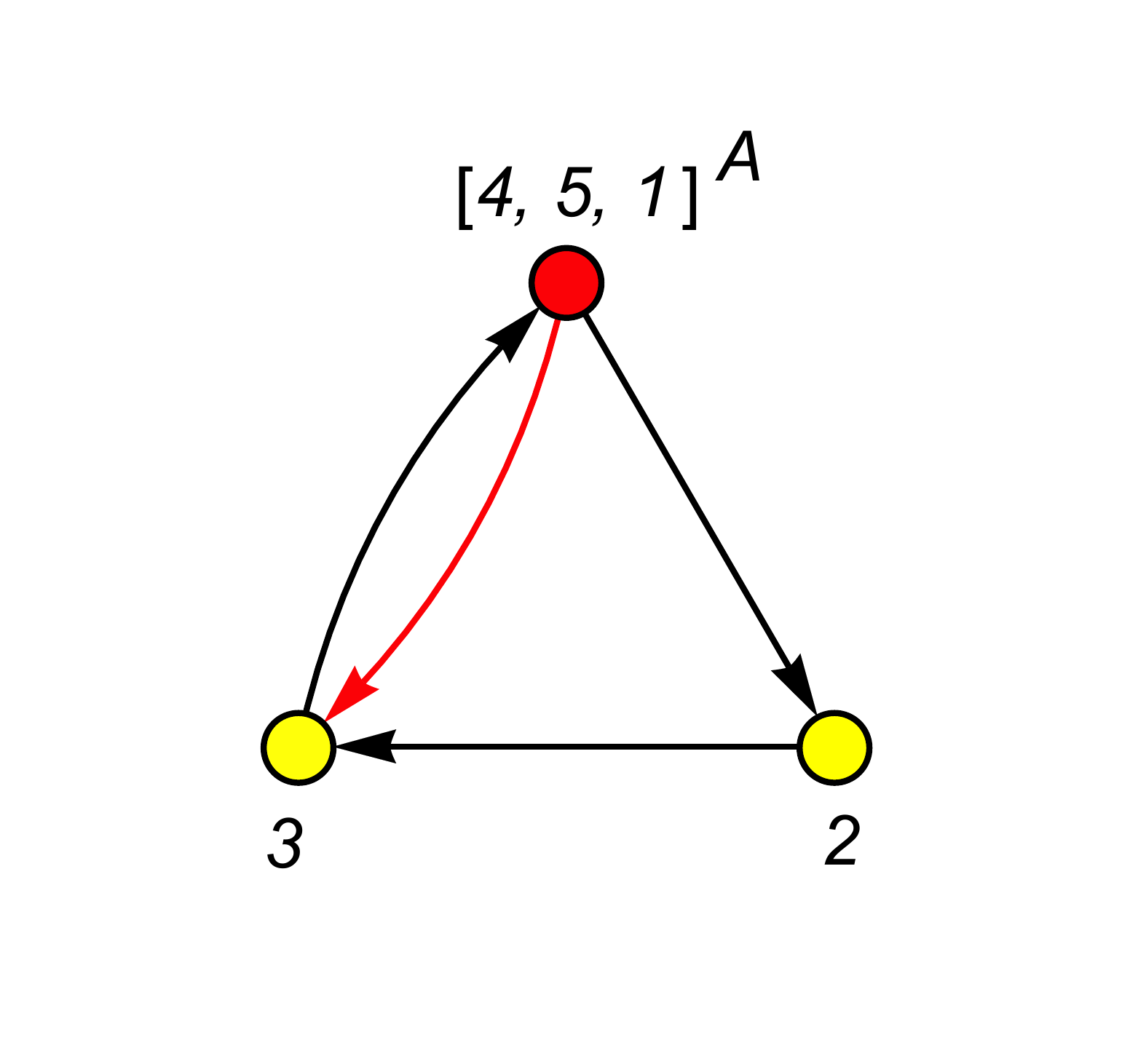}} 
\int d\mu_4^{\rm CHY} 
\hspace{-0.6cm}
\parbox[c]{5.7em}{\includegraphics[scale=0.17]{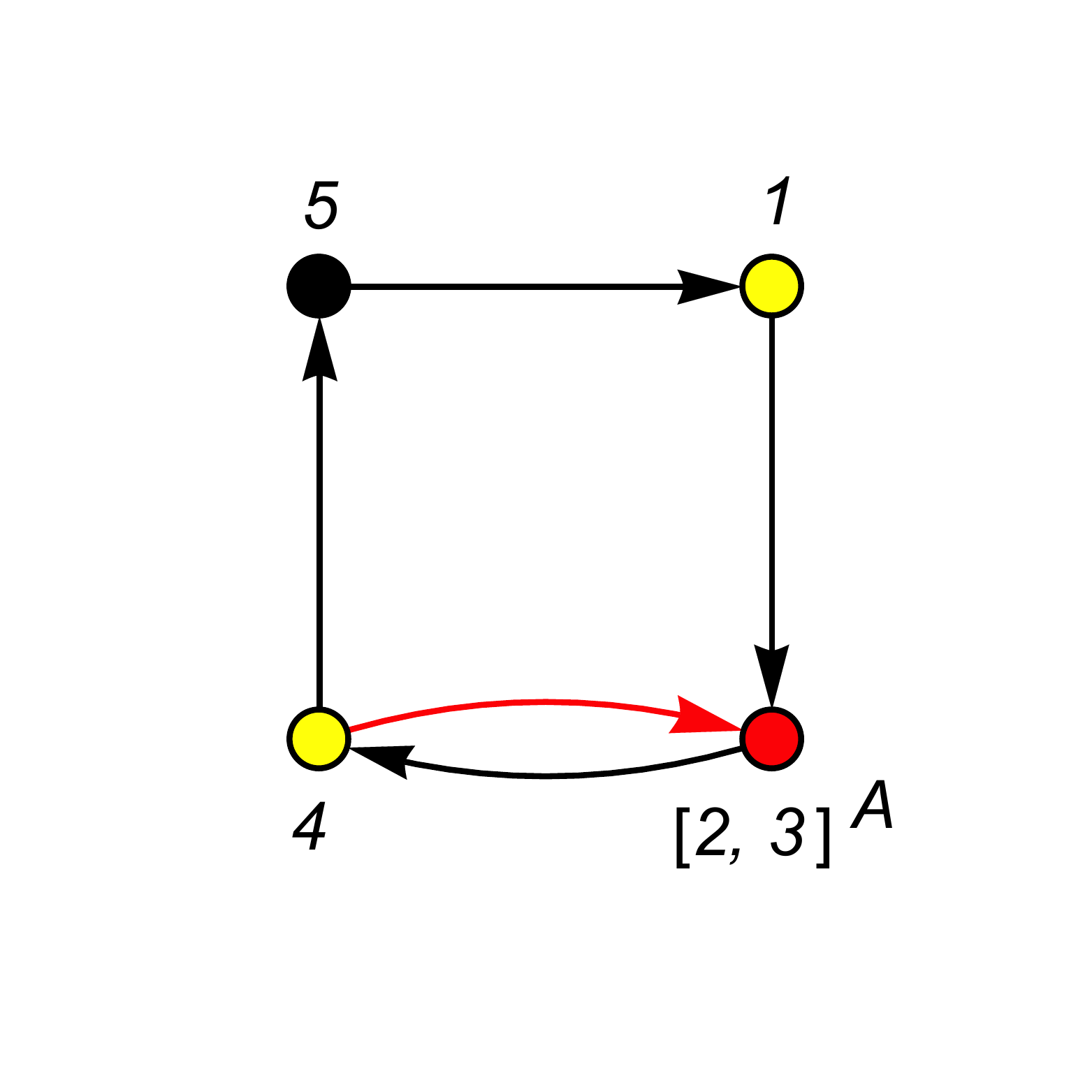}} ,\qquad 
 \nonumber \\
\end{eqnarray}
\vskip-0.5cm\noindent
where, $\sum_{r}\eps_{[4,5,1]}^{r,\mu}\, \eps_{[2,3]}^{r,\nu}=\eta^{\mu\nu}$ and   
$\sum_{A}\eps_{[4,5,1]}^{A,\mu}\, \eps_{[2,3]}^{A,\nu}=\eta^{\mu\nu}-\frac{2\,k_{[4,5,1]}^\mu k_{[2,3]}^\nu  }{k_{[4,5,1]}\cdot k_{[2,3]}   }$.  Clearly,  the {\bf integration  rules} do not work over the above four-point graphs. 

Fortunately, we have found two ways to face this issue, the first one is simple and intuitive, and the second one is more  systematic.

\begin{itemize}
\item{\bf First-method}
\end{itemize}

The idea of this method is to use  reverse engineering. First, we decompose the vectors, $\eps^{r,\mu}_{[i]}$, in two sectors, transverse  and longitudinal, i.e. $ \sum_{r}\eps_{[4,5,1]}^{r,\mu}\, \eps_{[2,3]}^{r,\nu} =  \sum_{T}\eps_{[4,5,1]}^{T,\mu}\, \eps_{[2,3]}^{T,\nu} +  \sum_{L}\eps_{[4,5,1]}^{L,\mu}\, \eps_{[2,3]}^{L,\nu}$, where, $ \sum_{T}\eps_{[4,5,1]}^{T,\mu}\, \eps_{[2,3]}^{T,\nu}  = \eta^{\mu\nu} -\frac{k_{[4,5,1]}^\mu k_{[2,3]}^\nu  }{k_{[4,5,1]}\cdot k_{[2,3]}   } $  and $\sum_{L}\eps_{[4,5,1]}^{L,\mu}\, \eps_{[2,3]}^{L,\nu} = \frac{k_{[4,5,1]}^\mu k_{[2,3]}^\nu  }{k_{[4,5,1]}\cdot k_{[2,3]} }$.  By the {\bf property III} in appendix \ref{appendix}, over the transverse sector, we can move the red arrow in the four-point graph from $(i,j)=(1,[2,3]^T) \rightarrow (i,j)=(1,4)$,  and now the {\bf integration rules} can be applied. 

On the other hand, although over the longitudinal sector the same trick doesn't work, we can make use reverse engineering with the help of \eqref{generalone} identity. To be more precise, from the {\bf properties-I,II} of the appendix \ref{appendix}, it is straightforward to see the equality,
\vspace{-0.4cm}
\begin{eqnarray}\label{method-1}
\hspace{-0.1cm}
\int d\mu_4^{\rm CHY} 
\hspace{-0.6cm}
\left.
\parbox[c]{5.6em}{\includegraphics[scale=0.17]{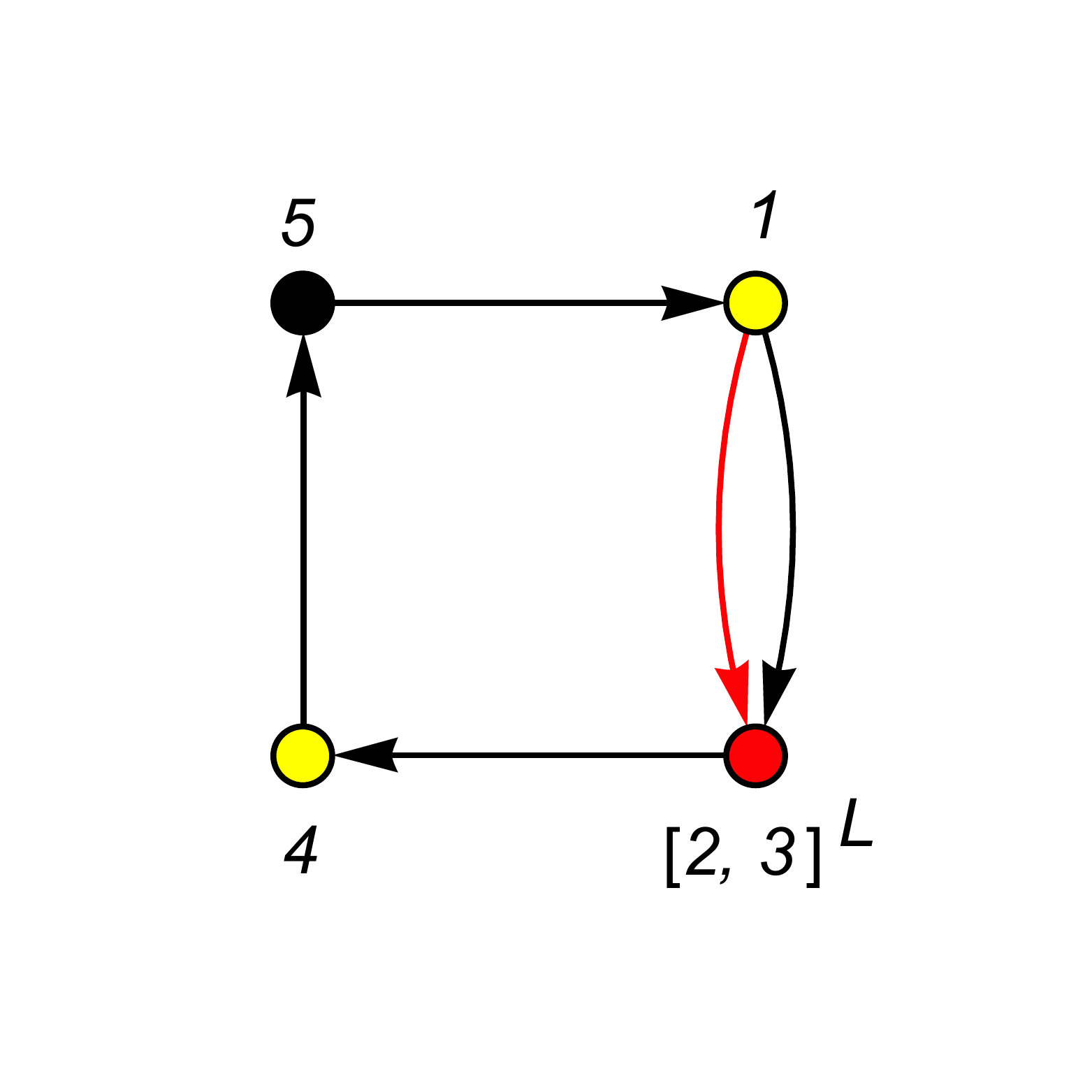}}\right|_{\eps^{L,\mu}_{[2,3]} \rightarrow k^\mu_{[2,3]}}
\hspace{-1.2cm}
=\frac{k_{[2,4]}\cdot k_{[1,4,5]} }{2}\times 
\int d\mu_4^{\rm CHY} 
\hspace{-0.6cm}
\parbox[c]{5.6em}{\includegraphics[scale=0.17]{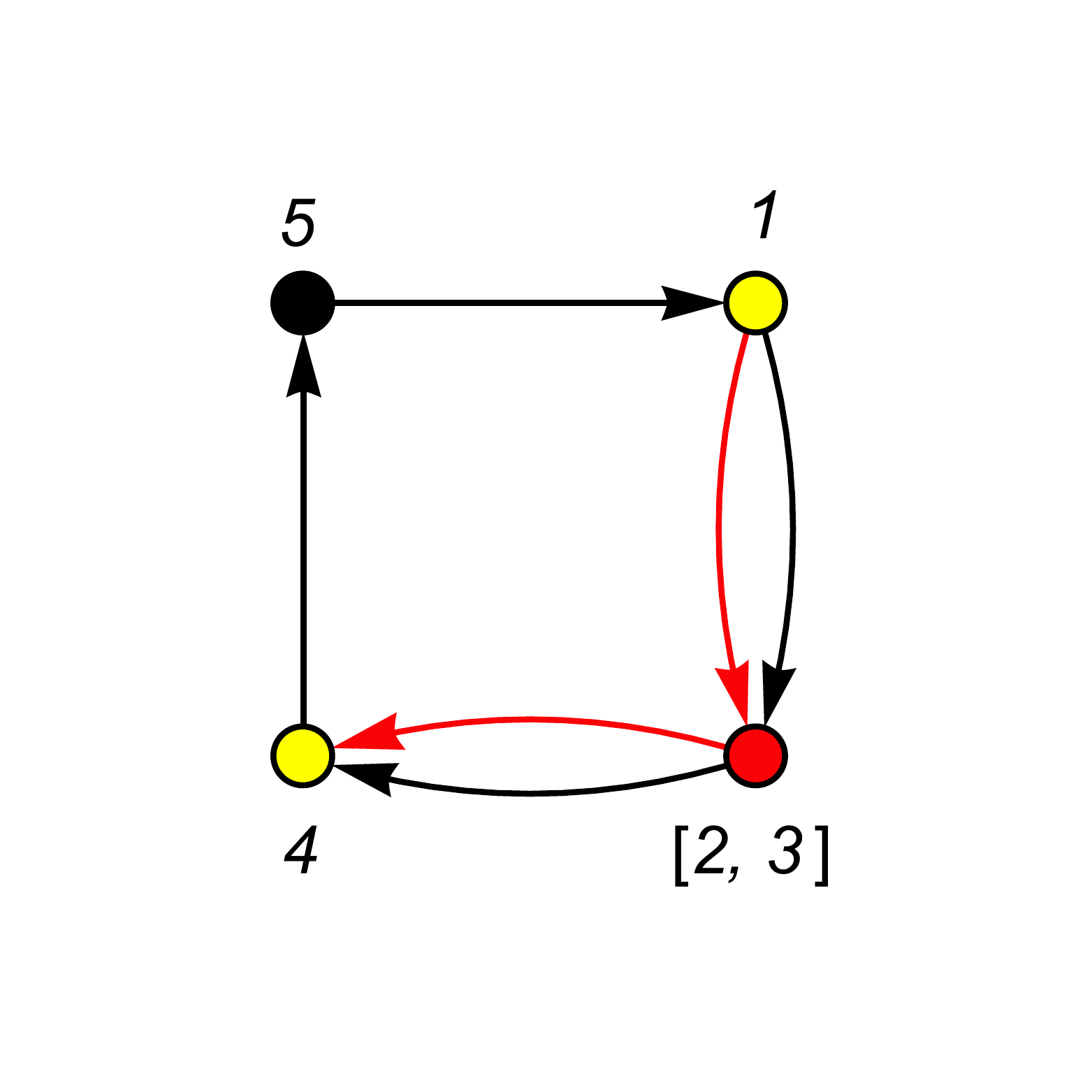}} .
\end{eqnarray}
\vskip-0.2cm\noindent
Clearly, the right-hand side graph is trivial to be computed via the  {\bf integration rules}, however, it has a strange shape. Furthermore, observe that the matrix associated with this graph is, $\left(\Psi_{\rm g,s:g}\right)^{1\,4\,[2,3]}_{1\,4\,[2,3]}$, where the gluon and scalar sets are given by, ${\rm g}=\{1,4,5  \}$ and ${\rm s}=\{ [2,3]  \}$.  In section \ref{sectionYMS} we are going to discuss  a little bit about this matrix. 

This  method can be extended to a higher number of particles or more off-shell vertices.

\begin{itemize}
\item{\bf Second-method}
\end{itemize}

The second method is based  on the cross-ratio identities \cite{Cardona:2016gon,Bjerrum-Bohr:2016juj}. For instance, in \eqref{example-CR} the  scattering equation, $S_5=\frac{\tilde s_{51}}{\s_{51}} + \frac{\tilde s_{5[2,3]}}{\s_{5[2,3]}}+ \frac{\tilde s_{54}}{\s_{54}} =0$, implies the cross-ratio identity, $ \frac{\tilde s_{[2,3]5}}{\tilde s_{45}} \left(  \frac{\s_{1[2,3]} \s_{45} }{\s_{[2,3]5} \s_{41}}   \right) =1$ or $\tilde s_{54} \, {\rm PT}_{(4,5,1,[2,3])}+   \tilde s_{5[4,1]}\,  {\rm PT}_{(4,1,5,[2,3])}=0$. Thus, the four-point graph in \eqref{example-CR} becomes
\vspace{-0.4cm}
\begin{eqnarray}\label{}
\hspace{-0.2cm}
\int d\mu_{4}^{\rm CHY}
\hspace{-0.5cm}
\parbox[c]{5.4em}{\includegraphics[scale=0.17]{cut3-5p-R2.pdf}} 
= \left(
\frac{\tilde s_{[2,3]5}}{\tilde s_{45}}
\right)
 \int d\mu_{4}^{\rm CHY}
\hspace{-0.5cm}
\parbox[c]{5.9em}{\includegraphics[scale=0.17]{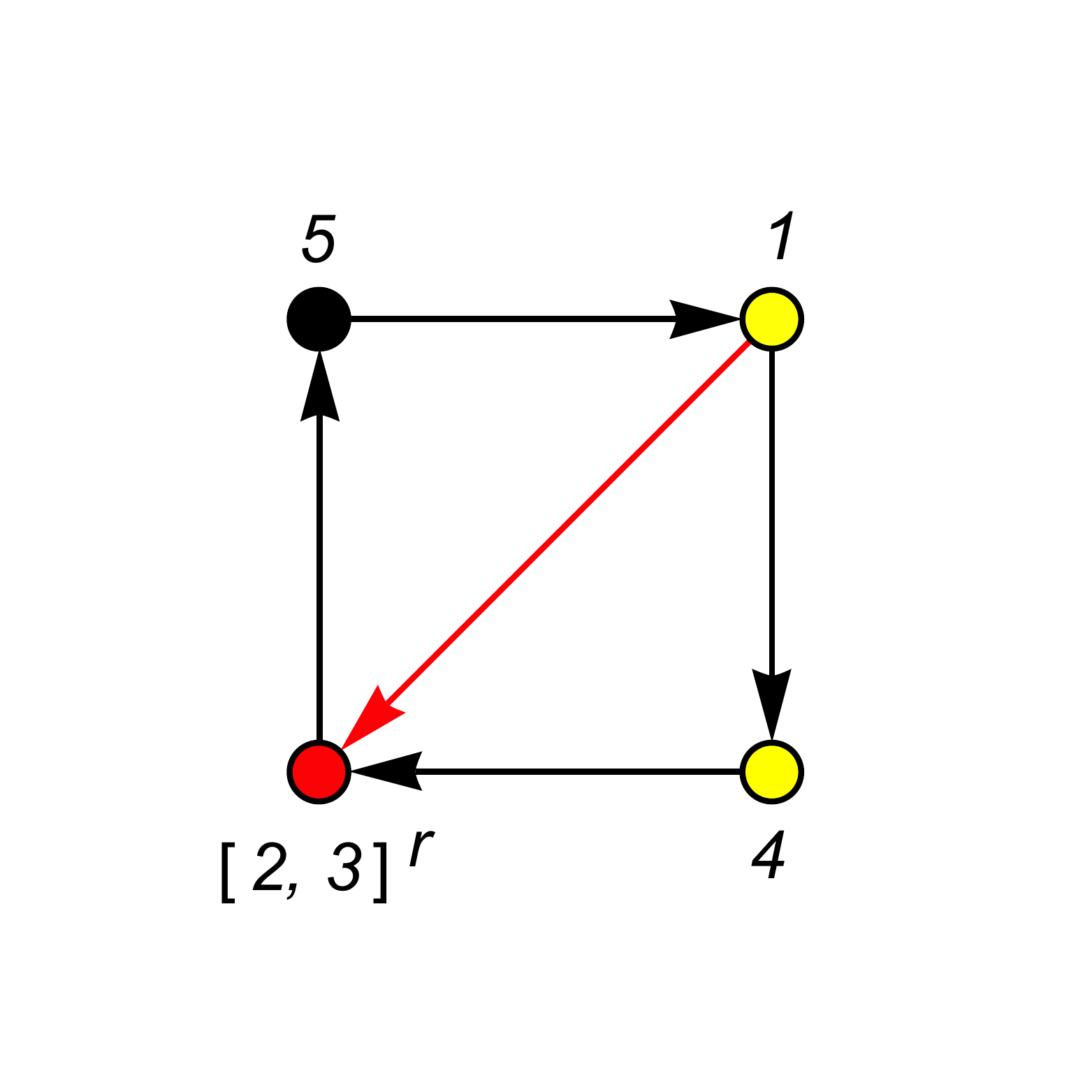}} ,
\end{eqnarray}
\vskip-0.4cm\noindent
now the  {\bf integration rules} can be applied easily. 

Let us consider one more example,  the five-point off-shell graph, 
\vspace{-0.4cm}
\begin{eqnarray}\label{five-off}
\hspace{-0.2cm}
\int d\mu_{5}^{\rm CHY}
\hspace{-0.2cm}
\parbox[c]{7.0em}{\includegraphics[scale=0.17]{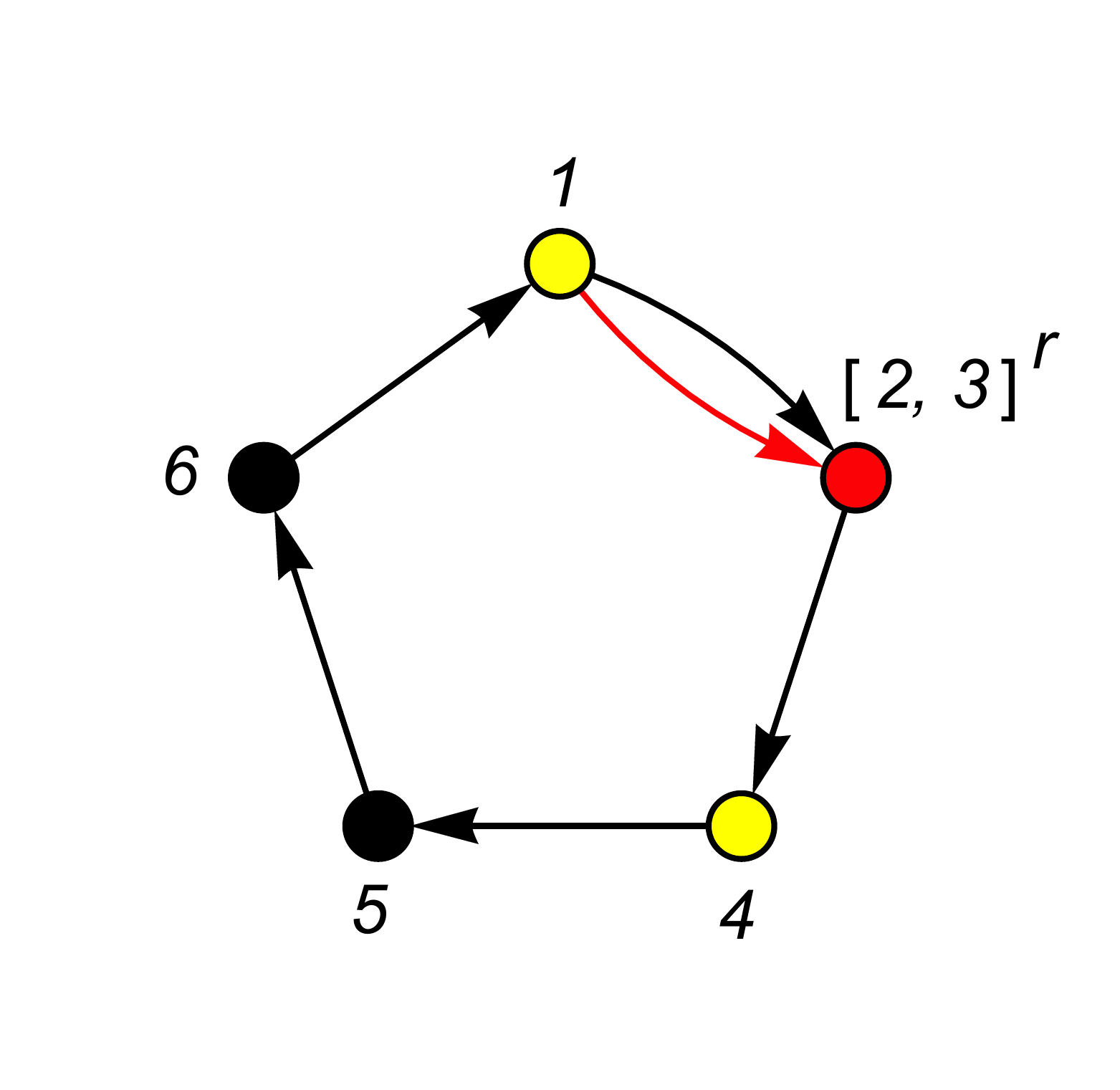}}  .
\end{eqnarray}
\vskip-0.3cm\noindent

On the support of the scattering equations, $S_5=\frac{\tilde s_{51}}{\s_{51}} + \frac{\tilde s_{5[2,3]}}{\s_{5[2,3]}}+ \frac{\tilde s_{54}}{\s_{54}} +\frac{\tilde s_{56}}{\s_{56}}  =0$ and $S_6=\frac{\tilde s_{61}}{\s_{61}} + \frac{\tilde s_{6[2,3]}}{\s_{6[2,3]}}+ \frac{\tilde s_{64}}{\s_{64}} +\frac{\tilde s_{65}}{\s_{65}}  =0$, it is straightforward to get the cross-ratio  and  the BCJ-like identity, $\tilde s_{456} + \tilde s_{5 [2,3]} \left(  \frac{\s_{[2,3]1}  \s_{54}  }{ \s_{5 [2,3] } \s_{41}}  \right)   +  \tilde s_{6 [2,3]} \left(  \frac{\s_{[2,3]1}  \s_{64}  }{ \s_{6 [2,3] } \s_{41}}  \right) =0$ and 
${\rm PT}_{(4,5,6,1,[2,3])}\tilde s_{456}+{\rm PT}_{(4,5,1,6,[2,3])}(\tilde s_{456}+\tilde s_{61}) +{\rm PT}_{(4,1,5,6,[2,3])}(\tilde s_{456}+\tilde s_{[5,6]1})=0$. So, by using these identities, the five-point graph in \eqref{five-off}  may be  rewritten as 
\vspace{-0.4cm}
\begin{eqnarray}\label{diffYMg}
\hspace{-0.5cm}
 \left(
\frac{\tilde s_{5 [2,3]}}{\tilde s_{456}}
\right)
\times
\int d\mu_{5}^{\rm CHY}
\hspace{-0.2cm}
\parbox[c]{6.3em}{\includegraphics[scale=0.17]{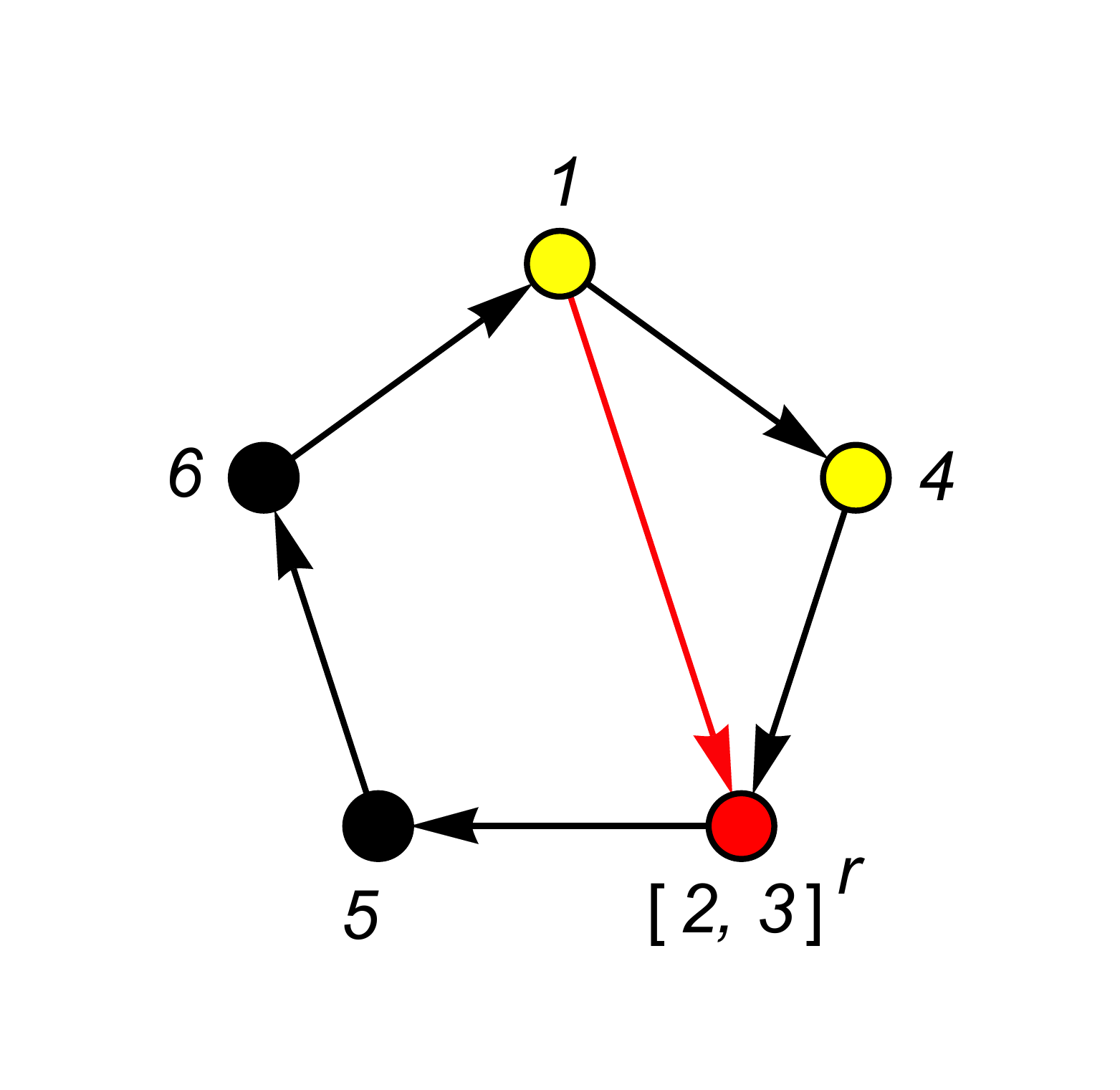}} 
+\,
\left(
\frac{\tilde s_{6 [2,3]}}{\tilde s_{456}}
\right)
\times
\int d\mu_{5}^{\rm CHY}
\hspace{-0.2cm}
\parbox[c]{6.4em}{\includegraphics[scale=0.18]{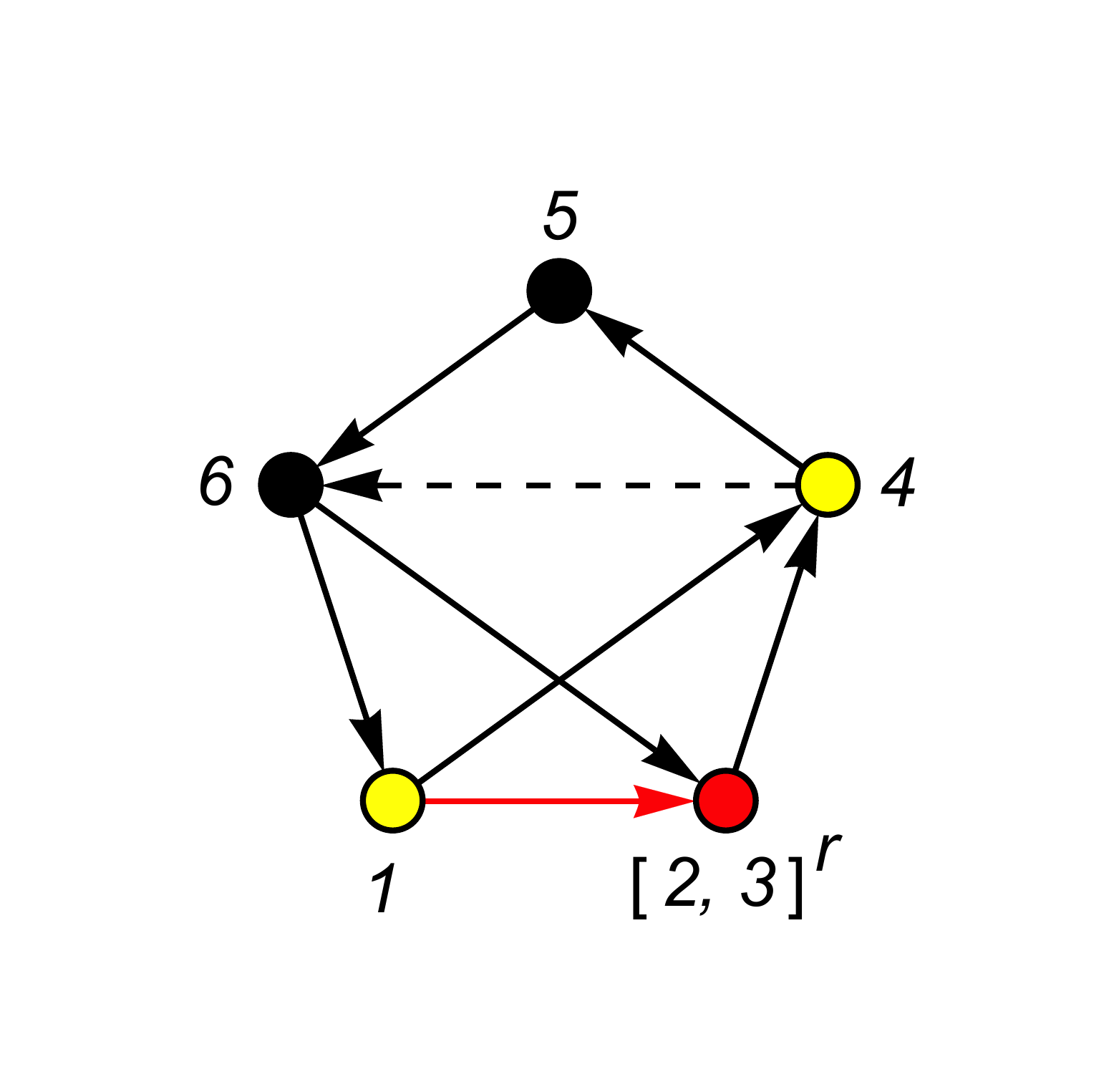}} ,
\end{eqnarray}
\vskip-0.5cm\noindent
\vspace{-0.6cm}
\begin{eqnarray}\label{}
\hspace{-0.2cm}
 \left(
\frac{\tilde s_{456} + \tilde s_{[5,6]1}}{\tilde s_{456}}
\right)
\,
\int d\mu_{5}^{\rm CHY}
\hspace{-0.2cm}
\parbox[c]{6.1em}{\includegraphics[scale=0.17]{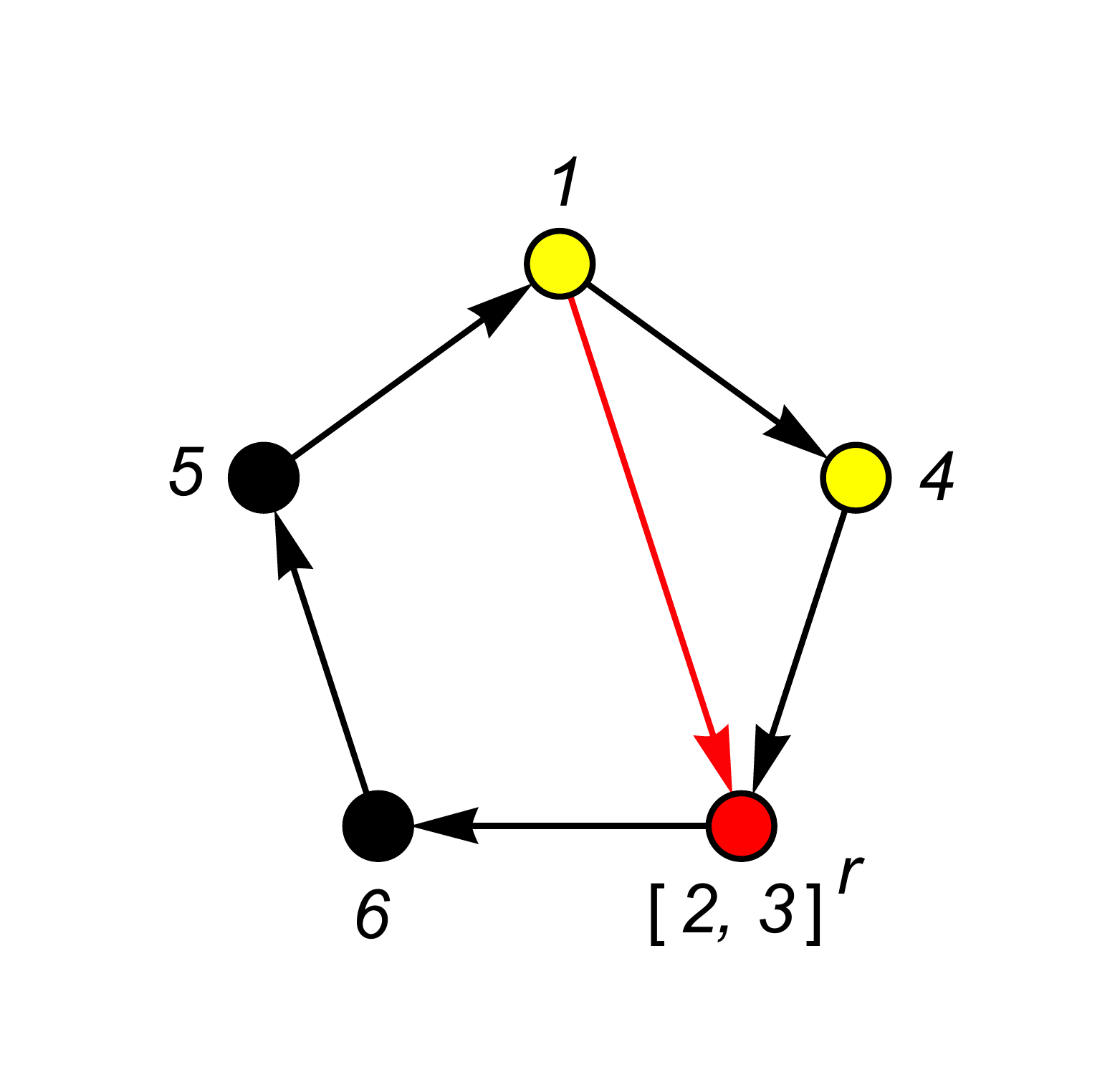}} -
\left(
\frac{\tilde s_{456} + \tilde s_{61}}{\tilde s_{456}}
\right)
\,
\int d\mu_{5}^{\rm CHY}
\hspace{-0.2cm}
\parbox[c]{6.0em}{\includegraphics[scale=0.17]{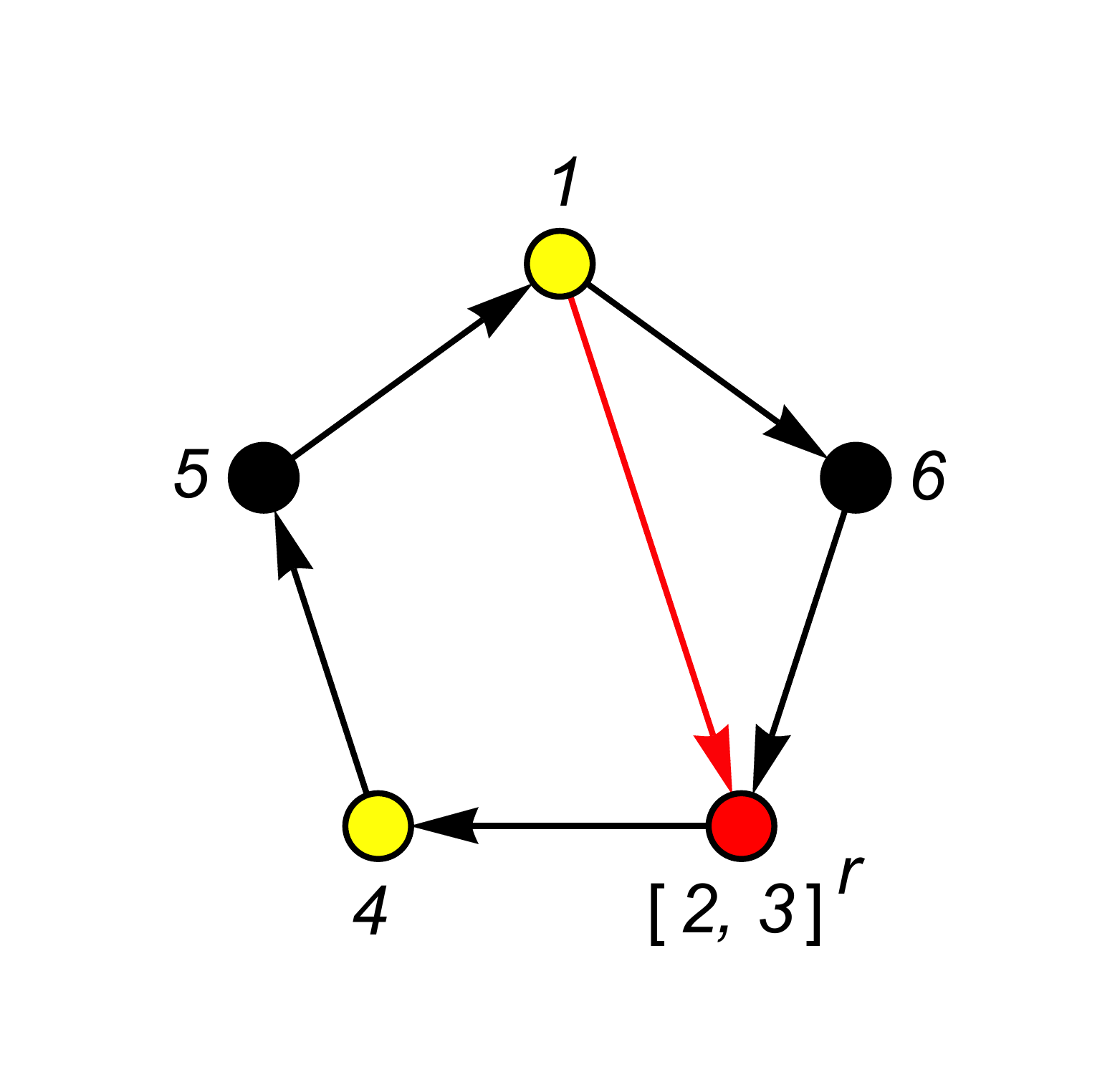}} ,\,\,\,~~
\end{eqnarray}
\vskip-0.2cm\noindent
respectively. On the second line, we obtained two YM graphs that can be computed by using the {\bf integration rules}. On the first line,  a new type of graph has arisen, and although we do not know its physical mean, the {\bf integration rules} work perfectly over it\footnote{Let us recall that the dashed arrow (anti-line) on this graph means the factor ``$\s_{46}$" is in the numerator. In addition, so as in \cite{Gomez:2016bmv}, when the integration rules are applied on this type of graphs, the anti-line subtracts in one the total number of arrows cut by a given configuration (dashed red line).}.

From simple examples, we have presented two more ways to deal with that kind of graphs. A generalization of  the graph in \eqref{five-off} is given by 
\vspace{-0.7cm}
\begin{eqnarray}\label{Goff-cr}
\hspace{-0.2cm}
\int d\mu_{p+3}^{\rm CHY}
\hspace{-0.5cm}
\parbox[c]{6.5em}{\includegraphics[scale=0.17]{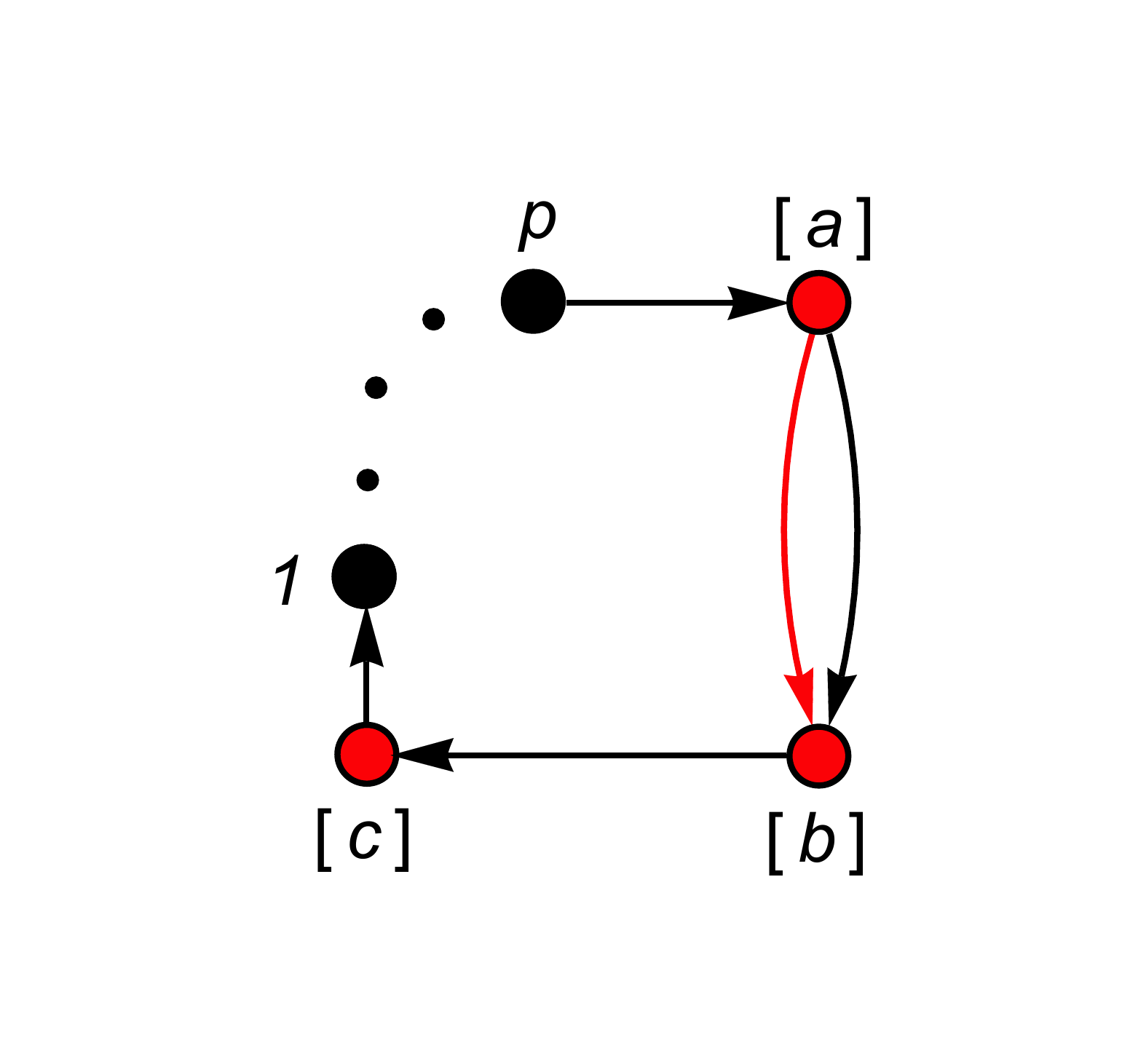}},
\end{eqnarray}
\vskip-0.4cm\noindent
where, $k_{[i]}^2\neq 0, \, \eps_{[i]}\cdot k_{[i]}\neq 0, \, [i]\in \{[a],[b]\} $, and $k_{[c]}^2\neq 0$, $\eps_{[c]}\cdot k_{[c]} =0$. Such as in above examples,  on the support, $S_1=\cdots =S_{p}=0$, the cross-ratio identity,  $\tilde s_{[c]1\ldots p}+ \cdots =0$, and the BCJ-like identity,  ${\rm PT}_{([c],1,\ldots,p,[a],[b])} \tilde s_{[c]1\ldots p}+
{\rm PT}_{([c],1,\ldots,[a],p,[b])} (\tilde s_{[c]1\ldots p} + \tilde s_{[a]p}) + \cdots +{\rm PT}_{([c],[a],1,\ldots,p,[b])} (\tilde s_{[c]1\ldots p} + \tilde s_{[a][p,...,1]})  =0$, are satisfied.  Thus, using one of these two identities, we may rewrite \eqref{Goff-cr} and apply the {\bf  integration rules}.

\section{Examples}\label{sectionEX5P}

As a final illustration, in this section we would like to apply the previous ideas to compute, explicitly, the five-point amplitude  $A_5^{\rm YM}(1,2,3,4,5)$. The plan is to write its five cuts in terms of the three-point building-block, $A_3^{\rm YM}$.  

Before computing the cuts obtained in \eqref{fivePcuts} from $A_5^{(1,3)}( 1,2,3,4,5)$, it is useful to  carry out the off-shell four-point amplitude,
\vspace{-0.5cm}
{\small
\begin{eqnarray}\label{}
\hspace{-0.05cm}
A_4^{([a],[c])}(  [a] ,b,[c],d)
=
\int d\mu_{4}^{\rm \L}
\hspace{-0.5cm}
\parbox[c]{5.4em}{\includegraphics[scale=0.17]{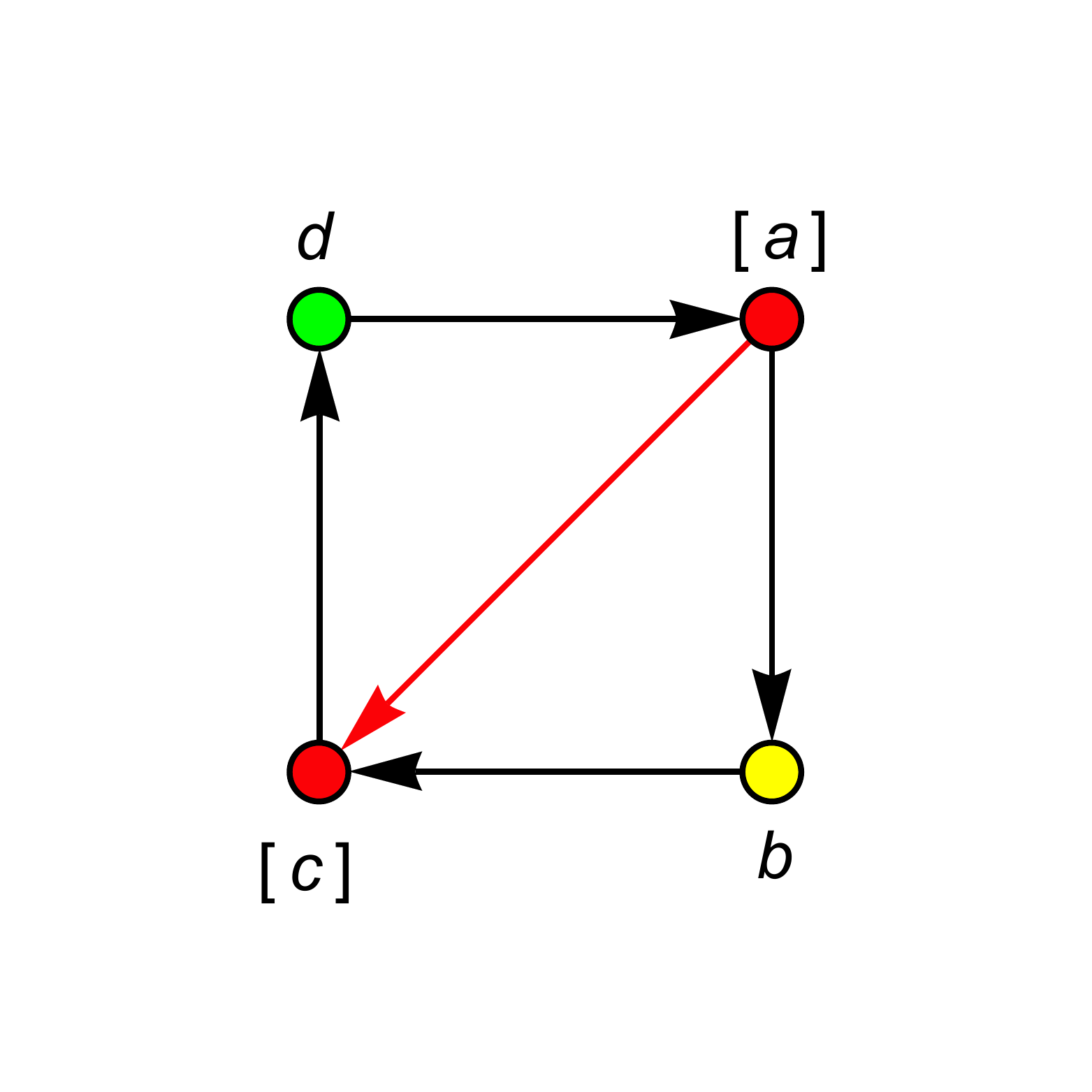}} =
 \nonumber 
\end{eqnarray}
}
\vskip-0.9cm\noindent
\vspace{-0.5cm}
{\small
\begin{eqnarray}\label{}
\hspace{-0.2cm}
\sum_r\left[\frac{A_{3}^{([a],[b,c])} (  [a] ,  [b,c]^r, d) 
\times A_{3}^{([c],[d,a])} (   [c] ,  [d,a]^r ,b)
}{\tilde s_{d[a]}} 
+\frac{A_{3}^{([a,b],[c])}( [a,b]^r , [c] , d)\, A_{3}^{([c,d],[a])}(  [c,d]^r , [a] , b )  }{\tilde s_{[c]d}}
 \right]
 \nonumber 
\end{eqnarray}
}
\vskip-0.9cm\noindent
{\small
 \begin{eqnarray}\label{}
\hspace{-0.1cm} 
 +2\,
 \sum_L
 \frac{A_{3}^{(d,[a,c])}(  d ,  [a,c]^L , b )}{\tilde s_{bd}}
 \times  
 \left.
 A_{3}^{([b,d],[a])}( [b,d]^L , [a] , [c]) 
 \right|_{\eps_{[a]}\cdot k_{[a]}=0 \atop
 \eps_{[c]}\cdot k_{[c]}=0 }, \quad\,\,\,
 \nonumber
\end{eqnarray}
}
\vskip-0.15cm\noindent
where, $\sum_r \eps_{[b,c]}^{r,\mu} \eps_{[d,a]}^{r,\nu} = \sum_r \eps_{[a,b]}^{r,\mu} \eps_{[c,d]}^{r,\nu} = \eta^{\mu\nu}$,  $\sum_L \eps_{[a,c]}^{L,\mu} \eps_{[b,d]}^{L,\nu} = -\frac{ k_{[a,c]}^\mu \, k_{[b,d]}^\nu }{k_{[b,d]}^2+ k_{[c]}^2 - k_{[a]}^2 }$, and the particles, ``$[a]$ and $[c]$", can be off-shell and non-transverse, i.e. $k_{[a]}^2\neq 0$, $k_{[c]}^2\neq 0$ and  $\eps_{[a]}\cdot k_{[a]} \neq 0$, $\eps_{[c]}\cdot k_{[c]}  \neq 0 $.  Let us keep in mind that  the momentum conservation condition is satisfies, $k_{[a]} + k_b + k_{[c]} + k_d=0$, and the particles, ``$b$" and ``$d$", are on-shell and transverse, $k_b^2=k_d^2=\eps_{b}\cdot k_b=\eps_{d}\cdot k_d =0 $. This four-point result  has been checked
numerically.

Using the off-shell amplitude, $A_4^{([a],[c])}(  [a] ,b, [c] ,d)$, and  the methods proposed in this work,  the five cuts obtained in \eqref{fivePcuts} for the amplitude, $A_5^{\rm YM}( 1,2,3,4,5)$, are given, explicitly, by the expressions 
\vspace{-0.4cm}
\begin{eqnarray}
\hspace{-2.8cm}
\parbox[c]{6.4em}{\includegraphics[scale=0.17]{5pt-cut1.pdf}} 
=\sum_r \frac{ A_3^{([5,1,2],3)}(   [5,1,2] ^r,  3, 4) \times A_4^{(1,[3,4] )}( 1, 2,  [3,4] ^r, 5) } {\tilde s_{34}}
\nonumber
\end{eqnarray}
\vskip-0.2cm\noindent
\vspace{-0.75cm}
\begin{eqnarray}
\hspace{-2.8cm}
\parbox[c]{6.4em}{\includegraphics[scale=0.17]{5pt-cut2.pdf}} 
=\sum_r \frac{ A_3^{(2,[3,4,5])}( 2,  [3,4,5]^r, 1) \times A_4^{([1,2],4)} ( [1,2] ^r ,3, 4, 5) } {\tilde s_{345}}
\nonumber
\end{eqnarray}
\vskip-0.1cm\noindent
\vspace{-0.85cm}
\begin{eqnarray}
\hspace{-0.8cm}
\parbox[c]{6.4em}{\includegraphics[scale=0.17]{5pt-cut3.pdf}} 
=
\hspace{-0.05cm}
\left( 
\frac{\tilde s_{[2,3]5}}{ \tilde s_{45}} 
\right)\times
\sum_r \frac{ A_3^{(3,[4,5,1])}( 3,  [4,5,1]^r, 2) \times  A_4^{(1,[2,3])}(1,4 ,  [2,3] ^r ,5) } {\tilde s_{451}}
\nonumber
\end{eqnarray}
\vskip-0.1cm\noindent
\vspace{-0.8cm}
\begin{eqnarray}
\hspace{-2.3cm}
\parbox[c]{6.4em}{\includegraphics[scale=0.17]{5pt-cut4.pdf}} 
=2\,\sum_L \frac{ A_3^{(4,[1,3,5])}(4,  [1,3,5]^L, 2) \times A_4^{(1,[2,4])}(1 , 3 ,  [2,4] ^L , 5)   } {\tilde s_{24}}
\nonumber
\end{eqnarray}
\vskip-0.1cm\noindent
\vspace{-0.8cm}
\begin{eqnarray}
\hspace{-2.3cm}
\parbox[c]{6.4em}{\includegraphics[scale=0.17]{5pt-cut5.pdf}} 
=2\,\sum_L \frac{ A_3^{([2,4,5],1)}( [2,4,5]^L, 1, 3)  \times A_4^{([1,3],4)}( [1,3]^L , 2 , 4 , 5) } {\tilde s_{245}}
\nonumber
\end{eqnarray}
\vskip-0.1cm\noindent
where\footnote{To compute the {\it cut-3} we used the second method developed in section \ref{sectionScuts}.}
$\sum_{r} \eps_{[a]}^{r,\mu} \, \eps_{[b]}^{r,\nu} = \eta^{\mu\nu}$ and
$\sum_L \eps_{[i]}^{L ,\mu} \eps_{[j]}^{L ,\nu} = \frac{ k_{[i]}^\mu \, k_{[j]}^\nu }{k_{[i]}\cdot k_{[j]} }$.  Finally, it is not hard to verify that,
$$
A_5^{\rm YM}(1,2,3,4,5) =-( \text{{\it cut-1}+{\it cut-2}+{\it cut-3}+{\it cut-4}+{\it cut-5}})
$$.

As a last point, notice the non conventional structure of the poles, for example, in the {\it cut-1} and {\it cut-2} one has,
$\tilde s_{5[3,4]}$ and $\tilde s_{5[1,2]} = -\tilde s_{5[3,4]} $, respectively.  This fact is a consequence from the scattering equations and the \eqref{fpL0} expansion.

\section{Special Yang-Mills-Scalar Theory}\label{sectionYMS}

After giving an extended analysis  and obtaining an alternative algorithm of the pure Yang-Mills theory in the CHY framework, the generalization to the special Yang-Mills-Scalar theory is simple.

The Lagrangian for this theory is given by the expression
{\small 
\begin{eqnarray}\label{YMSaction}
{\cal L}_{\rm YMS} =- {\rm Tr} \left( \frac{1}{4} F_{\mu\nu} F^{\mu\nu} + \frac{1}{2} D^\mu \phi^I D_\mu \phi^I - \frac{g^2}{4} \sum_{I\neq J}\left[  \phi^I, \phi^J \right]
\right) \, ,
\end{eqnarray}
}
where the gauge group is $U(N)$ and the scalars have a flavor index from a global symmetry group, $SO(M)$. 

In \cite{Cachazo:2014xea}, it was conjectured that the tree-level color-ordered amplitude for a set ``${\rm g}$"
of $p$ gluons (i.e. ${\rm g}=\{g_1, \ldots, g_p \}$) and a set ``${\rm s}$"
of $2m$ scalars\footnote{We apologize for the abuse of the notation. However, remember in this work we are calling the Mandelstam variables as $\tilde s_{a_1,...,a_p}$ (see appendix \ref{notation}).} (i.e. ${\rm s}=\{s_1, \ldots, s_{2m} \}$)  is given by the CHY integral
(note that, $p+2m=n$, where $n$ is the total number of particles)
{\small
\begin{eqnarray}
A^{\rm YMS}_{\rm g:s} = \int d\mu_n^{\rm CHY} \Delta(pqr)^2 \, \times \, {\cal I}^{\rm YMS}_{\rm g:s}(1,\ldots, n),
\end{eqnarray}
}
with 
{\small
\begin{eqnarray}\label{}
{\cal I}^{\rm YMS}_{\rm g:s} (1,\ldots, n)=  {\rm PT}_{(1,\ldots, n)} \times 
\hspace{-0.7cm}
\sum_{~~~\{a,b\}\in {\rm p.m. (s)}}
\hspace{-0.6cm}
\, \delta^{I_{a_1},I_{b_1}} \cdots \delta^{I_{a_m},I_{b_m}}
\frac{{\rm sgn}_{(\{a,b\})}}{  \s_{a_1b_1}\cdots \s_{a_m b_m} }
 \,
{\rm Pf}^{\prime} \Psi_{\rm g,s:g} \, ,  ~~ ~~~\label{}
\end{eqnarray}
}
where the $\Psi_{\rm g,s:g}$ matrix is given by the blocks
\begin{eqnarray}
\Psi_{\rm g,s:g} \, = \, 
\begin{blockarray}{cccc}
 b \in {\rm g}  &  b \in {\rm s}  &  b\in {\rm g}  \\
\begin{block}{(c|c|c)c}
{\cal A}_{ab} & {\cal A}_{ab} & \left(-{\cal C}^{\rm T}\right)_{ab} &  a\in {\rm g} \\
----& ----& ----\\
{\cal A}_{ab} & {\cal A}_{ab}  &\left(-{\cal C}^{\rm T}\right)_{ab} &   a\in {\rm s}  \\
---- & ----& ----\\
{\cal C}_{ab} & {\cal C}_{ab} & {\cal B}_{ab}   & a\in {\rm g}  \\
\end{block}
\end{blockarray} ~~~  .
\end{eqnarray}
It is useful to remind that  ``${\rm p.m.}$" means perfect matchings,  and note, $\{a_1,b_1,\ldots , a_m,b_m \} ={\rm s}$.  For instance, let us consider the punctures, $(\s_1)$, as  a gluon (${\rm g}=\{1 \}$), and $(\s_2, \s_3)$ as scalars (${\rm s}=\{2,3 \}$), then,  $\Psi_{\rm g,s:g} $ is a $4\times 4$ matrix given by
\vspace{-0.2cm}
{\small
\begin{eqnarray}\label{}
\Psi_{\rm g,s:g} \,
= \, \left(
\begin{array}[c]{c|cc|cc}
 0 & \frac{\tilde s_{12}}{\s_{12}} & \frac{\tilde s_{13}}{\s_{13}} & - {\cal C}_{11}   \\
 \hline
  \frac{\tilde s_{21}}{\s_{21}}  &0 &\frac{\tilde s_{23}}{\s_{23}} &- {\eps_1\cdot k_2 \over \s_{12} }  \\
   \frac{\tilde s_{31}}{\s_{31}} & {\tilde s_{32} \over \s_{32} } &0 & - {\eps_1\cdot k_3 \over \s_{13} }  \\ 
 \hline
{\cal C}_{11}  & {\eps_1\cdot k_2 \over \s_{12} } & {\eps_1\cdot k_3 \over \s_{13} } & 0  \\
\end{array}
\right)~~~ ,
\end{eqnarray}
}
\vskip-0.3cm\noindent
where, ${\cal C}_{11}= -\left(  \frac{\eps_1\cdot k_2}{\s_{12}} +\frac{\eps_1\cdot k_3}{\s_{13}}   \right)$.

As it was found in \eqref{PfYMS},  the double-cover version of the integrand, ${\cal I}^{\rm YMS}_{\rm g:s}$, is given by the expression 
\vspace{-0.1cm}
{\small
\begin{eqnarray}\label{}
{\cal I}^{\rm YMS}_{\rm g:s} (1,...,n) =  {\rm PT^\tau}_{(1,...,n)} \times 
\hspace{-0.7cm}
\sum_{~~~\{a,b\}\in {\rm p.m. (s)}}
\hspace{-0.6cm}
{\rm sgn}_{(\{a,b\})}\, \delta^{I_{a_1},I_{b_1}} \cdots \delta^{I_{a_m},I_{b_m}}
\, T_{a_1b_1}\cdots T_{a_m b_m} \,
{\bf Pf}^{\prime} \Psi^\L_{\rm g,s:g} \, ,\nonumber\\  ~~ ~~~\label{}
\end{eqnarray}
}
\vskip-0.5cm\noindent
where, ${\bf Pf}^{\prime} \Psi_{\rm g,s:g}^\L \equiv   
(-1)^{i+j}\, T_{ij}\left(  \left[ \prod_{a=1}^n   \frac{(y\s)_a}{y_a}    \right]
 {\rm Pf} \left[(\Psi^\L_{\rm g,s:g})^{ij}_{ij}\right] \right)$ and $\Psi^\L_{\rm g,s:g} \equiv \Psi_{\rm g,s:g}\Big|_{\frac{1}{\s_{ab}} \rightarrow T_{ab}}$. Therefore, a partial color-ordered amplitude among ${\rm g}$-gluons and ${\rm s}$-scalars is obtained from the integral, $A^{\rm YMS}_{\rm g:s} = \int d\mu_n^\L \, \frac{\Delta_{(pqr)} \Delta_{(pqr|m)}  }{ S^\tau_m} \times {\cal I}^{\rm YMS}_{\rm g:s} (1,...,n)$. 

Before to start with the examples, we define the partial color-flavor-ordered amplitude, $A^{\rm YMS}_{\rm g:s} (1,\ldots, n)_{(a_1,b_1:\cdots :a_m,b_m )}$, as
{\small
\begin{eqnarray}\label{color-flavorA}
\hspace{-0.2cm}
A^{\rm YMS}_{\rm g:s} (1,\ldots, n)_{(a_1,b_1:\cdots :a_m,b_m )} =  
\int d\mu_n^\L \, \frac{\Delta_{(pqr)} \Delta_{(pqr|m)}  }{ S^\tau_m} \,
{\rm PT}^\tau_{(1,\ldots, n)} \times 
 T_{a_1b_1}\cdots T_{a_m b_m} 
 \,
{\bf Pf}^{\prime} \Psi_{\rm g,s:g} \, . \nonumber\\
\end{eqnarray}
}
Visibly, $A^{\rm YMS}_{\rm g:s} (1,\ldots, n) =  
\sum\,
{\rm sgn}_{(\{a,b\})}\, \delta^{I_{a_1},I_{b_1}} \cdots \delta^{I_{a_m},I_{b_m}}\,
A^{\rm YMS}_{\rm g:s} (1,\ldots, n)_{(a_1,b_1:\cdots :a_m,b_m )} $, where $\sum=\sum_{\{a,b\}\in {\rm p.m. (s)}}$.

\subsection{Special Yang-Mills-Scalar Examples}\label{examplesYMS}

The ideas presented up to this point can be easily extrapolated to this special Yang-Mills-Scalar theory. Here, we show some simple examples to illustrate how  the {\bf integration rules} work over the amplitude, $A_{\rm g:s}^{\rm YMS}$.  Such as in Yang-Mills, we will introduce a superscript in order to indicate the red arrow, $(i,j)$. 

Let us consider the four-point example, ${\rm g}=\{ 1,2\}$ and ${\rm s}=\{ 3,4\}$. The color-ordered amplitude is given by the integral, $A^{\rm YMS}_{\rm g:s}(g_1,g_2,s_3,s_4)_{(3,4)} =  \int d\mu_4^\L \frac{\Delta_{(123)} \Delta_{(123|4)}  }{ S^\tau_4}$  $\times {\rm PT^\tau}_{(1,2,3,4)} \times 
\, T_{34}\,
{\bf Pf}^{\prime} \Psi^\L_{\rm g,s:g} $, where we have chosen, $(pqr|m)=(123|4)$.  To avoid singular configurations (see \eqref{Lbehavior}),  we pick out the red arrow to join the vertices, $(i,j)=(1,3)$. Therefore, 
following the graphical construction in section \ref{graphR}, the partial amplitude, $A^{\rm YMS}_{\rm g:s}(g_1,g_2,s_3,s_4)_{(3,4)}$, is represented by the graph (which we call as {\it YMS-graph})
\vspace{-0.5cm}
{\small
\begin{eqnarray}\label{}
A^{(1,3)}_{\rm g:s}( g_1, g_2,  s_3,s_4)_{(3,4)} \, =
\hspace{-0.1cm}
\int d\mu_4^{\L}
\hspace{-0.45cm}
\parbox[c]{5.8em}{\includegraphics[scale=0.17]{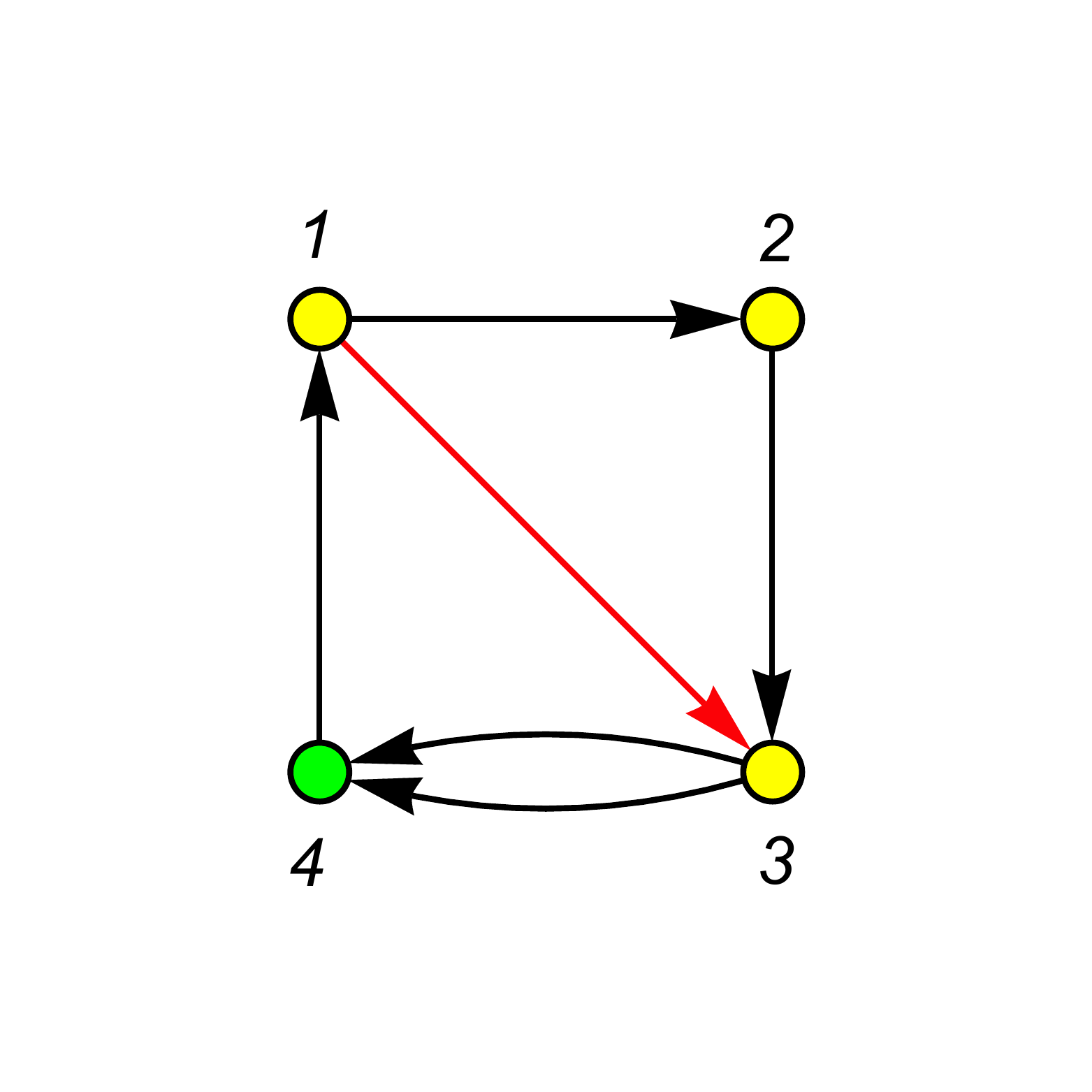}} =
\hspace{-0.45cm}
\parbox[c]{6.1em}{\includegraphics[scale=0.17]{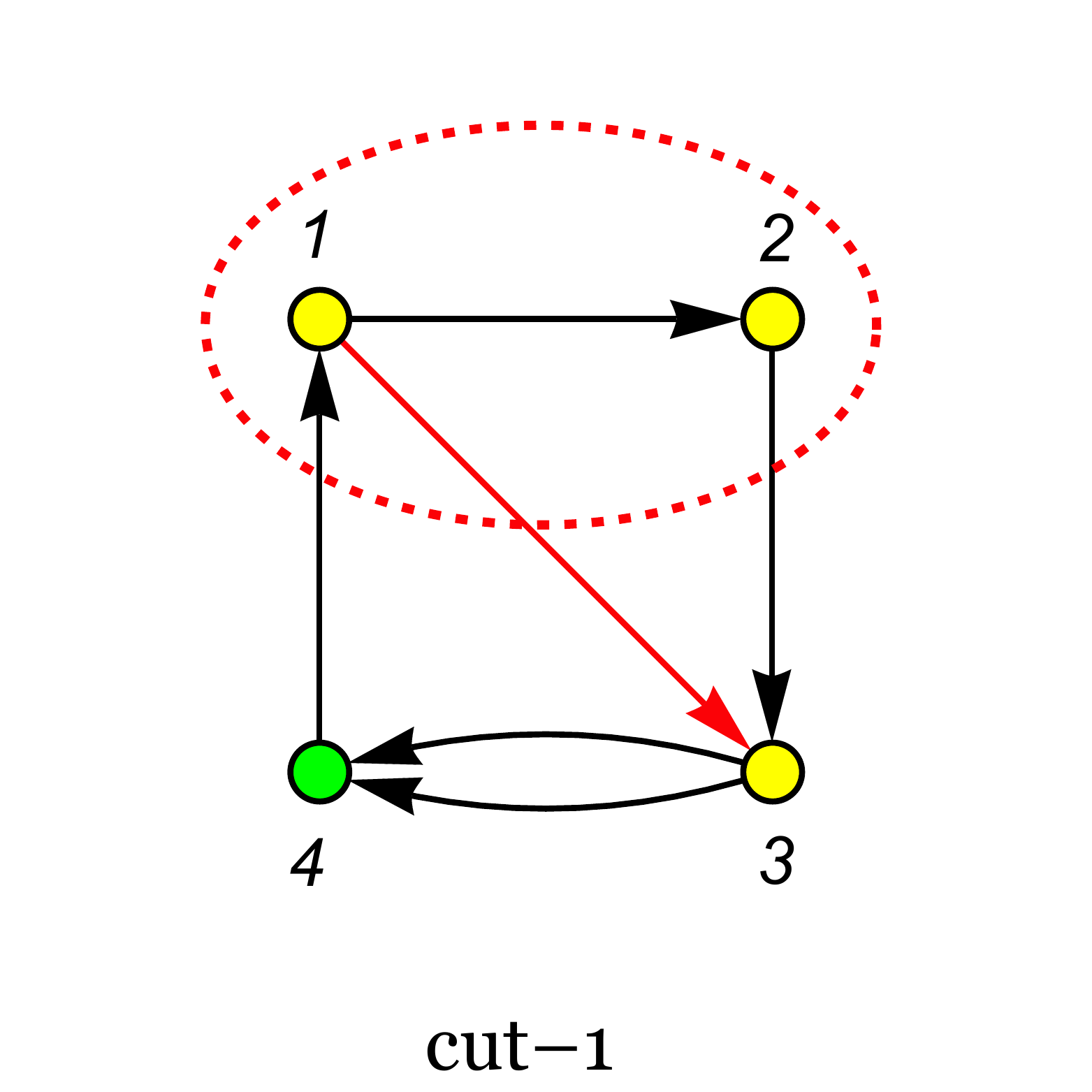}} +
\hspace{-0.45cm}
\parbox[c]{6.8em}{\includegraphics[scale=0.17]{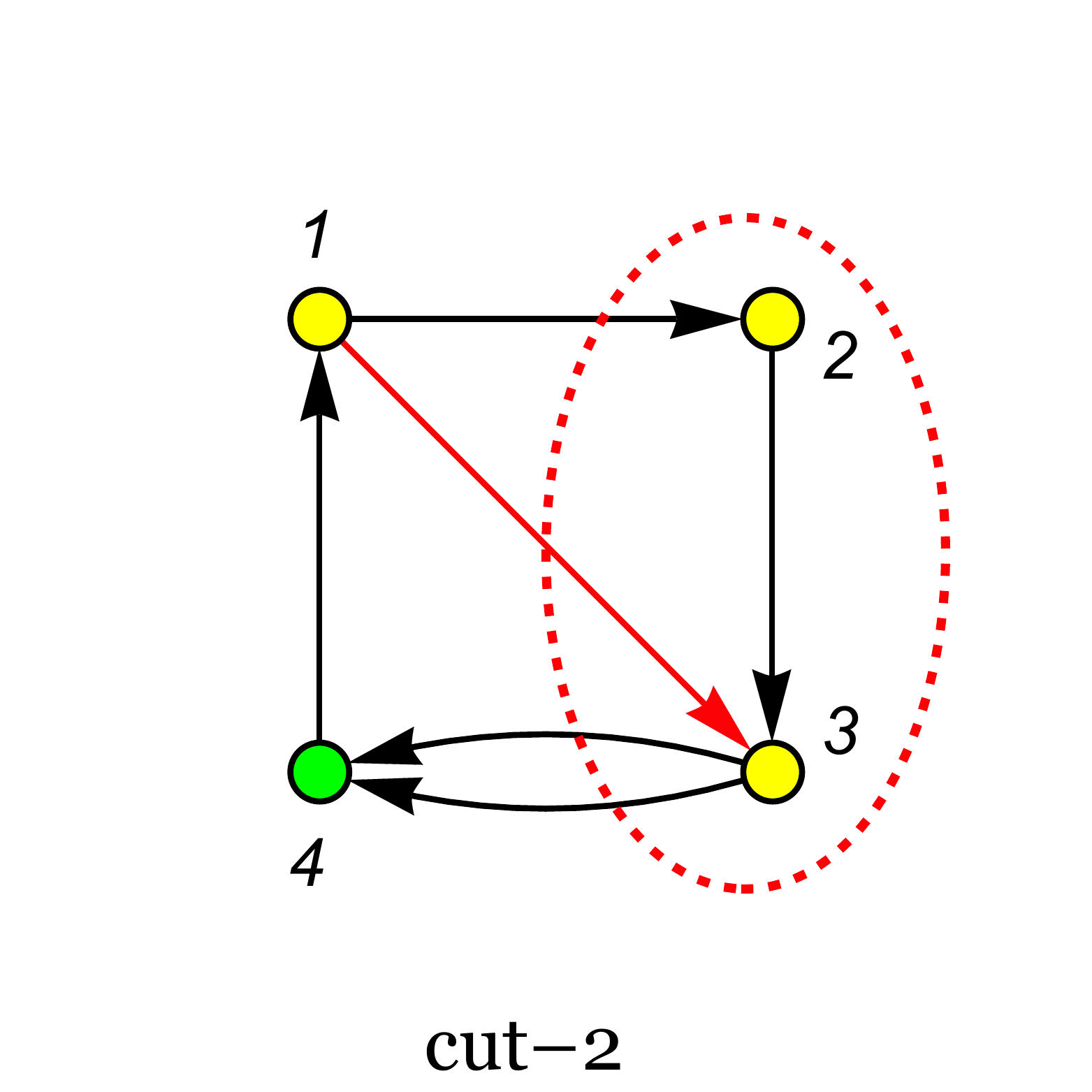}},
\end{eqnarray}
}
\vskip-0.5cm\noindent
where we have applied the {\bf rules-I,II}. It is very important to note that, the vertices with two black arrows are gluons, while the vertices with three black arrows are scalars. Additionally, from the ${\bf Pf}^\prime\Psi^\L_{\rm g,s:g}$ definition, the scalar vertex with four arrows (three black and one red) does not appear in the $\Psi^\L_{\rm g,s:g}$ matrix, i.e. all its associated rows/columns must be removed of $\Psi^\L_{\rm g,s:g}$ (such as it was said at the end of section \ref{specificrules}). Applying the {\bf rule-III} over the above cuts one has
\vspace{-0.5cm}
{\small
\begin{eqnarray}\label{}
&&\hspace{-0.85cm}
\parbox[c]{5.6em}{\includegraphics[scale=0.17]{YMS2g-2s-c1.pdf}} 
=\sum_r
\hspace{-0.7cm}
\parbox[c]{5.1em}{\includegraphics[scale=0.17]{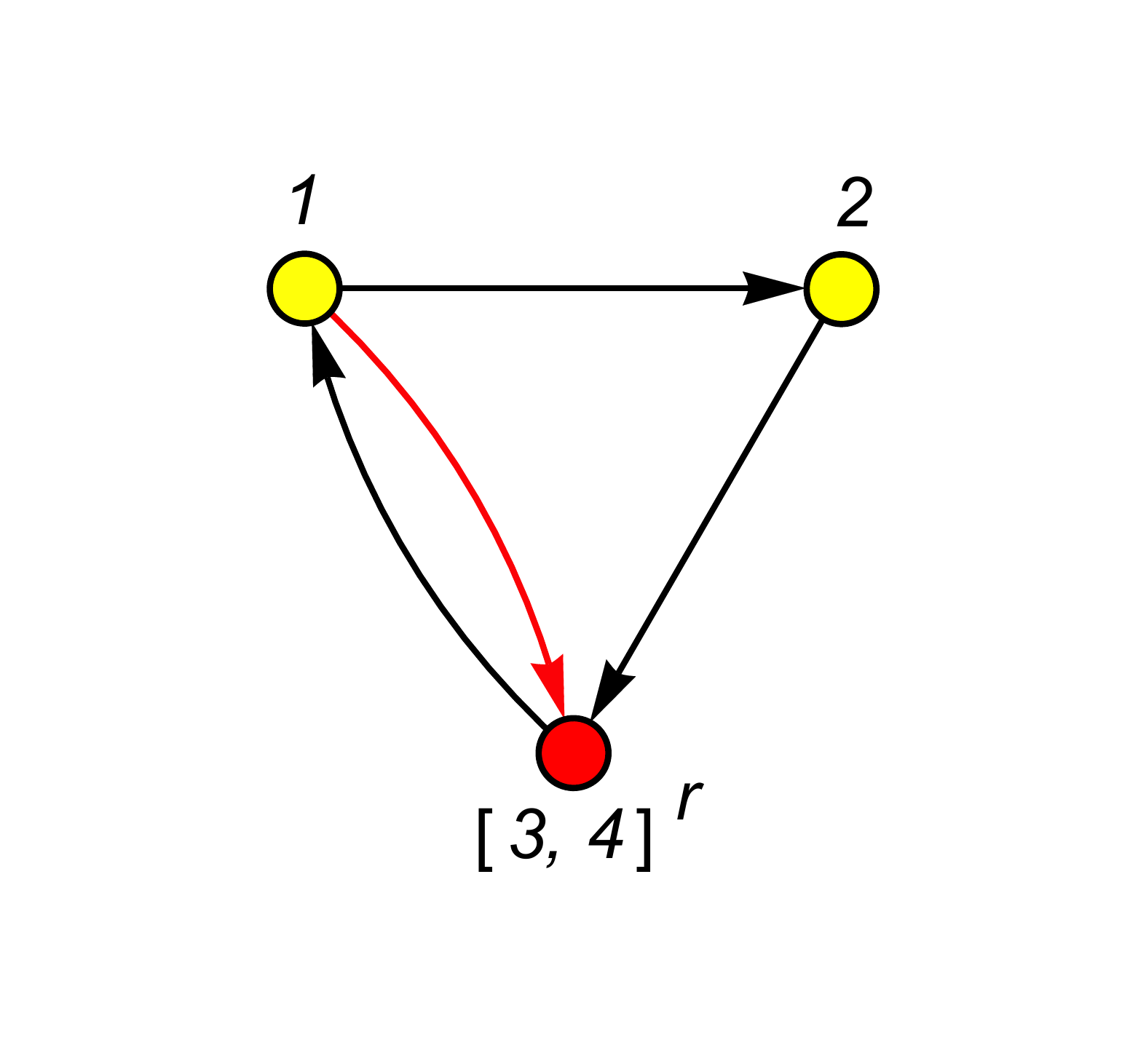}}
\times
\left( \frac{1}{\tilde s_{34} } \right) \times
\hspace{-0.8cm}
\parbox[c]{5.4em}{\includegraphics[scale=0.17]{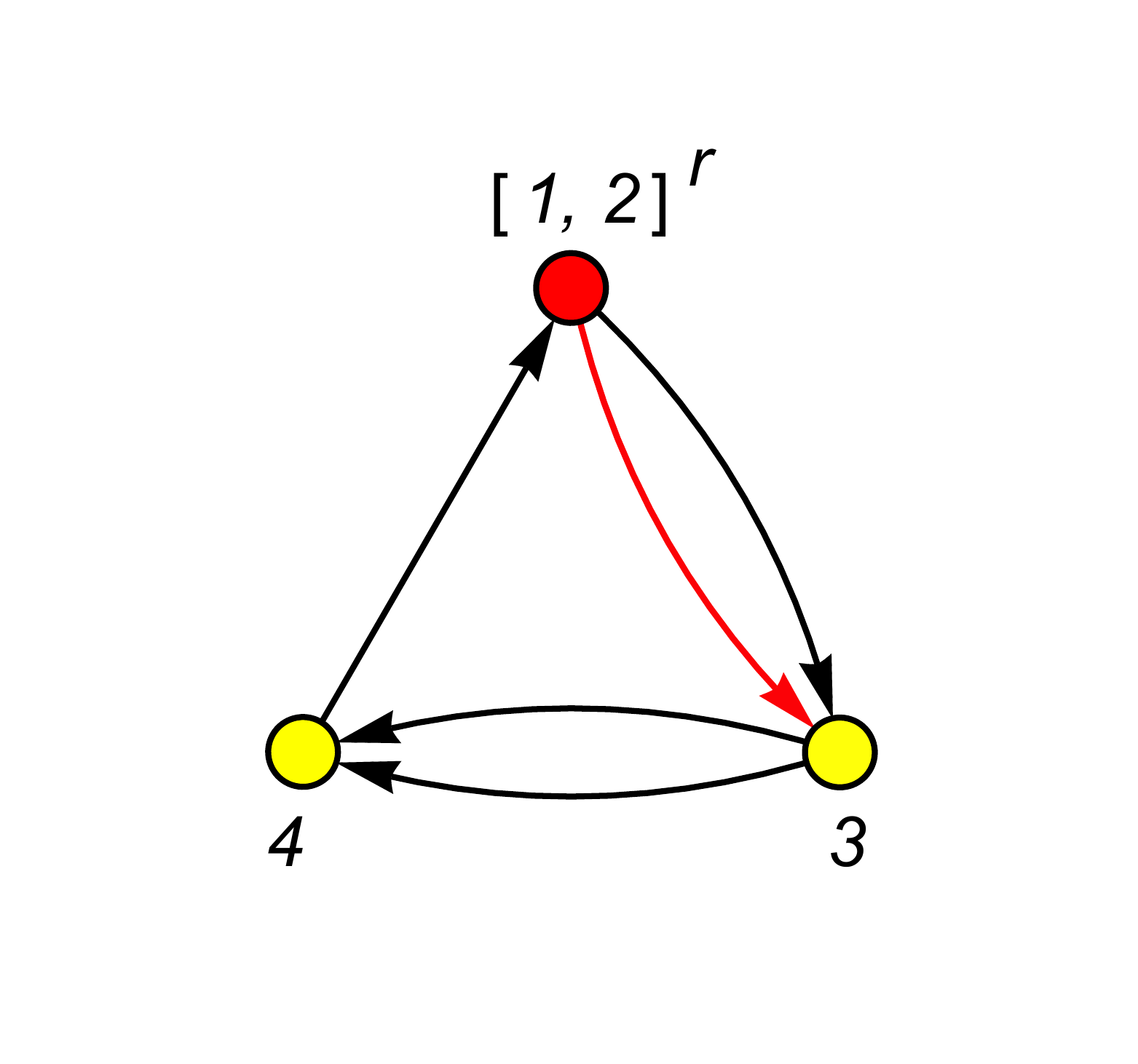}}
=
\frac{ \sum_r  A_3^{([3,4],1)} ([3,4]^r,1,2)  \,  A_{\rm g:s}^{([1,2],3)} ( g_{ [1,2]^r}, s_{3}, s_4)_{(3,4)} }{\tilde s_{34}}
\nonumber \\
&&\hspace{0.9cm}
=\frac{ (\eps_1\cdot k_3) (\eps_2\cdot k_1) - (\eps_1\cdot k_2) (\eps_2\cdot k_3) - (\eps_1\cdot \eps_2) \, \tilde s_{13} } {\tilde s_{12} }
,
\nonumber
\end{eqnarray}
}
\vskip-1.13cm\noindent
{\small
\begin{eqnarray}\label{2g-2s-YMS}
&&\hspace{-1.9cm}
\parbox[c]{6.2em}{\includegraphics[scale=0.17]{YMS2g-2s-c2.pdf}}
=   
\hspace{-0.75cm}
\parbox[c]{5.7em}{\includegraphics[scale=0.17]{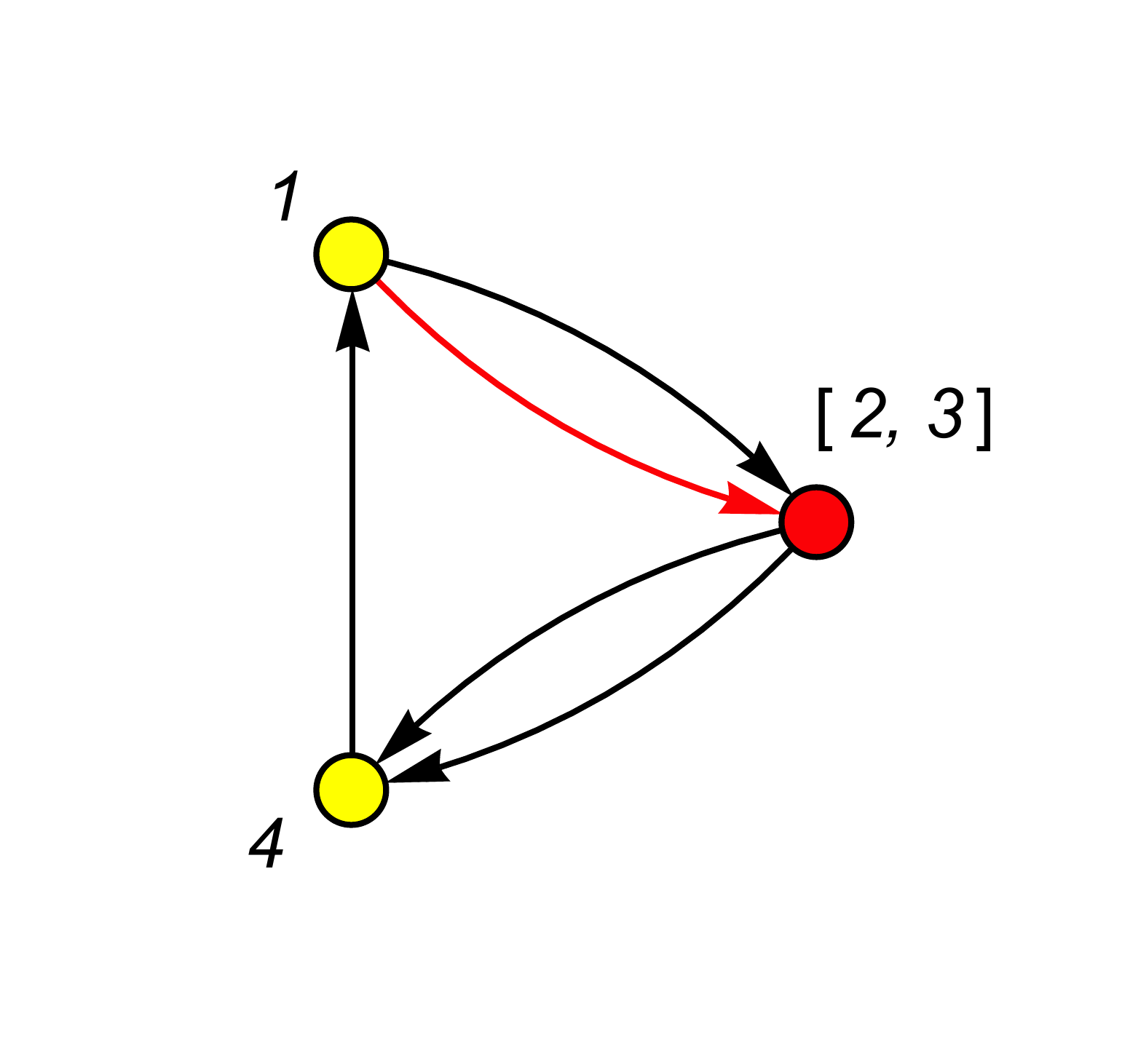}}
\times
\left( \frac{1}{\tilde s_{14} } \right) \times
\hspace{-0.4cm}
\parbox[c]{5.5em}{\includegraphics[scale=0.17]{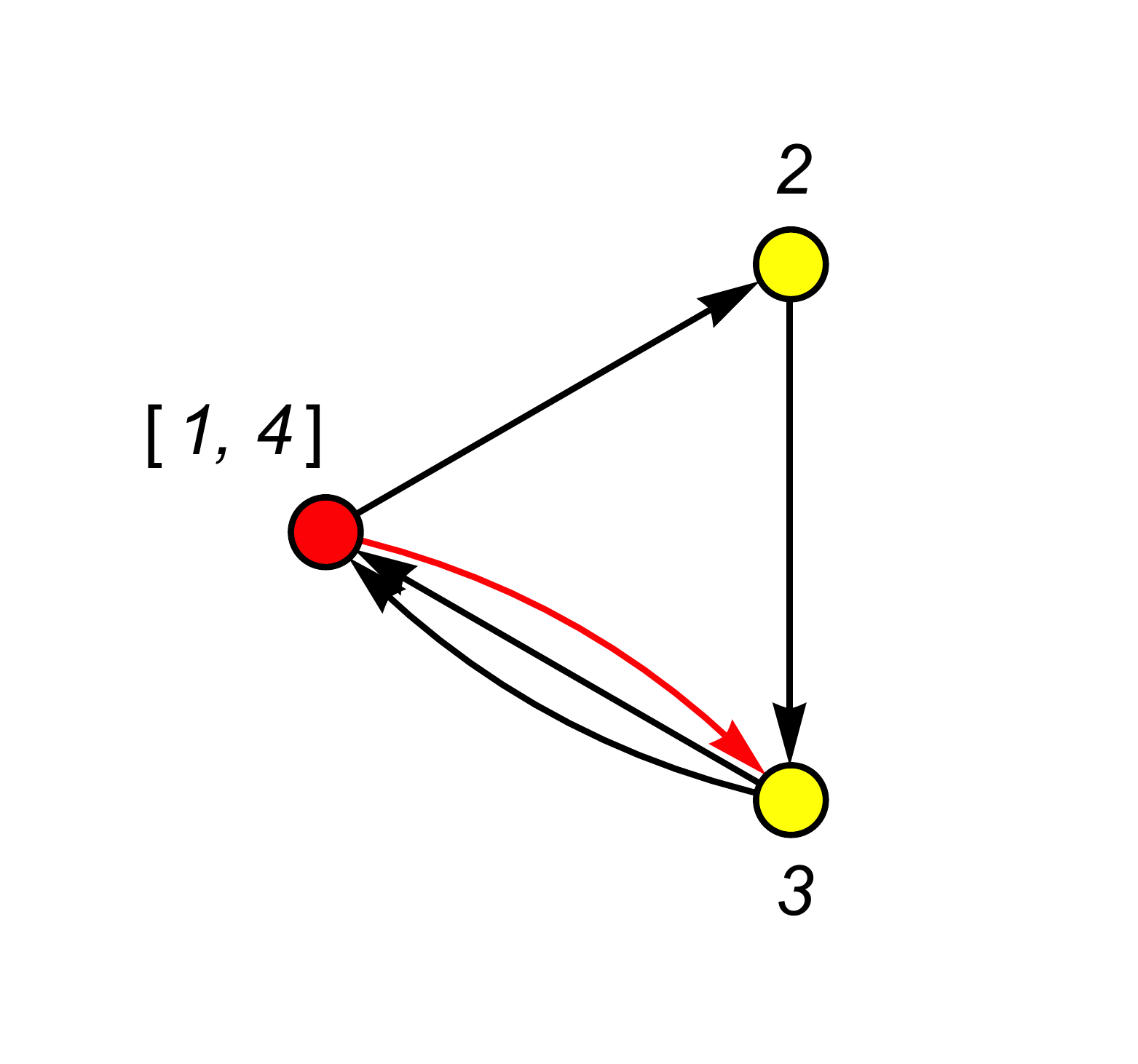}}
\nonumber \\
&&=
\frac{ A_{\rm g:s}^{(1,[2,3])} ( g_1, s_{[2,3]}, s_4)_{([2,3],4)}  \,  A_{\rm g:s}^{([1,4],3)} ( s_{ [1,4]}, g_2,s_{3})_{(3,[1,4])} }{\tilde s_{14}}\, =\, 
\frac{ (\eps_1\cdot k_4) \, (\eps_2\cdot k_3) }{\tilde s_{14}}, \nonumber\\
\end{eqnarray}
}
\vskip-0.7cm\noindent
where we have used the gluing identity  (by the {\bf rule-III}),  $\sum_{r} \eps^{r,\mu}_{[3,4]} \,  \eps^{r,\nu}_{[1,2]} = \eta^{\mu\nu}$, and the three-point building-block in \eqref{YM-BB} . Finally, it is not hard to show that, $A^{\rm YMS}_{\rm g:s}(g_1,g_2,s_3,s_4)_{(3,4)}=${\it cut-1} + {\it cut-2}, up to overall sign.

It is straightforward to note  that the three-point function, $A_{\rm g:s}^{([1,2],3)} (  g_{ [1,2]^r} , s_{3}, s_4)_{(3,4)} = (\eps_{[1,2]}^r \cdot k_{4} ) = \eps_{[1,2]}^{r,\mu}\times \left[ (k_4)_\mu - (k_3)_\mu \right]  \left( \frac{1}{2} \right)  - ( \eps_{[1,2]}^{r} \cdot k_{[1,2]} ) \left( \frac{1}{2} \right)$, is the Feynman vertex, ${\rm Tr} \left( A^\mu \left[\phi^I,\partial_\mu \phi^I\right]  \right)$  (Lagrangian \eqref{YMSaction}), plus an extra term that measures the transversality of the polarization vector  $\eps^{r,\mu}_{[1,2]}$ (like in pure Yang-Mills, expression \eqref{YM-BB}).

\subsubsection{Scalar Amplitudes,  from ${\cal A}$ to the $\Psi_{\rm g,s:g}$ Matrix}

In this section we consider the simplest examples, just scalar particles, i.e.  $\Psi^\L_{\rm g,s:g} = {\cal A}^\L$.  Let us begin with the four-point computations, $A^{\rm YMS}_{\rm g:s} (s_1,s_2,s_3,s_4)_{(1,3 :2,4)}$ and $A^{\rm YMS}_{\rm g:s} (s_1,s_2,s_3,s_4)_{(1,2: 3,4)}$. We set the gauge fixing, $(pqr | m)=(123 | 4)$ and, to avoid singular cuts, we choose the red arrow among, $(i,j)=(1,4)$.  So, for the amplitude $A^{(1,4)}_{\rm g:s} (s_1,s_2,s_3,s_4)_{(1,3 :2,4)}$, one has
\vspace{-0.5cm}
{\small
\begin{eqnarray}\label{4scalar-1}
A^{(1,4)}_{\rm g:s}( s_1, s_2,s_3, s_4)_{(1,3:2,4)} \, =
\hspace{-0.1cm}
\int d\mu_4^{\L}
\hspace{-0.45cm}
\parbox[c]{5.6em}{\includegraphics[scale=0.17]{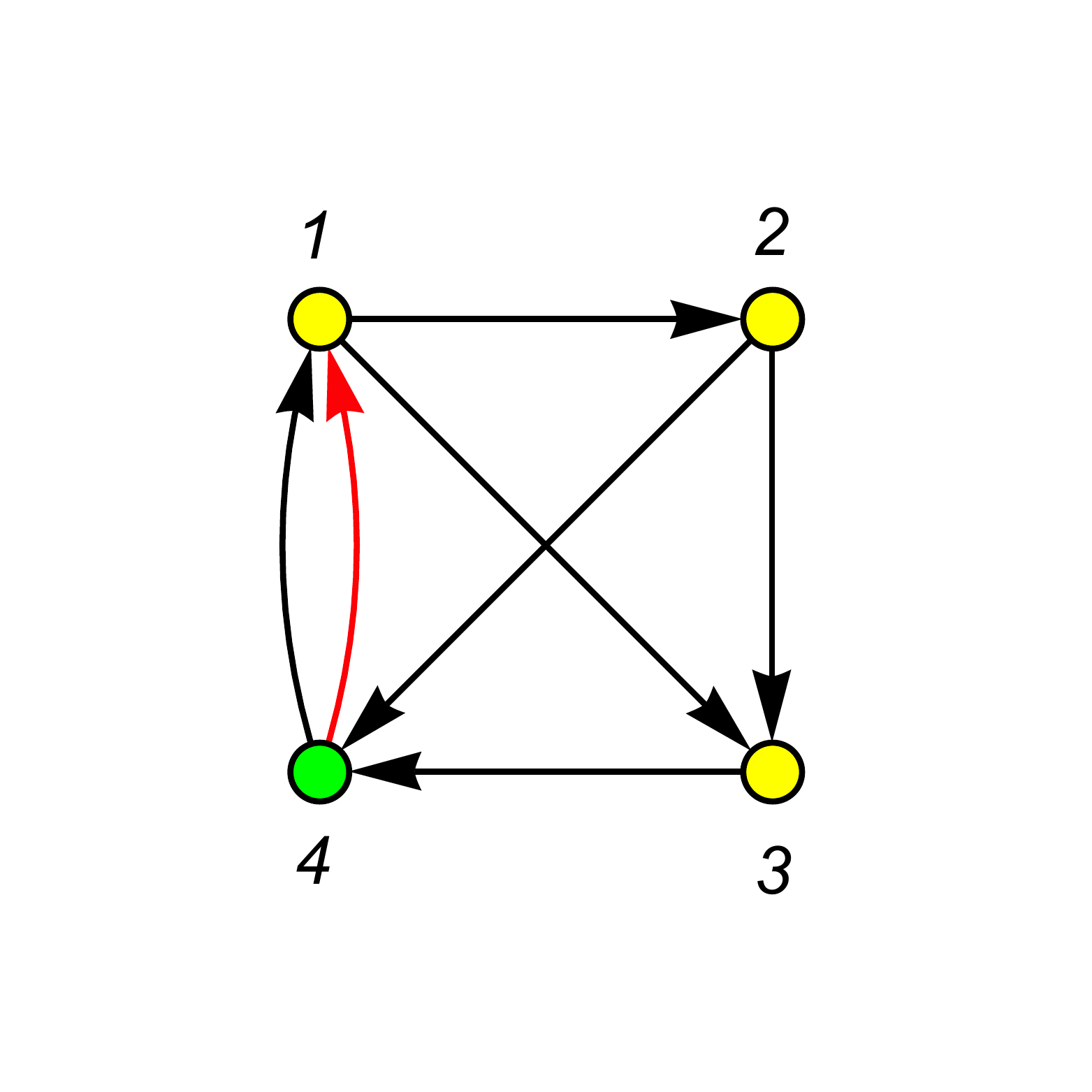}} =
\hspace{-0.5cm}
\parbox[c]{6.5em}{\includegraphics[scale=0.17]{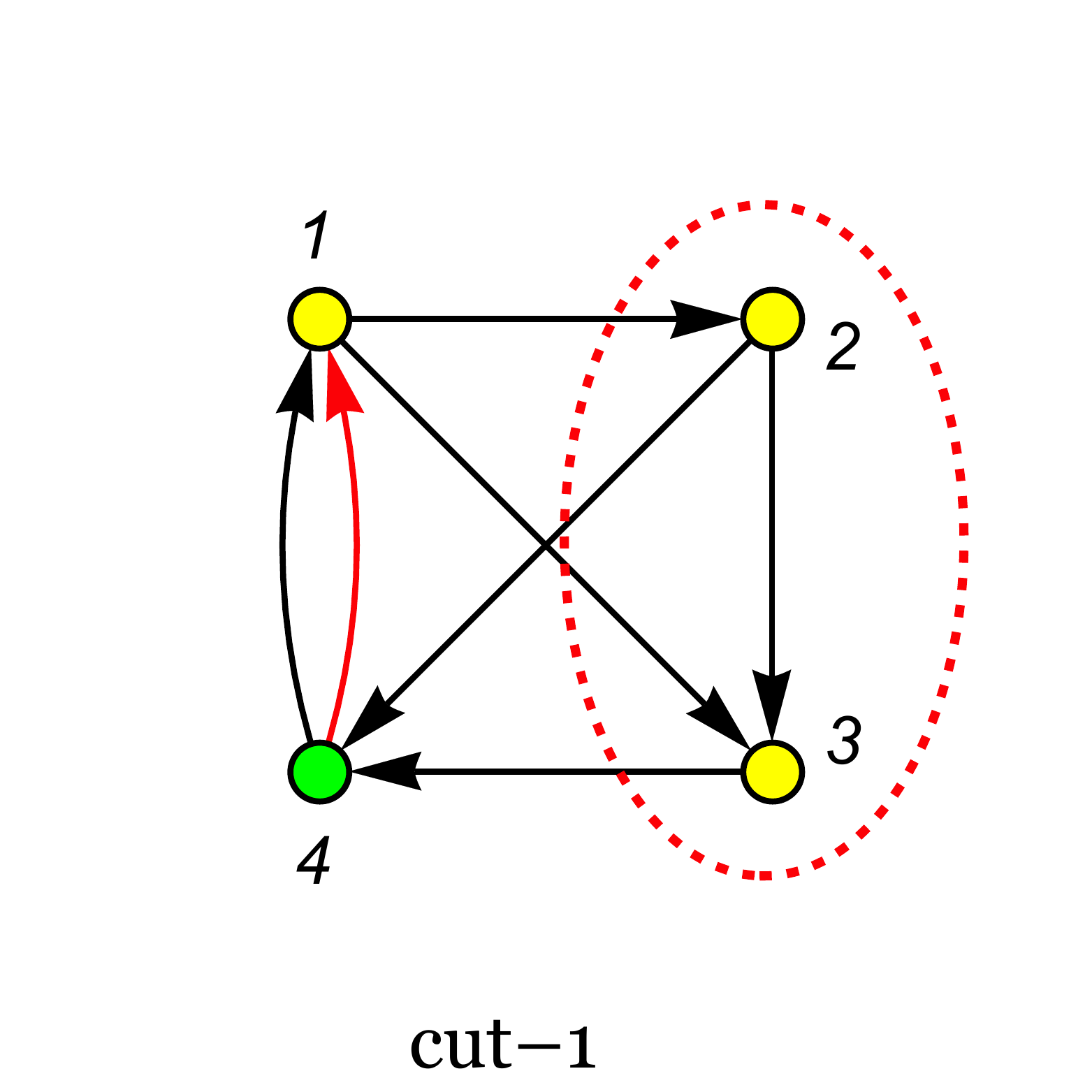}} \, ,
\quad~~
\end{eqnarray}
}
\vskip-0.5cm\noindent
where the {\bf rules-I,II} have been applied. By the {\bf rule-IIIb},  it is simple to carry out  {\it cut-1}
\vskip-0.4cm\noindent
{\small
\begin{eqnarray}\label{4scalar1-cuts}
&&
\hspace{0.5cm}
\parbox[c]{6.2em}{\includegraphics[scale=0.17]{4sc-1324-c1.pdf}}
=  
\hspace{-0.75cm}
\parbox[c]{5.7em}{\includegraphics[scale=0.17]{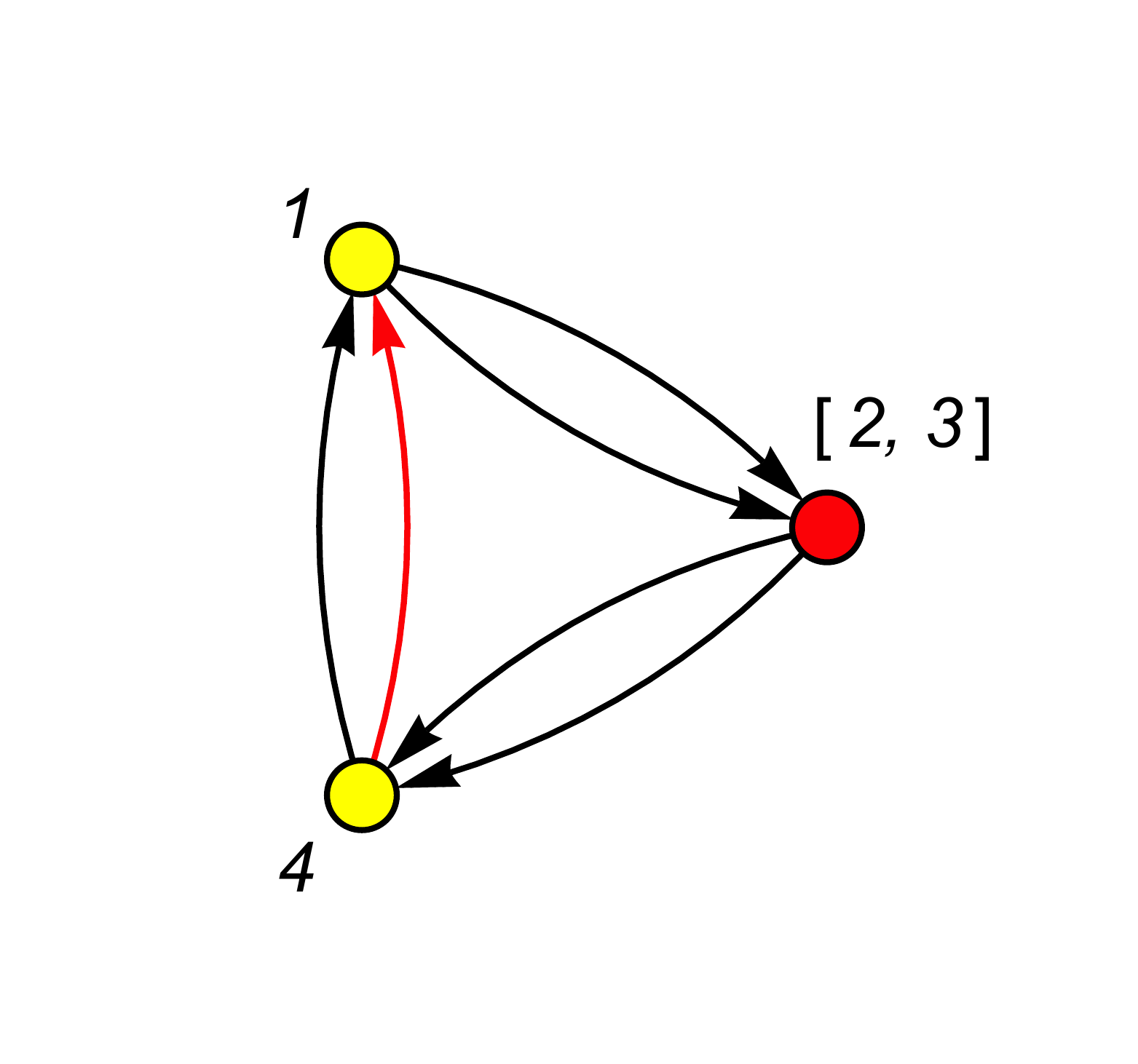}}
\times
\left( \frac{1}{\tilde s_{14} } \right) \times
\hspace{-0.4cm}
\parbox[c]{5.5em}{\includegraphics[scale=0.17]{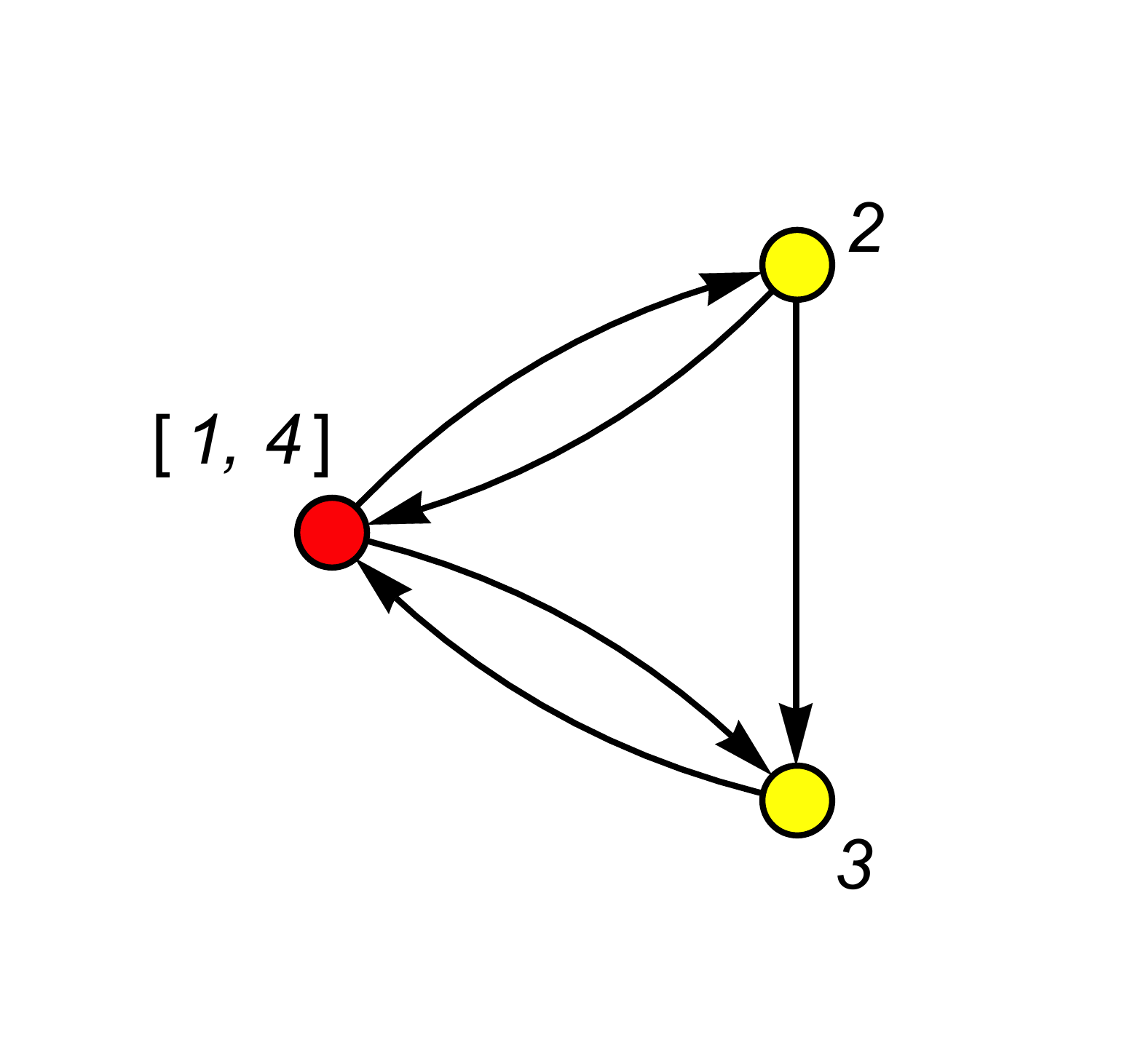}}
\nonumber \\
&&
\hspace{0.5cm}
=
1\times 
\left( \frac{1} {\tilde s_{14}} 
\right)
\times \frac{ (\s_{[1,4] 2} \, \s_{23} \, \s_{ 3[1,4] } )^2 }{ (\s_{[1,4] 2} \, \s_{23} \, \s_{ 3[1,4] } )\times (\s_{2 [1,4]} \,  \s_{ [1,4] 3 }) }
\times
{\rm Pf}
\left[
\begin{matrix}
0 & \frac{\tilde s_{23} }{\s_{23}} \\ 
\frac{\tilde s_{32} }{\s_{32}}  & 0
\end{matrix}
\right]  
=  \frac{1}{\tilde s_{14} } \times \tilde s_{23}
=1 , \qquad ~~~~ ~~
\end{eqnarray}
}
\vskip-0.4cm\noindent
where we have taken into account that a scalar vertex with four arrows means one must remove its row/column of the matrix. This simple result matched with the one given in \cite{Cachazo:2014xea} (equation (4.5)), in addition,  an interesting question arises, what is the physical meaning of the  three-point resulting graphs obtained in \eqref{4scalar1-cuts}? (there is no cubic scalar vertex\footnote{As one can note from the previous examples (YM and YMS), the number of arrows cut by a given configuration (red dashed line) is related with the Feynman vertex.  In the above example,  \eqref{4scalar-1}, the {\it YMS-graph}  has a configuration that cuts four arrows, therefore, this must be related with the Feynman vertex, ${\rm Tr}\left(\, \sum \, \left[\phi^I,\phi^J \right]^2 \,\right)$. Since this cut is the only one contribution, we could assume in advance that the final answer was 1.}  
in \eqref{YMSaction}).
 
Such as we learned in pure Yang-Mills  (section \ref{LongContributions}), the spurious pole contribution in \eqref{4scalar1-cuts} can be seen as a longitudinal gluon contribution. To be more precise, it is straightforward to check
\vspace{-0.2cm}
{\small
\begin{eqnarray}\label{phi4vertex}
A^{\rm YMS}_{\rm g:s} (s_1,s_2,s_3,s_4)_{(1,3 :2,4)} =
\frac{2\,\sum_L \, A^{([2,3],4)}_{\rm g,s:g} ( s_1, g_{[2,3]^L}, s_4)_{ (4,1)}  \times A^{([1,4],3)}_{\rm g,s:g} (g_{[1,4]^L},s_2, s_3)_{ (3,2)} }{\tilde s_{14}},
\qquad~
\end{eqnarray}
}
\vskip-0.3cm\noindent
with the gluing identity, $\sum_{L} \eps^{L,\mu}_{[2,3]} \eps^{L,\nu}_{[1,4]} = \frac{ k^\mu_{[2,3]}\, k^\nu_{[1,4]} }{k_{[2,3]} \cdot k_{[1,4]}  }$. This expression means that the $\phi^4$ vertex can be factorized as a product of two Yang-Mills-Scalar three-point amplitudes glue by a longitudinal off-shell gluon\footnote{This is the equivalent to the identity found in \eqref{generalone} for pure Yang-Mills.}.
We will give a non-trivial example later.

Let us consider the next example, , $A^{\rm YMS}_{\rm g:s} (s_1,s_2,s_3,s_4)_{(1,2: 3,4)}$. Its {\it YMS-graph} is given by 
\vspace{-0.5cm}
{\small
\begin{eqnarray}\label{}
A^{(4,1)}_{\rm g:s}(s_1, s_2, s_3, s_4)_{(1,2:3,4)} \, =
\hspace{-0.1cm}
\int d\mu_4^{\L}
\hspace{-0.45cm}
\parbox[c]{5.6em}{\includegraphics[scale=0.17]{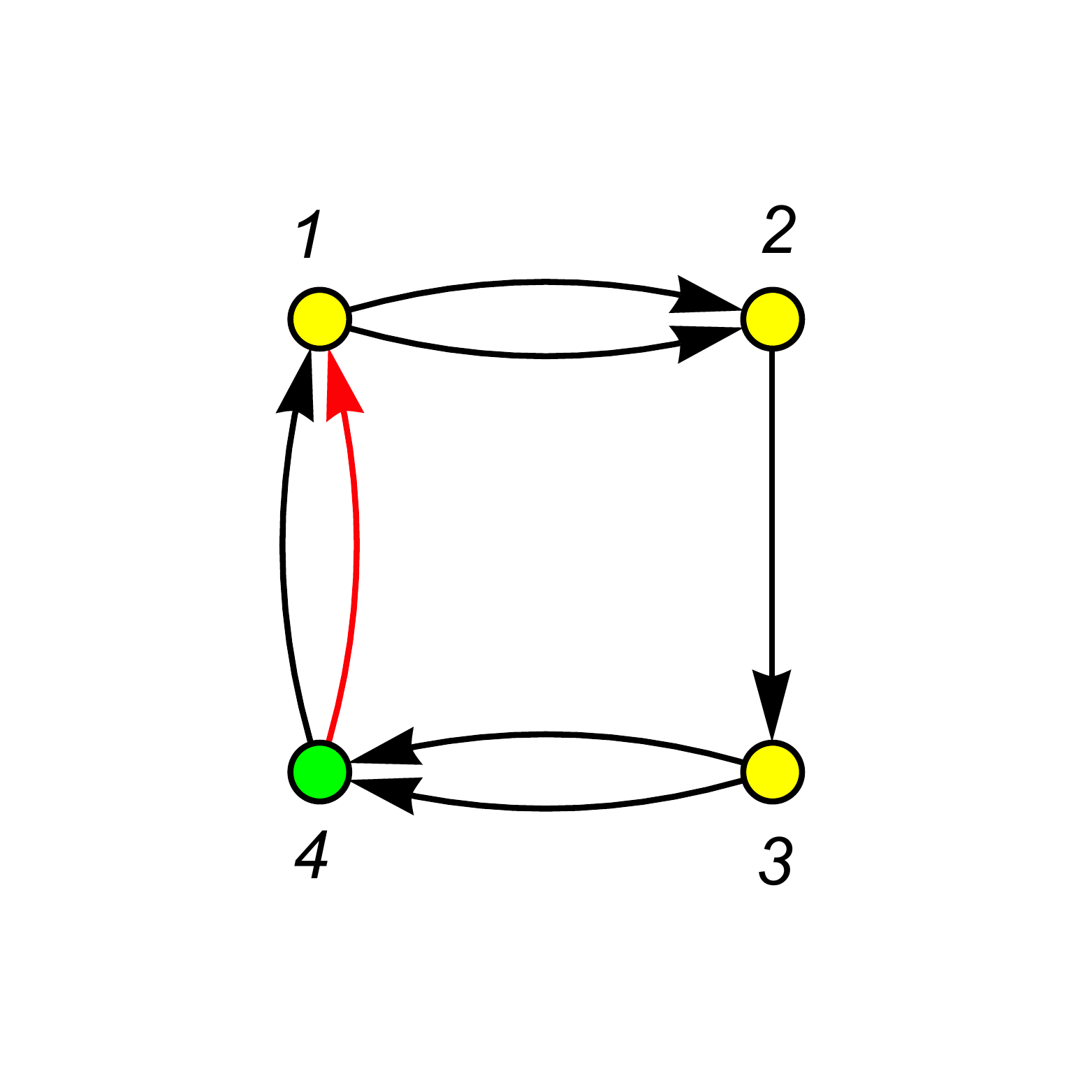}} =
\hspace{-0.45cm}
\parbox[c]{6.1em}{\includegraphics[scale=0.17]{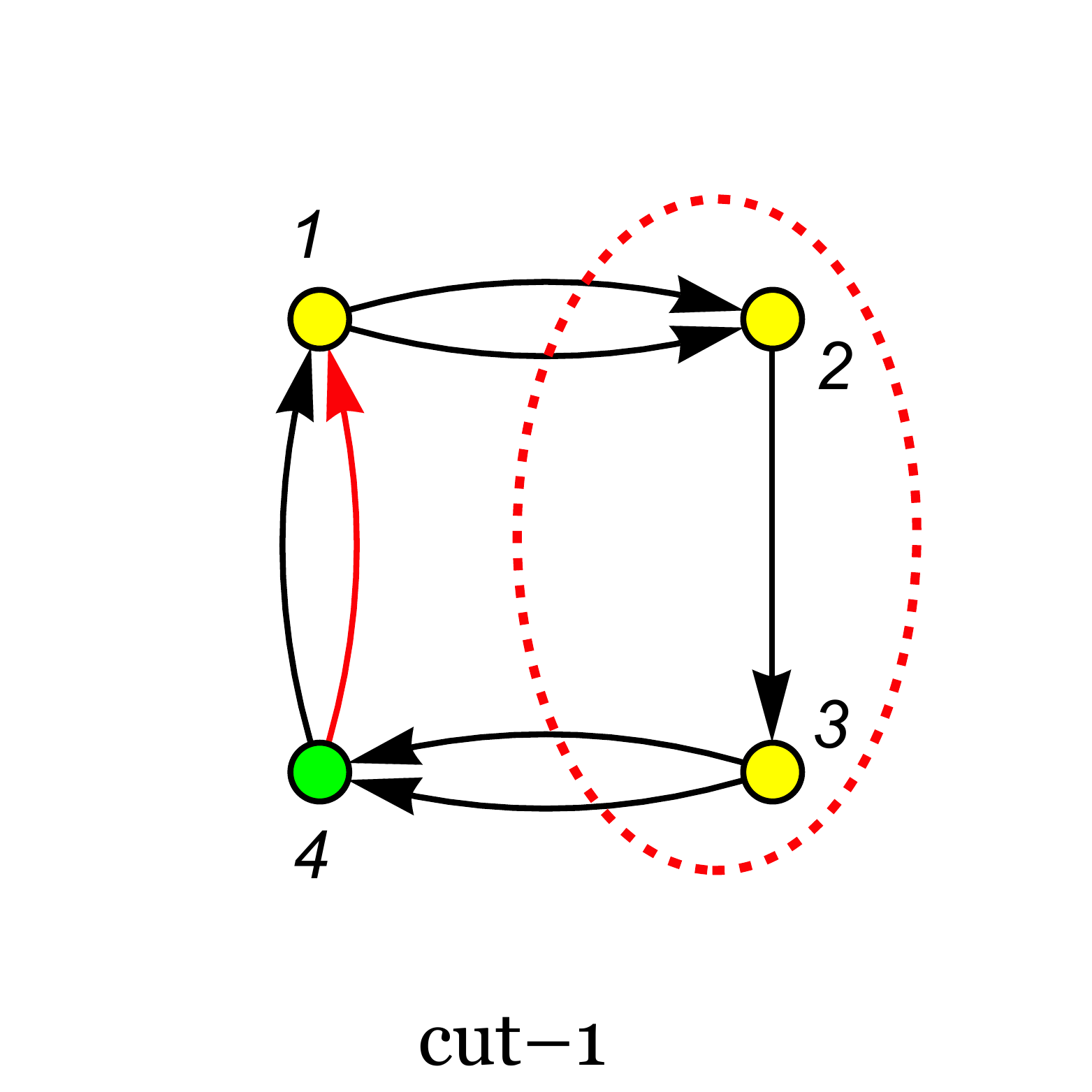}} +
\hspace{-0.45cm}
\parbox[c]{6.8em}{\includegraphics[scale=0.17]{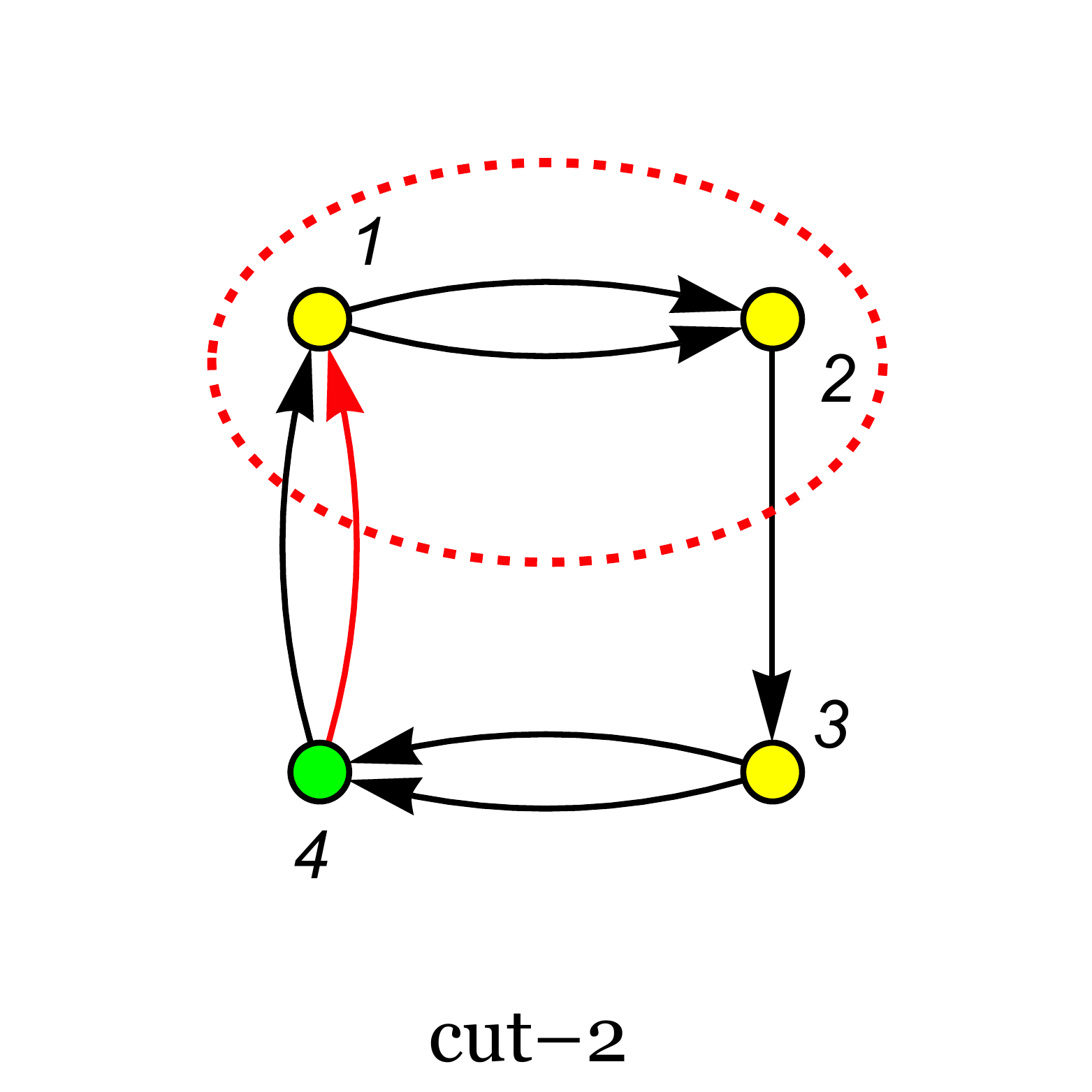}},
\quad~~
\end{eqnarray}
}
\vskip-0.5cm\noindent
where we have used the {\bf rules-I,II}.  While the {\it cut-1} was computed previously in \eqref{4scalar1-cuts},  {\it cut-1}$\,=1$, the {\it cut-2} is not obvious to carry out. If one applies the {\bf rule-IIIa}, the {\it cut-2} should be factorized  as
\vskip-0.6cm\noindent
{\small
\begin{eqnarray}\label{4scalar2-cut2}
\hspace{-0.5cm}
\parbox[c]{5.8em}{\includegraphics[scale=0.17]{4sc-1234-c2.pdf}}
=   \,
\sum_r
\hspace{-0.55cm}
\parbox[c]{5.5em}{\includegraphics[scale=0.17]{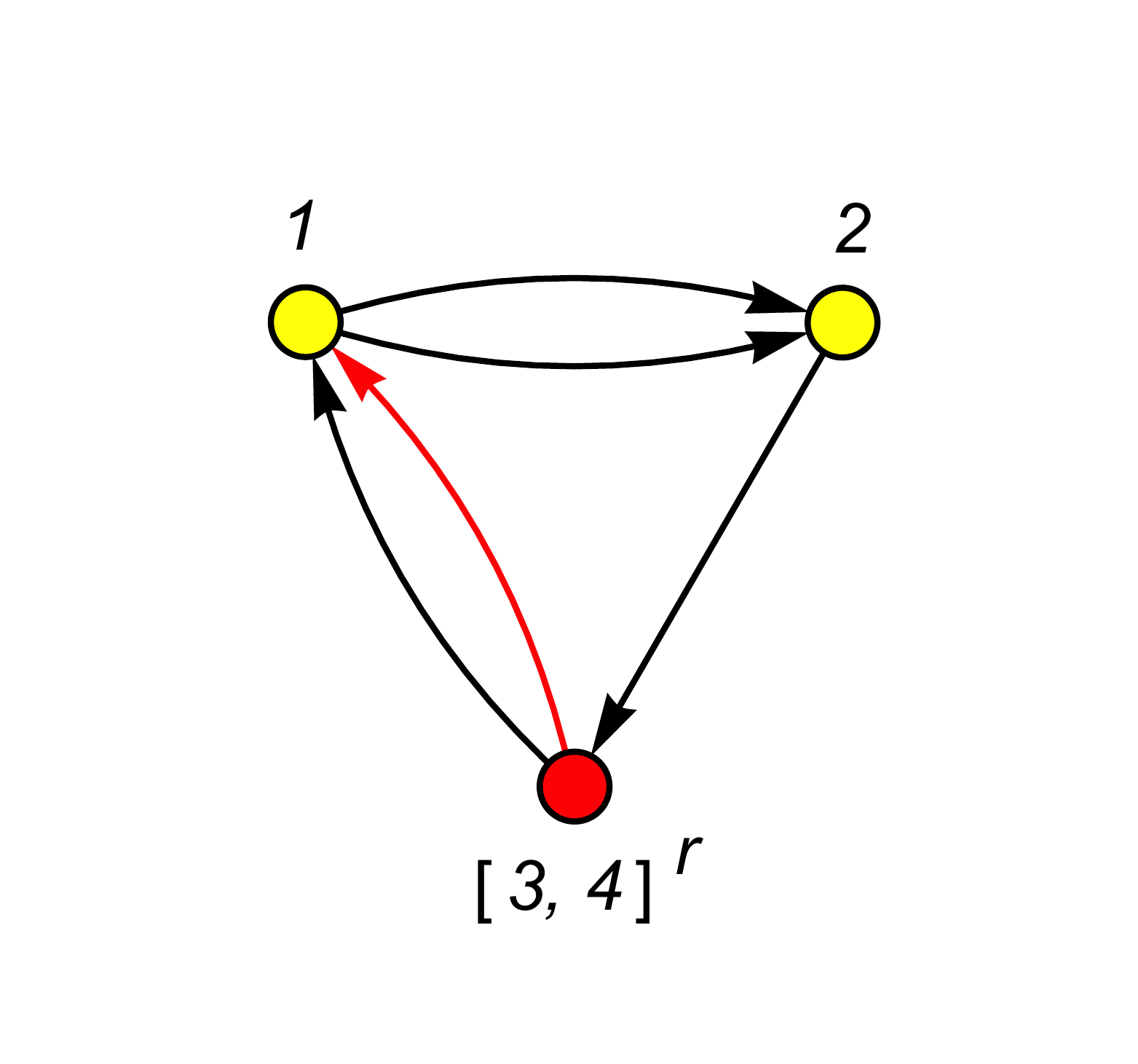}}
\times
\left( \frac{1}{\tilde s_{34} } \right) \times
\hspace{-0.6cm}
\parbox[c]{5.5em}{\includegraphics[scale=0.17]{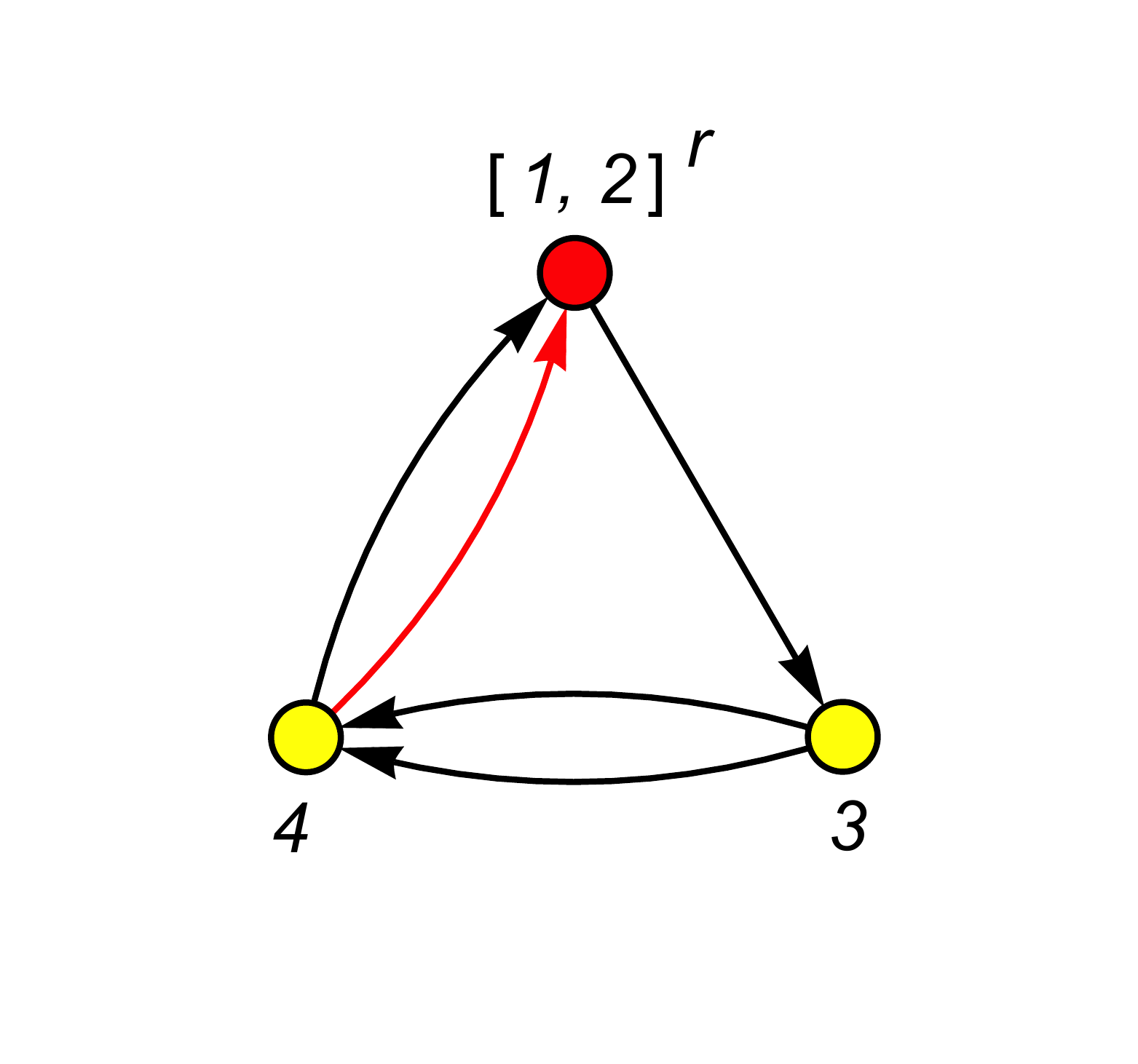}}
~~~ .
\end{eqnarray}
}
\vskip-0.4cm\noindent
Now the question is, How to read these resulting graphics?

Notice that these resulting graphs have an  identical structure as the second one obtained in \eqref{2g-2s-YMS}, and so, their meaning is the same. In  other words, each graphs in \eqref{4scalar2-cut2} have a off-shell gluon vertex with three arrows (i.e. its row/column must be removed of the $\Psi_{\rm g,s:g}$ matrix) and two on-shell scalar vertices, one of them with four-arrows (its row/column must be removed of the  $\Psi_{\rm g,s:g}$ matrix). Thus, the first graph becomes
\vskip-0.6cm\noindent
{\small
\begin{eqnarray}\label{}
\hspace{-0.55cm}
\parbox[c]{5.5em}{\includegraphics[scale=0.17]{4sc-1234-c2-r1.pdf}}
= A^{([3,4],1)}_{\rm g:s} ( g_{[3,4]^r}  ,  s_1 , s_2 )_{(1,2)}
=
(\s_{12}\,\s_{2[3,4]} \, \s_{[3,4]1})^2
\times {\rm PT}_{(1,2,[3,4])} \times \frac{1}{\s_{12}}  \nonumber 
\end{eqnarray}
}
\vskip-1.4cm\noindent
{\small
\begin{eqnarray}\label{}
\hspace{5.55cm}
\times \frac{1}{\s_{[3,4]1}} \, {\rm Pf}
\left[
\begin{matrix}
0 & -\frac{ \eps^r_{[3,4]}\cdot k_2 }{\s_{[3,4]2}} \\
\frac{ \eps^r_{[3,4]} \cdot k_2 }{\s_{[3,4] 2} }  & 0 \\ 
\end{matrix}
\right] =
(k_2\cdot \eps^r_{[3,4]}).
~~~
\end{eqnarray}
}
\vskip-0.2cm\noindent
In a similar way we compute the second resulting graph,  therefore the {\it cut-2} turns into, 
{\it cut-2} $=\frac{1}{\tilde s_{34}}\sum_r  A^{([3,4],1)}_{\rm g:s} (  g_{[3,4]^r } ,  s_1 , s_2 )_{(1,2)}\times  A^{(4,[1,2])}_{\rm g:s} (s_3, s_4,g_{[1,2]^r} )_{(3,4)} = \frac{\tilde s_{23}}{\tilde s_{34}}$. Finally,  the total result for $A^{\rm YMS}_{\rm g:s} (s_1,s_2,s_3,s_4)_{(1,2: 3,4)}$ is given by
\vskip-0.1cm\noindent
{\small
\begin{eqnarray}\label{scalar-YMS}
\hspace{-0.2cm}
A^{\rm YMS}_{\rm g:s} (s_1,s_2,s_3,s_4)_{(1,2: 3,4)}= 1 + \frac{\sum_r  A^{([3,4],1)}_{\rm g:s} (  g_{[3,4]^r } ,  s_1 , s_2 )_{(1,2)}\, A^{(4,[1,2])}_{\rm g:s} ( s_3,s_4,g_{[1,2]^r}  )_{(3,4)}  }{\tilde s_{34}}
= -\frac{\tilde s_{13}}{\tilde s_{12}}, \nonumber ~~\\
\end{eqnarray}
}
\vskip-0.8cm\noindent
which is the same aswer found in \cite{Cachazo:2014xea} (up to overall sign).

Roughly speaking, one of the interesting things here is that the ${\Psi}_{\rm g,s:g}$ matrix can emerge exponentially after factoring the ${\cal A}$ matrix,  to be more precise,  ${\rm Pf }^\prime \left[ {\cal A} \right] =\sum_L {\rm Pf }^\prime \left[ {\Psi}_{\rm g,s:g} \right] \times {\rm Pf }^\prime \left[ {\Psi}_{\rm g,s:g} \right]$ in \eqref{4scalar1-cuts} (longitudinal contributions), and  ${\rm Pf }^\prime \left[ {\cal A} \right] =\sum_r {\rm Pf }^\prime \left[ {\Psi}_{\rm g,s:g} \right] \times {\rm Pf }^\prime \left[ {\Psi}_{\rm g,s:g} \right]$
in \eqref{4scalar2-cut2}. This implies there are virtual gluons in a process that involves just scalar particles, which in the context of the special Yang-Mills-Scalar Lagrangian in \eqref{YMSaction} is a natural fact.  
However, it can also be seen in pure\footnote{The partial amplitude in \eqref{phi4vertex} represents the $\phi^4$ vertex.} $\phi^4$, and  additionally, a similar phenomenon appears in the effective field theories framework, which will be deeply studied in \cite{inpreparation}.

Finally, we consider the six-point amplitude, $A^{(6,1)}_{\rm g:s}(s_1, s_2, s_3,s_4,s_5,  s_6)_{(1,2:3,5:4,6)} $, 
\vspace{-0.4cm}
{\small
\begin{eqnarray}\label{}
\hspace{-0.1cm}
\int d\mu_6^{\L}
\hspace{-0.16cm}
\parbox[c]{6.6em}{\includegraphics[scale=0.17]{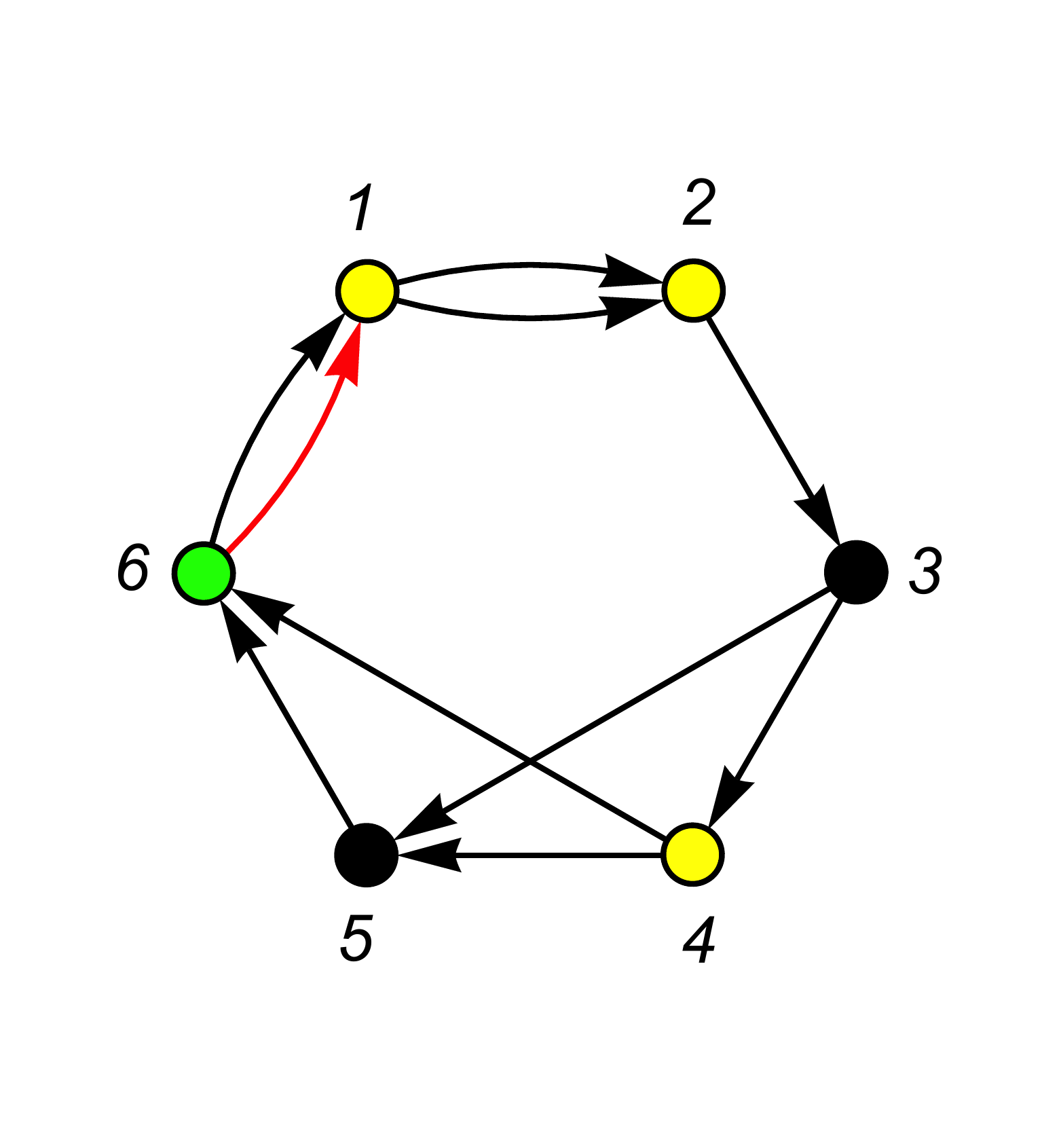}} =
\hspace{-0.2cm}
\parbox[c]{6.7em}{\includegraphics[scale=0.17]{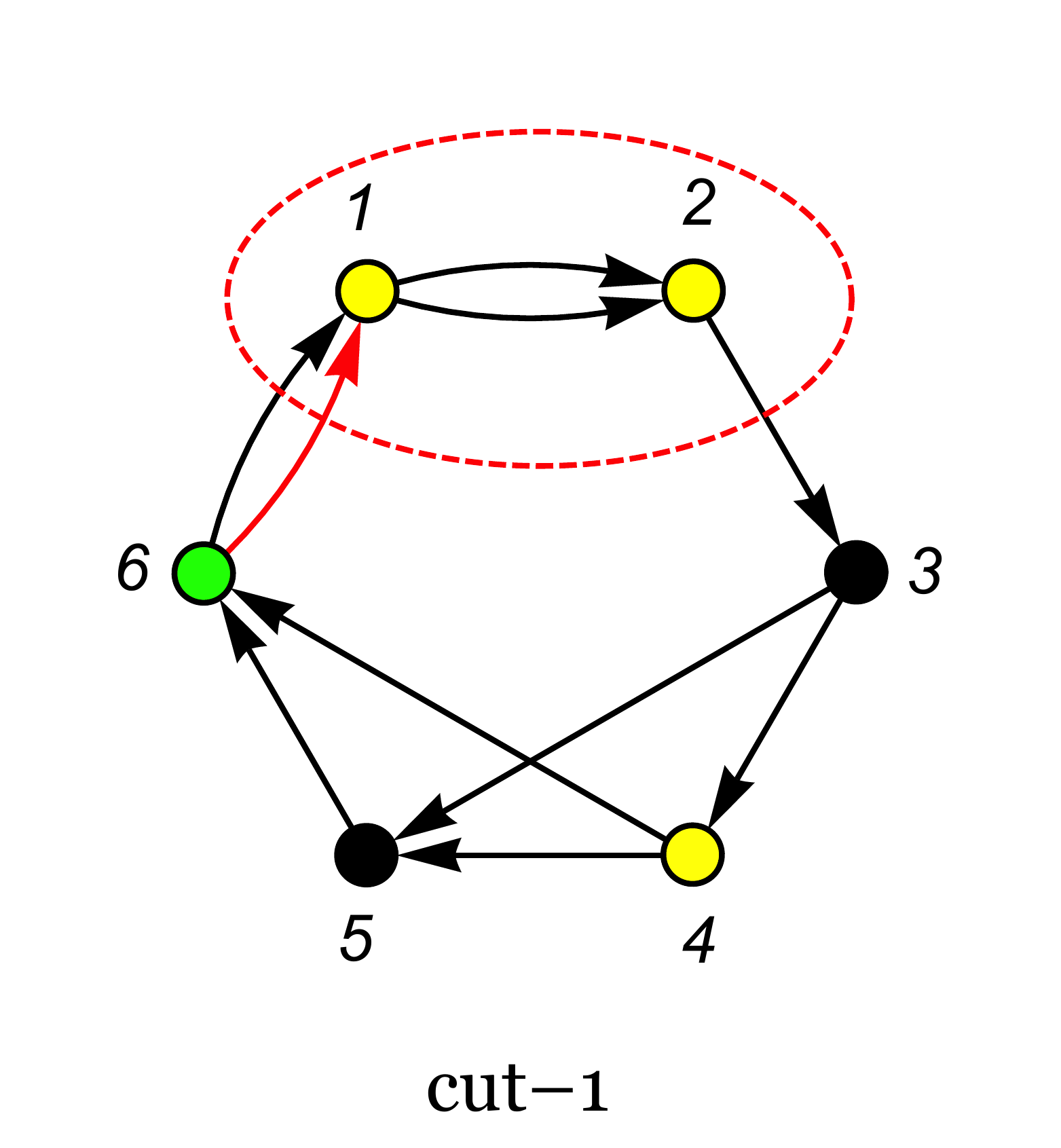}} +
\hspace{-0.2cm}
\parbox[c]{6.8em}{\includegraphics[scale=0.17]{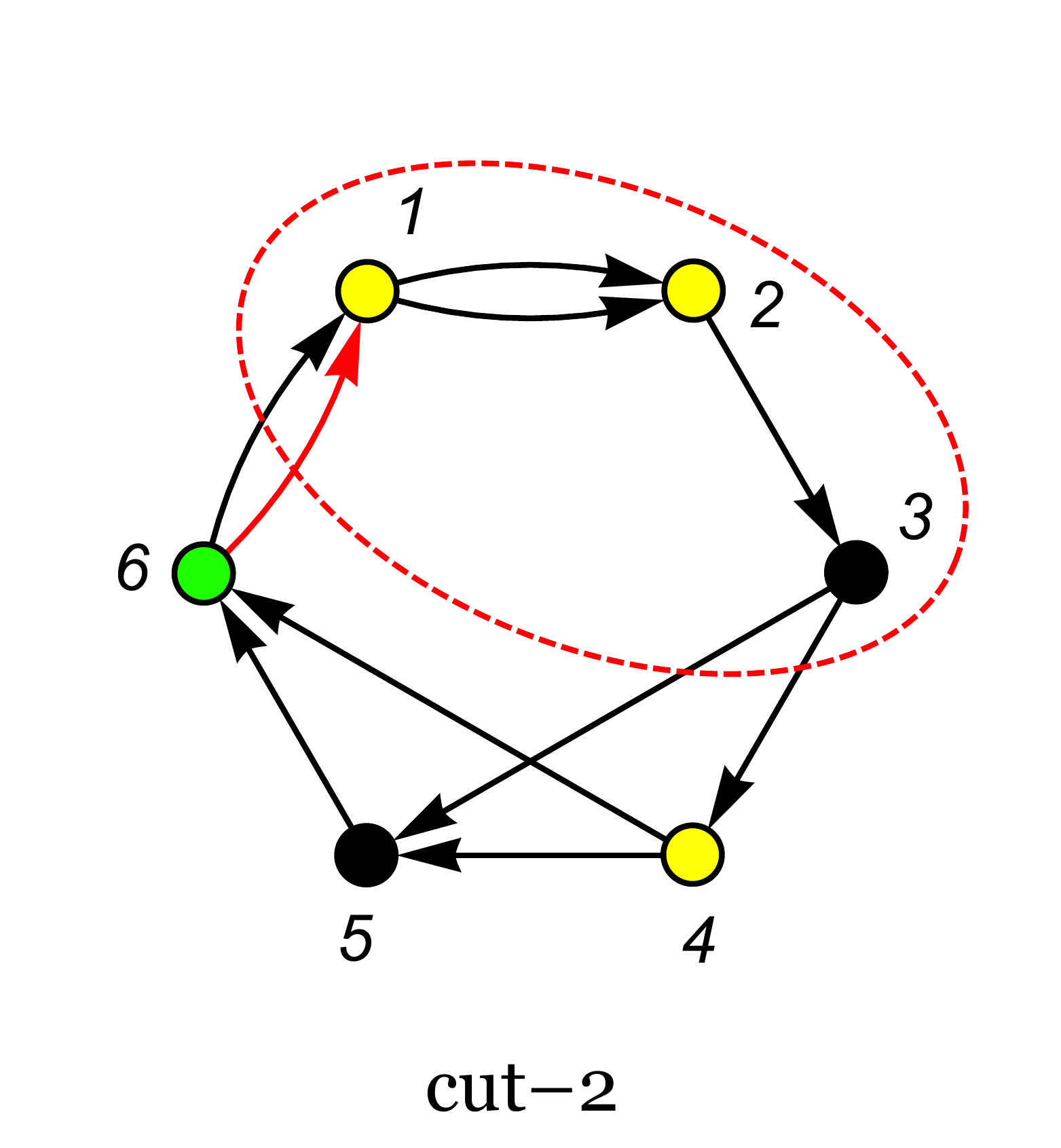}}
+
\hspace{-0.21cm}
\parbox[c]{6.8em}{\includegraphics[scale=0.17]{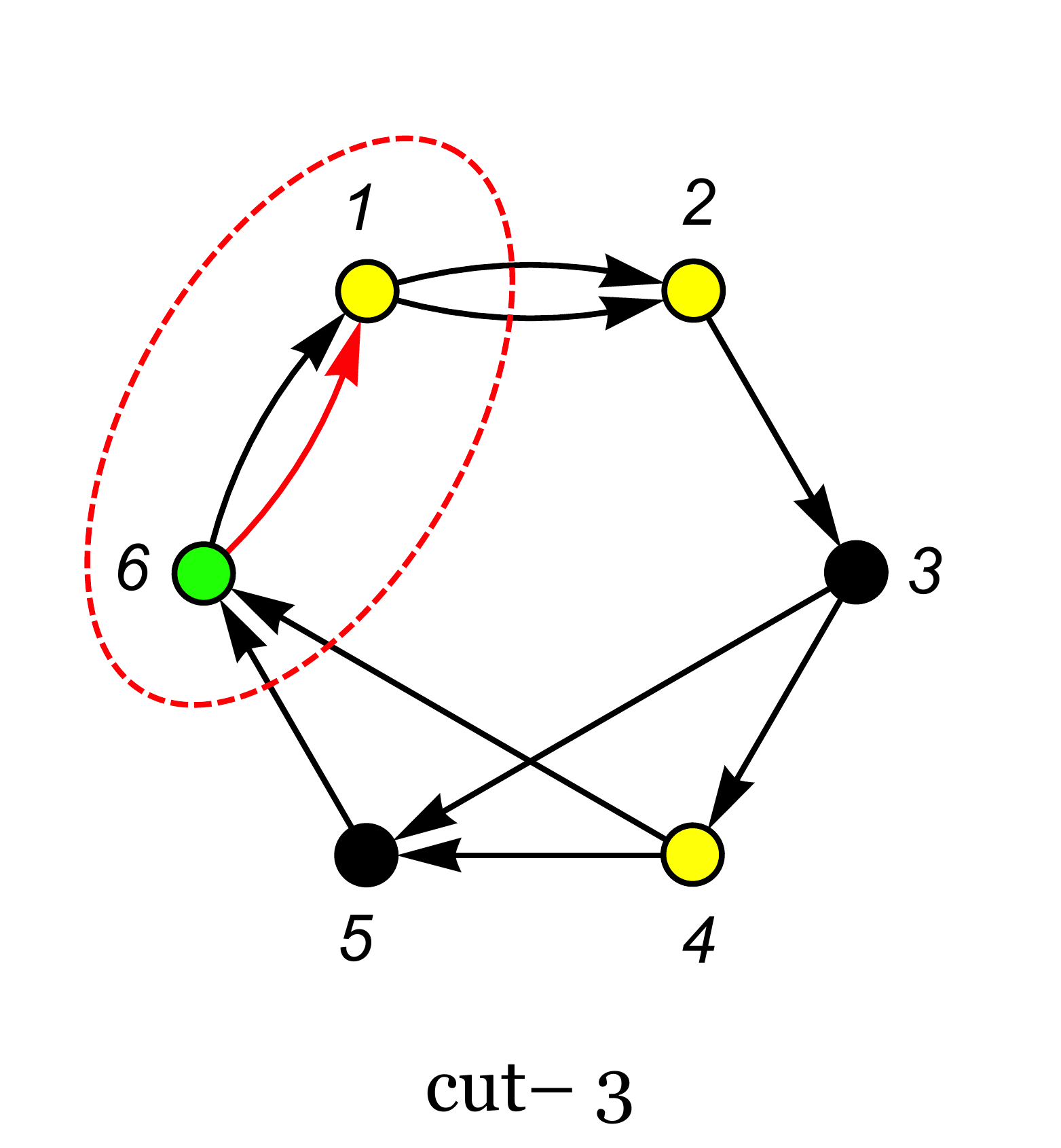}}\, ,
\quad~
\end{eqnarray}
}
\vskip-0.5cm\noindent
where we have chosen, $(pqr|m)=(124|6)$ and $(i,j)=(6,1)$.
It is simple to see that the {\it cut-1} is factorized by a vector field, while the {\it cut-2} by a scalar one, namely,
\vspace{-0.1cm}
{\small
\begin{eqnarray}\label{}
\hspace{-0.1cm}
\text{\it cut-1 } &=& \frac{\sum_r\,  A^{\rm ([3,4,5,6],1)}_{\rm g:s} ( g_{[3,4,5,6]^r},  s_1, s_2 )_{(1,2)} \times  A^{(6,[1,2])}_{\rm g:s} (s_6,g_{[1,2]^r}, s_3,s_4 ,s_5,)_{(3,5:4,6)} }{\tilde s_{3456}},\quad~~~~~~~ \\
\hspace{-0.1cm}
\text{\it cut-2 } &=& \frac{ A^{([4,5,6],1)}_{\rm g:s} ( s_3,s_{[4,5,6]},  s_1,s_2)_{(1,2:3,[4,5,6])} \times  A^{(6,[1,2,3])}_{\rm g:s} (s_4,s_5,s_6, s_{[1,2,3]})_{([1,2,3],5:4,6)} }{\tilde s_{456}},
\quad~~~~~~~
\end{eqnarray}
}
\vskip-0.5cm\noindent
with, $\sum_r\, \eps_{[4,5,6]}^{r,\mu}\, \eps_{[1,2]}^{r,\nu} = \eta^{\mu\nu}$. On the other hand, 
the resulting graphs obtained from the {\it cut-3} do not have an obvious physical interpretation,
\vspace{-0.5cm}
{\small
\begin{eqnarray}\label{}
\parbox[c]{6.6em}{\includegraphics[scale=0.17]{scalar-6pt-c3.pdf}} =
\hspace{-0.6cm}
\parbox[c]{5.9em}{\includegraphics[scale=0.17]{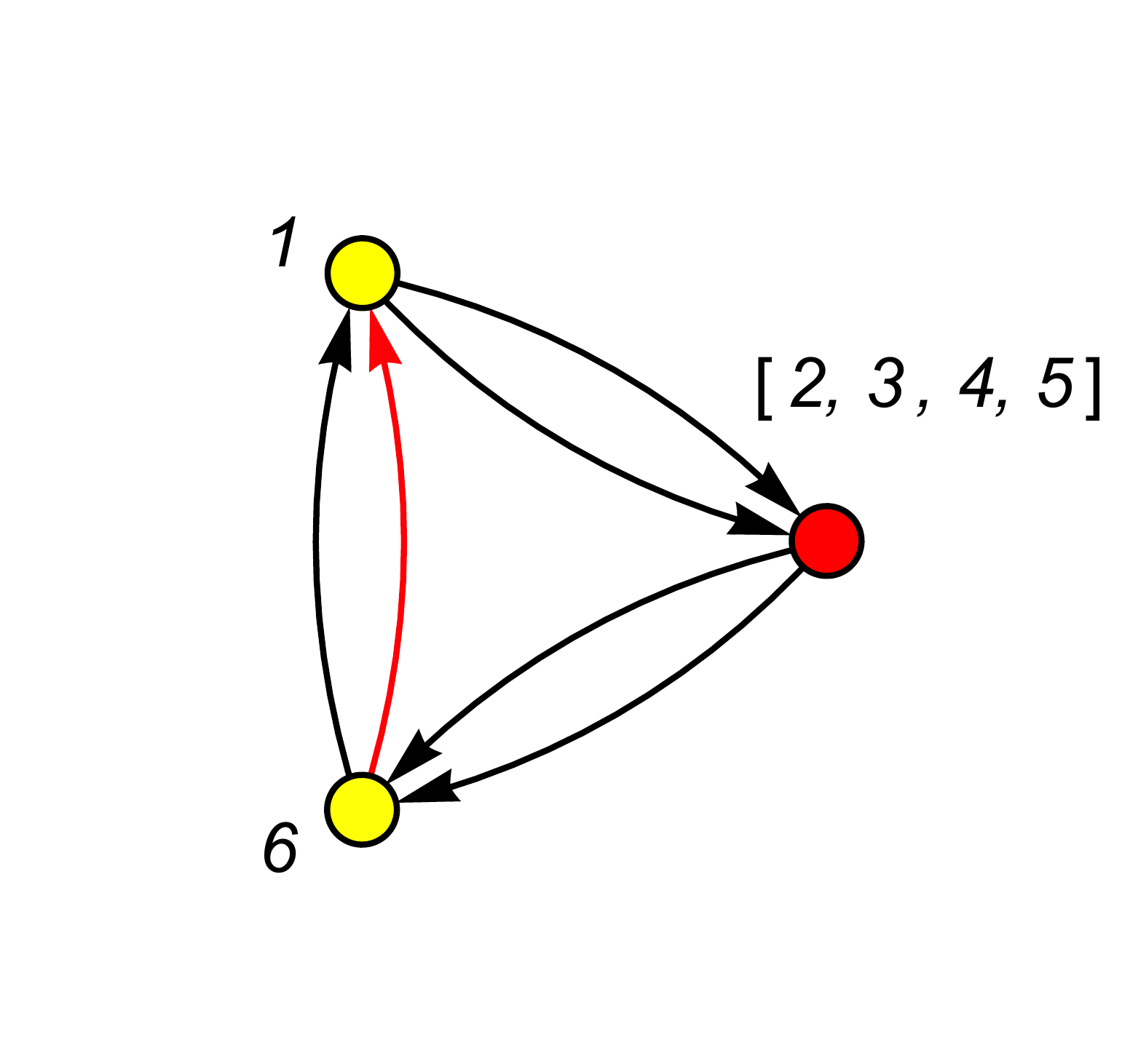}} \times
\left(\frac{1}{\tilde s_{16}} \right)
\times
\int d\mu_5^{\rm CHY}
\hspace{-0.3cm}
\parbox[c]{6.3em}{\includegraphics[scale=0.17]{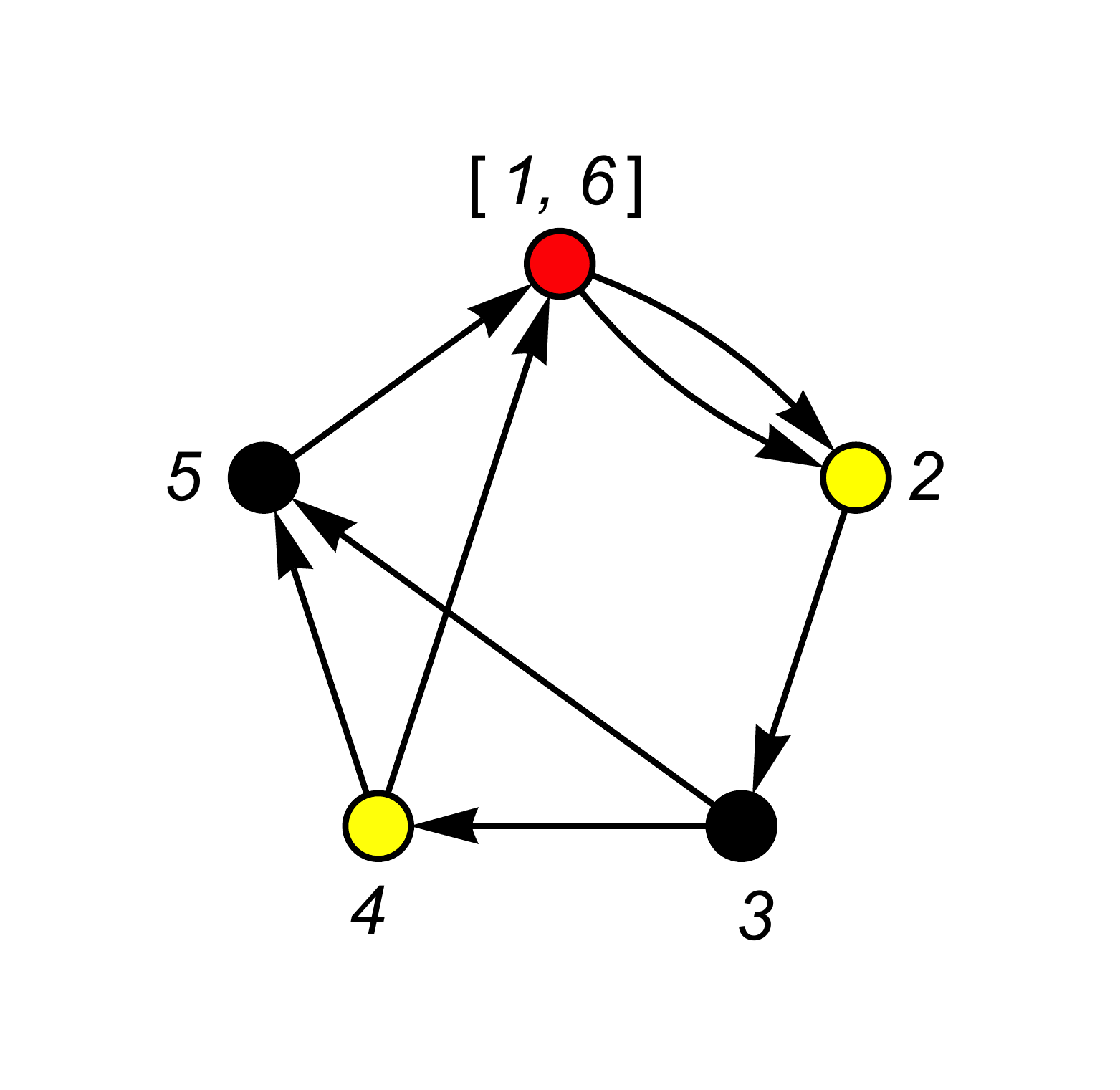}}
\, .
\qquad\qquad~
\end{eqnarray}
}
\vskip-0.5cm\noindent
Nevertheless, a similar example was shown in \eqref{} (the $\phi^4$ vertex), thus, it is simple to check that {\it cut-3} can be rewritten as a product of two {\it YMS-graphs} glued by a longitudinal gluon,
\vspace{-0.1cm}
{\small
\begin{eqnarray}\label{}
\hspace{-0.1cm}
\text{\it cut-3 } = 2\,  \frac{\sum_L\,  A^{\rm ([2,3,4,5],6)}_{\rm g:s} (s_6, s_1 , g_{[2,3,4,5]^L})_{(6,1)} \times  A^{(4,[1,6])}_{\rm g:s} (s_2,s_3, s_4,s_5,  g_{[1,6]^L})_{(2,4:3,5)} }{\tilde s_{16}}\, , \qquad ~~~~
\end{eqnarray}
}
\vskip-0.3cm\noindent
where, $\sum_L\, \eps_{[2,3,4,5]}^{L,\mu}\, \eps_{[1,6]}^{L,\nu} = \frac{ k^\mu_{[2,3,4,5]} \, k^\nu_{[1,6]} }{ k_{[2,3,4,5]} \cdot k_{[1,6]}}$.  This identity can be extended a higer numeber of points, such as it was done for pure Yang-Mills in section \ref{LongContributions}.

Finally, it is not hard to carry out each {\it cut} contribution
\vspace{-0.1cm}
{\small
\begin{eqnarray}\label{S6ptcuts}
\text{\it cut-1}= \frac{1}{\tilde s_{12}}\left( \frac{\tilde s_{25} + \tilde s_{24}}{\tilde s_{345}}
+ \frac{\tilde s_{23}}{\tilde s_{3[1,2]}}
+ \frac{\tilde s_{23}}{\tilde s_{345}}
\right),
~~
\text{\it cut-2}= -\frac{\tilde s_{23} - \tilde s_{3[1,2]}}{\tilde s_{123}\, \tilde s_{3[1,2]}}, 
~~
\text{\it cut-3}= \frac{1}{ \tilde s_{345} } .
\qquad ~~~
\end{eqnarray}
}
\vskip-0.4cm\noindent
These computations confirm that, $A^{\rm YMS}_{\rm g:s}( s_1, s_2, s_3,s_4,s_5, s_6)_{(1,2:3,5:4,6)} =$ {\it cut-1} +{\it cut-2}+{\it cut-3}, which is in agreement with the result  presented in \cite{Cachazo:2014xea}.

\section{Strange-cuts and special Yang-Mills-Scalar Amplitudes}\label{strange-YMS}

From the previous section, we learned that a vertex with four arrows may be interpreted as a scalar particle. Additionally, we saw as the $\Psi_{\rm g,s:g}$ matrix can emerge after factorizing the ${\cal A}$ matrix.

In another way, the resulting graphs obtained from a {\it strange-cut} in pure Yang-Mills have an off-shell vertex with four arrows, thus, in this section we come back to this point in order to understand its relationship with the scalar particles in a special Yang-Mills-Scalar theory.

The first observation  we would like to do is that there are two types of resulting graphs obtained from a {\it strange-cut}, which are related by \eqref{strangec-YMS1} ({\bf property-I} of the appendix \ref{appendix}). Therefore, it is enough just to  work  with one of them, to be more precise
\vspace{-0.6cm}
{\small
\begin{eqnarray}\label{strangec-YMS}
\int
d\mu_{(n-p+3)}^{\L}
\hspace{-0.4cm}
\parbox[c]{6.3em}{\includegraphics[scale=0.17]{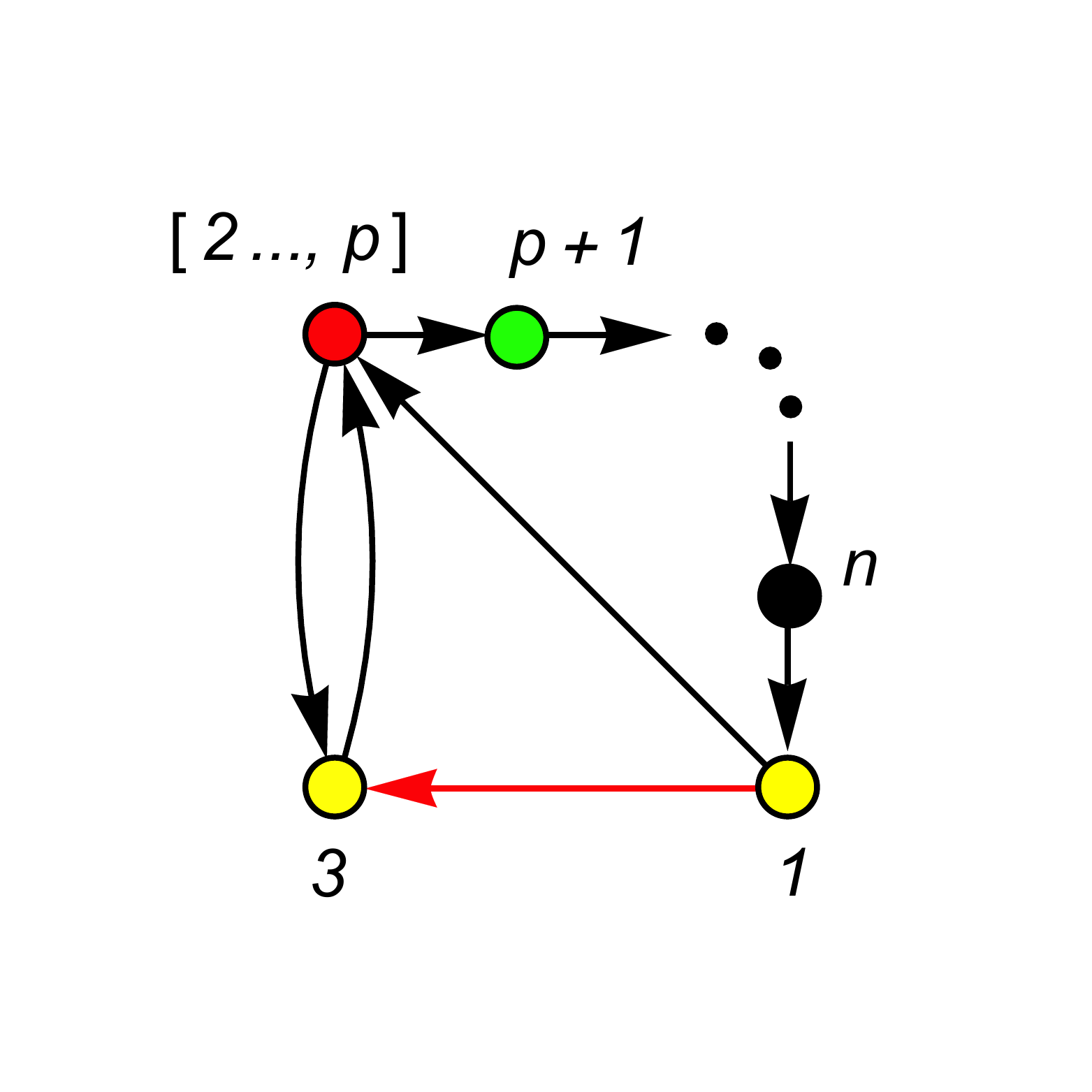}} .
\,\,~~~ 
\end{eqnarray}
}
\vskip-0.8cm\noindent
It is simple to note that the {\bf integration rules} can be applied over above graph.

Such as it was remarked in section \ref{LongContributions}, the  associated matrix to the above graph is given by, $\left(\Psi_{\rm g:s,g}\right)^{13[2,...,p]}_{13[2,...,p]}$, where ${\rm g}=\{ 1,3,p+1,p+2,...,n \}$ and ${\rm s}=\{ [2,...,p] \}$. So, from this point of view, the puncture ``$\s_{[2,3,...,p]}$" looks like an off-shell scalar particle, nevertheless, the CHY integrand of this graph does not fit over any theory known. It is important to note that the $\Psi_{\rm g:s,g}$ matrix has appeared after a factorization process, instead to squeeze or compactify the theory. For instance, in \eqref{strangeC1} we wrote the matrices, $\left(\Psi_{\rm g:s,g}\right)^{13[2,4]}_{13[2,4]}$ 
and $\left(\Psi_{\rm g:s,g}\right)^{[1,3]}_{[1,3]}$, for ${\rm g}=\{1,3\}$, ${\rm s}=\{ [2,4] \}$ and 
${\rm g}=\{2,4\}$, ${\rm s}=\{ [1,3] \}$, respectively.

In order to go beyond to the longitudinal gluon interpretation given in section \ref{LongContributions}, we are going to apply the {\bf integration rules} over the graph in \eqref{strangec-YMS}. Before carrying out the general case, we would like give a simple example. Let us consider the four-point computation,
\vspace{-0.5cm}
{\small
\begin{eqnarray}\label{}
\hspace{-4.8cm}
\int d\mu_4^{\L}
\hspace{-0.45cm}
\parbox[c]{5.8em}{\includegraphics[scale=0.17]{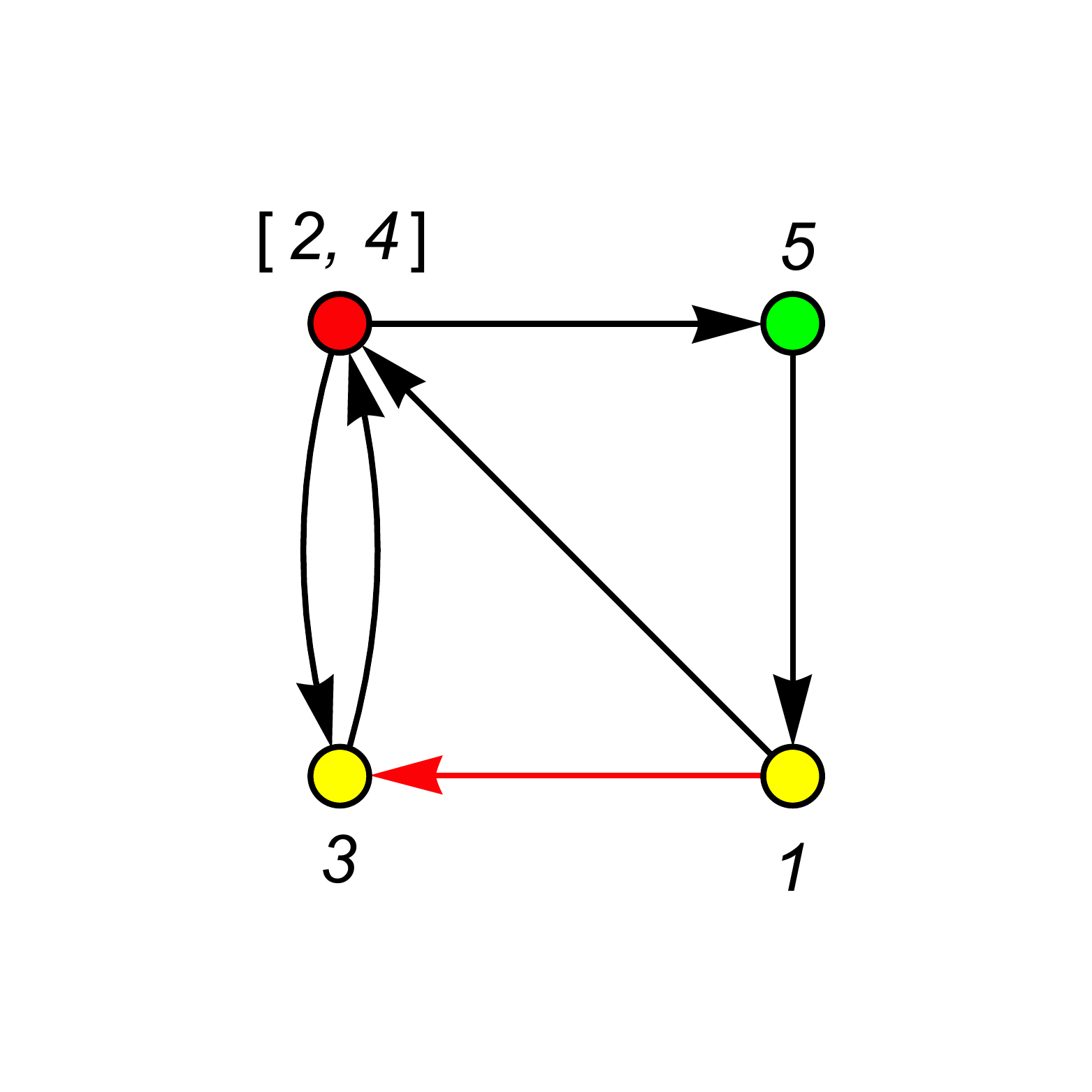}} =
\hspace{-0.5cm}
\parbox[c]{5.8em}{\includegraphics[scale=0.17]{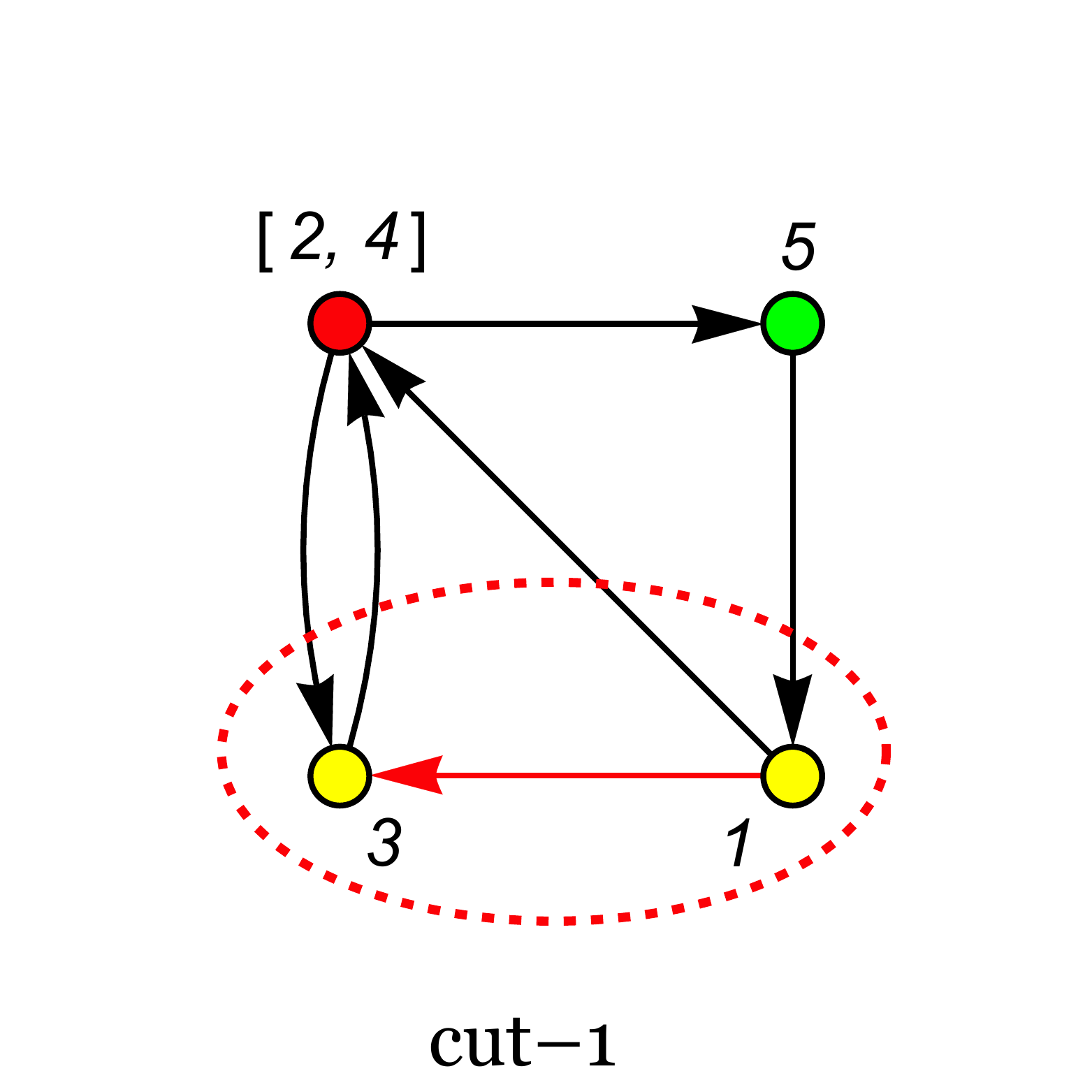}} 
+
\hspace{-0.5cm}
\parbox[c]{6.1em}{\includegraphics[scale=0.17]{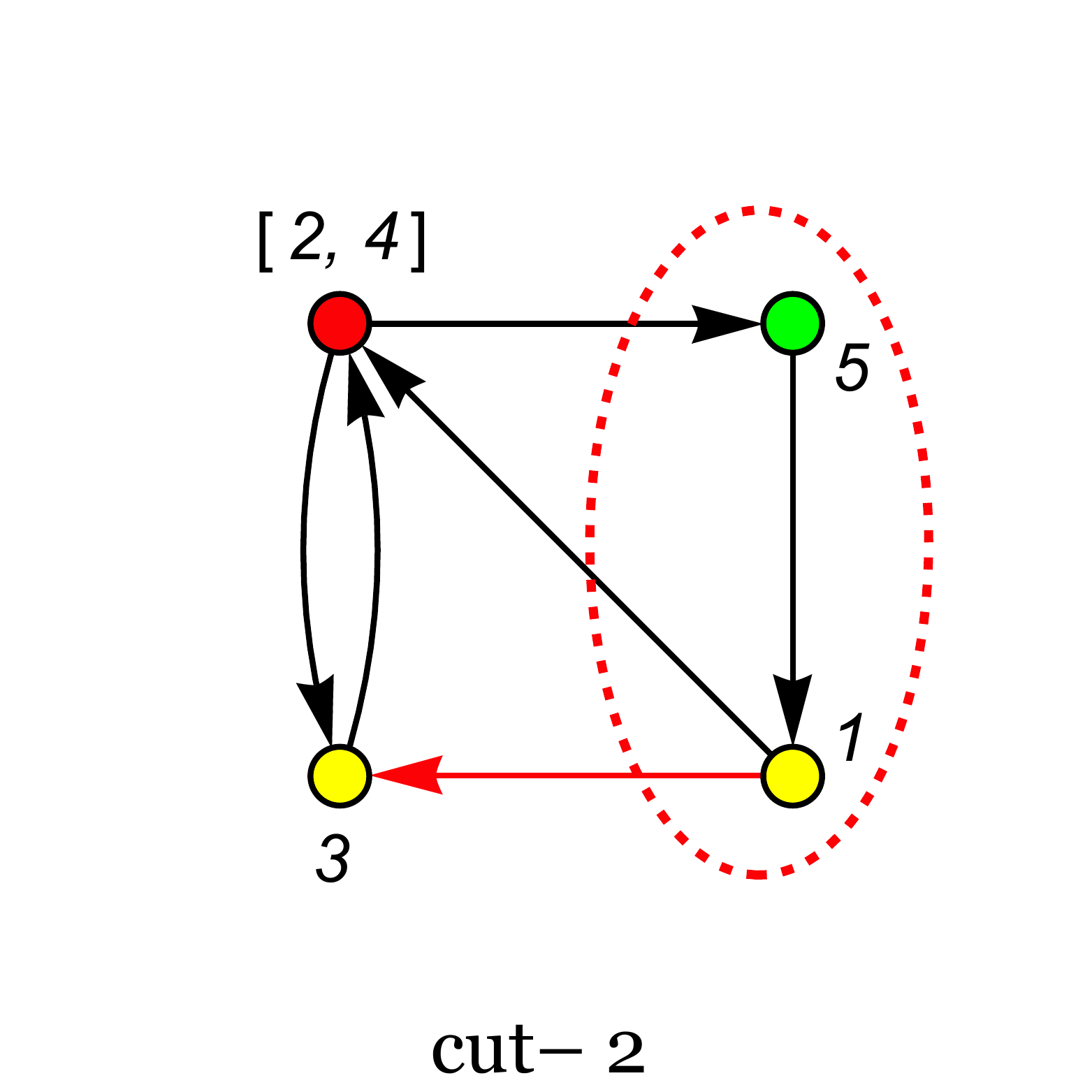}} 
\nonumber
\end{eqnarray}
}
\vskip-1.1cm\noindent
\vspace{-0.4cm}
\begin{eqnarray}\label{sc-YMS-4}
\hspace{-0.15cm}
=
\left\{
\hspace{-0.5cm}
\parbox[c]{5.2em}{\includegraphics[scale=0.17]{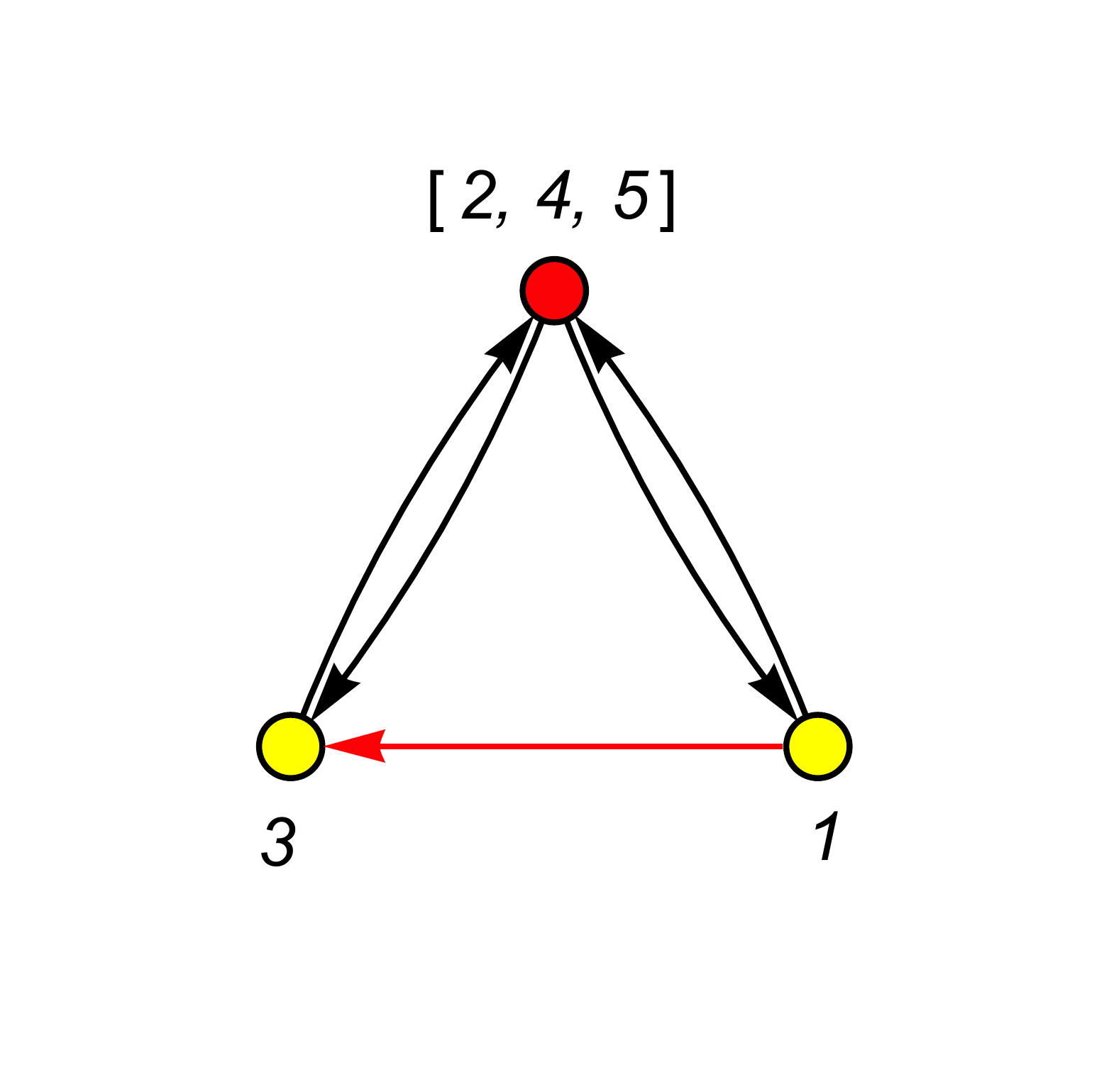}} 
\times \left(
\frac{1}{\tilde s_{5[2,4]}}
\right)
\times
\hspace{-0.6cm}
\parbox[c]{5.6em}{\includegraphics[scale=0.17]{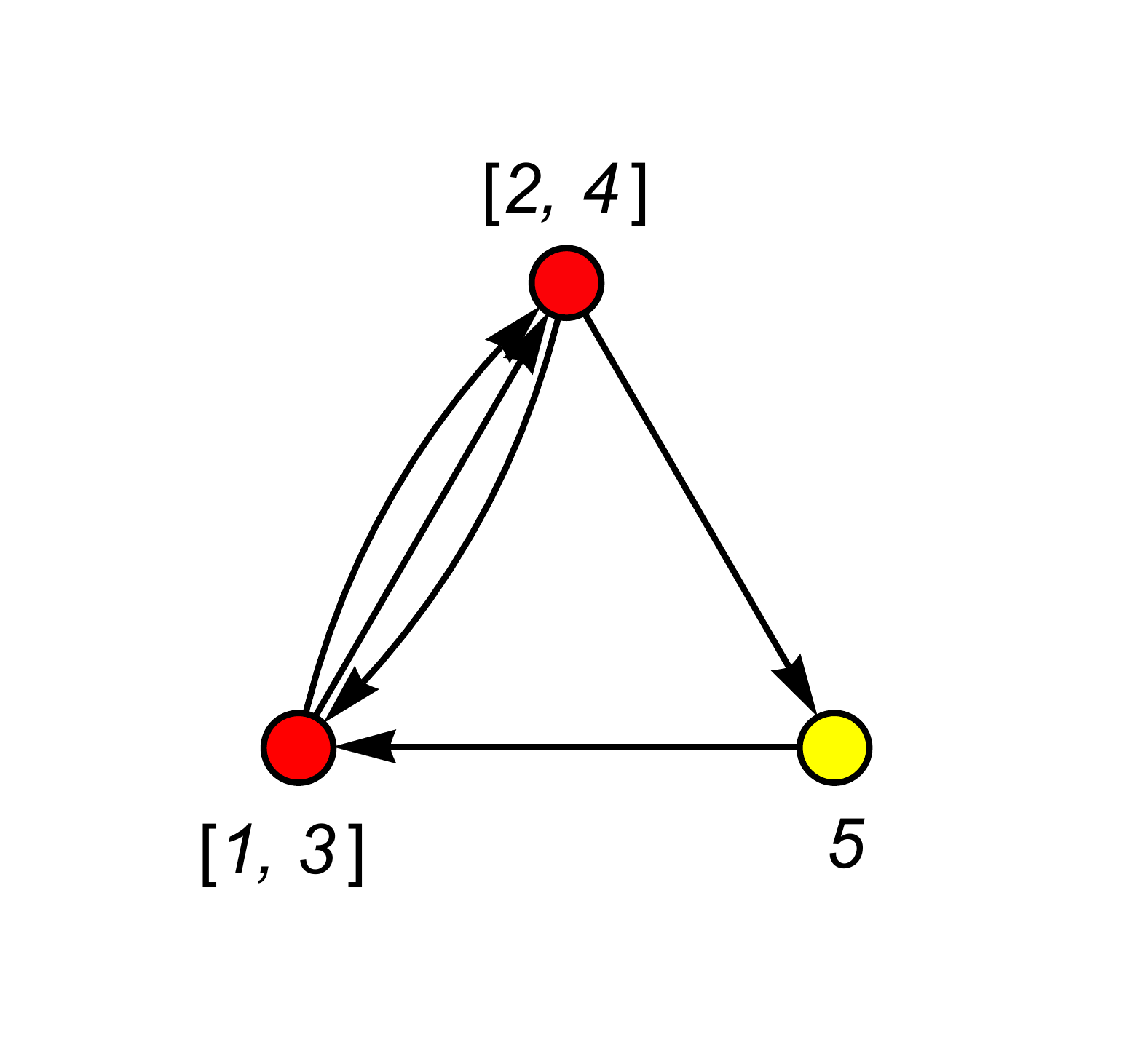}} 
\right\}~
+~
\left\{\sum_r
\hspace{-0.5cm}
\parbox[c]{5.2em}{\includegraphics[scale=0.17]{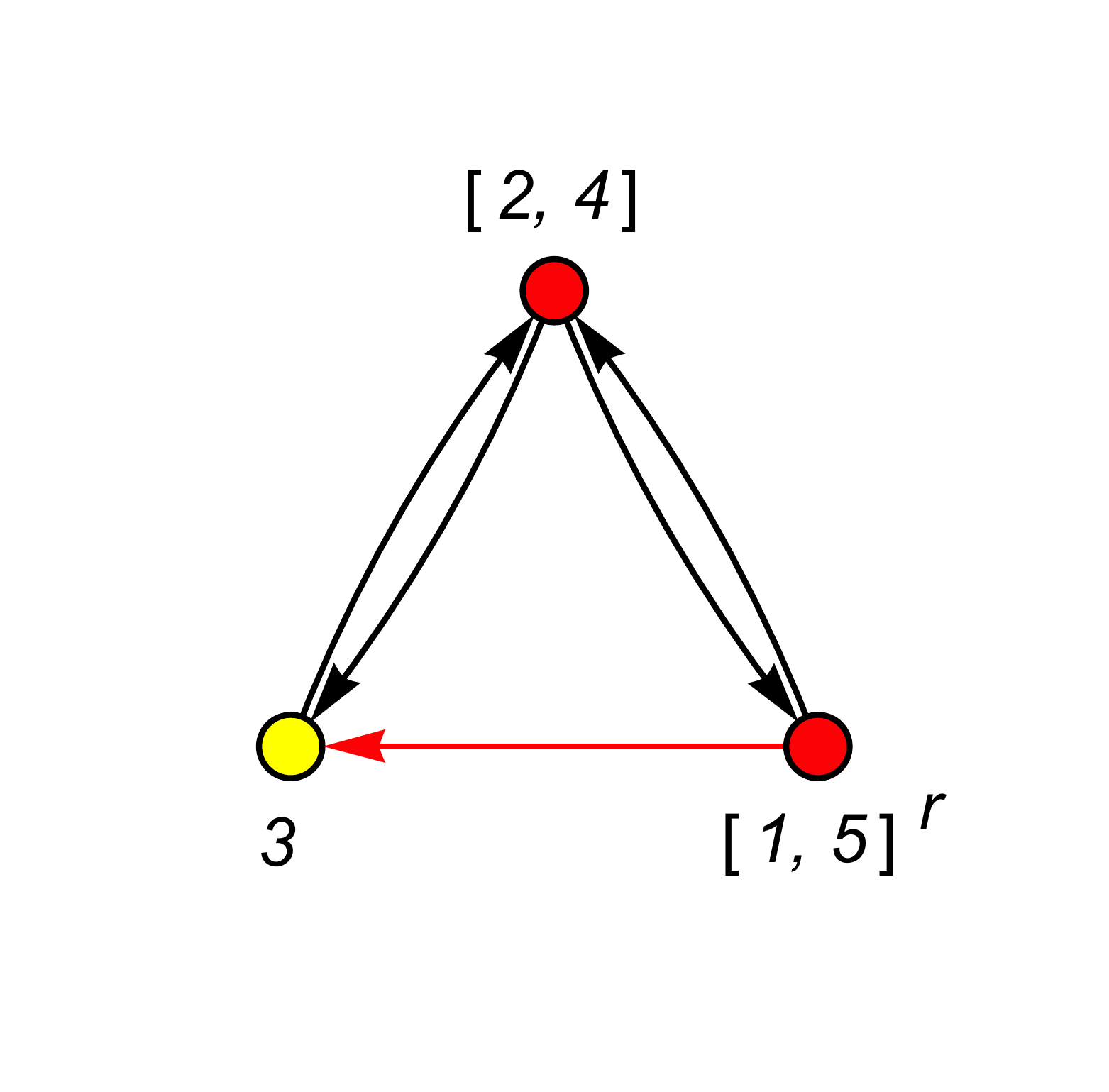}} 
\times \left(
\frac{1}{\tilde s_{51}}
\right)
\times
\hspace{-0.6cm}
\parbox[c]{5.6em}{\includegraphics[scale=0.17]{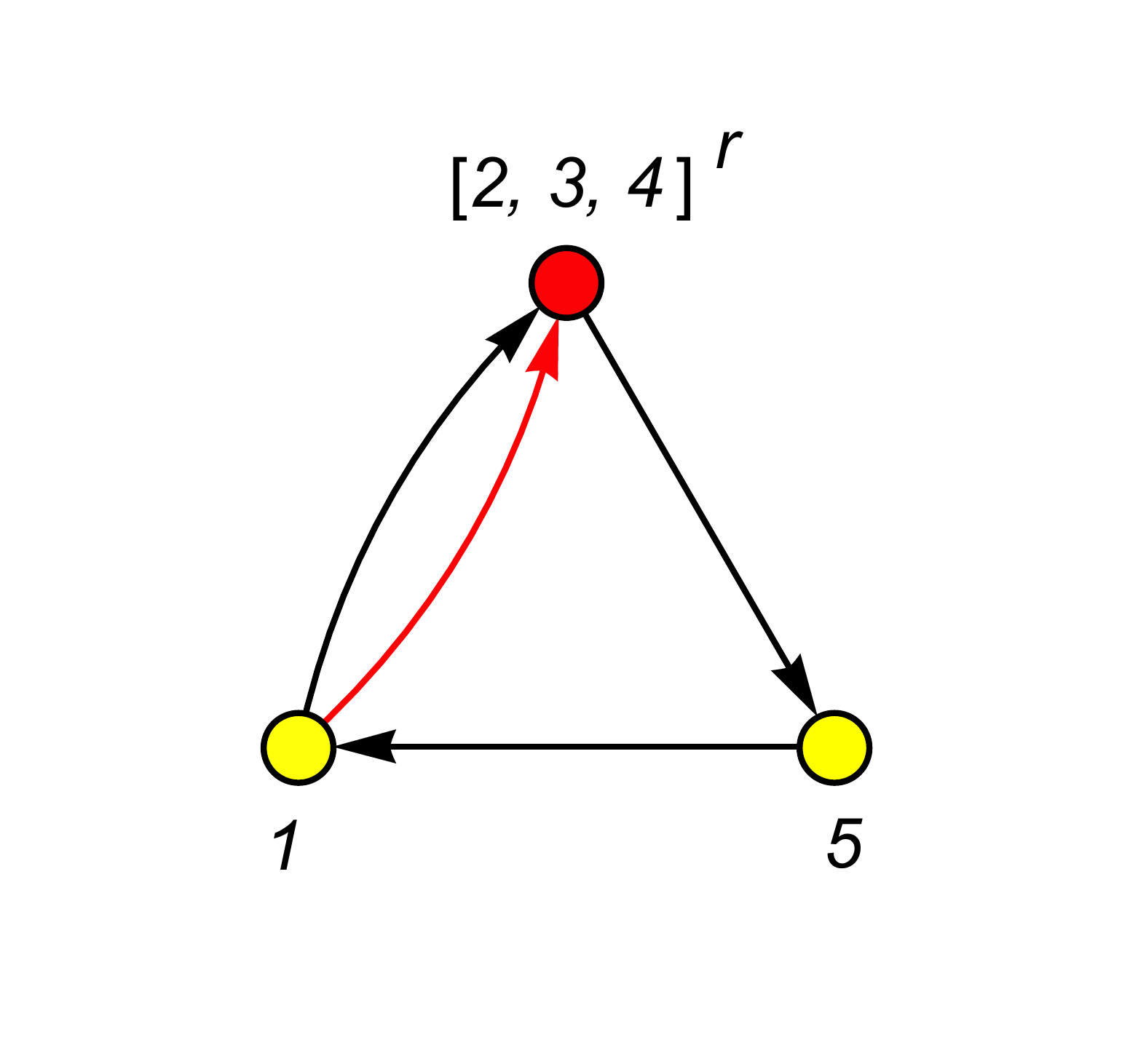}} 
\right\}\, , \quad ~~~
\end{eqnarray}
\vskip-0.2cm\noindent
where we have used the {\bf integration rules} and, $\sum_r \eps_{[1,5]}^{r,\mu} \eps_{[2,3,4]}^{r,\nu} = \eta^{\mu\nu}$. Notice that second resulting graph obtained in the first bracket was already found in the previous section, after factorizing a scattering among two gluons with two scalar particles,  equation \eqref{2g-2s-YMS}.  Thus, by writing \eqref{sc-YMS-4} in terms of the three-point amplitudes one has
\vspace{-0.2cm}
{\small
\begin{eqnarray}\label{sc-cuts-4pt}
&&\text{\it cut-1}= \frac{ A_{\rm g:s}^{([2,4],[1,3])} (s_{[2,4]},g_5,  s_{[1,3]})_{([1,3],[2,4])} }{\tilde s_{5[2,4]}}\times 2\,\left. A_3^{([2,4,5],1)} ([2,4,5]^L, 1 , 3)\right|_{\eps^{L,\mu}_{[2,4,5]} \rightarrow  \frac{- k^\mu_{[2,4,5]}}{k_{[2,4,5]} \cdot k_{[1,3]} } } ,
\nonumber\\
&&\text{\it cut-2}=\sum_r \frac{ A_3^{(1,[2,3,4])} (1,  [2,3,4]^r , 5)  }{\tilde s_{51}}\times 2\,\left. A_3^{([2,4],[1,5])} ([2,4]^L, [1,5]^r, 3)\right|_{ \hspace{-0.8cm}\eps^r_{[1,5]} \cdot k_{[1,5]} =0 \atop
\eps^{L,\mu}_{[2,4]} \rightarrow  \frac{k^\mu_{[2,4]}}{k_{[2,4]}^2 - k^2_{[1,5]} } }.
\qquad\quad ~~~
\end{eqnarray}
}
\vskip-0.2cm\noindent
Clearly, on the first line, the puncture ``$\s_{[2,4]}$" behaves as an off-shell scalar particle, which interacts with one more off-shell scalar and an on-shell gluon\footnote{Computing the cuts in \eqref{sc-cuts-4pt},
it is straightforward to check their results are given by the expressions, {\it cut-1} $=\frac{ (\eps_1\cdot \eps_3)\,  (\eps_5\cdot k_{[1,3]}) }{\tilde s_{5[2,4]}}$ and 
{\it cut-2} $= \frac{ (\eps_1\cdot \eps_3)\, (\eps_5\cdot k_1) +(\eps_5\cdot \eps_1)\, (\eps_3\cdot k_5) + (\eps_3\cdot \eps_5)\, (\eps_1\cdot k_{[2,3,4}]) }{\tilde s_{51}} $.  Observe the non-conventional pole, $\tilde s_{5[2,4]}$.}. Therefore, from the factorization method developed in this paper, a scattering of pure gluons is able to generate an off-shell scalar-like particle\footnote{Additionally, the replacement, $\eps^{L,\mu}_{[2,4,5]} \rightarrow  \frac{- k^\mu_{[2,4,5]}}{k_{[2,4,5]} \cdot k_{[1,3]} }$, also appears in \cite{Mizera:2018jbh}, in the context of the boundary conditions for the Berends-Giele currents obtained by using the new NLSM action proposed in \cite{Cheung:2016prv,Cheung:2017yef}.}. 

The above example is straightforward to generalize to higher number of points. Applying the {\bf integration rules} over the graph in \eqref{strangec-YMS}, we obtained two types of resulting graphs,  
\vspace{-0.3cm}
{\small
\begin{eqnarray}\label{twotypes}
\int
d\mu_{(n-p+3)}^{\L}
\hspace{-0.4cm}
\parbox[c]{6.0em}{\includegraphics[scale=0.17]{R1-cutpL.pdf}} 
=
\hspace{-0.3cm}
\parbox[c]{6.1em}{\includegraphics[scale=0.17]{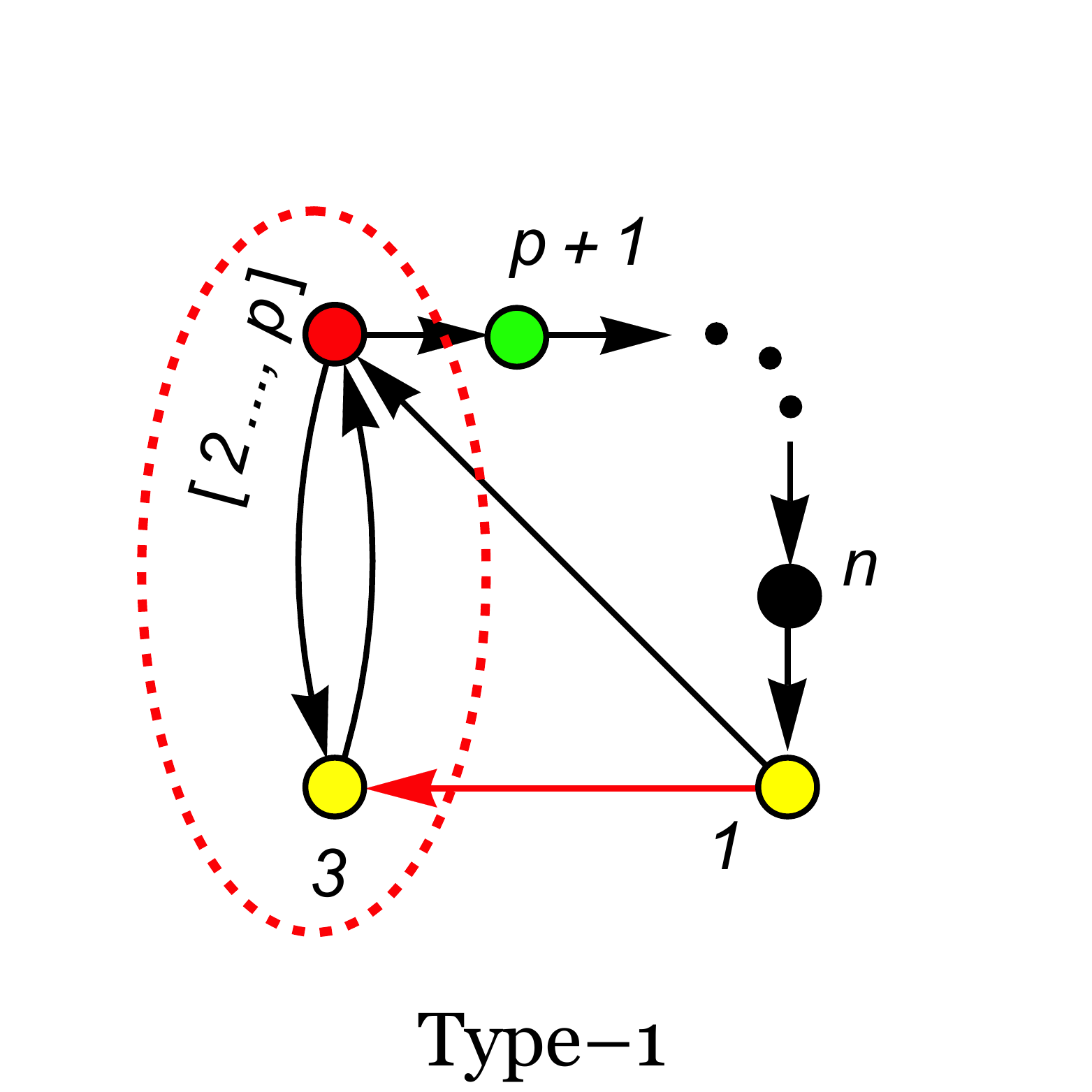}} 
+
\sum_{{\rm cut}\in {\bf C}}
\hspace{-0.4cm}
\parbox[c]{7.0em}{\includegraphics[scale=0.17]{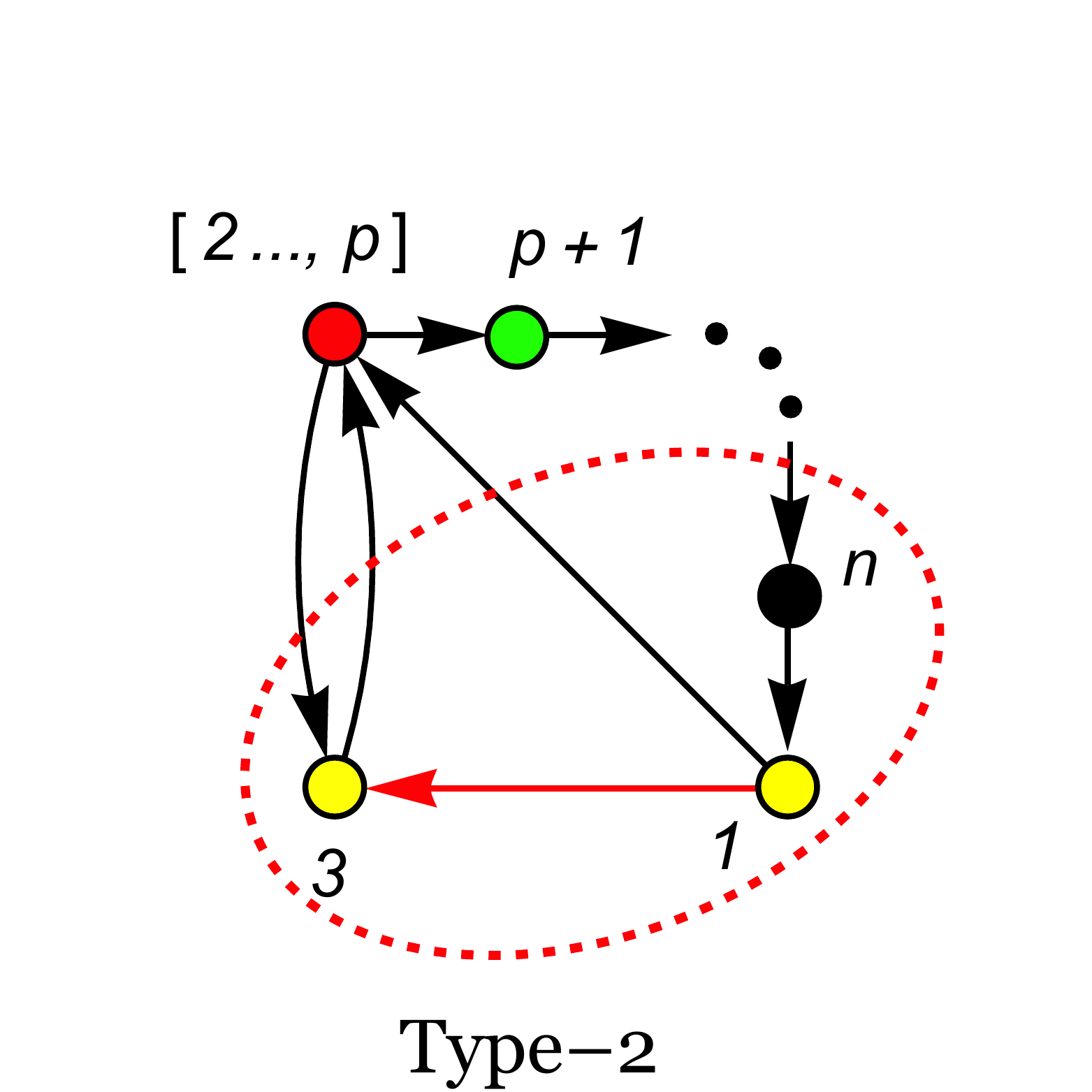}} ,
\,\,~~~ 
\end{eqnarray}
}
\vskip-0.5cm\noindent
where ${\bf C}$ is the set given by all possible punctures that the dashed red line can encircle,
 ${\bf C}=\{(1,3), (1,3,n), (1,3,n,n-1), ... , (1,3,n,...,p+2)   \}$. The first resulting graph is the generalization of  {\it cut-2} in \eqref{sc-cuts-4pt}, 
\vspace{-0.1cm}
{\small
\begin{eqnarray}\label{}
\text{\it type-1}&=&\sum_r \frac{ A_{n-p+2}^{(1,[3,...,p])} (1,  [3,2,4,...,p]^r , p+1,...,n)  }{\tilde s_{1p+1... n}} \nonumber\\
&&
\times \, 2\, \left. A_{3}^{([2,...,p],[1,...,n])} ([2,...,p]^L, [1,p+1,...,n]^r, 3)\right|_{ \bf Cond.}.
\qquad~
\end{eqnarray}
}
\vskip-0.2cm\noindent
where,  $\sum_{r} \eps^{r,\mu}_{[3,2,4,...,p]} \eps^{r,\nu}_{[1,p+1,...,n]}=\eta^{\mu\nu}$, and the conditions, ${\bf Cond}= \left\{\eps^r_{[1,p+1,...,n]} \cdot k_{[1,p+1,...,n]} =0,\right.$ $\left.
 \eps^{L,\mu}_{[2,...,p]} \rightarrow  \frac{ k^\mu_{[2,...,p]}}{k_{[2,...,p]}^2 - k^2_{[1,p+1,...,n]} }\right\}$. On the other hand, the cuts type-2 give arisen to off-shell scalar-like particles in the context of special Yang-Mills-Scalar theory. For instance, let us consider a generic element of the set ${\bf C}$, $(1,3,n,...,i)\in {\bf C}$, $i>p+1$, then, for this configuration the cut becomes
\vspace{-0.2cm}
{\small
\begin{eqnarray}\label{}
\text{\it cut}_{(1,3,n,...,i)}&=& \frac{ A_{\rm g:s}^{([1,3,...,i],[2,4,...,p])} ( s_{[1,3,n,..,i] }, s_{[2,4,...,p]} , g_{p+1},...,g_{i-1}    )_{([2,4,...,p],[1,3,n,..,i])  } }{\tilde s_{[2,4,...,p]p+1,...,i-1}} \nonumber
\end{eqnarray}
}
\vskip-1.0cm\noindent
\vspace{-0.9cm}
{\small
\begin{eqnarray}\label{}
\hspace{-2.9cm}
\times 
\hspace{-0.1cm}
\int
d\mu_{(n-i+4)}^{\rm CHY}
\hspace{-0.55cm}
\parbox[c]{6.0em}{\includegraphics[scale=0.17]{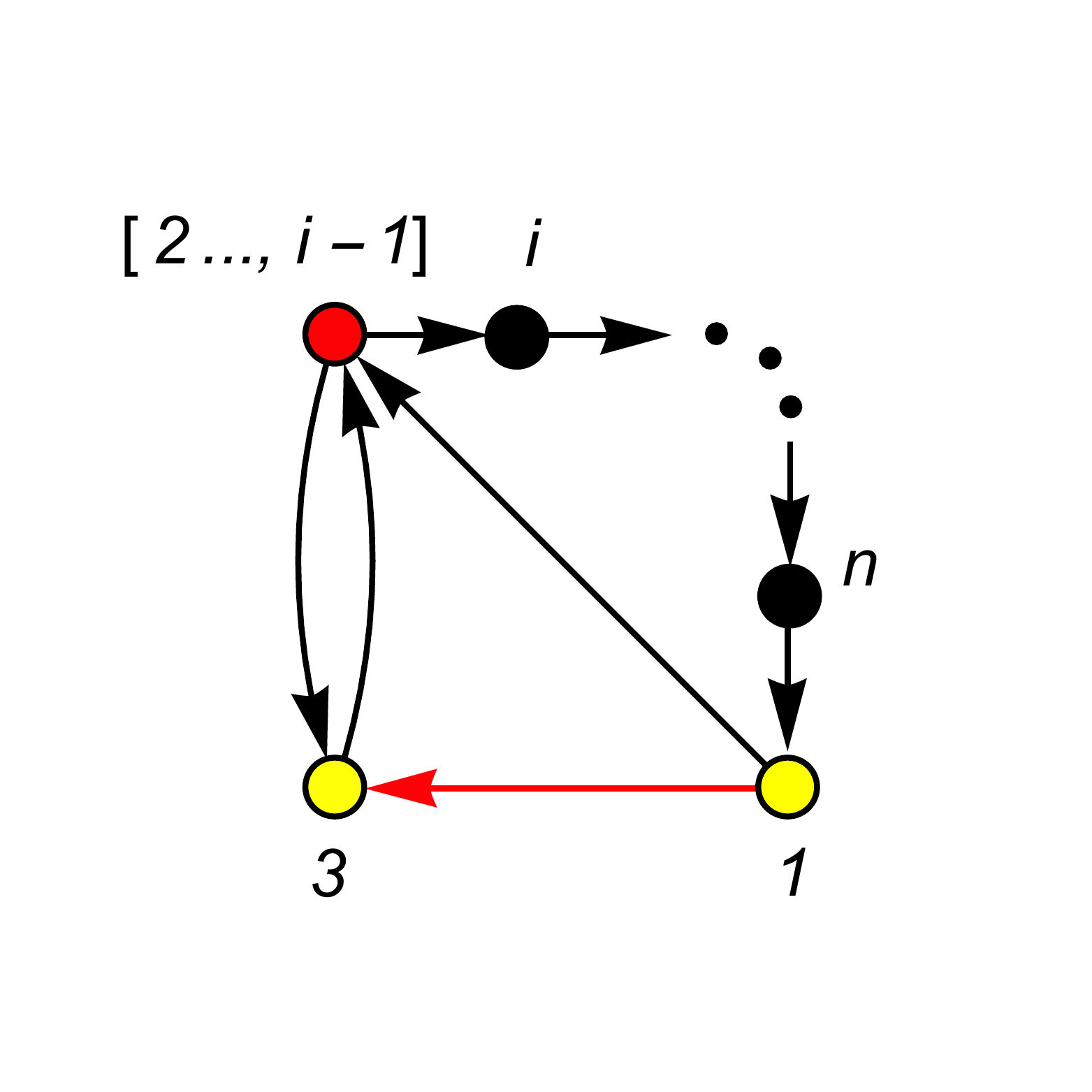}}. 
\end{eqnarray}
}
\vskip-0.4cm\noindent
Clearly, we have obtained an off-shell Yang-Mills-Scalar amplitude multiplied by a graph with the same structure as the original one, therefore, we can applied  \eqref{twotypes} again. 

One of the important things to remark here is to see as some off-shell scalar particles  (in a special Yang-Mills-Scalar theory) can emerge from the {\it strange cuts}.  This fact confirms the interpretation of seeing the off-shell puncture, ``$\s_{[2,...,p]}$'', as a scalar-like particle.

\section{Soft Limit}\label{sectionSoft}

As a final section, we would show how the soft limit in pure Yang-Mills theory \cite{Weinberg:1965nx},  at leading order,  can be computed  in a simple way by using the factorization method developed in this paper.

Before going to the general case, let us consider the four-point example,  $A_4^{\rm YM}(1,2,3,4)$.  Such as it was shown in section \ref{four-point}, this amplitude can be written in the following way
{\small
\begin{eqnarray}\label{}
\hspace{-3.0cm}
A_4^{\rm YM}(1,2,3,4)
=\sum_r\left[\frac{A_{3}^{([3,4],1)}([3,4]^r,1,2)\, A_{3}^{([1,2],3)}( [1,2]^r,3, 4)  }{\tilde s_{43}}  +
\right.~~~
 \nonumber 
\end{eqnarray}
}
\vskip-1.1cm\noindent
{\small
 \begin{eqnarray}\label{soft-one}
\left.
+
\frac{ A_{3}^{(1,[2,3])}(1, [2,3]^r,4) \,A_{3}^{(3,[4,1])}(3, [4,1]^r,2)}{\tilde s_{41}}
\right]
 +
\hspace{-0.5cm}
\parbox[c]{5.6em}{\includegraphics[scale=0.17]{R1-cut3.pdf}} 
\times
\left(
 \frac{1}{\tilde s_{42}} 
 \right)
 \times
\hspace{-0.55cm}
\parbox[c]{6.0em}{\includegraphics[scale=0.17]{R2-cut3.pdf}} ,
\end{eqnarray}
}
\vskip-0.55cm\noindent
where, $\sum_{r} \eps^{r,\mu}_{[3,4]} \eps^{r,\nu}_{[1,2]} = \sum_{r} \eps^{r,\mu}_{[2,3]} \eps^{r,\nu}_{[4,1]} = \eta^{\mu\nu} $. This important to recall that the setup used was,  $(pqr|m)=(123|4)$, and the red arrow by joining the punctures $(i,j)=(1,3)$. 

Now, we are taking the particle at the puncture ``$\s_4$" as the soft particle, i.e. $k^\mu_4=z\, q_4^\mu$, with $z\rightarrow 0$. Since that the second resulting graph on the second line in \eqref{soft-one} is proportional to $\tilde s_{42}$, the only leading order contribution comes from the first and second terms (with poles, $\tilde s_{43} = z(q_4\cdot k_3)$ and $\tilde s_{41} = z(q_4\cdot k_1)$, respectively). Notice that when, $z\rightarrow 0$, the off-shell momenta of the  particles at $\s_{[3,4]}$ and $\s_{[4,1]}$ become the on-shell momenta, $k_{[3,4]}^\mu\rightarrow k_3^\mu$ and $k_{[4,1]}^\mu\rightarrow k_1^\mu$, respectively. Thus, we only worry by the three-point functions,  $ A_{3}^{([1,2],3)}( [1,2]^r,3, 4)$ and $A_{3}^{(1,[2,3])}(1, [2,3]^r,4)$. Using the building block given in \eqref{YM-BB}, these functions turn into (at leading order in $z$.)
{\small
\begin{eqnarray}\label{}
\hspace{-0.1cm}
 A_{3}^{([1,2],3)}( [1,2]^r,3, 4) 
=  (\eps_{[1,2]}^r\cdot \eps_3)\, (\eps_4 \cdot k_{[1,2]})  , \quad
  A_{3}^{(1,[2,3])}(1, [2,3]^r, 4) 
  =  (\eps_{1}\cdot \eps^r_{[2,3]})\, (\eps_4 \cdot k_{1}) .
 \nonumber 
\end{eqnarray}
}
\vskip-0.4cm\noindent
Finally, from the gluing identities, $\sum_{r} \eps^{r,\mu}_{[3,4]} \eps^{r,\nu}_{[1,2]} = \sum_{r} \eps^{r,\mu}_{[2,3]} \eps^{r,\nu}_{[4,1]} = \eta^{\mu\nu}$, one has,  $\sum_{r} \eps^{r,\mu}_{[3,4]} (\eps^{r}_{[1,2]}\cdot \eps_3) = \eps_3^{\mu}$ and $\sum_{r} (\eps_1 \cdot \eps^{r}_{[2,3]}) \eps^{r,\nu}_{[4,1]} = \eps_1^{\nu}$.  Therefore, at leading order, the four-point amplitude is given by
{\small
\begin{eqnarray}\label{}
A_4^{\rm YM}(1,2,3,z\,q_4)
=\frac{1}{z}\left(\frac{\eps_4\cdot k_1  }{q_4\cdot k_1}  - \frac{\eps_4\cdot k_{3}  }{q_4\cdot k_3} 
\right) A_{3}^{\rm YM}(1,2,3)  \,+\, {\cal O}(z^0),
 \nonumber 
\end{eqnarray}
}
\vskip-0.3cm\noindent
where we have used\footnote{Notice that when, $k_{[3,4]}^\mu\rightarrow k_3^\mu$ and $\eps^{r,\mu}_{[3,4]} \rightarrow \eps_3^\mu$, we may relabel the puncture $\s_{[3,4]}$ by $\s_{3}$, in order to indicate its associated momentum and polarization vector. Of course, this is also applied for $\s_{[4,1]}$.},  $\eps_4\cdot k_{[1,2]}= -\eps_4\cdot k_{3}$ and $A_{3}^{(3,1)}(3,1,2) =A_{3}^{\rm YM}(1,2,3) $.  

This simple analysis can be generalized to a higher number of points, the main key is that the soft particle must be surrounded by fixed punctures (yellow vertices). In order to satisfy this requirement, we choose the gauge fixing, $(pqr|m) = ((n-1)12|n)$, the red arrow on, $(i,j)=(2,n)$, and the soft particle, $k^\mu_n = z\, q^\mu_n$, with $z\rightarrow 0$.  Under this setup, one can apply the {\bf integration rules} over the amplitude, $ A^{\rm YM}_n ( 1,\ldots , n) $, to obtain\footnote{There are some overall signs which do not affect our analysis. Nevertheless, the exact expressions for the $n$-point amplitudes can be found in section \ref{sectionConc}.}
\vspace{-0.1cm}
{\small
\begin{eqnarray}\label{Gen-soft}
\hspace{-0.4cm}
 A^{\rm YM}_n ( 1,\ldots , n)       
= \nonumber 
\end{eqnarray}
}
\vskip-0.4cm\noindent
\vspace{-1.1cm}
{\small
\begin{eqnarray}\label{Gen-soft}
\hspace{-0.4cm}
&& 
\sum_{i=0}^{n-4} \sum_r \frac{A_{i+3}^{(2,[ \mathbb{O}^\ast_i ,n-1,n])}(1, 2,\mathbb{O}_i, [ \mathbb{O}^\ast_i ,n-1,n]^r    )      \times 
A_{n-(i+1)}^{ ([1,2,\mathbb{O}_i],n)   }(
[1,2,\mathbb{O}_i]^r , \mathbb{O}^\ast_i,
  n-1,n ) }{\tilde s_{n\,n-1\,\mathbb{O}^\ast_i}}  \nonumber \\
\hspace{-0.4cm}
 &&
+ \sum_r \frac{A_{3}^{( [2,\mathbb{O}_{n-4},n-1],n  )}( [2,\mathbb{O}_{n-4},n-1]^r,n ,1  ) \times A_{n-1}^{(2,[n,1])}( 2,\mathbb{O}_{n-4}, n-1, [n,1]^r, ) }{\tilde s_{ n\,1}} 
+  \text{\it strange-cuts} \, ,\qquad\nonumber \\
\end{eqnarray}
}
\vskip-1.0cm\noindent
where we have defined,  $\mathbb{O}_0\equiv \emptyset, \, \mathbb{O}_i\equiv \{ 3,4,\ldots, i+2\},\, i=0,...,n-4$,  $\mathbb{O}^\ast_i\equiv \{3, 4,5,\ldots, n-2\}\diagdown \mathbb{O}_i,\,  i=0,...,n-4 $, and with the gluing identities, $\sum_{r} \eps_{[i]}^{r,\mu} \eps_{[j]}^{r,\nu} =\eta^{\mu \nu}$.  By the property in \eqref{strangec-YMS1}, the resulting graphs obtained from a strange-cut always cancel out the pole (spurious pole), therefore, those configurations do not contribute to the soft limit at leading order. On the other hand, the leading order contribution from the second line  in \eqref{Gen-soft} is given when $\mathbb{O}^\ast_i$ is an empty set, i.e. when $i=n-4$ ($\mathbb{O}^\ast_{n-4} = \emptyset$). Thus, when $z\rightarrow 0$ the leading order contribution of \eqref{Gen-soft}  is given by
\vspace{-0.1cm}
{\small
\begin{eqnarray}\label{}
\hspace{-0.4cm}
&& A^{\rm YM}_n ( 1,\ldots , z\,q_n)       
= \sum_r \frac{A_{n-1}^{(2,[n-1,n])}(1, 2,\mathbb{O}_{n-4}, [ n-1,n]^r  )      \times 
A_{3}^{(n, [1,2,\mathbb{O}_{n-4}])}( n,
 [1,2,\mathbb{O}_{n-4}]^r ,
  n-1 ) }{z\, q_n\cdot k_{n-1}}  \nonumber \\
\hspace{-0.4cm}
 &&
+ \sum_r \frac{A_{3}^{([2,\mathbb{O}_{n-4},n-1],n)}( [2,\mathbb{O}_{n-4},n-1]^r,n ,1  ) \times A_{n-1}^{(2,[n,1])}( 2,\mathbb{O}_{n-4}, n-1, [n,1]^r ) }{z\, q_n\cdot k_{1}}  + {\cal O}(z^0)\, .
\qquad
\end{eqnarray}
}
\vskip-0.3cm\noindent
Now, performing the same procedure as in the previous four-point example, it is trivial to see that
\vspace{-0.1cm}
{\small
\begin{eqnarray}\label{}
\hspace{-0.4cm}
 A^{\rm YM}_n ( 1,\ldots , z\,q_n)       
= \frac{1}{z}\left(
\frac{\eps_n \cdot k_1}{q_n\cdot k_1} -
\frac{\eps_n \cdot k_{n-1}}{q_n\cdot k_{n-1}}
\right)
A_{n-1}^{\rm YM}( 1,2,\ldots , n-1 )
+ {\cal O}(z^0)\, ,
\qquad
\end{eqnarray}
}
\vskip-0.3cm\noindent
which is the right answer \cite{Weinberg:1965nx}.

\section{Conclusions}\label{sectionConc} 

In this work, we have obtained a new graphical method to compute the scattering of $n$-gluons and interactions with scalar particles, which was developed under the double-cover formulation of the CHY-approach. The main idea was to define a graph representation for an ordered amplitude from the double-cover  (and single-cover) prescription. By integrating the extra parameter, ``$\L$", the double-cover breaks into two single-cover  and the {\bf integration  rules} arose naturally. This process generates smaller {\it YM/YMS-graphs} and, as a byproduct, some {\it strange-graphs}.  The {\it strange-graphs} did not seem to have a simple physical interpretation, however, we were able to conjecture that these are just longitudinal contributions from YM/YMS amplitudes. To be more precise, 
under the gauge fixing, $(pqr|m)=(123|4)$, and by choosing the red arrow from the vertices,  $(i,j)=(1,3)$, the
 {\bf integration rules} over the Yang-Mills amplitude, $A_n^{\rm YM} (1,...,n)$, 
produce the off-shell general formula\footnote{When the number of particles is even, there is an overall sign, $(-1)^{i}$, missing in the equations  (43) and (44)  in \cite{Bjerrum-Bohr:2018lpz}.}:
\begin{itemize}
\item
When $n=2m+1$
\vspace{-0.0cm}
{\small
\begin{eqnarray}\label{Gen-C-odd}
\hspace{-0.4cm}
&&A^{\rm YM}_n ( 1,2,\ldots , n)    
= (-1) \times \left\{   
 \sum_r \frac{A_{3}^{(3,  [\mathbb{O}_{n-3},1] )}( 3,  [\mathbb{O}_{n-3},1]^r,2) \times A_{n-1}^{(1,[2,3])}(1,[2,3]^r, \mathbb{O}_{n-3}) }{\tilde s_{\mathbb{O}_{n-3} 1}}  
  \right. \nonumber \\
&& 
+
\sum_{i=1}^{n-3} \sum_r \frac{A_{n-i}^{(1,[3,\mathbb{O}_i] )}(1,2,[3,\mathbb{O}_i]^r,\mathbb{O}^\ast_i) \times A_{i+2}^{( [\mathbb{O}^\ast_i,1,2],3)}( [\mathbb{O}^\ast_i,1,2]^r, 3, \mathbb{O}_i ) }{\tilde s_{3\mathbb{O}_i}} 
\nonumber \\
&&
-
2\times
\left. \left.
\sum_{i=1}^{n-3} \sum_L 
\frac{A_{n-i}^{(1,[3,\mathbb{O}_i] )}(1,2,[3,\mathbb{O}_i]^L,\mathbb{O}^\ast_i) \times A_{i+2}^{( [\mathbb{O}^\ast_i,1,2] , 3)}([\mathbb{O}^\ast_i,1,2]^L, 3, \mathbb{O}_i ) }{\tilde s_{3\mathbb{O}_i}} 
 \right|_{2 \leftrightarrow 3}  \right\} .
\nonumber\\
\end{eqnarray}
}
\vskip-0.2cm\noindent
\item
When $n=2m$
\vspace{-0.0cm}
{\small
\begin{eqnarray}\label{Gen-C}
\hspace{-0.4cm}
&&A^{\rm YM}_n ( 1,2,\ldots , n)    
=   
 \sum_r \frac{A_{3}^{(3,  [\mathbb{O}_{n-3},1] )}( 3,  [\mathbb{O}_{n-3},1]^r,2) \times A_{n-1}^{(1,[2,3])}(1,[2,3]^r, \mathbb{O}_{n-3}) }{\tilde s_{\mathbb{O}_{n-3} 1}}  
 \nonumber \\
&& 
+ 
\sum_{i=1}^{n-3} \sum_r (-1)^{ (i+1) } \times \frac{A_{n-i}^{(1,[3,\mathbb{O}_i] )}(1,2,[3,\mathbb{O}_i]^r,\mathbb{O}^\ast_i) \times A_{i+2}^{( [\mathbb{O}^\ast_i,1,2],3)}( [\mathbb{O}^\ast_i,1,2]^r, 3, \mathbb{O}_i ) }{\tilde s_{3\mathbb{O}_i}} 
\nonumber \\
&&
-
2\times
\left. 
\sum_{i=1}^{n-3} \sum_L  (-1)^{ (i+1) } \times
\frac{A_{n-i}^{(1,[3,\mathbb{O}_i] )}(1,2,[3,\mathbb{O}_i]^L,\mathbb{O}^\ast_i) \times A_{i+2}^{( [\mathbb{O}^\ast_i,1,2] , 3)}([\mathbb{O}^\ast_i,1,2]^L, 3, \mathbb{O}_i ) }{\tilde s_{3\mathbb{O}_i}} 
 \right|_{2 \leftrightarrow 3}   , \qquad
\nonumber\\
\end{eqnarray}
}
\vskip-0.2cm\noindent
\end{itemize} 
where we have defined the sets,  $\mathbb{O}_0\equiv \emptyset, \, \mathbb{O}_i\equiv \{ 4,5,\ldots, i+3\},\, i=1,...,n-3$,  $\mathbb{O}^\ast_{n-3}\equiv \emptyset, \, \mathbb{O}^\ast_i\equiv \{ 4,5,\ldots, n\}\diagdown \mathbb{O}_i,\,  i=1,...,n-3 $, and with the gluing identities, $\sum_{r} \eps_{[3, \mathbb{O}_i ]}^{r,\mu} \eps_{[\mathbb{O}^\ast_i,1,2]}^{r,\nu} =\sum_{r} \eps_{[ \mathbb{O}_{n-3},1 ]}^{r,\mu} \eps_{[2,3]}^{r,\nu} =\eta^{\mu\nu}$,  $\sum_{L} \eps_{[2, \mathbb{O}_i ]}^{L,\mu} \eps_{[1,3,\mathbb{O}^\ast_i]}^{L,\nu} =\frac{ k_{[2, \mathbb{O}_i ]}^{\mu}  k_{ [1,3,\mathbb{O}^\ast_i]}^{\nu} }{ k_{[2, \mathbb{O}_i ]}\cdot  k_{ [1,3,\mathbb{O}^\ast_i]} }$. Let us remind ourselves the three punctures which must be fixed in the smaller off-shell Yang-Mills amplitudes are given by the set in \eqref{fixed-punctures}, namely,  $\{\textbf{Fixed  punctures} \} =\left( \{\textbf{All  punctures  in the graph}\} \cap \{1,2,3,4 \} \right) \cup\{\textbf{off-shell punctures}\}$.

Notice that the poles, $\tilde s_{2\mathbb{O}_i}$, are not physical and these must cancel out. Although we do not have a formal proof, we have carried out several examples to verify that, in effect, the amplitudes, 
$A_{n-i}^{(1,[2,\mathbb{O}_i])}(1,3,[2,\mathbb{O}_i]^L,\mathbb{O}^\ast_i)$ and $A_{i+2}^{([1,3,\mathbb{O}^\ast_i], 4)}([1,3,\mathbb{O}^\ast_i]^L, 2,  \mathbb{O}_i )$, are proportional to  $\tilde s_{2\mathbb{O}_i}$ when the off-shell gluons are longitudinal,       
$\eps_{[2,\mathbb{O}_i]}^{L,\mu} = k^\mu_{[2,\mathbb{O}_i]}$  
and $\eps_{ [1,3,\mathbb{O}^\ast_i] }^{L,\mu} = k^\mu_{ [1,3,\mathbb{O}^\ast_i] }$, respectively. Additionally, we are looking for formal proof (it is enough just to prove the off-shell Pfaffian properties presented in appendix \ref{appendix}.)
 
One of the more amazing things we would like to remark is the computational information that there is behind of the identities obtained in section \ref{LongContributions} ({\bf properties-I, II} in appendix).  For example, in the identity obtained in \eqref{strangec-YMS1},  we related a graph with a matrix $2(p-2)\times 2(p-2)$ to another  graph with a matrix $2(p-2)-2\times 2(p-2)-2$, and the relationship is given just for an overall factor. Computationally, this is very important,  let us suppose that $p = 5$, we are simplifying the problem from a matrix $6 \times 6$ to a new one $4\times 4$.

On the other hand, although the {\bf integration rules} emerged  from the pure Yang-Mills amplitudes, these can be applied naturally over a large spectrum of graphs, such as special Yang-Mills-Scalar theory, effective field theories \cite{inpreparation}, and an others strange graphs. In addition to this fact, it  was very interesting to see as the ${\cal A}$ and $\Psi_n$ matrices can be factorized in terms of the off-shell $\Psi_{\rm g,s:g}$ matrix\footnote{Let us remind that in \cite{Cachazo:2014xea}, the $\Psi_{\rm g,s:g}$ matrix arose after compactifying.}
\\

{\bf Feynman diagrams and BCJ numerators:}\\
A direct relation between the CHY and Feynman diagrams is unknown. However, since that the method developed in this paper is a factorization-like approach, perhaps, this is the closest bridge among CHY amplitudes and Feynman rules. Let us observe how to obtain the four-point Feynman diagrams since our method.

From the general formula in \eqref{Gen-C}, one has
\vspace{-0.1cm}
{\small
\begin{eqnarray}\label{}
\hspace{-0.05cm}
A_4^{\rm YM}(1,2,3,4)
=
 \nonumber 
\end{eqnarray}
}
\vspace{-1.1cm}
{\small
\begin{eqnarray}\label{}
\sum_r
\left[
\frac{   A_{3}^{([1,2],3)}( [1,2]^r , 3 ,4) \, 
A_{3}^{([3,4],1)}( [3,4]^r,1 , 2)  }{\tilde s_{12}}
+
\frac{ A_{3}^{(1,[2,3])}( 1 , [2,3]^r, 4) \, 
 A_{3}^{(3,[4,1])}(3 , [4,1]^r , 2)   }{\tilde s_{14}} 
 \right]
 \nonumber 
\end{eqnarray}
}
\vskip-0.7cm\noindent
{\small
 \begin{eqnarray}\label{Feynman-one} 
 \hspace{-0.2cm}
+ 
 \sum_L 
 2\, \frac{  A_{3}^{(4,[1,3])}(4,[1,3]^L , 2) \times  A_{3}^{([2,4],1)}( [2,4]^L,1 ,3)  }{\tilde s_{13}}\, .
 \qquad~~
\end{eqnarray}
}
\vskip-0.6cm\noindent
As it was shown in  \eqref{YM-BB}, to obtain the three-point Feynman vertex,   
the polarization vectors on the first line, ``$\eps_{[i]}^r$", must be transverse, thus, we  use the identity, $\sum_{r}\eps_{[1,2]}^{r,\mu}\eps_{[3,4]}^{r,\nu}= \sum_{T}\eps_{[1,2]}^{T,\mu}\eps_{[3,4]}^{T,\nu}  + \sum_{L}\eps_{[1,2]}^{L,\mu}\eps_{[3,4]}^{L,\nu} $, where,
$\sum_{T}\eps_{[1,2]}^{T,\mu}\eps_{[3,4]}^{T,\nu} = \eta^{\mu\nu} - \frac{k^\mu_{[1,2]} k^\nu_{[3,4]}  }{k_{[1,2]} \cdot k_{[3,4]} }$ and $\sum_{L}\eps_{[1,2]}^{L,\mu}\eps_{[3,4]}^{L,\nu} =  \frac{ k^\mu_{[1,2]} k^\nu_{[3,4]}  }{k_{[1,2]} \cdot k_{[3,4]} }$ (in a similar way for $\eps^r_{[2,3]}$ and $\eps^r_{[4,1]}$).  Therefore, \eqref{Feynman-one} becomes
\vspace{-0.1cm}
{\small
\begin{eqnarray}\label{}
\hspace{-0.05cm}
A_4^{\rm YM}(1,2,3,4)
=
 \nonumber 
\end{eqnarray}
}
\vspace{-1.0cm}
{\small
\begin{eqnarray}\label{}
\hspace{-0.05cm}
\sum_T
\left[
\frac{   A_{3}^{([1,2],3)}( [1,2]^T , 3 ,4)
A_{3}^{([3,4],1 )}( [3,4]^T,1 , 2)  }{\tilde s_{12}}
+
\frac{ A_{3}^{(1,[2,3])}(  1 ,  [2,3]^T, 4) 
 A_{3}^{(3,[4,1] )}(3 , [4,1]^T , 2)   }{\tilde s_{14}} 
 \right]
 \nonumber 
\end{eqnarray}
}
\vskip-0.7cm\noindent
{\small
 \begin{eqnarray}\label{} 
 \hspace{-0.2cm}
+ 
 \sum_L \left[
 2\,\frac{  A_{3}^{(4,[1,3])}(4 ,[1,3]^L , 2) \times  A_{3}^{([2,4],1)}( [2,4]^L,1 ,3)  }{\tilde s_{13}}
+
\frac{  A_{3}^{([1,2],3)}( [1,2]^L , 3 , 4)  \times  A_{3}^{([3,4],1)}(  [3,4]^L ,1  ,2)  }{ \tilde s_{12}} 
\right.
\nonumber
\end{eqnarray}
}
\vskip-0.6cm\noindent
{\small
 \begin{eqnarray}\label{fourP-Fey} 
 \hspace{-0.2cm}
  \left.
+ 
\frac{ A_{3}^{(1,[2,3])}( 1 , [2,3]^L, 4) \times  A_{3}^{( 3,[4,1] )}( 3 , [4,1]^L, 2)   }{\tilde s_{14}} 
\right]. \quad\, 
\end{eqnarray}
}
\vskip-0.15cm\noindent
It is trivial to check that the two terms on the second line are just the Feynman diagrams,  $\parbox[c]{2.4em}{\includegraphics[scale=0.062]{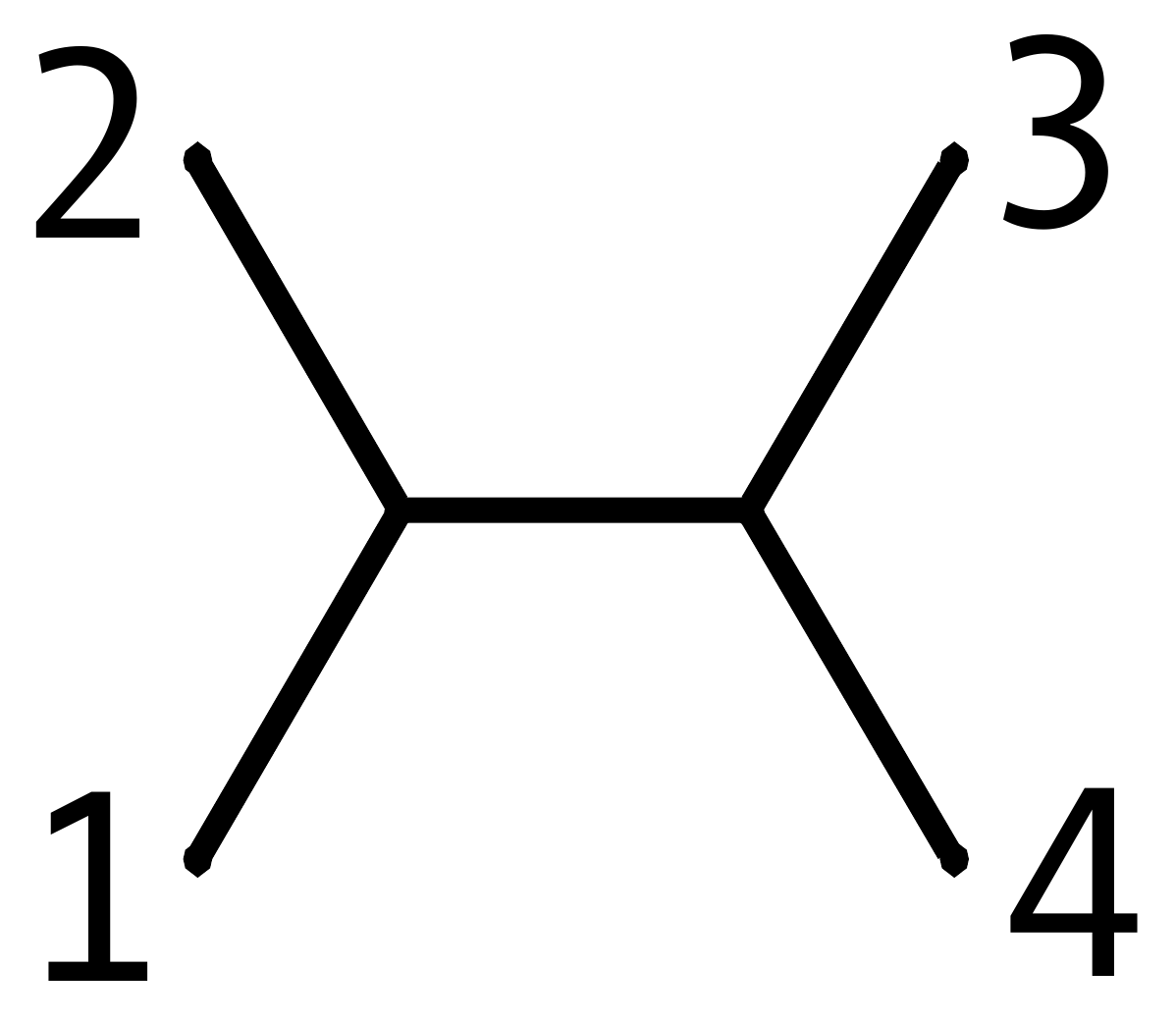}} $ and $\parbox[c]{2.6em}{\includegraphics[scale=0.062]{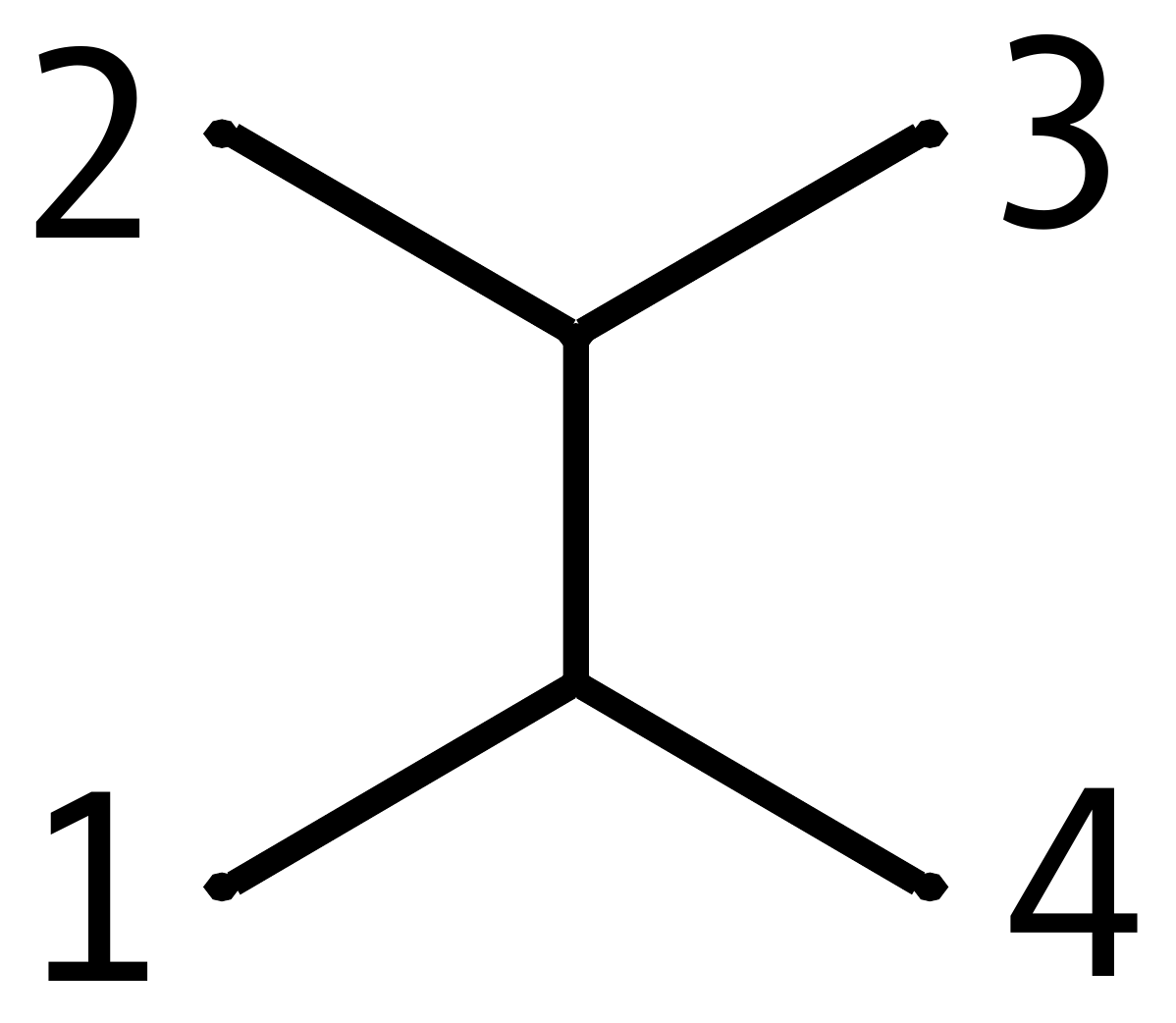}}$, respectively, and the last three terms give us the color-ordered quartic vertex $\parbox[c]{2.6em}{\includegraphics[scale=0.062]{FeyQ.pdf}}$, \cite{Dixon:1996wi}.  

To obtain the BCJ numerators, it is necessary to  reorganize \eqref{fourP-Fey} in the following way
\vspace{-0.0cm}
{\small
\begin{eqnarray}\label{}
\hspace{-2.6cm}
A_4^{\rm YM}(1,2,3,4)
=\frac{{\bf n}_{s_{12}}  }{\tilde s_{12}} + \frac{{\bf n}_{s_{14}}  }{\tilde s_{14}}, \quad {\rm with},
\nonumber
\end{eqnarray}
}
\vskip-0.6cm\noindent
{\small
\begin{eqnarray}\label{}
\hspace{-6.1cm}
{\bf n}_{s_{12}}
=
 \sum_T  A_{3}^{([1,2],3)}(  [1,2]^T , 3 ,4)\times  A_{3}^{([3,4],1)}( [3,4]^T,1 , 2) \,+ 
 \nonumber 
\end{eqnarray}
}
\vskip-0.6cm\noindent
{\small
 \begin{eqnarray}\label{}
\tilde s_{12} 
\times
\sum_L
\left[
 \frac{  A_{3}^{( 4,[1,3] )}(  4 ,  [1,3]^L  , 2) \,  A_{3}^{([2,4],1)}( [2,4]^L, 1 , 3)  }{\tilde s_{13}}
 +
\frac{ A_{3}^{(1,[2,3])}(1 ,[2,3]^L, 4) \,  A_{3}^{(3,[4,1])}(3 , [4,1]^L,2)   }{\tilde s_{14}} \right]
\nonumber
\end{eqnarray}
}
\vskip-0.5cm\noindent
{\small
\begin{eqnarray}\label{}
\hspace{-6.1cm}
{\bf n}_{s_{14}}
=
 \sum_T  A_{3}^{(1,[2,3])}(  1 ,  [2,3]^T , 4)\times  A_{3}^{(3,[4,1] )}( 3,   [4,1]^T , 2) \,+ 
 \nonumber 
\end{eqnarray}
}
\vskip-0.6cm\noindent
{\small
 \begin{eqnarray}\label{}
\tilde s_{14} 
\times 
\sum_L
\left[
 \frac{  A_{3}^{(4,[1,3])}(  4 , [1,3]^L , 2) \,  A_{3}^{([2,4],1)}(  [2,4]^L , 1 , 3)  }{\tilde s_{13}}
 +
\frac{ A_{3}^{([1,2],3)}( [1,2]^L, 3, 4)   A_{3}^{([3,4],1 )}( [3,4]^L,   1 , 2)            }{\tilde s_{12}}
\right]\, .
\nonumber
\end{eqnarray}
}
\vskip-0.2cm\noindent
From the above ${\bf n}_{s_{ij}}$ definitions, it is trivial to check the identity,     
${\bf n}_{s_{12}} - {\bf n}_{s_{14}} = {\bf n}_{s_{13}}$, where ${\bf n}_{s_{13}}$ is obtained from ${\bf n}_{s_{12}}$ under the permutation, $(1,2,3,4) \rightarrow (1,3,2,4)$, i.e.   ${\bf n}_{s_{13}} = {\bf n}_{s_{12}}\Big|_{(1,2,3,4) \rightarrow (1,3,2,4)}$. 

Nevertheless, the generalization of these ideas (using this approach) to higher number of points is unknown and it would be very interesting to be studied.\\

{\bf Effective Field Theories and final remarks:}\\
In \cite{inpreparation}, we applied the ideas presented in this paper to effective field theories, such as NLSM, multi-trace and special Galileon theory.  Additionally,  we are going to present a new NLSM prescription in order to obtain a relationship among the CHY approach and the gauge theory version for NLSM. found in \cite{Cheung:2016prv,Cheung:2017yef}. 

Although, the soft limit, at leading order, was trivially obtained using the technology developed in this paper, we would like to look for the sub-leading order contributions, and the extension to others theories.

As a final point, we would like to understand the connection among our graphic method, the ambitwistor string \cite{Mason:2013sva} and the gluing operator in \cite{Roehrig:2017gbt}.

\subsection*{Acknowledgements}

We are in debt to N. E. J. Bjerrum-Bohr, J. Bourjaily and P.H. Damgaard, for discussions and their collaboration during the initial stage of this project. We are very grateful 
with N. Ahmadiniaz, C. Lopez-Arcos, Song He, C. Cardona and A. Helset for useful discussions and comments.
We especially thank to J. Bourjaily for all his comments and observations, we are in debt to him.
Finally, we thank the referee of JHEP for many comments and corrections.
The work of  H.G.  is supported by 
the Niels Bohr Institute - University of Copenhagen and the Santiago de Cali University (USC).

\appendix

\section{Notation}\label{notation}

For convenience, in this paper we use the following notation
\begin{equation}
k_{ a_1,\ldots, a_m}\equiv\sum_{i=1}^m k_{a_i}= [a_1,\ldots, a_m], \quad
s_{a_1\ldots a_m}\equiv k_{ a_1,\ldots, a_m}^2, 
\quad \tilde s_{a_1\ldots a_m}\equiv \sum_{a_i<a_j}^m k_{a_i}\cdot k_{a_j}.
\end{equation}
Clearly, when $k_i^2=0$, then, $s_{a_1\ldots a_m}=2\,\tilde s_{a_1\ldots a_m}$

\begin{figure}[!h]
	\centering
	\includegraphics[width=5.2in]{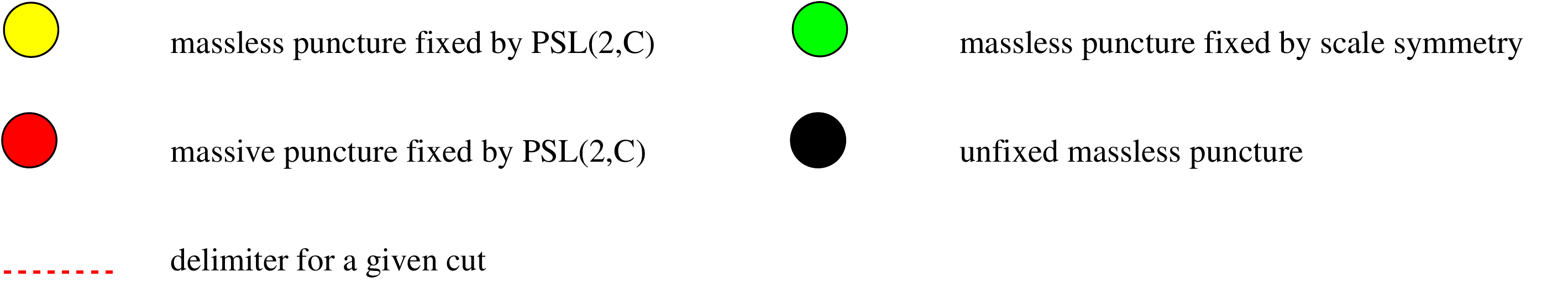}
	\caption{Vertex Color code in the CHY-graphs.}\label{color_codV}
\end{figure}

To have a graphical description for the CHY integrands, it is useful to represent each puncture, $\s_a$ (or $(y_a,\s_a)$ in double-cover prescription), by a vertex, the factor ${1 \over \s_{ab}}$ or $T_{ab}$ by an arrow and the numerator $\s_{ab}$ or $T_{ab}^{-1}$ by a dashed arrow that we call as {\it anti-line}. Additionally,   in Fig. \ref{color_codV} we give the color code for a  mnemonic understanding.
\noindent


\section{Off-shell Pfaffian Properties}\label{appendix}

In this appendix we are going to give some properties of the Pfaffian when there is an off-shell particle. 
These properties involve the matrices, $\Psi_n$ and $\Psi_{\rm g,s:g}$.

This is very important to remark that those properties are supported on the solution of the scattering equations, and, although we do not have a formal proof, they have been checked up to eight points.

First of all, we remind the notation, $\left( \Psi_n  \right)^{i_1...i_p}_{{j_1...j_p}}$ and $\left( \Psi_{\rm g,s:g}  \right)^{i_1...i_p}_{{j_1...j_p}}$, which means that the rows, $(i_1,...,i_p)$, and the columns, $(j_1,...,j_p)$, must be removed from, $ \Psi_n$ and $ \Psi_{\rm g,s:g}$.

Let us consider the total number of particles is $n$, i.e. $k_1+\cdots + k_n=0$. Additionally, since we are interested to show some off-shell properties, we choose a configuration where the off-shell particle has momentum, $k_{[1,...,p]}=k_1+\cdots +k_p$, and its puncture is fixed\footnote{Usually we fix the puncture, $\s_{[1,2,\ldots, p]}=0$.},  i.e. $\s_{[1,2,\ldots, p]}=\text{c}_1\in \mathbb{C}$.  In order to complete the setup,  we fix the punctures, $(\s_{p+1},\, \s_{p+2})=(\text{c}_2,\text{c}_3) \in \mathbb{C}^2$, where $\text{c}_1\neq \text{c}_2 \neq \text{c}_3$. Therefore, we are going to work on the support of the ``$n-(p+2)$" scattering equations
{\small
\begin{eqnarray}
S_a=\sum_{b=p+1\atop a\neq b  }^n \frac{k_a\cdot k_b}{\s_{ab}} + \frac{k_a\cdot k_{[1,\ldots , p]}}{\s_{a[1,\ldots, p]}}=0, \qquad  {\rm with\,\, } a=p+3, \ldots , n.
\end{eqnarray}
}

{\bf Properties:} \\
Under the previous setup, we have the following  properties
\begin{itemize}
\item[\bf I.] 
\begin{equation}\label{firstP}
{\rm Pf}\left[\left(\Psi_{\rm g,s: g}\right)^{[1,...,p]}_{[1,...,p]}   \right] = \frac{ k^2_{[1,...,p]} }{2} \times \frac{1}{\s_{p+1\, p+2}} \, {\rm Pf}\left[\left(\Psi_{\rm g,s: g}\right)^{p+1\, p+2\, [1,...,p]}_{p+1\, p+2\, [1,...,p]}   \right] ,
\end{equation}
where the gluon and scalar sets are given by particles,  ${\rm g}=\{ p+1,...,n  \}$ and ${\rm s}= \{ [1,...., p]\}$. Notice that if  all particles are on-shell, $k_i^2=0$, the right hand side vanishes trivially by the overall factor, $k^2_{[1,...,p]}$.  

\item[\bf II.] 
\begin{eqnarray}\label{secondP}
\frac{1}{\s_{p+2[1,...,p]}}\times {\rm Pf}\left[\left(\Psi_{\rm g,s: g}\right)^{[1,...,p]}_{[1,...,p]}   \right] =  \frac{1}{\s_{p+1\, p+2}} \times \left.{\rm Pf}\left[\left(\Psi_{n-p+1}\right)^{p+1\, [1,...,p]}_{p+1\,[1,...,p]}   \right] \right|_{\eps^{L,\mu}_{[1,...,p]} \rightarrow k^\mu_{[1,...,p]} }  ,\nonumber\\
\end{eqnarray}
with the gluons and scalars given by the sets,  ${\rm g}=\{ p+1,...,n  \}$ and ${\rm s}= \{ [1,...., p]\}$, on the left-hand side.  On the right hand side, the polarization vector, $\eps^{L,\mu}_{[1,...,p]}$, is longitudinal. Clearly, if all particle are on-shell the Pfaffian on the right-hand side vanishes trivially, since the replacement, $\eps^{L,\mu}_{[1,...,p]} \rightarrow k^\mu_{[1,...,p]}$, becomes a gauge transformation. This fact agrees with the {\bf property-I}.

\item[\bf III.] 
\begin{eqnarray}\label{thirdP}
&&\frac{1}{\s_{p+1[1,...,p]}} \times {\rm Pf}\left[\left(\Psi_{n-p+1}\right)^{p+1\, [1,...,p]}_{p+1\,[1,...,p]}   \right] = \frac{(-1)}{\s_{p+2[1,...,p]}} \times {\rm Pf}\left[\left(\Psi_{n-p+1}\right)^{p+2\, [1,...,p]}_{p+2\,[1,...,p]}   \right]   \nonumber\\
&&
= \frac{(-1)}{\s_{p+1p+2}} \times {\rm Pf}\left[\left(\Psi_{n-p+1}\right)^{p+1\, p+2}_{p+1\,p+2}   \right]\, ,
\end{eqnarray}
where the polarization vector of the off-shell puncture, which we denote by $\eps^{T,\mu}_{[1,...,p]}$, must be transverse, i.e. $\eps^T_{[1,...,p]} \cdot k_{[1,...,p]}=0$. This property just involves the fixed punctures, $(\s_{p+1},\, \s_{p+2}, \,\s_{[1,...,p]} )$, of course, when all particles are on-shell this property works for any couple of punctures.

\item[\bf IV.] 
\begin{eqnarray}\label{fourP}
\frac{ \s_{p+2[1,...,p]} }{\s_{p+1[1,...,p]} } \,\left. {\rm Pf}\left[\left(\Psi_{n-p+1}\right)^{p+1\, [1,...,p]}_{p+1\,[1,...,p]}   \right]\right|_{\eps^{L,\mu}_{[1,...,p]}  \rightarrow k^\mu_{[1,...,p]}} 
\hspace{-0.7cm}
= 
\left.  {\rm Pf}\left[\left(\Psi_{n-p+1}\right)^{p+2\, [1,...,p]}_{p+2\,[1,...,p]}   \right]\right|_{\eps^{L,\mu}_{[1,...,p]}  \rightarrow k^\mu_{[1,...,p]}} .\nonumber\\
\end{eqnarray}
It is important to remark that the overall signs in \eqref{thirdP} and \eqref{fourP} are different.

\end{itemize}


\bibliographystyle{JHEP}
\bibliography{new-YM-v1.bbl}

\end{document}